\documentclass[%
  seceqn       
]{elsart}      
\usepackage{setup}
\begin{document}
\begin{frontmatter}
\title{%
  Transverse Polarisation
\break
  of Quarks in Hadrons
}
\author{Vincenzo Barone}
\address{%
  Di.S.T.A., Universit{\`a} del Piemonte Orientale ``A. Avogadro'',
\break
  15100 Alessandria, Italy,
\break
  INFN---sezione di Torino, 10125 Torino, Italy
\break
  and Dipartimento di Fisica Teorica, Universit{\`a} di Torino,
\break
  10125 Torino, Italy
}
\author{Alessandro Drago}
\address{%
  Dipartimento di Fisica, Universit{\`a} di Ferrara, Italy
\break
  and INFN---sezione di Ferrara, 44100 Ferrara, Italy
}
\author{Philip G. Ratcliffe}
\address{%
  Dipartimento di Scienze CC.FF.MM., Universit{\`a} degli Studi dell'Insubria,
\break
  sede di Como, 22100 Como, Italy
\break
  and INFN---sezione di Milano, 20133 Milano, Italy
}

\begin{abstract}
  We review the present state of knowledge regarding the transverse
  polarisation (or transversity) distributions of quarks. After some
  generalities on transverse polarisation, we formally define the
  transversity distributions within the framework of a classification of all
  leading-twist distribution functions. We describe the QCD evolution of
  transversity at leading and next-to-leading order. A comprehensive
  treatment of non-perturbative calculations of transversity distributions
  (within the framework of quark models, lattice QCD and QCD sum rules) is
  presented. The phenomenology of transversity (in particular, in Drell--Yan
  processes and semi-inclusive leptoproduction) is discussed in some detail.
  Finally, the prospects for future measurements are outlined.
\end{abstract}


\end{frontmatter}
\newpage
\tableofcontents
\newpage
\section{Introduction}
\label{introduction}

There has been, in the past, a common prejudice that all transverse spin
effects should be suppressed at high energies. While there is some basis to
such a belief, it is far from the entire truth and is certainly misleading as a
general statement. The main point to bear in mind is the distinction between
transverse polarisation \emph{itself} and its \emph{measurable effects}. As
well-known, even the ultra-relativistic electrons and positrons of the LEP
storage ring are significantly polarised in the transverse plane
\cite{Blondel:1997rg} due to the Sokolov--Ternov effect \cite{Sokolov:1964zn}.
Thus, the real problem is to identify processes sensitive to such polarisation:
while this is not always easy, it is certainly not impossible.

Historically, the first extensive discussion of transverse spin effects in
high-energy hadronic physics followed the discovery in 1976 that $\Lambda^0$
hyperons produced in $pN$ interactions even at relatively high $p_T$ exhibit an
anomalously large transverse polarisation \cite{Bunce:1976yb}.\footnote{An 
issue related to hadronic transverse spin, and investigated theoretically in 
the same period, is the $g_2$ spin structure function \cite{Hey:1972pp,
Heimann:1974bv}; we shall discuss its relation to transversity later.} This
result implies a non-zero imaginary part of the off-diagonal elements of the
fragmentation matrix of quarks into $\Lambda^0$ hyperons. It was soon pointed
out that this is forbidden in leading-twist \QCD, and arises only as a
$\Ord(1/p_T)$ effect \cite{Kane:1978nd, Efremov:1981sh, Efremov:1984ip}. It 
thus took a while to fully realise that transverse spin phenomena are sometimes
unsuppressed and observable.\footnote{This was pointed out in the pioneering
paper of Ralston and Soper \cite{Ralston:1979ys} on longitudinally and
transversely polarised Drell--Yan processes, but the idea remained almost
unnoticed for a decade, see below.}

The subject of this report is the transverse polarisation of quarks. This is an
elusive and difficult to observe property that has escaped the attention of
physicists for many years. Transverse polarisation of quarks is not, in fact,
probed in the cleanest hard process, namely \DIS, but is measurable in other
hard reactions, such as semi-inclusive leptoproduction or Drell--Yan dimuon
production.

At leading-twist level, the quark structure of hadrons is described by three
distribution functions: the number density, or unpolarised distribution,
$f(x)$; the longitudinal polarisation, or helicity, distribution $\DL{f}(x)$;
and the transverse polarisation, or transversity, distribution $\DT{f}(x)$.

The first two are well-known quantities: $f(x)$ is the probability of finding a
quark with a fraction $x$ of the longitudinal momentum of the parent hadron,
regardless of its spin orientation; $\DL{f}(x)$ measures the net helicity of a
quark in a longitudinally polarised hadron, that is, the number density of
quarks with momentum fraction $x$ and spin parallel to that of the hadron minus
the number density of quarks with the same momentum fraction but spin
antiparallel. If we call $f_{\pm}(x)$ the number densities of quarks with helicity
${\pm}1$, then we have
\begin{subequations}
\begin{eqnarray}
  f(x) &=& f_+(x) + f_-(x) \, ,
  \label{i1}
\\
  \DL{f}(x) &=& f_+(x) - f_-(x) \, .
  \label{i2}
\end{eqnarray}
\end{subequations}

The third distribution function, $\DT{f}(x)$, although less familiar, also has
a very simple meaning. In a transversely polarised hadron $\DT{f}(x)$ is the
number density of quarks with momentum fraction $x$ and polarisation parallel
to that of the hadron, minus the number density of quarks with the same
momentum fraction and antiparallel polarisation, \ie,\footnote{Throughout this
paper the subscripts ${\pm}$ will denote helicity whereas the subscripts
$\uparrow\downarrow$ will denote transverse polarisation.}
\begin{equation}
  \DT{f}(x) = f_\uparrow(x) - f_\downarrow(x) \, .
  \label{i3}
\end{equation}
In a basis of transverse polarisation states $\DT{f}$ too has a probabilistic
interpretation. In the helicity basis, in contrast, it has no simple meaning,
being related to an off-diagonal quark-hadron amplitude.

Formally, quark distribution functions are light-cone Fourier transforms of
connected matrix elements of certain quark-field bilinears. In particular, as
we shall see in detail (see Sec.~\ref{leading}), $\DT{f}$ is given by (we take
a hadron moving in the $z$ direction and polarised along the $x$-axis)
\begin{equation}
  \DT{f}(x) = \int \! \frac{\d\xi^-}{4\pi} \, \e^{\I xP^+ \xi^-}
  \langle PS |
    \anti\psi(0) \I \sigma^{1+} \gamma_5 \psi(0,\xi^-,\Vec{0}_\perp)
  | PS \rangle \, .
  \label{i4}
\end{equation}
In the parton model the quark fields appearing in (\ref{i4}) are free fields.
In \QCD they must be renormalised (see Sec.~\ref{renormalisation}). This
introduces a renormalisation-scale dependence into the parton distributions:
\begin{equation}
  f(x) \, , \, \DL{f}(x) \, , \, \DT{f}(x)
  \to
  f(x, \mu^2) \, , \, \DL{f}(x, \mu^2) \, , \, \DT{f}(x, \mu^2) \, ,
  \label{i4b}
\end{equation}
which is governed by the \DGLAP equations (see Sec.~\ref{transvqcd}).

It is important to appreciate that $\DT{f}(x)$ is a \emph{leading-twist}
quantity. Hence it enjoys the same status as $f(x)$ and $\DL{f}(x)$ and,
\emph{a priori}, there is no reason that it should be much smaller than its
helicity counterpart. In fact, model calculations show that $\DT{f}(x)$ and
$\DL{f}(x)$ are typically of the same order of magnitude, at least at low $Q^2$,
where model pictures hold (see Sec.~\ref{models}).

The \QCD evolution of $\DT{f}(x)$ and $\DL{f}(x)$ is, however, quite different
(see Sec.~\ref{comparison}). In particular, at low $x$, $\DT{f}(x)$ turns out
to be suppressed with respect to $\DL{f}(x)$. As we shall see, this behaviour
has important consequences for some observables. Another peculiarity of
$\DT{f}(x)$ is that it has no gluonic counterpart (in spin-$\half$ hadrons):
gluon transversity distributions for nucleons do not exist (Sec.~\ref{helamp}).
Thus $\DT{f}(x)$ does not mix with gluons in its evolution, and evolves as a
non-singlet quantity.

\begin{figure}[htbp]
  \centering
  \begin{picture}(380,140)(0,0)
    \SetOffset(-50,0)
    \SetWidth{2}
    \ArrowLine(80,20)(120,30)
    \ArrowLine(180,30)(220,20)
    \SetWidth{0.5}
    \ArrowLine(120,50)(120,80)
    \ArrowLine(180,80)(180,50)
    \GOval(150,40)(20,40)(0){0.5}
    \DashLine(150,10)(150,80){4}
    \Text(80,30)[]{$+$}
    \Text(220,30)[]{$-$}
    \Text(110,80)[]{$-$}
    \Text(190,80)[]{$+$}
    \Text(150,0)[]{(a)}
    \SetOffset(130,0)
    \SetWidth{2}
    \ArrowLine(80,20)(120,30)
    \ArrowLine(180,30)(220,20)
    \SetWidth{0.5}
    \ArrowLine(120,50)(120,80)
    \ArrowLine(180,80)(180,50)
    \ArrowLine(120,80)(180,80)
    \Photon(90,100)(120,80){3}{4.5}
    \Photon(210,100)(180,80){3}{4.5}
    \GOval(150,40)(20,40)(0){0.5}
    \DashLine(150,10)(150,77){4}
    \DashLine(150,83)(150,100){4}
    \Text(80,30)[]{$+$}
    \Text(220,30)[]{$-$}
    \Text(110,65)[]{$-$}
    \Text(190,65)[]{$+$}
    \Text(150,0)[]{(b)}
  \end{picture}
  \caption{(a) Representation of the chirally-odd distribution $\DT{f}(x)$.
           (b) A handbag diagram forbidden by chirality conservation.}
  \label{chiflip}
\end{figure}
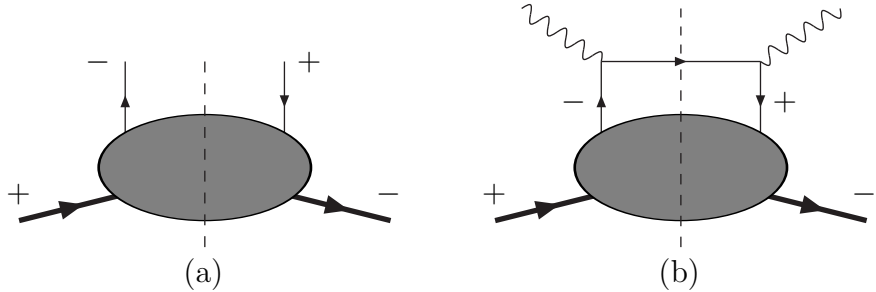

One may wonder why the transverse polarisation distributions are so little
known, if they are quantitatively comparable to the helicity distributions. No
experimental information on $\DT{f}(x)$ is indeed available at present (see,
however, Sec.~\ref{pionlepto}, where mention is made of some preliminary data
on pion leptoproduction that might involve $\DT{f}(x)$). The reason has already
been mentioned: transversity distributions are not observable in fully
inclusive \DIS, the process that has provided most of the information on the
other distributions. Examining the operator structure in (\ref{i4}) one can see
that $\DT{f}(x)$, in contrast to $f(x)$ and $\DL{f}(x)$, which contain
$\gamma^+$ and $\gamma^+\gamma_5$ instead of $\I\sigma^{1+}\gamma_5$, is a
chirally-odd quantity (see Fig.~\ref{chiflip}a). Now, fully inclusive \DIS
proceeds via the so-called handbag diagram which cannot flip the chirality of
the probed quark (see Fig.~\ref{chiflip}b). In order to measure $\DT{f}$ the
chirality must be flipped twice, so one needs either two hadrons in the initial
state (hadron--hadron collisions), or one hadron in the initial state and one
in the final state (semi-inclusive leptoproduction), and at least one of these
two hadrons must be transversely polarised. The experimental study of these
processes has just started and will provide in the near future a great wealth
of data (Sec.~\ref{phenomenology}).

So far we have discussed the distributions $f(x)$, $\DL{f}(x)$ and $\DT{f}(x)$.
If quarks are perfectly collinear, these three quantities exhaust the
information on the internal dynamics of hadrons. If we admit instead a finite
quark transverse momentum $\Vec{k}_\perp$, the number of distribution functions
increases (Sec.~\ref{transvmot}). At leading twist, assuming time-reversal
invariance, there are six $\Vec{k}_\perp$-dependent distributions. Three of
them, called in the Jaffe--Ji--Mulders classification scheme
\cite{Jaffe:1992ra, Mulders:1996dh} $f_1(x,\Vec{k}_\perp^2)$,
$g_{1L}(x,\Vec{k}_\perp^2)$ and $h_1(x,\Vec{k}_\perp^2)$, upon integration over
$\Vec{k}_\perp^2$, yield $f(x)$, $\DL{f}(x)$ and $\DT{f}(x)$, respectively. The
remaining three distributions are new and disappear when the hadronic tensor is
integrated over $\Vec{k}_\perp$, as is the case in \DIS. Mulders has called
them $g_{1T}(x,\Vec{k}_\perp^2)$, $h_{1L}^\perp(x,\Vec{k}_\perp^2)$ and
$h_{1T}^\perp(x,\Vec{k}_\perp^2)$. If time-reversal invariance is not applied
(for the physical motivation behind this, see Sec.~\ref{todd}), two more,
$T$-odd, $\Vec{k}_\perp$-dependent distribution functions appear
\cite{Boer:1998nt}: $f_{1T}^\perp(x,\Vec{k}_\perp^2)$ and
$h_1^\perp(x,\Vec{k}_\perp^2)$. At present the existence of these distributions
is merely conjectural.

To summarise, here is an overall list of the leading-twist quark distribution
functions:
\begin{equation}
  \overbrace{f \, , \; \DL{f} \, , \; \DT{f}\vphantom{^\perp}}
           ^{\text{no $\Vec{k}_\perp$}} \, , \,
  \overbrace{g_{1T} \, , \; h_{1L}^\perp \, , \; h_{1T}^\perp}
           ^{\!\text{$\Vec{k}_\perp$-dependent}\!} \, , \,
  \underbrace{f_{1T}^\perp \, , \; h_1^\perp\downstrut{1ex}}
            _{\text{$T$-odd}} \, .
  \nonumber
\end{equation}

At higher twist the proliferation of distribution functions continues
\cite{Jaffe:1992ra, Mulders:1996dh}. Although, for the sake of completeness, we
shall also briefly discuss the $\Vec{k}_\perp$-dependent and the twist-three
distributions, most of our attention will be directed to $\DT{f}(x)$. Less
space will be dedicated to the other transverse polarisation distributions,
many of which have, at present, only an academic interest.

In hadron production processes, which, as mentioned above, play an important
r\^{o}le in the study of transversity, there appear other dynamical quantities:
fragmentation functions. These are in a sense specular to distribution
functions, and represent the probability for a quark in a given polarisation
state to fragment into a hadron carrying some momentum fraction $z$. When the
quark is transversely polarised and so too is the produced hadron, the process
is described by the leading-twist fragmentation function $\DT{D}(z)$, which is
the analogue of $\DT{f}(x)$ (see Sec.~\ref{systfrag}). A $T$-odd fragmentation
function, usually called $H_1^\perp(z)$, describes instead the production of
unpolarised (or spinless) hadrons from transversely polarised quarks, and
couples to $\DT{f}(x)$ in certain semi-inclusive processes of great relevance
for the phenomenology of transversity (the emergence of $\DT{f}$ via its
coupling to $H_1^\perp$ is known as the Collins effect \cite{Collins:1993kk}).
The fragmentation of transversely polarised quarks will be described in detail
in Sec.~\ref{semiinclusive} and Sec.~\ref{transvhad}.

\subsection{History}
\label{history}

The transverse polarisation distributions were first introduced in 1979 by
Ralston and Soper in their seminal work on Drell--Yan production with polarised
beams \cite{Ralston:1979ys}. In that paper $\DT{f}(x)$ was called $h_T(x)$.
This quantity was apparently forgotten for about a decade, until the beginning
of nineties, when it was rediscovered by Artru and Mekhfi \cite{Artru:1990zv},
who called it $\Delta_1q(x)$ and studied its \QCD evolution, and also by Jaffe
and Ji \cite{Jaffe:1991kp, Jaffe:1992ra}, who renamed it $h_1(x)$ in the
framework of a general classification of all leading-twist and higher-twist
parton distribution functions. At about the same time, other important studies
of the transverse polarisation distributions exploring the possibility of
measuring them in hadron--hadron or lepton-hadron collisions were carried out
by Cortes, Pire and Ralston \cite{Cortes:1992ja}, and by Ji \cite{Ji:1992ev}.

The last few years have witnessed a great revival of interest in the transverse
polarisation distributions. A major effort has been devoted to investigating
their structure using more and more sophisticated model calculations and other
non-perturbative tools (\QCD sum rules, lattice \QCD \etc). Their \QCD
evolution has been calculated up to \NLO. The related phenomenology has been
explored in detail: many suggestions for measuring (or at least detecting)
transverse polarisation distributions have been put forward and a number of
predictions for observables containing $\DT{f}$ are now available. We can say
that our \emph{theoretical} knowledge of the transversity distributions is by
now nearly comparable to that of the helicity distributions. What is really
called for is an \emph{experimental} study of the subject.

On the experimental side, in fact, the history of transverse polarisation
distributions is readily summarised: (almost) no measurements of $\DT{f}$ have
been performed as yet. Probing quark transverse polarisation is among the goals
of a number of ongoing or future experiments. At \theRHIC $\DT{f}$ can be
extracted from the measurement of the double-spin transverse asymmetry in
Drell--Yan dimuon production with two transversely polarised hadron beams
\cite{Bunce:2000uv} (Sec.~\ref{ppexp}). Another important class of reactions
that can probe transverse polarisation distributions is semi-inclusive \DIS.
The HERMES collaboration at HERA \cite{Airapetian:1999tv} and the SMC
collaboration at CERN \cite{Bravar:1999rq} have recently presented results on
single-spin transverse asymmetries, which could be related to the transverse
polarisation distributions via the hypothetical Collins mechanism
\cite{Collins:1993kk} (Sec.~\ref{pionlepto}). The study of transversity in
semi-inclusive \DIS is one of the aims of the upgraded HERMES experiment and of
the COMPASS experiment at the CERN SPS collider, which started taking data in
2001 \cite{Baum:1996yv}. It also represents a significant part of other
projects (see Sec.~\ref{lnexp}). We may therefore say that the experimental
study of transverse polarisation distributions, which is right now only at the
very beginning, promises to have an exciting future.

\subsection{Notation and terminology}
\label{notations}

Transverse polarisation of quarks is a relatively young and still unsettled
subject, hence it is not surprising that the terminology is rather confused.
Notation that has been used in the past for the transverse polarisation of
quarks comprises
\begin{equation}
  \begin{array}{r@{\qquad}l}
  h_T (x)       & \text{(Ralston \& Soper),}
\\
  \Delta_1 q(x) & \text{(Artru \& Mekhfi),}
\\
  h_1(x)        & \text{(Jaffe \& Ji),}
  \end{array}
  \nonumber
\end{equation}
The first two forms are now obsolete while the third is still widely employed.
This last was introduced by Jaffe and Ji in their classification of all
twist-two, twist-three and twist-four parton distribution functions. In the
Jaffe--Ji scheme, $f_1(x)$, $g_1(x)$ and $h_1(x)$ are the unpolarised,
longitudinally polarised and transversely polarised distribution functions,
respectively, with the subscript 1 denoting leading-twist quantities. The main
disadvantage of this nomenclature is the use of $g_1$ to denote a
\emph{leading-twist distribution function} whereas the same notation is
universally adopted for one of the two polarised \emph{structure functions}.
This is a serious source of confusion. In the most recent literature the
transverse polarisation distributions are often called
\begin{displaymath}
  \delta{q}(x) \quad \text{or} \quad \DT{q}(x) \, .
\end{displaymath}
Both forms appear quite natural as they emphasise the parallel between the
longitudinal and the transverse polarisation distributions.

In this report we shall use $\DT{f}$, or $\DT{q}$, to denote the
\emph{transverse polarisation distributions}, reserving $\delta{q}$ for the
\emph{tensor charge} (the first moment of $\DT{q}$).

The Jaffe--Ji classification scheme has been extended by Mulders and
collaborators \cite{Mulders:1996dh, Boer:1998nt} to all twist-two and
twist-three $\Vec{k}_\perp$-dependent distribution functions. The letters $f$,
$g$ and $h$ denote unpolarised, longitudinally polarised, and transversely
polarised quark distributions, respectively. A subscript $1$ labels the
leading-twist quantities. Subscripts $L$ and $T$ indicate that the parent
hadron is longitudinally or transversely polarised. Finally, a superscript
$\perp$ signals the presence of transverse momenta with uncontracted Lorentz
indices.

In the present paper we adopt a hybrid terminology. We use the traditional
notation for the $\Vec{k}_\perp$-integrated distribution functions: $f(x)$, or
$q(x)$, for the number density, $\DL{f}(x)$, or $\DL{q}(x)$, for the helicity
distributions, $\DT{f}(x)$, or $\DT{q}(x)$, for the transverse polarisation
distributions, and Mulders' notation for the additional
$\Vec{k}_\perp$-dependent distribution functions: $g_{1T}$, $h_{1L}^\perp$,
$h_{1T}^\perp$, $f_{1T}^\perp$ and $h_1^\perp$.

We make the same choice for the fragmentation functions. We call the
$\Vec\kappa_\perp$-integrated fragmentation functions $D(z)$ (unpolarised),
$\DL{D}(z)$ (longitudinally polarised) and $\DT{D}(z)$ (transversely
polarised). For the $\Vec\kappa_\perp$-dependent functions we use Mulders'
terminology.

Occasionally, other notation will be introduced, for the sake of clarity, or to
maintain contact with the literature on the subject. In particular, we shall
follow these rules:
\begin{itemize}
\item
the \emph{subscripts} $0,L,T$ in the distribution and fragmentation functions
denote the polarisation state of the \emph{quark} ($0$ indicates unpolarised,
and the subscript $L$ is actually omitted in the familiar helicity distribution
and fragmentation functions);
\item
the \emph{superscripts} $0,L,T$ denote the polarisation state of the parent
\emph{hadron}.
\end{itemize}
Thus, for instance, $\DT^L{f}$ represents the distribution function of
transversely polarised quarks in a longitudinally polarised hadron (it is
related to Mulders' $h_{1L}^\perp$). The Jaffe--Ji--Mulders terminology is
compared to ours in Table~\ref{table_intr}. The correspondence with other
notation encountered in the literature \cite{Anselmino:1999pw,
Anselmino:1999gd} is
\begin{eqnarray}
  \Delta^N f_{q/N^\uparrow} & \equiv & \Delta_0^T f \, ,
  \nonumber
\\
  \Delta^N f_{q^\uparrow/N} & \equiv & \DT^0 f \, ,
  \nonumber
\\
  \Delta^N D_{h/q^\uparrow} & \equiv & 2 \DT^0 D_{h/q} \, .
  \nonumber
\end{eqnarray}

Finally, we recall that the name \emph{transversity}, as a synonym for
\emph{transverse polarisation}, was proposed by Jaffe and Ji
\cite{Jaffe:1991kp}. In \cite{Cortes:1991ka, Anselmino:1995gn} it was noted
that ``transversity'' is a pre-existing term in spin physics, with a different
meaning, and that its use therefore in a different context might cause
confusion. In this report we shall ignore this problem, and use both terms,
``transverse polarisation distributions'' and ``transversity distributions'' with
the same meaning.
\begin{table}[htbp]
  \centering
  \caption{Notation for the distribution and the fragmentation functions (JJM
           denotes the Jaffe--Ji--Mulders classification.}
  \label{table_intr}
  \vspace{1ex}
  \begin{tabular}{cc}
   \hline
   \multicolumn{2}{c}{Distribution functions} \\
   \hline\hline
   JJM            & This paper \\
   \hline\hline
   $f_1$          & $f$, $q$ \\
   $g_1$          & $\DL{f}$, $\DL{q}$ \\
   $h_1$          & $\DT{f}$, $\DT{q}$ \\
   $g_{1T}$       & $g_{1T}$ \\
   $h_{1L}^\perp$ & $h_{1L}^\perp$ \\
   $h_{1T}^\perp$ & $h_{1T}^\perp$ \\
   $f_{1T}^\perp$ & $f_{1T}^\perp$, $\Delta_0^T f$ \\
   $h_1^\perp$    & $h_1^\perp$, $\DT^0f$ \\
   \hline
  \end{tabular}
  \hspace{1cm}
  \begin{tabular}{cc}
   \hline
   \multicolumn{2}{c}{Fragmentation functions} \\
   \hline\hline
   JJM            & This paper \\
   \hline\hline
   $D_1$          & $D$ \\
   $G_1$          & $\DL{D}$ \\
   $H_1$          & $\DT{D}$ \\
   $G_{1T}$       & $G_{1T}$ \\
   $H_{1L}^\perp$ & $H_{1L}^\perp$ \\
   $H_{1T}^\perp$ & $H_{1T}^\perp$ \\
   $D_{1T}^\perp$ & $D_{1T}^\perp$ \\
   $H_1^\perp$    & $H_1^\perp$, $\DT^0D$ \\
   \hline
  \end{tabular}
  \vspace{2ex}
\end{table}

\subsection{Conventions}
\label{conventions}

We now list some further conventions adopted throughout the paper.

The metric tensor is
\begin{equation}
  g^{\mu\nu} = g_{\mu\nu} = \diag (+1, -1, -1, -1) \, .
  \label{conv1}
\end{equation}
The totally antisymmetric tensor $\varepsilon^{\mu\nu\rho\sigma}$ is normalised
so that
\begin{equation}
  \varepsilon^{0123} = - \varepsilon_{0123} = +1 \,.
  \label{conv1b}
\end{equation}
A generic four-vector $A^\mu$ is written, in Cartesian contravariant
components, as
\begin{equation}
  A^\mu = (A^0, A^1, A^2, A^3) = ( A^0, \Vec{A}) \, ,
  \label{conv2}
\end{equation}
The light-cone components of $A^\mu$ are defined as
\begin{equation}
  A^{\pm} = \textfrac{1}{\sqrtno2} \, (A^0 \pm A^3) \, ,
  \label{conv3}
\end{equation}
and in these components $A^\mu$ is written as
\begin{equation}
  A^\mu = (A^+, A^-, \Vec{A}_\perp) \, .
  \label{conv4}
\end{equation}
The norm of $A^\mu$ is given by
\begin{equation}
  A^2 = {(A^0)}^2 - \Vec{A}^2 = 2 A^+ A^- - \Vec{A}_\perp^2 \, ,
  \label{conv5}
\end{equation}
and the scalar product of two four-vectors $A^\mu$ and $B^\mu$ is
\begin{equation}
  A{\cdot}B
  =
  A^0 B^0 - \Vec{A}{\cdot}\Vec{B}
  =
  A^+ B^- + A^- B^+ - \Vec{A}_\perp{\cdot}\Vec{B}_\perp \, .
  \label{conv6}
\end{equation}

Our fermionic states are normalised as
\begin{equation}
  \langle p | p' \rangle
  =
  (2\pi)^3 \, 2E \, \delta^3(\Vec{p} - \Vec{p}')
  =
  (2\pi)^3 \, 2p^+ \, \delta(p^+ - {p'}^+) \,
  \delta(\Vec{p}_\perp - \Vec{p}_\perp') \,  ,
  \label{conv7}
\end{equation}
\begin{equation}
  \anti{u}(p,s) \gamma^\mu u(p,s') = 2 p^\mu \, \delta_{ss'} \, ,
  \label{conv8}
\end{equation}
with $E=(\Vec{p}^2+m^2)^{1/2}$. The creation and annihilation operators satisfy
the anticommutator relations
\begin{equation}
  \left  \{ b(p,s), b^\dagger(p',s') \right \} =
  \left  \{ d(p,s), d^\dagger(p',s') \right \} =
  (2\pi)^3 \, 2E\, \delta_{ss'} \, \delta^3 (\Vec{p} - \Vec{p}') \, .
  \label{conv9}
\end{equation}

\section{Longitudinal and transverse polarisation}
\label{polarisation}

The representations of the Poincar{\'e} group are labelled by the eigenvalues of
two Casimir operators, $P^2$ and $W^2$ (see \eg, \cite{Itzykson:1980rh}). $P^\mu$
is the energy--momentum operator, $W^\mu$ is the Pauli--Lubanski operator,
constructed from $P^\mu$ and the angular-momentum operator $J^{\mu\nu}$
\begin{equation}
  W^\mu =
  - \half \, \varepsilon^{\mu\nu\rho\sigma} J_{\nu\rho} P_\sigma \, .
  \label{tp1}
\end{equation}
The eigenvalues of $P^2$ and $W^2$ are $m^2$ and $-s(s+1)m^2$ respectively, where
$m$ is the mass of the particle and $s$ its spin.

The states of a Dirac particle ($s=1/2$) are eigenvectors of $P^\mu$ and of the
polarisation operator $\Pi\equiv-W{\cdot}s/m$
\begin{eqnarray}
  P^\mu \, | p,s \rangle
  &=& p^\mu \, | p,s \rangle \, ,
  \label{tp2}
\\
  - \frac{W{\cdot}s}{m} \, | p,s \rangle
  &=& \pm \frac12 \, | p,s \rangle \, ,
  \label{tp3}
\end{eqnarray}
where $s^\mu$ is the spin (or polarisation) vector of the particle, with the
properties
\begin{equation}
  s^2 = -1, \quad s{\cdot}p = 0 \, .
\end{equation}
In general, $s^\mu$ may be written as
\begin{equation}
  s^\mu =
  \left(
    \frac{\Vec{p}{\cdot}\Vec{n}}{m}, \;
    \Vec{n} + \frac{(\Vec{p}{\cdot}\Vec{n}) \, \Vec{p}}{m (m+p^0)}
  \right) ,
  \label{tp4}
\end{equation}
where $\Vec{n}$ is a unit vector identifying a generic space direction.

The polarisation operator $\Pi$ can be re-expressed as
\begin{equation}
  \Pi = \frac1{2m} \, \gamma_5 \, \slashed{s} \, \slashed{p} \, ,
  \label{tp6}
\end{equation}
and if we write the plane-wave solutions of the free Dirac equation in the form
\begin{equation}
  \psi(x) =
  \left\{
    \begin{array}{ll}
      \e^{- \I p{\cdot}x} \, u(p) & \qquad \text{(positive energy),}
    \\
      \e^{+ \I p{\cdot}x} \, v(p) & \qquad \text{(negative energy),}
    \end{array}
  \right.
  \label{tp6b}
\end{equation}
with the condition $p^0 >0$, $\Pi$ becomes
\begin{subequations}
\begin{align}
  \Pi &= + \half \, \gamma_5 \, \slashed{s} &
  & \text{(positive-energy states),}
  \label{tp7}
\intertext{when acting on positive-energy states, $(\slashed{p}-m)\,u(p)=0$,
  and}
  \Pi &= - \half \, \gamma_5 \, \slashed{s} &
  & \text{(negative-energy states),}
  \label{tp8}
  \sublabel{subtp8}
\end{align}
\end{subequations}
when acting on negative-energy states, $(\slashed{p}+m)\,v(p)=0$. Thus the
eigenvalue equations for the polarisation operator read ($\alpha=1,2$)
\begin{equation}
  \begin{array}{rccl}
    \Pi \, u_{(\alpha)} =
    +\half \, \gamma_5 \, \slashed{s} \, u_{(\alpha)}
    &=&
    \pm \, \half \, u_{(\alpha)} & \quad \text{(positive energy),}
  \\
    \Pi \, v_{(\alpha)} =
    -\half \, \gamma_5 \, \slashed{s} \, v_{(\alpha)}
    &=&
    \pm \, \half \, v_{(\alpha)} & \quad \text{(negative energy).}
  \end{array}
  \label{tp9&10}
\end{equation}

Let us consider now particles that are at rest in a given frame. The spin
$s^\mu$ is then (set $\Vec{p}=0$ in eq.~(\ref{tp4}))
\begin{equation}
  s^\mu = (0, \Vec{n}) \, ,
  \label{tp5}
\end{equation}
and in the Dirac representation we have the operator
\begin{equation}
  \half \, \gamma_5 \, \slashed{s}
  =
  \left(
    \begin{array}{cc}
      \Vec\sigma{\cdot}\Vec{n} & 0
    \\
      0 & - \Vec\sigma{\cdot}\Vec{n}
  \end{array}
  \right),
  \label{tp10b}
\end{equation}
acting on
\begin{equation}
  u_{(\alpha)} =
  \left(
    \begin{array}{c}
      \varphi_{(\alpha)} \\ 0
    \end{array}
  \right), \qquad
  v_{(\alpha)} =
  \left(
    \begin{array}{c}
      0 \\ \chi_{(\alpha)}
    \end{array}
  \right).
  \label{tp10c}
\end{equation}
Hence, the spinors $u_{(1)}$ and $v_{(1)}$ represent particles with spin
$\half\Vec\sigma{\cdot}\Vec{n}=+\half$ in their rest frame whereas the spinors
$u_{(2)}$ and $v_{(2)}$ represent particles with spin
$\half\Vec\sigma{\cdot}\Vec{n}=-\half$ in their rest frame. Note that the
polarisation operator in the form (\ref{tp7},~\ref{subtp8}) is also well
defined for massless particles.

\subsection{Longitudinal polarisation}
\label{longpol}

For a longitudinally polarised particle ($\Vec{n}=\Vec{p}/|\Vec{p}|$), the spin
vector reads
\begin{equation}
  s^\mu =
  \left(
    \frac{|\Vec{p}|}{m}, \frac{p^0}{m} \frac{\Vec{p}}{|\Vec{p}|}
  \right) ,
\end{equation}
and the polarisation operator becomes the \emph{helicity} operator
\begin{equation}
  \Pi = \frac{\Vec\Sigma{\cdot}\Vec{p}}{2 | \Vec{p} |} \, ,
  \label{tp11}
\end{equation}
with $\Vec\Sigma=\gamma_5\gamma^0\Vec\gamma$. Consistently with
eq.~(\ref{tp6b}), the \emph{helicity states} satisfy the equations
\begin{equation}
  \begin{split}
    \frac{\Vec\Sigma{\cdot}\Vec{p}}{|\Vec{p}|} \, u_{\pm}(p) &= \pm \, u_{\pm}(p) \, ,
  \\
    \frac{\Vec\Sigma{\cdot}\Vec{p}}{|\Vec{p}|} \, v_{\pm}(p) &= \mp \, v_{\pm}(p) \, .
  \end{split}
  \label{tp12&13}
\end{equation}
Here the subscript $+$ indicates positive helicity, that is spin parallel to
the momentum ($\Vec\Sigma{\cdot}\Vec{p}>0$ for positive-energy states,
$\Vec\Sigma{\cdot}\Vec{p}<0$ for negative-energy states); the subscript $-$ indicates
negative helicity, that is spin antiparallel to the momentum
($\Vec\Sigma{\cdot}\Vec{p}<0$ for positive-energy states, $\Vec\Sigma{\cdot}\Vec{p}>0$ for
negative-energy states). The correspondence with the spinors $u_{(\alpha)}$ and
$v_{(\alpha)}$ previously introduced is: $u_+=u_{(1)}$, $u_-=u_{(2)}$,
$v_+=v_{(2)}$, $v_-=v_{(1)}$.

In the case of massless particles one has
\begin{equation}
  \Pi = \frac{\Vec\Sigma{\cdot}\Vec{p}}{2|\Vec{p}|}
  = \half \, \gamma_5 \, .
  \label{tp14}
\end{equation}
Denoting again by $u_{\pm},v_{\pm}$ the helicity eigenstates, eqs.~(\ref{tp12&13})
become for zero-mass particles
\begin{equation}
  \begin{split}
    \gamma_5 \, u_{\pm}(p) &= \pm \, u_{\pm}(p) \, ,
  \\
    \gamma_5 \, v_{\pm}(p) &= \mp \, v_{\pm}(p) \, .
  \end{split}
  \label{tp15&16}
\end{equation}
Thus \emph{helicity} coincides with \emph{chirality} for positive-energy
states, while it is opposite to chirality for negative-energy states. The
helicity projectors for massless particles are then
\begin{equation}
  \mathcal{P}_{\pm} =
  \left\{
  \begin{array}{l@{\qquad}l}
    \half \, (1 \pm \gamma_5) & \text{positive-energy states,}
  \\
    \half \, (1 \mp \gamma_5) & \text{negative-energy states.}
  \end{array}
  \right.
  \label{tp18}
\end{equation}

\subsection{Transverse polarisation}
\label{transvpol}

Let us come now to the case of transversely polarised particles. With
$\Vec{n}{\cdot}\Vec{p}=0$ and assuming that the particle moves along the $z$
direction, the spin vector (\ref{tp4}) becomes, in Cartesian components
\begin{equation}
  s^\mu = s_\perp^\mu = (0, \Vec{n}_\perp, 0) \, ,
  \label{tp19}
\end{equation}
where $\Vec{n}_\perp$ is a transverse two-vector. The polarisation operator
takes the form
\begin{equation}
  \Pi=
  \left\{
  \begin{array}{l@{\quad}l}
    -
    \half \, \gamma_5 \Vec\gamma_\perp{\cdot}\Vec{n}_\perp =
    \hphantom{-}
    \half \, \gamma_0 \Vec\Sigma_\perp{\cdot}\Vec{n}_\perp
    & \text{(positive-energy states),}
  \\
    \hphantom{-}
    \half \, \gamma_5 \Vec\gamma_\perp{\cdot}\Vec{n}_\perp = -
    \half \, \gamma_0 \Vec\Sigma_\perp{\cdot}\Vec{n}_\perp
    & \text{(negative-energy states),}
  \end{array}
  \right.
  \label{tp20}
\end{equation}
and its eigenvalue equations are
\begin{equation}
  \begin{split}
    \half \, \gamma_5 \, \slashed{s}_\perp \, u_{\uparrow\downarrow}
    &=
    \pm \, \half \, u_{\uparrow\downarrow} \, ,
  \\
    \half \, \gamma_5 \, \slashed{s}_\perp \, v_{\uparrow\downarrow}
    &=
    \mp \, \half \, v_{\uparrow\downarrow} \, .
  \end{split}
  \label{tp21&tp22}
\end{equation}
The transverse polarisation projectors along the directions $\Vec{\hat{x}}$ and
$\Vec{\hat{y}}$ are
\begin{equation}
  \begin{split}
    \mathcal{P}_{\uparrow\downarrow}^{(x)}
    &= \half \, (1 \pm \gamma^1\gamma_5) \, ,
  \\
    \mathcal{P}_{\uparrow\downarrow}^{(y)}
    &= \half \, (1 \pm \gamma^2\gamma_5) \, ,
  \end{split}
  \label{tp23&tp24}
\end{equation}
for positive-energy states, and
\begin{equation}
  \begin{split}
    \mathcal{P}_{\uparrow\downarrow}^{(x)}
    &= \half \, (1 \mp \gamma^1\gamma_5) \, ,
  \\
    \mathcal{P}_{\uparrow\downarrow}^{(y)}
    &= \half \, (1 \mp \gamma^2\gamma_5) \, ,
  \end{split}
  \label{tp25&tp26}
\end{equation}
for negative-energy states.

The relations between transverse polarisation states and helicity states are
(for positive-energy wave functions)
\begin{equation}
  \left\{
  \begin{array}{c}
    u_\uparrow^{(x)}   = \frac1{\sqrtno2} \, (u_+ + u_-)
  \\
    u_\downarrow^{(x)} = \frac1{\sqrtno2} \, (u_+ - u_-)
  \end{array}
  \right. \qquad
  \left\{
  \begin{array}{c}
    u_\uparrow^{(y)}   = \frac1{\sqrtno2} \, (u_+ + \I u_-)
  \\
    u_\downarrow^{(y)} = \frac1{\sqrtno2} \, (u_+ - iu_-)
  \end{array}
  \right.
\end{equation}

\subsection{Spin density matrix}
\label{densmat}

The spinor $u(p,s)$ for a particle with polarisation vector $s^\mu$ satisfies
\begin{equation}
  u(p,s) \, \anti{u}(p,s) =
  (\slashed{p} + m) \, \half(1 + \gamma_5 \slashed{s})  \, .
  \label{dm1}
\end{equation}
If the particle is at rest, then
$s^{\mu}=(0,\Vec{s})=(0,\Vec{s}_\perp,\lambda)$ and (\ref{dm1}) gives
\begin{equation}
  \frac1{2m} \, u(p,s) \, \anti{u}(p,s) =
  \left(
  \begin{array}{c@{\quad}c}
    \half (1 + \Vec\sigma{\cdot}\Vec{s}) & 0
  \\
    0 & 0
  \end{array}
  \right).
  \label{dm2}
\end{equation}
Here one recognises the spin density matrix for a spin-half particle
\begin{equation}
  \rho = \half \, (1 + \Vec\sigma{\cdot}\Vec{s}) \, .
  \label{dm3}
\end{equation}
This matrix provides a general description of the spin structure of a system
that is also valid when the system is not in a pure state. The polarisation
three-vector $\Vec{s}=(\lambda,\Vec{s}_\perp)$ is, in general, such that
$\Vec{s}^2\le1$: in particular, $\Vec{s}^2=1$ for pure states and $\Vec{s}^2<1$
for mixtures. Explicitly, $\rho$ reads
\begin{equation}
  \rho = \half
  \left(
    \begin{array}{cc}
      1 + \lambda & s_x - \I s_y
    \\
      s_x + \I s_y & 1 - \lambda
    \end{array}
  \right) .
  \label{dm4}
\end{equation}
The entries of the spin density matrix have an obvious probabilistic
interpretation. If we call $P_m(\hat{\Vec{n}})$ the probability that the spin
component in the $\hat{\Vec{n}}$ direction is $m$, we can write
\begin{eqnarray}
  \lambda &=& P_{1/2} (\hat{z}) - P_{-1/2}(\hat{z}) \, ,
  \nonumber
\\
  s_x &=& P_{1/2} (\hat{x}) - P_{-1/2}(\hat{x}) \, ,
  \label{dm5}
\\
  s_y &=& P_{1/2} (\hat{y}) - P_{-1/2}(\hat{y}) \, .
  \nonumber
\end{eqnarray}

In the high-energy limit the polarisation vector is
\begin{equation}
  s^\mu = \lambda \, \frac{p^\mu}{m} + s_\perp^\mu,
\end{equation}
where $\lambda$ is (twice) the helicity of the particle. Thus we have
\begin{equation}
  (1 + \gamma_5 \slashed{s}) \, (m \pm \slashed{p})
  =
  (1 \pm \lambda \gamma_5 + \gamma_5 \slashed{s}_\perp) \,
  (m \pm \slashed{p}),
\end{equation}
and the projector (\ref{dm1}) becomes (with $m\to0$)
\begin{equation}
  u(p,s) \, \anti{u}(p,s)
  =
  \half \, \slashed{p} \,
  ( 1 - \lambda \gamma_5 + \gamma_5 \slashed{s}_\perp ) \, .
  \label{dm6}
\end{equation}
If $u_\lambda(p)$ are helicity spinors, calling $\rho_{\lambda\lambda'}$ the
elements of the spin density matrix, one has
\begin{equation}
  \half \, \slashed{p} \,
  (1 - \lambda \gamma_5 + \gamma_5 \slashed{s}_\perp )
  =
  \rho_{\lambda\lambda'} \, u_{\lambda'}(p) \, \anti{u}_\lambda(p),
  \label{dm7}
\end{equation}
where the r.h.s.\ is a trace in helicity space.

\section{Quark distributions in DIS}
\label{quarkdistr}

Although the transverse polarisation distributions cannot be probed in fully
inclusive \DIS for the reasons mentioned in the Introduction, it is convenient
to start from this process to illustrate the field-theoretical definitions of
quark (and antiquark) distribution functions. In this manner, we shall see why
the transversity distributions $\DT{f}$ decouple from \DIS even when quark
masses are taken into account (which would in principle allow chirality-flip
distributions). We start by reviewing some well-known features of \DIS (for an
exhaustive treatment of the subject see \eg, \cite{Leader:1996hk}).

\subsection{Deeply-inelastic scattering}
\label{deepinelastic}

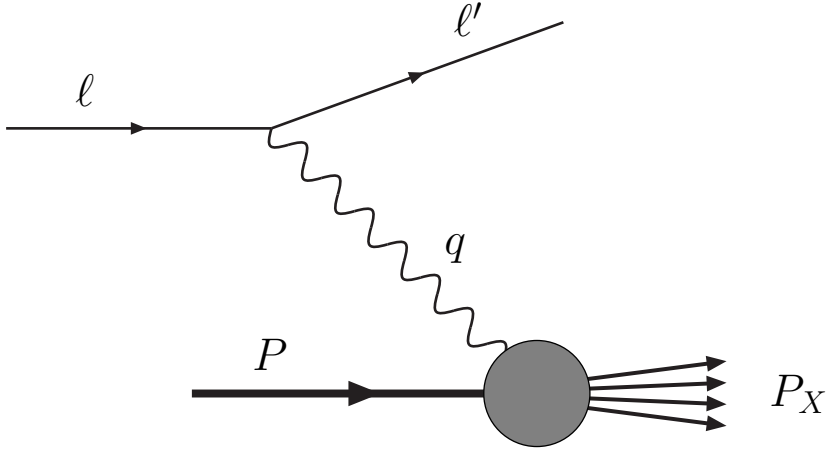
\begin{figure}
  \centering
  \begin{picture}(370,170)(20,30)
    \SetWidth{1}
    \ArrowLine(50,150)(150,150)
    \Text(80,165)[]{\Large$\ell$}
    \ArrowLine(150,150)(260,190)
    \Text(225,192)[]{\Large$\ell'$}
    \Photon(150,150)(250,50){-5}{8}
    \SetWidth{3}
    \ArrowLine(120,50)(250,50)
    \Text(220,105)[]{\Large$q$}
    \Text(150,65)[]{\Large$P$}
    \SetWidth{1.5}
    \LongArrow(250,53)(320,62)
    \LongArrow(250,51)(320,54)
    \LongArrow(250,49)(320,46)
    \LongArrow(250,47)(320,38)
    \Text(350,50)[]{\Large$P_X$}
    \SetWidth{0.5}
    \GCirc(250,50){20}{0.5}
  \end{picture}
  \caption{Deeply-inelastic scattering.}
  \label{fig_dis}
\end{figure}

Consider inclusive lepton--nucleon scattering (see Fig.~\ref{fig_dis}, where
the dominance of one-photon exchange is assumed)
\begin{equation}
  l (\ell) \, + \, N (P) \to l' (\ell') \, + X (P_X) \, ,
  \label{dis0}
\end{equation}
where $X$ is an undetected hadronic system (in brackets we put the four-momenta
of the particles). Our notation is as follows: $M$ is the nucleon mass,
$m_\ell$ the lepton mass, $s_\ell(s'_\ell)$ the spin four-vector of the
incoming (outgoing) lepton, $S$ the spin four-vector of the nucleon, while
$\ell=(E,\Vec\ell)$, and $\ell'=(E',\Vec\ell')$ are the lepton four-momenta.

Two kinematic variables (besides the centre-of-mass energy $s=(\ell+P)^2$, or,
alternatively, the lepton beam energy $E$) are needed to describe reaction
(\ref{dis0}). They can be chosen among the following invariants (unless
otherwise stated, we neglect lepton masses):
\begin{align*}
  q^2 &= (\ell - \ell')^2 = -2 E E' (1 - \cos\vartheta) \, ,
\\
  \nu &= \frac{P{\cdot}q}{M} &
  & \text{(the lab-frame photon energy),}
\\
  x &= \frac{Q^2}{2P{\cdot}q} = \frac{Q^2}{2M\nu} &
  & \text{(the Bjorken variable),}
\\
  y &= \frac{P{\cdot}q}{P{\cdot}\ell} &
  & \text{(the inelasticity),}
\end{align*}
where $\vartheta$ is the scattering angle. The photon momentum $q$ is a
spacelike four-vector and one usually introduces the positive quantity
$Q^2\equiv-q^2$. Both the Bjorken variable $x$ and the inelasticity $y$ take on
values between 0 and 1. They are related to $Q^2$ by $xy=Q^2/(s-M^2)$.

The \DIS cross-section is
\begin{equation}
  \d\sigma =
  \frac1{4 \, \ell{\cdot}P} \, \frac{e^4}{Q^4} \, L_{\mu\nu} \,
  W^{\mu\nu} \, 2\pi \, \frac{\d^3\Vec{\ell}'}{(2\pi)^3 \, 2E'} \, ,
  \label{dis4b}
\end{equation}
where the leptonic tensor $L_{\mu\nu}$ is defined as (lepton masses are
retained here)
\begin{eqnarray}
  L_{\mu\nu}
  &=& \sum_{s_{l'}}
  \left[ \strut
    \anti{u}_{l'}(\ell', s_{l'}) \gamma_\mu u_l(\ell, s_l)
  \right]^*
  \left[ \strut
    \anti{u}_{l'}(\ell', s_{l'}) \gamma_\nu u_l(\ell, s_l)
  \right]
  \nonumber
\\
  &=& \Tr
  \left[
    (\slashed\ell + m_l) \, \half(1 + \gamma_5 \slashed{s}_l) \,
    \gamma_\mu \, (\slashed\ell' + m_l) \, \gamma_\nu
  \right] .
  \label{dis4}
\end{eqnarray}
and the hadronic tensor $W^{\mu\nu}$ is
\begin{eqnarray}
  W^{\mu\nu}
  &=&
  \frac1{2\pi} \, \sum_X \int \! \frac{\d^3\Vec{P}_X}{(2\pi)^3\,2E_X} \,
  (2\pi)^4 \, \delta^4(P + q - P_X)
  \nonumber
\\
  && \hspace{5em} \null \times
  \langle PS | J^\mu (0) | X  \rangle
  \langle X  | J^\nu (0) | PS \rangle \, .
  \label{dis2b}
\end{eqnarray}
Using translational invariance this can be also written as
\begin{equation}
  W^{\mu\nu} = \frac1{2\pi} \int \! \d^4 \xi \, \e^{\I q{\cdot}\xi} \,
  \langle PS | J^\mu(\xi) J^\nu(0) | PS \rangle \, .
  \label{dis3}
\end{equation}
It is important to recall that the matrix elements in (\ref{dis3}) are
connected. Therefore, vacuum transitions of the form
$\langle0|J^\mu(\xi)J^\nu(0)|0\rangle\,\langle{PS}|{PS}\rangle$ are excluded.

Note that in (\ref{dis4}) and (\ref{dis2b}) we summed over the final lepton
spin but did not average over the initial lepton spin, nor sum over the hadron
spin. Thus we are describing, in general, the scattering of polarised leptons
on a polarised target, with no measurement of the outgoing lepton polarisation
(for comprehensive reviews on polarised \DIS see \cite{Anselmino:1995gn,
Lampe:1998eu, Filippone:2001ux}).

In the target rest frame, where $\ell{\cdot}P=ME$, (\ref{dis4b}) reads
\begin{equation}
  \frac{\d\sigma}{\d{E}' \, \d\Omega } =
  \frac{\alpha_\text{em}^2}{2MQ^4} \, \frac{E'}{E} \,
  L_{\mu\nu} W^{\mu\nu} \, ,
  \label{dis5}
\end{equation}
where $\d\Omega=\d\cos\vartheta\,\d\varphi$.

The leptonic tensor $L_{\mu\nu}$ can be decomposed into a symmetric and an
antisymmetric part under $\mu\leftrightarrow\nu$ interchange
\begin{equation}
  L_{\mu\nu} =
  L_{\mu\nu}^\text{(S)} (\ell, \ell') + \I \,
  L_{\mu\nu}^\text{(A)} (\ell, s_l; \ell') \, ,
  \label{dis7}
\end{equation}
and, computing the trace in (\ref{dis4}), we obtain
\begin{subequations}
\begin{eqnarray}
  L_{\mu\nu}^\text{(S)} &=&
  2 (\ell_\mu \ell'_\nu + \ell_\nu \ell'_\mu - g_{\mu\nu} \, \ell{\cdot}\ell') \, ,
  \label{dis8}
\\
  L_{\mu\nu}^\text{(A)} &=& 2 m_l \,
  \varepsilon_{\mu\nu\rho\sigma} s_\ell^\rho (\ell - \ell')^\sigma \, .
  \label{dis9}
\end{eqnarray}
\end{subequations}
If the incoming lepton is longitudinally polarised, its spin vector is
\begin{equation}
  s_l^\mu = \frac{\lambda_l}{m_l} \, \ell^\mu, \quad
  \lambda_l = \pm 1 \, ,
  \label{dis15c}
\end{equation}
and (\ref{dis9}) becomes
\begin{equation}
  L_{\mu\nu}^\text{(A)} =
  2 \lambda_l \, \varepsilon_{\mu\nu\rho\sigma} \ell^\rho q^\sigma \, .
  \label{dis9b}
\end{equation}
Note that the lepton mass $m_l$ appearing in (\ref{dis9}) has been cancelled by
the denominator of (\ref{dis15c}). In contrast, if the lepton is transversely
polarised, that is $s_l^\mu=s_{l\perp}^\mu$, no such cancellation occurs and
the process is suppressed by a factor $m_l/E$. In what follows we shall
consider only unpolarised or longitudinally polarised lepton beams.

The hadronic tensor $W_{\mu\nu}$ can be split as
\begin{equation}
  W_{\mu\nu} =
  W_{\mu\nu}^\text{(S)}(q, P) + \I \, W_{\mu\nu}^\text{(A)}(q;P,S) \, ,
  \label{dis10}
\end{equation}
where the symmetric and the antisymmetric parts are expressed in terms of two
pairs of structure functions, $F_1$, $F_2$ and $G_1$, $G_2$, as
\begin{subequations}
\begin{eqnarray}
  \frac1{2M} \, W_{\mu\nu}^\text{(S)}
  &=&
  \left(- g_{\mu\nu} + \frac{q_\mu q_\nu}{q^2} \right) W_1(P{\cdot}q,q^2)
  \nonumber
\\
  && \null + \frac1{M^2}
  \left[
    \left( P_\mu - \frac{P{\cdot}q}{q^2} \, q_\mu \right)
    \left( P_\nu - \frac{P{\cdot}q}{q^2} \, q_\nu \right)
  \right] W_2(P{\cdot}q,q^2) \, , \hspace{4em}
  \label{dis11}
\\
  \frac1{2M} \, W_{\mu\nu}^\text{(A)}
  &=&
  \varepsilon_{\mu\nu\rho\sigma} \, q^\rho
  \left\{ M S^\sigma \, G_1(P{\cdot}q,q^2) \Strut \right.
  \nonumber
\\
  && \null +
  \left.
    \frac1{M} \left[ \strut
P{\cdot}q \, S^\sigma - S{\cdot}q \, P^\sigma \right] G_2(P{\cdot}q,q^2)
  \right\}.
  \label{dis12}
  \sublabel{subdis12}
\end{eqnarray}
\end{subequations}
Eqs.~(\ref{dis11},~\ref{subdis12}) are the most general expressions compatible
with the requirement of gauge invariance, which implies
\begin{equation}
  q_\mu W^{\mu\nu} = 0 = q_\nu W^{\mu\nu}.
\end{equation}
Using (\ref{dis7}, \ref{dis10}) the cross-section (\ref{dis5}) becomes
\begin{equation}
  \frac{\d\sigma}{\d{E}' \, \d\Omega } =
  \frac{\alpha_\text{em}^2}{2MQ^4} \,
  \frac{E'}{E}
  \left[
    L_{\mu\nu}^\text{(S)} W^{\mu\nu\,\text{(S)}} -
    L_{\mu\nu}^\text{(A)} W^{\mu\nu\,\text{(A)}}
  \right].
  \label{dis13}
\end{equation}

The unpolarised cross-section is then obtained by averaging over the spins of
the incoming lepton ($s_l$) and of the nucleon ($S$) and reads
\begin{equation}
  \frac{\d\sigma^\text{unp}}{\d{E}' \, \d\Omega }
  =
  \frac12 \sum_{s_l} \,
  \frac12 \sum_S \frac{\d\sigma}{\d{E}' \, \d\Omega }
  =
  \frac{\alpha_\text{em}^2}{2MQ^4} \, \frac{E'}{E} \,
  L_{\mu\nu}^\text{(S)} W^{\mu\nu \, \text{(S)}} \, .
  \label{dis14}
\end{equation}
Inserting eqs.~(\ref{dis8}) and (\ref{dis11}) into (\ref{dis14}) one obtains
the well-known expression
\begin{equation}
  \frac{\d\sigma^\text{unp}}{\d{E}' \, \d\Omega }
  =
  \frac{4\alpha_\text{em}^2{E'}^2}{Q^4}
  \left[
    2 W_1 \, \sin^2\frac{\vartheta}{2}
    + W_2 \, \cos^2\frac{\vartheta}{2}
  \right] .
  \label{dis14b}
\end{equation}

Differences of cross-sections with opposite target spin probe the antisymmetric
part of the leptonic and hadronic tensors
\begin{equation}
  \frac{\d\sigma(+S)}{\d{E}'\,\d\Omega} -
  \frac{\d\sigma(-S)}{\d{E}'\,\d\Omega}
  =
  - \frac{\alpha_\text{em}^2}{2M Q^4} \, \frac{E'}{E} \,
  2 L_{\mu\nu}^\text{(A)}
    W^{\mu\nu\,\text{(A)}} \, .
  \label{dis15}
\end{equation}
In the target rest frame the spin of the nucleon can be parametrised as
(assuming $|\Vec{S}|=1$)
\begin{equation}
  S^\mu =
  (0, \Vec{S}) =
  (0, \, \sin\alpha \cos\beta, \, \sin\alpha \sin\beta, \, \cos\alpha) \, .
  \label{dis15b}
\end{equation}
Taking the direction of the incoming lepton to be the $z$-axis, we have
\begin{equation}
  \begin{split}
    \ell^\mu  &= E (1, 0, 0, 1) \, ,
  \\
    {\ell'}^\mu &= E'
    (1, \,
     \sin\vartheta \cos\varphi, \,
     \sin\vartheta \sin\varphi, \,
     \cos\vartheta
    ) \, .
  \end{split}
  \label{dis15d&e}
\end{equation}

Inserting (\ref{dis9}) and (\ref{dis12}) in eq.~(\ref{dis15}), with the above
parametrisations for the spin and the momentum four-vectors, for the
cross-section asymmetry in the target rest frame we now obtain
\begin{eqnarray}
  \frac{\d\sigma(+S)}{\d{E}' \, \d\Omega }
  &-&
  \frac{\d\sigma(-S)}{\d{E}' \, \d\Omega } =
  - \frac{4\alpha_\text{em}^2 \, E'}{Q^2 \, E}
  \nonumber
\\
  && \null \times
  \left\{ \strut
    \left[
      E \cos\alpha +
      E'(\sin\vartheta \sin\alpha \cos\phi + \cos\vartheta \cos\alpha)
    \right] MG_1
  \right.
  \nonumber
\\
  && \quad \null
  \left. \strut
    + 2EE'
    \left[
      \sin\vartheta \sin\alpha \cos\phi +
      \cos\vartheta \cos\alpha -
      \cos\alpha
    \right] G_2
  \right\} , \hspace{4em}
  \label{dis16}
\end{eqnarray}
where $\phi=\beta-\varphi$ is the azimuthal angle between the lepton plane and
the $(\hat{\Vec\ell}, \hat{\Vec{S}})$ plane.

In particular, when the target nucleon is \emph{longitudinally polarised} (that
is, \emph{polarised along the incoming lepton direction}), one has $\alpha=0$
and the spin asymmetry becomes
\begin{subequations}
\begin{equation}
  \frac{\d\sigma^{\Rightarrow}}{\d{E}' \, \d\Omega} -
  \frac{\d\sigma^{\Leftarrow}} {\d{E}' \, \d\Omega} =
  - \frac{4\alpha_\text{em}^2 E'}{Q^2 E}
  \left[
    (E + E' \cos\vartheta) \, M \, G_1 - Q^2 \, G_2
  \right].
  \label{dis17}
\end{equation}
When the target nucleon is \emph{transversely polarised} (that is,
\emph{polarised orthogonally to the incoming lepton direction}), one has
$\alpha=\pi/2$ and the spin asymmetry is
\begin{equation}
  \frac{\d\sigma^{\Uparrow}}  {\d{E}' \, \d\Omega} -
  \frac{\d\sigma^{\Downarrow}}{\d{E}' \, \d\Omega} =
  - \frac{4\alpha_\text{em}^2 {E'}^2}{Q^2 E} \,
  \sin\vartheta \left[ M \, G_1 + 2E \, G_2 \strut \right] .
  \label{dis18}
\end{equation}
\end{subequations}

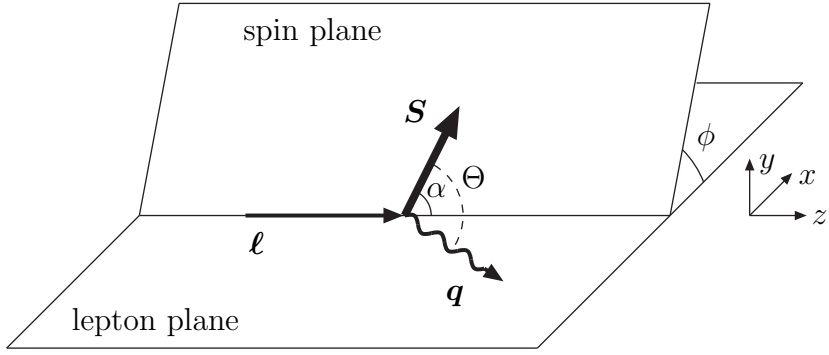
\begin{figure}
  \centering
  \begin{picture}(350,140)(80,50)
    \Line(100,50)(300,50)
    \Line(150,100)(350,100) 
    \Line(100,50)(150,100)
    \Line(300,50)(350,100)
    \Text(125,60)[l]{lepton plane}
    \Line(150,100)(165,180)
    \Line(350,100)(365,180)
    \Line(165,180)(365,180)
    \Text(190,170)[l]{spin plane}
    \Line(350,100)(400,150)
    \Line(400,150)(360,150)
    \CArc(330,100)(35,21,45)
    \Text(365,130)[]{$\phi$}
    \SetWidth{1.5}
    \LongArrow(190,100)(248,100)
    \Text(195,90)[]{$\Vec\ell$}
    \Photon(250,100)(280,80){2}{3}
    \LongArrow(280,80)(286,76)
    \Text(270,70)[]{$\Vec{q}$}
    \SetWidth{0.5}
    \CArc(250,100)(10,0,60)
    \Text(263,110)[]{$\alpha$}
    \DashCArc(250,100)(22,330,60){3}
    \Text(277,115)[]{$\Theta$}
    \SetWidth{3}
    \LongArrow(250,100)(270,140)
    \Text(255,140)[]{$\Vec{S}$}
    \SetWidth{0.5}
    \LongArrow(380,100)(400,100)
    \LongArrow(380,100)(380,120)
    \LongArrow(380,100)(395,115)
    \Text(405,100)[l]{$z$}
    \Text(385,120)[l]{$y$}
    \Text(400,115)[l]{$x$}
  \end{picture}
  \caption{Lepton and spin planes. The lepton plane is taken here to coincide
           with the $xz$ plane, \ie, $\varphi=0$.}
  \label{plane}
\end{figure}

A remark on the terminology is in order here. The terms ``longitudinal'' and
``transverse'' are somewhat ambiguous, insofar as a reference axis is not
specified. From an experimental point of view, the ``longitudinal'' and
``transverse'' polarisations of the nucleon are in reference to the lepton beam
axis. Thus ``longitudinal'' (``transverse'') indicates parallel (orthogonal) to
this axis. We use the large arrows $\Rightarrow$ and $\Uparrow$ to denote these
two cases respectively. From a theoretical point of view, it is simpler to
refer to the direction of motion of the virtual photon. One then speaks of the
``longitudinal'' ($\SL$) and ``transverse'' ($\ST$) spin of the nucleon, meaning by
this spin parallel and perpendicular, respectively, to the photon axis. When
the target is ``longitudinally'' or ``transversely'' polarised in this sense, we
shall make explicit reference to $\SL$ and $\ST$ in the cross-section. Later,
it will be shown how to pass from $\d\sigma^\Rightarrow-\d\sigma^\Leftarrow$
and $\d\sigma^\Uparrow-\d\sigma^\Downarrow$ to $\d\sigma(+\SL)-\d\sigma(-\SL)$
and $\d\sigma(+\ST)-\d\sigma(-\ST)$. Note that, in general,
$\d\sigma^\Rightarrow$ is a combination of $\d\sigma(\SL)$ and $\d\sigma(\ST)$.
We shall return to this point in Sec.~\ref{transvlepnuc}.

It is customary to introduce the dimensionless structure functions
\begin{equation}
\begin{split}
  F_1(x,Q^2) &\equiv M \, W_1(\nu,Q^2) \, ,
\\
  F_2(x,Q^2) &\equiv \nu \, W_2(\nu,Q^2) \, ,
\\
  g_1(x,Q^2) &\equiv M^2 \nu \, G_1(\nu,Q^2) \, ,
\\
  g_2(x,Q^2) &\equiv M \nu^2 \, G_2(\nu,Q^2) \, .
\end{split}
\end{equation}
In the Bjorken limit
\begin{equation}
  \nu, Q^2 \to \infty \, , \quad
  x = \frac{Q^2}{2M\nu} \quad \text{fixed},
  \label{dis23}
\end{equation}
$F_1$, $F_2$, $g_1$ and $g_2$ are expected to scale approximately, that is, to
depend only on $x$. In terms of $F_1$, $F_2$, $g_1$ and $g_2$, the hadronic
tensor reads
\begin{subequations}
\begin{eqnarray}
  W_{\mu\nu}^\text{(S)} &=&
  2
  \left(- g_{\mu\nu} +
  \frac{q_\mu q_\nu}{q^2}\right) F_1(x,Q^2)
  \nonumber
\\
  && \null
  + \frac2{P{\cdot}q}
  \left[
    \left( P_\mu - \frac{P{\cdot}q}{q^2} \, q_\mu \right)
    \left( P_\nu - \frac{P{\cdot}q}{q^2} \, q_\nu \right)
  \right] F_2(x,Q^2) \, ,
  \label{dis24}
\\
  W_{\mu\nu}^\text{(A)}
  &=&
  \frac{2M \, \varepsilon_{\mu\nu\rho\sigma} \, q^\rho}{P{\cdot}q}
  \left\{
    S^\sigma \, g_1(x,Q^2) +
    \left[ S^\sigma - \frac{S{\cdot}q}{P{\cdot}q} \, P^\sigma \right] g_2(x,Q^2)
  \right\} . \hspace{4em}
  \label{dis25}
\end{eqnarray}
\end{subequations}
The unpolarised cross-section then becomes (as a function of $x$ and $y$)
\begin{equation}
  \frac{\d\sigma}{\d{x} \, \d{y}}
  =
  \frac{4\pi\alpha_\text{em}^2s}{Q^4}
  \left\{
    x y^2 \, F_1(x,Q^2) +
    \left( 1 - y - \frac{xy m_N^2}{s} \right) F_2(x,Q^2)
  \right\} ,
  \label{dis25b}
\end{equation}
whereas the spin asymmetry (\ref{dis16}), in terms of $g_1$ and $g_2$, takes on
the form
\begin{eqnarray}
  &&
  \frac{\d\sigma(+S)}{\d{x} \, \d{y} \, \d\varphi}
  -
  \frac{\d\sigma(-S)}{\d{x} \, \d{y} \, \d\varphi}
  =
  \frac{4\alpha_\text{em}^2}{Q^2}
  \left\{(2-y) \, g_1(x,Q^2) \, \cos\alpha
  \STRUT \right.
  \nonumber
\\
  && \qquad \left. \null
  + \frac{2Mx}{Q} \, \sqrt{1-y}
  \left[
    y \, g_1(x,Q^2) + 2 g_2(x,Q^2)
  \right] \sin\alpha \, \cos\phi \right\} , \qquad
  \label{dis26}
\end{eqnarray}
where we have neglected contributions of order $M^2/Q^2$. Note that the term
containing $g_2$ is suppressed by one power of $Q$. This renders the
measurement of $g_2$ quite a difficult task.

It is useful at this point to re-express the cross-section asymmetry
(\ref{dis26}) in terms of the angle $\Theta$ between the spin of the nucleon
$\Vec{S}$ and the photon momentum $\Vec{q}=\Vec{l}-\Vec{l}'$. The relation
between $\alpha$, $\phi$ and $\Theta$, ignoring terms $\Ord(M^2/Q^2)$, is
\begin{equation}
  \begin{split}
    \cos\alpha &=
    \cos\Theta + \frac{2Mx}{Q} \, \sqrt{1-y} \, \cos\phi \, \sin\Theta \, ,
  \\
    \sin\alpha &=
    \sin\Theta - \frac{2Mx}{Q} \, \sqrt{1-y} \, \cos\phi \, \cos\Theta \, ,
  \end{split}
  \label{dis26b&c}
\end{equation}
and hence we obtain
\begin{subequations}
\begin{eqnarray}
  \frac{\d\sigma(+S)}{\d{x} \, \d{y} \, \d\varphi} -
  \frac{\d\sigma(-S)}{\d{x} \, \d{y} \, \d\varphi}
  &=&
  - \frac{4\alpha_\text{em}^2}{Q^2}
  \left[ \STRUT
    (2-y) \, g_1 \, \cos\Theta
  \right.
  \nonumber
\\
  && \quad \left. \null +
    \frac{4Mx}{Q} \, \sqrt{1-y} \, (g_1 + g_2) \, \sin\Theta \cos\phi
  \right] , \hspace{4em}
  \label{dis27}
\end{eqnarray}
which demonstrates that when the target spin is perpendicular to the photon
momentum ($\Theta=\pi/2$) \DIS probes the combination $g_1+g_2$; and
\begin{equation}
  \frac{\d\sigma(+\ST)}{\d{x} \, \d{y} \, \d\varphi} -
  \frac{\d\sigma(-\ST)}{\d{x} \, \d{y} \, \d\varphi} =
  - \frac{4\alpha_\text{em}^2}{Q^2}
  \left[ \frac{4Mx}{Q} \, \sqrt{1-y} \, (g_1 + g_2) \right] \cos\phi \, .
  \label{dis27a}
\end{equation}
\end{subequations}
This result can be obtained in another, more direct, manner. Splitting the spin
vector of the nucleon into a longitudinal and transverse part (with respect to
the photon axis):
\begin{equation}
  S^\mu = \SL^\mu + \ST^\mu \, ,
  \label{dis27b}
\end{equation}
where $\lambda_N=\pm1$ is (twice) the helicity of the nucleon, the
antisymmetric part of the hadronic tensor becomes
\begin{equation}
  W_{\mu\nu}^\text{(A)} =
  \frac{2M \, \varepsilon_{\mu\nu\rho\sigma} \, q^\rho}{P{\cdot}q}
  \left[ \SL^\sigma \, g_1 + \ST^\sigma \, (g_1 + g_2) \right] .
  \label{dis27c}
\end{equation}
Thus, if the nucleon is longitudinally polarised the \DIS cross-section depends
only on $g_1$; if it is transversely polarised (with respect to the photon
axis) what is measured is the sum of $g_1$ and $g_2$. We shall use expression
(\ref{dis27c}) when studying the quark content of structure functions in the
parton model, to which we now turn.

\subsection{The parton model}
\label{parton}

In the parton model the virtual photon is assumed to scatter incoherently off
the constituents of the nucleon (quarks and antiquarks). Currents are treated
as in free field theory and any interaction between the struck quark and the
target remnant is ignored.

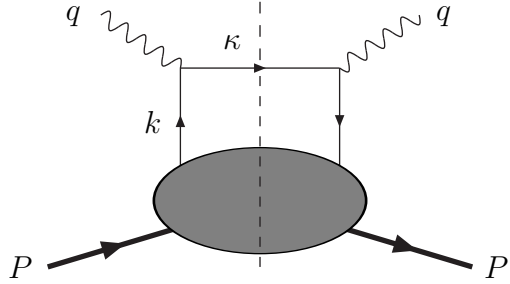
\begin{figure}
  \centering
  \begin{picture}(300,110)(0,20)
    \SetWidth{2}
    \ArrowLine(70,25)(120,40)
    \ArrowLine(180,40)(230,25)
    \Text(60,25)[]{$P$}
    \Text(240,25)[]{$P$}
    \SetWidth{0.5}
    \ArrowLine(120,60)(120,100)
    \ArrowLine(180,100)(180,60)
    \ArrowLine(120,100)(180,100)
    \Text(110,80)[]{$k$}
    \Text(140,110)[]{$\kappa$}
    \Photon(90,120)(120,100){3}{4.5}
    \Photon(210,120)(180,100){3}{4.5}
    \Text(80,120)[]{$q$}
    \Text(220,120)[]{$q$}
    \GOval(150,50)(20,40)(0){0.5}
    \DashLine(150,25)(150,97){4}
    \DashLine(150,103)(150,125){4}
  \end{picture}
  \caption{The so-called handbag diagram.}
  \label{handbag}
\end{figure}

The hadronic tensor $W^{\mu\nu}$ is then represented by the handbag diagram
shown in Fig.~\ref{handbag} and reads (to simplify the presentation, for the
moment we consider only quarks, the extension to antiquarks being rather
straight-forward)
\begin{eqnarray}
  W^{\mu\nu}
  &=&
  \frac1{(2\pi)} \, \sum_a e_a^2 \, \sum_X
  \int \! \frac{\d^3\Vec{P}_X}{(2\pi)^3 \, 2E_X}
  \int \! \frac{\d^4k}{(2\pi)^4}
  \int \! \frac{\d^4\kappa}{(2\pi)^4} \,
  \delta(\kappa^2)
  \nonumber
\\
  && \qquad \null \times
  \left[ \strut \anti{u}(\kappa) \gamma^\mu \phi(k;P,S) \right]^*
  \left[ \strut \anti{u}(\kappa) \gamma^\nu \phi(k;P,S) \right]
  \nonumber
\\
  && \qquad \null \times
  (2\pi)^4 \, \delta^4(P - k - P_X) \;
  (2\pi)^4 \, \delta^4(k + q - \kappa) \, ,
  \label{pm1}
\end{eqnarray}
where $\sum_a$ is a sum over the flavours, $e_a$ is the quark charge in units
of $e$, and we have introduced the matrix elements of the quark field between
the nucleon and its remnant
\begin{equation}
  \phi_i(k,P,S) = \langle X | \psi_i(0) | PS \rangle \, .
  \label{pm2}
\end{equation}
We define the quark--quark correlation matrix $\Phi_{ij}(k,P,S)$ as
\begin{eqnarray}
  \Phi_{ij}(k,P,S)
  &=& \sum_X
  \int \! \frac{\d^3\Vec{P}_X}{(2\pi)^3 \, 2E_X} \,
  (2\pi)^4 \, \delta^4(P - k - P_X)
  \nonumber
\\
  && \hspace{5em} \null \times
  \langle PS | \psi_j(0) | X \rangle \langle X | \psi_i(0) | PS \rangle \, .
  \label{pm3}
\end{eqnarray}
Using translational invariance and the completeness of the $|X\rangle$ states
this matrix can be re-expressed in the more synthetic form
\begin{equation}
  \Phi_{ij}(k,P,S) = \int \! \d^4 \xi \, \e^{\I k{\cdot}\xi} \,
  \langle PS | \anti\psi_j(0) \psi_i(\xi) | PS \rangle \, .
  \label{pm5}
\end{equation}

With the definition (\ref{pm3}) the hadronic tensor becomes
\begin{eqnarray}
  W^{\mu\nu}
  &=&
  \sum_a e_a^2
  \int \! \frac{\d^4k}{(2\pi)^4}
  \int \! \frac{\d^4\kappa}{(2\pi)^4} \,
  \delta(\kappa^2) \, (2\pi)^4  \, \delta^4(k + q - \kappa) \,
  \Tr \left[ \Phi \gamma^\mu \slashed\kappa \gamma^\nu \right]
  \nonumber
\\
  &=&
  \sum_a e_a^2
  \int \! \frac{\d^4k}{(2\pi)^4} \,
  \delta \! \left( (k + q)^2 \right)
  \Tr \left[ \Phi \gamma^\mu (\slashed{k} + \slashed{q}) \gamma^\nu \right] .
  \label{pm4}
\end{eqnarray}

\begin{figure}
  \centering
  \begin{picture}(340,110)(80,50)
    \Line(100,50)(300,50)
    \Line(100,50)(200,150)
    \Line(300,50)(400,150)
    \Line(200,150)(400,150)
    \SetWidth{2}
    \LongArrow(330,100)(252,100)
    \SetWidth{1}
    \Photon(200,100)(243,100){3}{3.5}
    \LongArrow(243,100)(250,100)
    \LongArrow(200,100)(180,115)
    \LongArrow(160,85)(200,100)
    \SetWidth{0.5}
    \Text(280,90)[]{$\Vec{P}$}
    \Text(290,60)[r]{lepton plane}
    \Text(225,90)[]{$\Vec{q}$}
    \Text(160,76)[]{$\Vec\ell$}
    \Text(188,123)[l]{$\Vec\ell'$}
    \LongArrow(380,100)(365,100)
    \LongArrow(380,100)(380,85)
    \LongArrow(380,100)(390,110)
    \Text(365,97)[t]{$z$}
    \Text(377,83)[r]{$y$}
    \Text(395,110)[l]{$x$}
  \end{picture}
  \caption{The $\gamma^*N$ collinear frame (note our convention for the axes).}
  \label{plane3}
\end{figure}
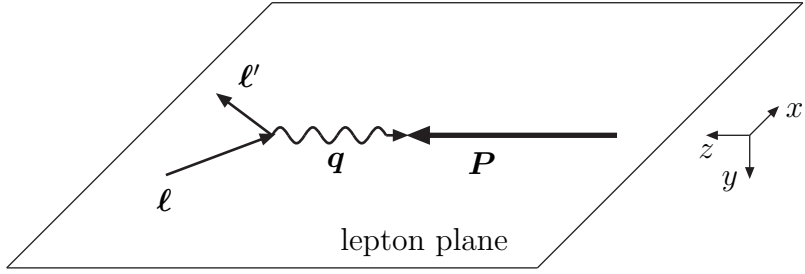

In order to calculate $W^{\mu\nu}$, it is convenient to use a Sudakov
parametrisation of the four-momenta at hand (the Sudakov decomposition of
vectors is described in Appendix~\ref{sudakov}). We introduce the null vectors
$p^\mu$ and $n^\mu$, satisfying
\begin{equation}
  p^2 = 0 = n^2 \, , \qquad p{\cdot}n = 1 \, , \qquad n^+ = 0 = p^- \, .
  \label{pm4b}
\end{equation}
and we work in a frame where the virtual photon and the proton are collinear.
As is customary, the proton is taken to be directed along the positive $z$
direction (see Fig.~\ref{plane3}). In terms of $p^\mu$ and $n^\mu$ the proton
momentum can be parametrised as
\begin{equation}
  P^\mu = p^\mu + \frac{M^2}{2} n^\mu \simeq p^\mu \, .
  \label{pm6}
\end{equation}
Note that, neglecting the mass $M$, $P^\mu$ coincides with the Sudakov vector
$p^\mu$. The momentum $q^\mu$ of the virtual photon can be written as
\begin{equation}
  q^\mu \simeq P{\cdot}q \, n^\mu - x \, p^\mu  \, ,
  \label{pm7}
\end{equation}
where we are implicitly ignoring terms $\Ord(M^2/Q^2)$. Finally, the Sudakov
decomposition of the quark momentum is
\begin{equation}
  k^\mu =
  \alpha \, p^\mu +
  \frac{(k^2 + \Vec{k}_\perp^2)}{2\alpha} \, n^\mu +
  k_\perp^\mu \, .
  \label{pm8}
\end{equation}

In the parton model one assumes that the handbag-diagram contribution to the
hadronic tensor is dominated by small values of $k^2$ and $\Vec{k}_\perp^2$. This
means that we can write $k^\mu$ approximately as
\begin{equation}
  k^\mu \simeq \alpha \, p^\mu \, .
  \label{pm9}
\end{equation}
The on-shell condition of the outgoing quark then implies
\begin{equation}
  \delta \! \left( (k + q)^2 \right)
  \simeq \delta(- Q^2 + 2\alpha \, P{\cdot}q)
  = \frac1{2P{\cdot}q} \, \delta(\alpha - x) \, ,
  \label{pm10}
\end{equation}
that is, $k^\mu\simeq{x}\,P^\mu$. Thus the Bjorken variable
$x\equiv{Q^2}/(2P{\cdot}q)$ is the fraction of the longitudinal momentum of the
nucleon carried by the struck quark: $x=k^+/P^+$. (In the following we shall
also consider the possibility of retaining the quark transverse momentum; in
this case (\ref{pm8}) will be approximated as
$k^\mu\simeq{x}P^\mu+k_\perp^\mu$.)

Returning to the hadronic tensor (\ref{pm4}), the identity
\begin{equation}
  \gamma^\mu \gamma^\rho \gamma^\nu
  =
  \left[
    g^{\mu\rho} g^{\nu\sigma} +
    g^{\mu\sigma} g^{\nu\rho} -
    g^{\mu\nu} g^{\rho\sigma} +
    \I \varepsilon^{\mu\rho\nu\sigma} \gamma_5
  \right]
  \gamma_\sigma \, ,
  \label{pm12}
\end{equation}
allows us to split $W^{\mu\nu}$ into symmetric (S) and antisymmetric (A) parts
under $\mu\leftrightarrow\nu$ interchange. Let us first consider
$W_{\mu\nu}^\text{(S)}$ (\ie, unpolarised \DIS):
\begin{eqnarray}
  W_{\mu\nu}^\text{(S)}
  &=&
  \frac1{2P{\cdot}q} \, \sum_a e_a^2 \int \! \frac{\d^4k}{(2\pi)^4} \;
  \delta \! \left(x - \frac{k^+}{P^+}\right)
  \nonumber
\\
  && \hspace{5em} \null \times
  \left[ \strut
    (k_\mu + q_\mu) \Tr(\Phi \gamma_\nu) +
    (k_\nu + q_\nu) \Tr(\Phi \gamma_\mu)
  \right.
  \nonumber
\\
  && \hspace{6em} \null -
  \left. \strut
    g_{\mu\nu} (k^\rho + q^\rho) \Tr(\Phi \gamma_\rho)
  \right] .
  \label{pm13}
\end{eqnarray}
From (\ref{pm7}) and (\ref{pm8}) we have $k_\mu+q_\mu\simeq(P{\cdot}q)\,n_\mu$ and
(\ref{pm13}) becomes
\begin{eqnarray}
  W_{\mu\nu}^\text{(S)}
  &=&
  \frac12 \, \sum_a e_a^2 \int \! \frac{\d^4k}{(2\pi)^4} \;
  \delta\!\left(x - \frac{k^+}{P^+}\right)
  \nonumber
\\
  && \hspace{3em} \null \times
  \left[ \strut
    n_\mu \Tr(\Phi \gamma_\nu) +
    n_\nu \Tr(\Phi \gamma_\mu) -
    g_{\mu\nu} n^\rho \Tr(\Phi \gamma_\rho)
  \right] . \hspace{4em}
  \label{pm14}
\end{eqnarray}
Introducing the notation
\begin{eqnarray}
  \langle \Gamma \rangle
  &\equiv&
  \int \! \frac{\d^4k}{(2\pi)^4}\,
  \delta\!\left(x - \frac{k^+}{P^+}\right) \Tr(\Gamma \, \Phi)
  \nonumber
\\
  &=&
  P^+ \int \! \frac{\d\xi^-}{2\pi} \, \e^{\I xP^+ \xi^-} \,
  \langle PS |
    \anti\psi(0) \, \Gamma \, \psi(0, \xi^-, \Vec{0}_\perp)
  | PS \rangle
  \nonumber
\\
  &=&
  \int \! \frac{\d\tau}{2\pi} \, \e^{\I \tau x} \,
  \langle PS | \anti\psi(0) \, \Gamma \, \psi(\tau n) | PS \rangle \, ,
  \label{pm15}
\end{eqnarray}
where $\Gamma$ is a Dirac matrix, $W_{\mu\nu}^\text{(S)}$ is written as
\begin{equation}
  W_{\mu\nu}^\text{(S)} =
  \half \sum_a \, e_a^2
  \left[ \strut
    n_\mu \, \langle \gamma_\nu \rangle +
    n_\nu \, \langle \gamma_\mu \rangle -
    g_{\mu\nu} \, n^\rho \, \langle \gamma_\rho \rangle
  \right] .
  \label{pm16}
\end{equation}
We have now to parametrise $\langle\gamma^\mu\rangle$, which is a vector
quantity containing information on the quark dynamics. At leading twist, \ie,
considering contributions $\Ord(P^+)$ in the infinite momentum frame, the only
vector at our disposal is $p^\mu\simeq{P}^\mu$ (recall that $n^\mu=\Ord(1/P^+)$
and $k^\mu\simeq{x}P^\mu$). Thus we can write
\begin{eqnarray}
  \langle \gamma^\mu \rangle
  &\equiv&
  \int \! \frac{\d^4k}{(2\pi)^4}\, \delta \! \left(x - \frac{k^+}{P^+}\right)
  \Tr(\gamma^\mu \, \Phi)
  \nonumber
\\
  &=&
  \int \! \frac{\d\tau}{2\pi} \, \e^{\I \tau x} \,
  \langle PS | \anti\psi(0) \, \gamma^\mu \, \psi(\tau n) | PS \rangle
  =
  2 f(x) \, P^\mu \, ,
  \label{pm17}
\end{eqnarray}
where the coefficient of $P^\mu$, which we called $f(x)$, is the \emph{quark
number density}, as will become clear later on (see Secs.~\ref{leading} and
\ref{probab}). From (\ref{pm17}) we obtain the following expression for $f(x)$
\begin{equation}
  f(x) =
  \int \! \frac{\d\xi^-}{4\pi} \, \e^{\I xP^+ \xi^-} \,
  \langle PS | \anti\psi(0) \gamma^+ \psi(0, \xi^-, \Vec{0}_\perp)| PS \rangle
  \, .
  \label{pm17b}
\end{equation}
Inserting (\ref{pm17}) into (\ref{pm16}) yields
\begin{equation}
  W_{\mu\nu}^\text{(S)} =
  \sum_a e_a^2 \, ( n_\mu P_\nu  + n_\nu P_\mu - g_{\mu\nu} ) \, f_a(x) \, .
  \label{pm18}
\end{equation}
The structure functions $F_1$ and $F_2$ can be extracted from $W_{\mu\nu}$ by
means of two projectors (terms of relative order $1/Q^2$ are neglected)
\begin{subequations}
\begin{eqnarray}
  F_1
  &=&
  \mathcal{P}_1^{\mu\nu} W_{\mu\nu} =
  \frac1{4} \left( \frac{4x^2}{Q^2} \, P^\mu P^\nu - g^{\mu\nu} \right)
  W_{\mu\nu} \, ,
  \label{pm19}
\\
  F_2
  &=&
  \mathcal{P}_2^{\mu\nu} W_{\mu\nu} =
  \frac{x}{2} \left( \frac{12x^2}{Q^2} \, P^\mu P^\nu - g^{\mu\nu} \right)
  W_{\mu\nu} \, .
  \label{pm20}
\end{eqnarray}
\end{subequations}
Since $(P^{\mu}P^\nu/Q^2)\,W_{\mu\nu}=\Ord(M^2/Q^2)$, we find that $F_1$ and $F_2$
are proportional to each other (the so-called Callan--Gross relation) and are
given by
\begin{equation}
  F_2(x) = 2 x \, F_1(x)  =
  - \frac{x}{2} \, g^{\mu\nu} W_{\mu\nu}^\text{(S)} =
  \sum_a e_a^2 \, x \, f_a(x) \, ,
  \label{pm21}
\end{equation}
which is the well-known parton model expression for the unpolarised structure
functions, restricted to quarks. To obtain the full expressions for $F_1$ and
$F_2$, one must simply add to (\ref{pm20}) the antiquark distributions
$\anti{f}_a$, which were left aside in the above discussion. They read (the
r\^{o}le of $\psi$ and $\anti\psi$ is interchanged with respect to the quark
distributions: see Sec.~\ref{leading} for a detailed discussion)
\begin{equation}
  \anti{f}(x)
  =
  \int \! \frac{\d\xi^-}{4\pi} \, \e^{\I xP^+\xi^-} \,
  \langle PS |
  \Tr \left[ \gamma^+ \psi(0) \anti\psi(0, \xi^-, \Vec{0}_\perp) \right]
  | PS \rangle \, .
  \label{pm22}
\end{equation}
and the structure functions $F_1$ and $F_2$ are
\begin{equation}
  F_2(x)
  = 2 x \, F_1(x)
  = \sum_a e_a^2 \, x \left[ f_a(x) + \anti{f}_a(x) \right] .
  \label{pm23}
\end{equation}

\subsection{Polarised DIS in the parton model}
\label{poldis}

Let us turn now to polarised \DIS. The parton-model expression of the
antisymmetric part of the hadronic tensor is
\begin{equation}
  W_{\mu\nu}^\text{(A)}
  =
  \frac1{2P{\cdot}q} \, \sum_a e_a^2 \int \! \frac{\d^4k}{(2\pi)^4} \,
  \delta \! \left(x - \frac{k^+}{P^+}\right)
  \varepsilon_{\mu\nu\rho\sigma} (k + q)^\rho \,
  \Tr(\gamma^\sigma \gamma_5 \Phi) \, .
  \label{pm25}
\end{equation}
With $k^\mu=xP^\mu$ this becomes, using the notation (\ref{pm15})
\begin{equation}
  W_{\mu\nu}^\text{(A)} =
  \varepsilon_{\mu\nu\rho\sigma} \, n^\rho \, \sum_a \frac{e^2_a}{2} \,
  \langle \gamma^\sigma \gamma_5 \rangle \, .
  \label{pm26}
\end{equation}
At leading twist the only pseudovector at hand is $\SL^\mu$ (recall that
$\SL^\mu=\Ord(P^+)$ and $\ST^\mu=\Ord(1)$) and
$\langle\gamma^\sigma\gamma_5\rangle$ is parametrised as (a factor $M$ is
inserted for dimensional reasons)
\begin{equation}
  \langle \gamma^\sigma \gamma_5 \rangle =
  2M \, \DL{f}(x) \, \SL^\sigma =
  2 \lambda_N \, \DL{f}(x) \, P^\sigma \, .
  \label{pm27}
\end{equation}
Here $\DL{f}(x)$, given explicitly by
\begin{equation}
  \DL{f}(x) =
  \int \! \frac{\d\xi^-}{4\pi} \, \e^{\I xP^+ \xi^-} \,
  \langle PS |
    \anti\psi(0) \gamma^+ \gamma_5 \psi(0, \xi^-, \Vec{0}_\perp)
  | PS \rangle \, ,
  \label{pm28}
\end{equation}
is the longitudinal polarisation (\ie, helicity) distribution of quarks. In
fact, inserting (\ref{pm27}) in (\ref{pm26}), we find
\begin{equation}
  W_{\mu\nu}^\text{(A)} =
  2 \lambda_N \, \varepsilon_{\mu\nu\rho\sigma} \,
  n^\rho p^\sigma \, \sum_a \frac{e^2_a}{2} \, \DL{f}_a(x) \, .
  \label{pm29}
\end{equation}
Comparing with the longitudinal part of the hadronic tensor (\ref{dis27c}),
which can be rewritten as
\begin{equation}
  W^\text{(A)}_{\mu\nu, \; \text{long}} =
  2 \lambda_N \,
  \varepsilon_{\mu\nu\rho\sigma} \, n^\rho p^\sigma \, g_1 \, ,
  \label{pm30}
\end{equation}
we obtain the usual parton model expression for the polarised structure
function $g_1$
\begin{equation}
  g_1(x) = \half \, \sum_a e_a^2 \, \DL{f}_a(x) \, .
  \label{pm31}
\end{equation}
Again, antiquark distributions $\DL{f}_{\anti{q}}$ should be added to
(\ref{pm31}) to obtain the full parton model expression for $g_1$
\begin{equation}
  g_1(x) =
  \half \, \sum_a e_a^2 \left[ \DL{f}_a(x) + \DL\anti{f}_a(x) \right] .
  \label{pm31b}
\end{equation}
The important lesson we learned is that, at leading twist, only longitudinal
polarisation contributes to \DIS.

\subsection{Transversely polarised targets}
\label{transvtarget}

\begin{figure}
  \centering
  \begin{picture}(300,150)(0,0)
    \SetWidth{2}
    \ArrowLine(70,25)(120,40)
    \ArrowLine(180,40)(230,25)
    \SetWidth{0.5}
    \ArrowLine(120,60)(120,100)
    \ArrowLine(180,100)(180,60)
    \ArrowLine(120,100)(180,100)
    \Photon(90,120)(120,100){3}{4.5}
    \Photon(210,120)(180,100){3}{4.5}
    \Gluon(135,100)(135,68){-3}{5}
    \GOval(150,50)(20,40)(0){0.5}
    \DashLine(150,25)(150,97){4}
    \DashLine(150,103)(150,120){4}
  \end{picture}
  \caption{Higher-twist contribution to DIS involving quark--gluon correlation.}
  \label{nonhand}
\end{figure}
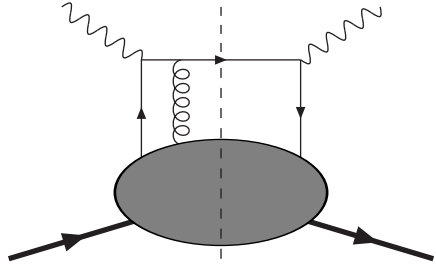

Since $\ST$ is suppressed by a power of $P^+$ with respect to its longitudinal
counterpart $\SL$, transverse polarisation effects in \DIS manifest themselves
at twist-three level. Including subdominant contributions, eq.~(\ref{pm27})
becomes
\begin{equation}
  \langle \gamma^\sigma \gamma_5 \rangle =
  2M \, \DL{f}(x) \, \SL\sigma +
  2M \, g_T(x) \, \ST^\sigma \, ,
  \label{pm32}
\end{equation}
where we have introduced a new, twist-three, distribution function $g_T$,
defined as (we take the nucleon spin to be directed along the $x$-axis)
\begin{eqnarray}
  g_T(x) &=& \frac{P^+}{2M} \, \langle \gamma^i \gamma_5 \rangle
  \nonumber
\\
  &=&
  \frac{P^+}{M} \int \! \frac{\d\xi^-}{4\pi} \, \e^{\I xP^+\xi^-} \,
  \langle PS |
    \anti\psi(0) \, \gamma^i \gamma_5 \, \psi(0, \xi^-, \Vec{0}_\perp)
  | PS \rangle \, ,
  \label{pm33}
\end{eqnarray}
As we are working at twist 3 (that is, with quantities suppressed by $1/P^+$)
we take into account the transverse components of the quark momentum,
$k^\mu\simeq{x}p^\mu+k_\perp^\mu$. Moreover, quark mass terms cannot be
ignored. Reinstating these terms in the hadronic tensor, we have
\begin{eqnarray}
  W_{\mu\nu}^\text{(A)}
  &=&
  \frac1{2P{\cdot}q} \, \sum_a e_a^2 \int \! \frac{\d^4k}{(2\pi)^4} \,
  \delta \! \left( x - \frac{k^+}{P^+} \right)
  \nonumber
\\
  && \null \times
  \varepsilon_{\mu\nu\rho\sigma}
  \left\{
    (k + q)^\rho \, \Tr\left[ \gamma^\sigma \gamma_5 \Phi \right] -
    \frac12 \, m_q \,
    \Tr\left[ \I \, \sigma^{\rho\sigma} \, \gamma_5 \, \Phi \right]
  \right\} . \hspace{2em}
  \label{pm34}
\end{eqnarray}
Notice that now we cannot simply set $k^\rho+q^\rho\simeq{P{\cdot}q}\,n^\mu$, as we
did in the case of longitudinal polarisation. Let us rewrite eq.~(\ref{pm34})
as
\begin{equation}
  W_{\mu\nu}^\text{(A)} =
  \frac1{2P{\cdot}q} \, \varepsilon_{\mu\nu\rho\sigma} \, q^\rho \,
  \sum_a e_a^2 \, \langle \gamma^\sigma \gamma_5 \rangle +
  \Delta W_{\mu\nu}^\text{(A)} \, ,
  \label{pm35}
\end{equation}
where
\begin{equation}
  \Delta W_{\mu\nu}^\text{(A)} =
  \frac1{2P{\cdot}q} \,
  \varepsilon_{\mu\nu\rho\sigma} \, \sum_a e_a^2
  \left(
    \langle \I \gamma^\sigma \gamma_5 \partial^\rho \rangle -
    \frac12 m_q \, \langle \I \sigma^{\rho\sigma} \gamma_5 \rangle
  \right) .
  \label{pm36}
\end{equation}
If we could neglect the term $\Delta{W}_{\mu\nu}^\text{(A)}$ then, for a
transversely polarised target, we should have, using eq.~(\ref{pm32})
\begin{equation}
  W_{\mu\nu}^\text{(A)}
  = \frac{2M \varepsilon_{\mu\nu\rho\sigma} q^\rho \, \ST^\sigma}{P{\cdot}q} \,
  \sum_a \frac{e_a^2}{2} \, g_T^a(x) \, .
  \label{pm37}
\end{equation}
Comparing with eq.~(\ref{dis27c}) yields the parton-model expression for the
polarised structure function combination $g_1+g_2$:
\begin{equation}
  g_1(x) + g_2(x) = \half \, \sum_a e_a^2 \, g_T^a(x) \, ,
  \label{pm38}
\end{equation}
This result has been obtained by ignoring the term
${\Delta}W_{\mu\nu}^\text{(A)}$ in the hadronic tensor, rather a strong
assumption, which seems lacking in justification. Surprisingly enough, however,
eq.~(\ref{pm36}) turns out to be correct. The reason is that at twist 3 one has
to add an extra term $W_{\mu\nu}^\text{(A)g}$ into (\ref{pm35}), arising from
non-handbag diagrams with gluon exchange (see Fig.~\ref{nonhand}) and which
exactly cancel out ${\Delta}W_{\mu\nu}^\text{(A)}$. Referring the reader to the
original papers \cite{Efremov:1983eb} (see also \cite{Ratcliffe:1985mp}) for a
detailed proof, we limit ourselves to presenting the main steps.

For the sum ${\Delta}W_{\mu\nu}^\text{(A)}+W_{\mu\nu}^{\text{(A)}g}$ one
obtains
\begin{eqnarray}
  \Delta W_{\mu\nu}^\text{(A)} + W_{\mu\nu}^{\text{(A)}g}
  &=&
  \frac1{4P{\cdot}q} \, \sum_a e_a^2
  \left\{ \varepsilon_{\mu\nu\rho\sigma}
    \left(
      \langle \I \gamma^\sigma \gamma_5 D^\rho(\tau n) \rangle -
      \frac12 m_q \, \langle \I \sigma^{\rho\sigma} \gamma_5 \rangle
    \right)
  \right.
  \nonumber
\\
  && \null
  \left. \Strut -
    \langle \gamma_\mu D_\nu(\tau n) - \gamma_\nu D_\mu(\tau n) \rangle +
    \dots
  \right\} ,
  \label{pm39}
\end{eqnarray}
where $D_\mu=\partial_\mu-{\I}g\,A_\mu$ and the ellipsis denotes terms with the
covariant derivative acting to the left and the gluon field evaluated at the
space--time point 0. We now resort to the identity
\begin{equation}
  \half \, \varepsilon_{\mu\nu\tau\sigma}
  \sigma^{\tau\sigma} \gamma_5 \gamma_\rho =
  g_{\rho\nu} \gamma_\mu -
  g_{\mu\rho} \gamma_\nu -
  \I \varepsilon_{\mu\nu\rho\sigma} \gamma^\sigma \gamma_5 \, .
  \label{pm40}
\end{equation}
Contracting with $D^\rho$ and using the equations of motion
$(\I\slashed{D}-m_q)\psi=0$ we ultimately obtain
\begin{equation}
  \half \, m_q \, \varepsilon_{\mu\nu\tau\sigma}
  \langle \sigma^{\tau\sigma} \gamma_5 \rangle =
  \langle \gamma_\nu D_\mu \rangle -
  \langle \gamma_\mu D_\nu \rangle +
  \varepsilon_{\mu\nu\rho\sigma}
  \langle \I \gamma^\sigma \gamma_5 D^\rho \rangle \, ,
  \label{pm41}
\end{equation}
which implies the vanishing of (\ref{pm39}). Concluding, \DIS with transversely
polarised nucleon (where transverse refers to the photon axis) probes a
twist-three distribution function, $g_T(x)$, which, as we shall see, has no
probabilistic meaning and is \emph{not} related in a simple manner to
transverse quark polarisation.

\subsection{Transverse polarisation distributions of quarks in DIS}
\label{transvpolqrk}

Let us focus now on the quark mass term appearing in the antisymmetric hadronic
tensor -- see eqs.~(\ref{pm36}) and (\ref{pm39}). We have shown that actually
it cancels out and does not contribute to \DIS. Its structure, however, is
quite interesting. It contains, in fact, the transverse polarisation
distribution of quarks, $\DT{f}$, which is the main subject of this report. The
decoupling of the quark mass term thus entails the absence of $\DT{f}$ from
\DIS, even at higher-twist level.

The matrix element $\langle\I\sigma^{\rho\sigma}\gamma_5\rangle$ admits a
unique leading-twist parametrisation in terms of a tensor structure containing
the transverse spin vector of the target $\ST^\mu$ and the dominant Sudakov
vector $p^\mu$
\begin{equation}
  \langle \I \sigma^{\rho\sigma} \gamma_5 \rangle
  =
  2 (p^\sigma \ST^\rho - p^\rho \ST^\sigma) \, \DT{f}(x) \, .
  \label{pm41b}
\end{equation}
The coefficient $\DT{f}(x)$ is indeed the transverse polarisation distribution.
It can be singled out by contracting (\ref{pm41b}) with $n_\rho$, which gives
(for definiteness, we take the spin vector directed along $x$)
\begin{eqnarray}
  \DT{f}(x)
  &=&
  \half \, \langle \I n_\rho \sigma^{1\rho} \gamma_5 \rangle
  \nonumber
\\
  &=& \int \! \frac{\d\xi^-}{4\pi} \, \e^{\I xP^+ \xi^-} \,
  \langle PS |
    \anti\psi(0) \I \sigma^{1+} \gamma_5 \psi(0, \xi^-, \Vec{0}_\perp)
  | PS \rangle \, ,
  \label{pm41c}
\end{eqnarray}
Eq.~(\ref{pm41}) can be put in the form of a constraint between $\DT{f}(x)$ and
other twist-three distributions embodied in $\langle\gamma_{\mu}D_\nu\rangle$
and $\langle\I\gamma^\sigma\gamma_5D^\rho\rangle$. Let us consider the partonic
content of the last two quantities. The gluonic (non-handbag) contribution
$W_{\mu\nu}^{\text{(A)}g}$ to the hadronic tensor involves traces of a
quark--gluon--quark correlation matrix. We introduce the following two
quantities:
\begin{subequations}
\begin{eqnarray}
  \langle \! \langle \gamma^\mu D^\nu (\tau_2 n) \rangle \! \rangle
  &\equiv&
  \int \! \frac{\d\tau_1}{2\pi} \int \! \frac{\d\tau_2}{2\pi} \,
  \e^{\I \tau_1 x_2} \, \e^{\I \tau_2 (x_1 - x_2)}
  \nonumber
\\
  && \hspace{2.5em} \null \times
  \langle PS |
    \anti\psi(0) \gamma^\mu D^\nu(\tau_2 n) \, \psi(\tau_1 n)
  | PS \rangle \, ,
  \label{pm42}
\\[1ex]
  \langle \! \langle \I \gamma^\mu \gamma_5 D^\nu (\tau_2 n) \rangle \! \rangle
  &\equiv&
  \int \! \frac{\d\tau_1}{2\pi} \int \! \frac{\d\tau_2}{2\pi}
  \e^{\I \tau_1 x_2} \, \e^{\I \tau_2(x_1 - x_2)}
  \nonumber
\\
  && \hspace{2.5em} \null \times
  \langle PS |
   \anti\psi(0) \I \gamma^\mu \gamma_5 D^\nu(\tau_2 n) \, \psi(\tau_1 n)
  | PS \rangle \, . \qquad
  \label{pm43}
\end{eqnarray}
\end{subequations}
These matrix elements are related to those appearing in (\ref{pm39}) by
\begin{subequations}
\begin{eqnarray}
  \int \! \d{x}_2 \,
  \langle \! \langle \gamma^\mu D^\nu(\tau_2 n) \rangle \! \rangle
  &=&
  \langle \gamma^\mu D^\nu(\tau_2 n) \rangle \, ,
  \label{pm44}
\\
  \int \! \d{x}_2 \,
  \langle \! \langle \I \gamma^\mu \gamma_5 D^\nu(\tau_1 n) \rangle \! \rangle
  &=&
  \langle \I \gamma^\mu \gamma_5 D^\nu(\tau_1 n) \rangle \, .
  \label{pm45}
\end{eqnarray}
\end{subequations}
At leading order (which for the quark--gluon--quark correlation functions
implies twist 3) the possible Lorentzian structures of
$\langle\!\langle\gamma^{\mu}D^\nu\rangle\!\rangle$ and
$\langle\!\langle\I\gamma^\mu\gamma_5D^\nu\rangle\!\rangle$ are
\begin{subequations}
\begin{eqnarray}
  \langle \! \langle \gamma^\mu D^\nu \rangle \! \rangle
  &=&
  2M \, G_D(x_1, x_2) \,
  p^\mu \, \varepsilon^{\nu\alpha\beta\rho} \,
  p_\alpha \, n_\beta \, S_{\perp\rho} \, ,
  \label{pm46}
\\
  \langle \! \langle \I \gamma^\mu \gamma_5 D^\nu \rangle \! \rangle
  &=&
  2M \, \widetilde{G}_D(x_1, x_2)  \, p^\mu \ST^\nu +
  2M \, \widetilde{G}'_D(x_1, x_2) \, p^\nu \ST^\mu \, .
  \label{pm47}
\end{eqnarray}
\end{subequations}
Here three multiparton distributions, $G_D(x_1,x_2)$,
$\widetilde{G}_D(x_1,x_2)$ and $\widetilde{G}'_D(x_1,x_2)$, have been
introduced. One of them, $\widetilde{G}'(x_1,x_2)$, is only apparently a new
quantity. Contracting eq.~(\ref{pm47}) with $n_\nu$ and exploiting the gauge
choice $A^+=0$, it is not difficult to derive a simple connection between
$\widetilde{G}'(x_1,x_2)$ and the twist-three distribution function $g_T(x_2)$
\cite{Efremov:1983eb}
\begin{equation}
  \widetilde{G}'_D(x_1, x_2) = x_2 \, \delta(x_1 - x_2) \, g_T(x_2) \, ,
  \label{pm48}
\end{equation}
Hence $\widetilde{G}'_D(x_1,x_2)$ can be eliminated in favour of the more
familiar $g_T(x_2)$. We are now in the position to translate eq.~(\ref{pm41})
into a relation between quark and multiparton distribution functions. Using
(\ref{pm41b}) and (\ref{pm44}--\ref{pm48}) in (\ref{pm41}) we find
\begin{equation}
  \int \! \d{y} \left[ G_D(x,y) + \widetilde{G}_D(x,y) \right] =
  x \, g_T (x) - \frac{m_q}{M} \, \DT{f}(x) \, .
  \label{pm49}
\end{equation}
By virtue of this constraint, the transverse polarisation distributions of
quarks, that one could na\"{\i}vely expect to be probed by \DIS at a subleading
level, turn out to be completely absent from this process.

\section{Systematics of quark distribution functions}
\label{syst}

In this section we present in detail the systematics of quark and antiquark
distribution functions. Our focus will be on leading-twist distributions. For
the sake of completeness, however, we shall also sketch some information on the
higher-twist distributions.

\subsection{The quark--quark correlation matrix}
\label{qqcorr}

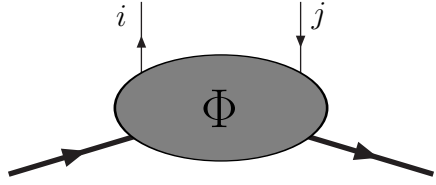
\begin{figure}
  \centering
  \begin{picture}(300,100)(0,0)
    \SetWidth{2}
    \ArrowLine(70,15)(120,30)
    \ArrowLine(180,30)(230,15)
    \SetWidth{0.5}
    \ArrowLine(120,50)(120,80)
    \ArrowLine(180,80)(180,50)
    \Text(115,75)[r]{$i$}
    \Text(185,75)[l]{$j$}
    \GOval(150,40)(20,40)(0){0.5}
    \Text(150,40)[]{\LARGE $\Phi$}
  \end{picture}
  \caption{The quark--quark correlation matrix $\Phi$.}
  \label{phi}
\end{figure}

Let us consider the quark--quark correlation matrix introduced in
Sec.~\ref{parton} and represented in Fig.~\ref{phi},
\begin{equation}
  \Phi_{ij}(k, P, S) = \int \! \d^4\xi \, \e^{\I k{\cdot}\xi} \,
  \langle PS | \anti\psi_j(0) \psi_i(\xi) | PS \rangle \, .
  \label{df1}
\end{equation}
Here, we recall, $i$ and $j$ are Dirac indices and a summation over colour is
implicit. The quark distribution functions are essentially integrals over $k$
of traces of the form
\begin{equation}
  \Tr(\Gamma\Phi) = \int \! \d^4\xi \, \e^{\I k{\cdot}\xi} \,
  \langle PS | \anti\psi(0) \, \Gamma \, \psi(\xi) | PS \rangle \, ,
  \label{df1bis}
\end{equation}
where $\Gamma$ is a Dirac matrix structure.

In Sec.~\ref{parton}, $\Phi$ was defined within the na\"{\i}ve parton model. In
\QCD, in order to make $\Phi$ gauge invariant, a path-dependent link operator
\begin{equation}
  \mathcal{L} (0,\xi) =
  \mathcal{P} \, \exp \left( - \I g \int_0^\xi \d{s}_\mu \, A^\mu(s) \right),
\end{equation}
where $\mathcal{P}$ denotes path-ordering, must be inserted between the quark
fields. It turns out that the distribution functions involve separations $\xi$
of the form $[0,\xi^-,\Vec{0}_\perp]$, or $[0,\xi^-,\Vec\xi_\perp]$. Thus, by
working in the axial gauge $A^+=0$ and choosing an appropriate path,
$\mathcal{L}$ can be reduced to unity. Hereafter we shall simply assume that
the link operator is unity, and just omit it.

The $\Phi$ matrix satisfies certain relations arising from hermiticity, parity
invariance and time-reversal invariance \cite{Mulders:1996dh}:
\begin{subequations}
\begin{align}
  \Phi^\dagger(k, P, S) &=
  \gamma^0 \, \Phi(k, P, S) \, \gamma^0 &
  & \text{(hermiticity),}
  \label{df1b}
\\
  \Phi(k, P, S) &=
  \gamma^0 \, \Phi(\widetilde{k}, \widetilde{P}, -\widetilde{S}) \, \gamma^0 &
  & \text{(parity),}
  \label{df1c}
\\
  \Phi^{*}(k, P, S) &=
  \gamma_5 C \, \Phi(\widetilde{k}, \widetilde{P}, \widetilde{S}) \, C^\dagger \gamma_5 &
  & \text{(time-reversal),}
  \label{df1d}
  \sublabel{subdf1d}
\end{align}
\end{subequations}
where $C=\I\gamma^2\gamma^0$ and the tilde four-vectors are defined as
$\widetilde{k}^\mu=(k^0,-\Vec{k})$. As we shall see, the time-reversal
condition (\ref{df1d}) plays an important r\^{o}le in the phenomenology of
transverse polarisation distributions. It is derived in a straight-forward
manner by using $T\,\psi(\xi)\,T^\dagger=-\I\gamma_5C\,\psi(-\widetilde\xi)$
and $T\,|PS\rangle=(-1)^{S-S_z}\,|\widetilde{P}\widetilde{S}\rangle$, where $T$
is the time-reversal operator. The crucial point, to be kept in mind, is the
transformation of the nucleon state, which is a free particle state. Under $T$,
this goes into the same state with reversed $\Vec{P}$ and $\Vec{S}$.

The most general decomposition of $\Phi$ in a basis of Dirac matrices,
\begin{equation}
  \Gamma =
  \left\{
  \one, \;
  \gamma^\mu, \;
  \gamma^\mu \gamma_5, \;
  \I \gamma_5, \;
  \I \sigma^{\mu\nu} \gamma_5
  \right\} ,
  \label{df1e}
\end{equation}
is (we introduce a factor $\half$ for later convenience)
\begin{equation}
  \Phi (k, P, S) =
  \half
  \left\{
    \mathcal{S} \, \one +
    \mathcal{V}_\mu \, \gamma^\mu +
    \mathcal{A}_\mu \gamma_5 \gamma^\mu +
    \I \mathcal{P}_5 \gamma_5 +
    \half \I \, \mathcal{T}_{\mu\nu}  \, \sigma^{\mu\nu} \gamma_5
  \right\} .
  \label{df2}
\end{equation}
The quantities $\mathcal{S}$, $\mathcal{V}^\mu$, $\mathcal{A}^\mu$,
$\mathcal{P}_5$ and $\mathcal{T}^{\mu\nu}$ are constructed with the vectors
$k^\mu$, $P^\mu$ and the pseudovector $S^\mu$. Imposing the constraints
(\ref{df1b}--\ref{subdf1d}) we have, in general,
\begin{subequations}
\begin{eqnarray}
  \mathcal{S} &=&
  \textfrac12  \, \Tr(\Phi)
  = C_1 \, ,
  \label{df2.1}
\\
  \mathcal{V}^\mu &=&
  \textfrac12 \, \Tr(\gamma^\mu \Phi)
  = C_2 \, P^\mu
  + C_3 \, k^\mu \, ,
  \label{df2.2}
\\
  \mathcal{A}^\mu &=&
  \textfrac12 \, \Tr(\gamma^\mu \gamma_5 \Phi)
  = C_4 \, S^\mu
  + C_5 \, k{\cdot}S \, P^\mu
  + C_6 \, k{\cdot}S \, k^\mu \, ,
  \label{df2.3}
  \sublabel{subdf2.3}
\\
  \mathcal{P}_5 &=&
  \textfrac1{2\I} \, \Tr(\gamma_5 \Phi)
  = 0 \, ,
  \label{df2.4}
\\
  \mathcal{T}^{\mu\nu} &=&
  \textfrac1{2\I} \, \Tr(\sigma^{\mu\nu} \gamma_5 \Phi)
  = C_7 \, P^{[\mu} S^{\nu]}
  + C_8 \, k^{[\mu} S^{\nu]}
  + C_9 \, k{\cdot}S \, P^{[\mu} k^{\nu]} \, , \qquad
  \label{df2.5}
  \sublabel{subdf2.5}
\end{eqnarray}
\end{subequations}
where the coefficients $C_i=C_i(k^2,k{\cdot}P)$ are real functions, owing to
hermiticity.

If we relax the constraint (\ref{df1d}) of time-reversal invariance (for the
physical relevance of this, see Sec.~\ref{todd} below), three more terms
appear:
\begin{align}
  \mathcal{V}^\mu &= \dots +
  C_{10} \, \varepsilon^{\mu\nu\rho\sigma} S_\nu P_\rho k_\sigma \, ,
  \tag{\ref{df2.2}$'$}
  \label{df2.6}
\\
  \mathcal{P}_5 &=
  C_{11} \, k{\cdot}S \, ,
  \tag{\ref{df2.4}$'$}
  \label{df2.7}
\\
  \mathcal{T}^{\mu\nu} &=  \dots +
  C_{12} \, \varepsilon^{\mu\nu\rho\sigma} P_\rho k_\sigma \, .
  \tag{\ref{df2.5}$'$}
  \label{df2.8}
\end{align}

\subsection{Leading-twist distribution functions}
\label{leading}

We are mainly interested in the leading-twist contributions, that is the terms
in eqs.~(\ref{df2.1}--\ref{subdf2.5}) that are of order $\Ord(P^+)$ in the
infinite momentum frame.

The vectors at our disposal are $P^\mu$, $k^\mu\simeq{}xP^\mu$, and
$S^\mu\simeq\lambda_NP^\mu/M+\ST^\mu$, where the approximate equality signs
indicate that we are neglecting terms suppressed by $(P^+)^{-2}$. Remember that
the transverse spin vector $\ST^\mu$ is of order $(P^+)^0$. For the time being
we ignore quark transverse momentum $k_\perp^\mu$ (which in \DIS is integrated
over). We shall see later on how the situation becomes more complicated when
$k_\perp^\mu$ enters the game.

At leading order in $P^+$ only the vector, axial, and tensor terms in
(\ref{df2}) survive and eqs.~(\ref{df2.2}, \ref{subdf2.3}, \ref{subdf2.5})
become
\begin{subequations}
\begin{eqnarray}
  \mathcal{V}^\mu &=&
  \textfrac12 \int \! \d^4\xi \, \e^{\I k{\cdot}\xi} \,
  \langle PS | \anti\psi(0) \gamma^\mu \psi(\xi) | PS \rangle =
  A_1 \, P^\mu \, ,
  \label{df5}
\\
  \mathcal{A}^\mu &=&
  \textfrac12 \int \! \d^4\xi \, \e^{\I k{\cdot}\xi} \,
  \langle PS | \anti\psi(0) \gamma^\mu \gamma_5 \psi(\xi) | PS \rangle =
  \lambda_N \,
  A_2 \, P^\mu \, ,
  \label{df6}
\\
  \mathcal{T}^{\mu\nu} &=&
  \textfrac1{2\I} \int \! \d^4 \xi \, \e^{\I k{\cdot}\xi} \,
  \langle PS | \anti\psi(0) \sigma^{\mu\nu} \gamma_5 \psi(\xi) | PS \rangle =
  A_3 \, P_{\vphantom\perp}^{[\mu} \ST^{\nu]} \, ,
  \label{df7}
  \sublabel{subdf7}
\end{eqnarray}
\end{subequations}
where we have introduced new real functions $A_i(k^2,k{\cdot}P)$. The leading-twist
quark correlation matrix (\ref{df2}) is then (we use
$P_{\vphantom\perp}^{[\mu}\ST^{\nu]}\sigma_{\mu\nu}=
2\I\,\slashed{P}\,\slashed{S}_\perp$)
\begin{equation}
  \Phi(k, P, S) =
  \half
  \left\{
    A_1 \, \slashed{P} +
    A_2 \, \lambda_N \, \gamma_5  \, \slashed{P} +
    A_3 \, \slashed{P} \, \gamma_5 \, \slashed{S}_\perp
  \right\} .
  \label{df9}
\end{equation}
From (\ref{df5}--\ref{subdf7}) we obtain
\begin{subequations}
\begin{eqnarray}
  A_1
  &=& \frac1{2P^+} \, \Tr(\gamma^+ \Phi) \, ,
  \label{df13}
\\
  \lambda_N \, A_2
  &=& \frac1{2P^+} \, \Tr(\gamma^+ \gamma_5 \Phi) \, ,
  \label{df14}
\\
  \ST^i \, A_3
  &=& \frac1{2P^+} \, \Tr(\I \sigma^{i+} \gamma_5 \Phi)
  = \frac1{2P^+} \, \Tr(\gamma^+ \gamma^i \gamma_5 \Phi) \, .
  \label{df15}
  \sublabel{subdf15}
\end{eqnarray}
\end{subequations}
The leading-twist distribution functions $f(x)$, $\DL{f}(x)$ and $\DT{f}(x)$
are obtained by integrating $A_1$, $A_2$ and $A_3$, respectively, over $k$,
with the constraint $x=k^+/P^+$, that is,
\begin{equation}
  \left\{
  \begin{array}{c}
    f(x) \\ \DL{f}(x) \\ \DT{f}(x)
  \end{array}
  \right\}
  = \int \! \frac{\d^4k}{(2\pi)^4}
  \left\{
  \begin{array}{c}
    A_1(k^2, k{\cdot}P) \\ A_2(k^2, k{\cdot}P) \\ A_3(k^2, k{\cdot}P)
  \end{array}
  \right\} \,
  \delta \! \left(x - \frac{k^+}{P^+} \right) ,
  \label{df15b}
\end{equation}
that is, using (\ref{df13}--\ref{subdf15}) and setting for definiteness
$\lambda_N=+1$ and $\ST^i=(1,0)$
\begin{subequations}
\begin{eqnarray}
  f(x) &=&
  \frac12 \int \! \frac{\d^4k}{(2\pi)^4} \,
  \Tr(\gamma^+ \Phi) \, \delta(k^+ - xP^+) \, ,
  \label{df16}
\\
  \DL{f}(x) &=&
  \frac12 \int \! \frac{\d^4k}{(2\pi)^4} \,
  \Tr(\gamma^+ \gamma_5 \Phi) \, \delta(k^+ -xP^+) \, ,
  \label{df17}
\\
  \DT{f}(x) &=&
  \frac12 \int \! \frac{\d^4k}{(2\pi)^4} \,
  \Tr(\gamma^+ \gamma^1 \gamma_5 \Phi) \, \delta(k^+ - xP^+) \, .
  \label{df18}
  \sublabel{subdf18}
\end{eqnarray}
\end{subequations}
Finally, inserting the definition (\ref{df1}) of $\Phi$ in
(\ref{df16}--\ref{subdf18}), we obtain the three leading-twist distribution
functions as light-cone Fourier transforms of expectation values of quark-field
bilinears \cite{Collins:1982uw}:
\begin{subequations}
\begin{eqnarray}
  f(x) &=&
  \int \! \frac{\d\xi^-}{4\pi} \e^{\I xP^+ \xi^-}
  \langle PS |
    \anti\psi(0) \gamma^+ \psi(0, \xi^-, \Vec{0}_\perp)
  | PS \rangle \, ,
  \label{df19}
\\
  \DL{f}(x) &=&
  \int \! \frac{\d\xi^-}{4\pi} \e^{\I xP^+ \xi^-}
  \langle PS |
    \anti\psi(0) \gamma^+ \gamma_5 \psi(0, \xi^-, \Vec{0}_\perp)
  | PS \rangle \, ,
  \label{df20}
\\
  \DT{f}(x) &=&
  \int \! \frac{\d\xi^-}{4\pi} \e^{\I xP^+ \xi^-}
  \langle PS |
    \anti\psi(0) \gamma^+ \gamma^1 \gamma_5 \psi(0, \xi^-, \Vec{0}_\perp)
  | PS \rangle \, . \hspace{2em}
  \label{df21}
  \sublabel{subdf21}
\end{eqnarray}
\end{subequations}

The quark--quark correlation matrix $\Phi$ integrated over $k$ with the
constraint $x=k^+/P^+$
\begin{eqnarray}
  \Phi_{ij}(x)
  &=&
  \int \! \frac{\d^4k}{(2\pi)^4} \, \Phi_{ij}(k, P, S) \, \delta(x - k^+/P^+)
  \nonumber
\\
  &=&
  \int \! \frac{\d\tau}{2\pi} \, \e^{\I \tau x} \,
  \langle PS | \anti\psi_j(0) \psi_i(\tau n) | PS \rangle \, ,
  \label{df22}
\end{eqnarray}
in terms of the three leading-twist distribution functions, reads
\begin{equation}
  \Phi(x) = \half
  \left\{
    f(x) \, \slashed{P} +
    \lambda_N \, \DL{f}(x) \, \gamma_5  \, \slashed{P} +
    \DT{f}(x) \, \slashed{P} \, \gamma_5 \, \slashed{S}_\perp
  \right\} .
  \label{df23}
\end{equation}

Let us now complete the discussion introducing the antiquarks. Their
distribution functions are obtained from the correlation matrix
\begin{equation}
  \overline\Phi_{ij}(k,P,S)
  = \int \! \d^4\xi \, \e^{\I k{\cdot}\xi} \,
  \langle PS | \psi_i(0) \anti\psi_j(\xi) | PS \rangle \, .
  \label{adf1}
\end{equation}
Tracing $\overline\Phi$ with the Dirac matrices $\Gamma$ gives
\begin{equation}
  \Tr(\Gamma \Phi)
  = \int \! \d^4 \xi \, \e^{\I k{\cdot}\xi} \,
  \langle PS |
    \Tr \left[ \Gamma \, \psi(0) \anti\psi(\xi) \right]
  | PS \rangle \, ,
  \label{adf2}
\end{equation}
In deriving the expressions for $\anti{f}$, $\DL\anti{f}$, $\DT\anti{f}$ care
is needed with the signs. By charge conjugation, the field bilinears in
(\ref{df1b}) transform as
\begin{equation}
  \anti\psi(0) \, \Gamma \, \psi(\xi)
  \to
  \pm \, \Tr \left[ \Gamma \, \psi(0) \anti\psi(\xi) \right] ,
  \label{adf3}
\end{equation}
where the $+$ sign is for $\Gamma=\gamma^\mu$, $\I\sigma^{\mu\nu}\gamma_5$ and
the $-$ sign for $\Gamma=\gamma^\mu\gamma_5$. We thus obtain the antiquark
density number:
\begin{eqnarray}
  \anti{f}(x)
  &=&
  \frac12 \int \! \frac{\d^4k}{(2\pi)^4} \,
  \Tr(\gamma^+ \overline\Phi) \,
  \delta(k^+ - xP^+)
  \nonumber
\\
  &=&
  \int \! \frac{\d\xi^-}{4\pi} \e^{\I xP^+ \xi^-}
  \langle PS |
    \Tr \left[ \gamma^+ \psi(0) \anti\psi(0, \xi^-, \Vec{0}_\perp) \right]
  | PS \rangle \, ,
  \label{adf4}
\end{eqnarray}
the antiquark helicity distribution
\begin{eqnarray}
  \DL\anti{f}(x)
  &=& - \frac12 \int \! \frac{\d^4k}{(2\pi)^4} \,
  \Tr \left(\gamma^+ \gamma_5 \overline\Phi\right)
  \delta(k^+ - xP^+)
  \nonumber
\\
  &=& \int \! \frac{\d\xi^-}{4\pi} \e^{\I xP^+ \xi^-}
  \langle PS |
    \Tr \left[
          \gamma^+ \gamma_5 \psi(0) \anti\psi(0, \xi^-, \Vec{0}_\perp)
        \right]
  | PS \rangle \, , \hspace{2em}
  \label{adf5}
\end{eqnarray}
and the antiquark transversity distribution
\begin{eqnarray}
  \DT\anti{f}(x)
  &=& \frac12 \int \! \frac{\d^4k}{(2\pi)^4} \,
  \Tr \left( \gamma^+ \gamma^1 \gamma_5 \overline\Phi \right)
  \delta(k^+ - xP^+)
  \nonumber
\\
  &=& \int \! \frac{\d\xi^-}{4\pi} \e^{\I xP^+ \xi^-}
  \langle PS |
    \Tr \left[
          \gamma^+ \gamma^1 \gamma_5 \psi(0) \anti\psi(0, \xi^-, \Vec{0}_\perp)
        \right]
  | PS \rangle \, .  \hspace{4em}
  \label{adf6}
\end{eqnarray}
Note the minus sign in the definition of the antiquark helicity distribution.

If we adhere to the definitions of quark and antiquark distributions,
eqs.~(\ref{df19}--\ref{subdf21}) and (\ref{adf4}--\ref{adf6}), the variable
$x\equiv{}k^+/P^+$ is not \emph{a priori} constrained to be positive and to
range from 0 to 1 (we shall see in Sec.~\ref{probab} how the correct support
for $x$ comes out, hence justifying its identification with the Bjorken
variable). It turns out that there is a set of symmetry relations connecting
quark and antiquark distribution functions, which are obtained by continuing
$x$ to negative values. Using anticommutation relations for the fermion fields
in the connected matrix elements
\begin{equation}
   \langle PS | \anti\psi(\xi) \psi(0) | PS \rangle_c
  =
  -\langle PS | \psi(0) \anti\psi(\xi) | PS \rangle_c \, ,
\end{equation}
one easily obtains the following relations for the three distribution functions
\begin{subequations}
\begin{eqnarray}
  \anti{f}(x)    &=&           -      f (-x) \, ,
  \label{adf8}
\\
  \DL\anti{f}(x) &=& \hphantom{-} \DL{f}(-x) \, ,
  \label{adf9}
\\
  \DT\anti{f}(x) &=&           -  \DT{f}(-x) \, .
  \label{adf10}
  \sublabel{subadf10}
\end{eqnarray}
\end{subequations}
Therefore, antiquark distributions are given by the continuation of the
corresponding quark distributions into the negative $x$ region.

\subsection{Probabilistic interpretation of distribution functions}
\label{probab}

Distribution functions are essentially the probability densities for finding
partons with a given momentum fraction and a given polarisation inside a
hadron. We shall now see how this interpretation comes about from the
field-theoretical definitions of quark (and antiquark) distribution functions
presented above.

Let us first of all decompose the quark fields into ``good'' and ``bad''
components:
\begin{equation}
  \psi = \psi_{(+)} + \psi_{(-)} \, ,
  \label{good1}
\end{equation}
where
\begin{equation}
  \psi_{({\pm})} = \half \, \gamma^{\mp} \, \gamma^{\pm} \, \psi \, .
  \label{good2}
\end{equation}
The usefulness of this procedure lies in the fact that ``bad'' components are not
dynamically independent: using the equations of motion, they can be eliminated
in favour of ``good'' components and terms containing quark masses and gluon
fields. Since in the $P^+\to\infty$ limit $\psi_{(+)}$ dominates over
$\psi_{(-)}$, the presence of ``bad'' components in a parton distribution
function signals higher twists. Using the relations
\begin{subequations}
\begin{eqnarray}
  \anti\psi \, \gamma^+ \, \psi &=&
  \sqrt2 \, \psi_{(+)}^\dagger \, \psi_{(+)} \, ,
  \label{good6}
\\
  \anti\psi \, \gamma^+ \gamma_5 \, \psi &=&
  \sqrt2 \, \psi_{(+)}^\dagger \, \gamma_5 \, \psi_{(+)} \, ,
  \label{good7}
\\
  \anti\psi \, \I \sigma^{i+} \gamma_5 \, \psi &=&
  \sqrt2 \, \psi_{(+)}^\dagger \, \gamma^i \gamma_5 \, \psi_{(+)} \, .
  \label{good8}
\end{eqnarray}
\end{subequations}
the leading-twist distributions (\ref{df19}--\ref{subdf21}) can be re-expressed
as \cite{Collins:1982uw}
\begin{subequations}
\begin{eqnarray}
  f(x) &=&
  \int \! \frac{\d\xi^-}{2\sqrt2 \, \pi} \e^{\I xP^+ \xi^-}
  \langle PS |
    \psi_{(+)}^\dagger(0) \gamma^+ \psi_{(+)} (0, \xi^-, \Vec{0}_\perp)
  | PS \rangle \, ,
  \label{good9}
\\
  \DL{f}(x) &=&
  \int \! \frac{\d\xi^-}{2\sqrt2 \, \pi} \e^{\I xP^+ \xi^-}
  \langle PS |
    \psi_{(+)}^\dagger(0) \gamma_5 \psi_{(+)} (0, \xi^-, \Vec{0}_\perp)
  | PS \rangle \, ,
  \label{good10}
\\
  \DT{f}(x) &=&
  \int \! \frac{\d\xi^-}{2\sqrt2 \, \pi} \e^{\I xP^+ \xi^-}
  \langle PS |
    \psi_{(+)}^\dagger(0) \gamma^1 \gamma_5 \psi_{(+)} (0, \xi^-, \Vec{0}_\perp)
  | PS \rangle \, , \hspace{2em}
  \label{good11}
  \sublabel{subgood11}
\end{eqnarray}
\end{subequations}
Note that, as anticipated, only ``good'' components appear. It is the peculiar
structure of the quark-field bilinears in eqs.~(\ref{good9}--\ref{subgood11})
that allows us to put the distributions in a form that renders their
probabilistic nature transparent.

\begin{figure}
  \centering
  \begin{picture}(360,110)(0,0)
    \SetWidth{3}
    \ArrowLine(10,60)(35,60)
    \ArrowLine(145,60)(170,60)
    \SetWidth{0.5}
    \ArrowLine(60,51)(120,51)
    \ArrowLine(60,57)(120,57)
    \ArrowLine(60,63)(120,63)
    \ArrowLine(60,69)(120,69)
    \ArrowLine(50,75)(50,95)
    \ArrowLine(130,95)(130,75)
    \Text(90,80)[]{$\vert n \rangle$}
    \GCirc(50,60){15}{0.5}
    \GCirc(130,60){15}{0.5}
    \Text(90,20)[]{(a)}
    \SetWidth{3}
    \ArrowLine(190,60)(215,60)
    \ArrowLine(230,50)(340,25)
    \SetWidth{0.5}
    \ArrowLine(240,51)(300,51)
    \ArrowLine(240,57)(300,57)
    \ArrowLine(240,63)(300,63)
    \ArrowLine(240,69)(300,69)
    \ArrowLine(230,75)(230,95)
    \ArrowLine(310,95)(310,75)
    \GCirc(230,60){15}{0.5}
    \GCirc(310,60){15}{0.5}
    \Text(270,80)[]{$\vert n \rangle$}
    \Text(270,20)[]{(b)}
  \end{picture}
  \caption{(a) A connected matrix element with the insertion of a complete set
           of intermediate states and (b) a semi-connected matrix element.}
  \label{interm}
\end{figure}
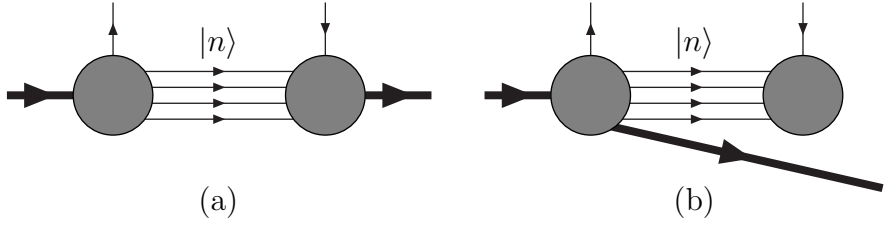

A remark on the support of the distribution functions is now in order. We
already mentioned that, according to the definitions of the quark
distributions, nothing constrains the ratio $x\equiv{}k^+/P^+$ to take on
values between 0 and 1. The correct support of the distributions emerges, along
with their probabilistic content, if one inserts into
(\ref{good9}--\ref{subgood11}) a complete set of intermediate states
$\{|n\rangle\}$ \cite{Jaffe:1983hp} (see Fig.~\ref{interm}). Considering, for
instance, the unpolarised distribution we obtain from (\ref{good9})
\begin{equation}
  f(x) =
  \textfrac1{\sqrt2} \, \sum_n \, \delta \! \left((1-x)P^+ - P_n^+\right)
  | \langle PS | \psi_{(+)}(0)| n \rangle|^2 \, ,
  \label{good12}
\end{equation}
where $\sum_n$ incorporates the integration over the phase space of the
intermediate states. Eq.~(\ref{good12}) clearly gives the probability of
finding inside the nucleon a quark with longitudinal momentum $k^+/P^+$,
irrespective of its polarisation. Since the states $|n\rangle$ are physical we
must have $P_n^+\ge0$, that is $E_n\ge|\Vec{P}_n|$, and therefore $x\le1$.
Moreover, if we exclude semi-connected diagrams like that in
Fig.~\ref{interm}b, which correspond to $x<0$, we are left with the connected
diagram of Fig.~\ref{interm}a and with the correct support $0\le{}x\le1$. A
similar reasoning applies to antiquarks.

Let us turn now to the polarised distributions. Using the Pauli--Lubanski
projectors $\mathcal{P}_{\pm}=\half\,(1\pm\gamma_5)$ (for helicity) and
$\mathcal{P}_{\uparrow\downarrow}=\half(1\pm\gamma^1\gamma_5)$ (for transverse
polarisation), we obtain
\begin{subequations}
\begin{eqnarray}
  \DL{f}(x)
  &=& \textfrac1{\sqrtno2} \, \sum_n \,
      \delta \! \left( (1-x)P^+ - P_n^+ \right)
  \nonumber
\\
  && \hspace{2.5em} \null \times
  \left\{
    \left|
      \langle PS | \mathcal{P}_+ \psi_{(+)}(0) | n \rangle
    \right|^2 -
    \left|
      \langle PS | \mathcal{P}_- \psi_{(+)}(0) | n \rangle
    \right|^2
  \right\} ,  \hspace{4em}
  \label{good15}
\\
  \DT{f}(x)
  &=& \textfrac1{\sqrtno2} \, \sum_n \, \delta\left( (1-x)P^+ - P_n^+ \right)
  \nonumber
\\
  && \hspace{2.5em} \null \times
  \left\{
    \left|
      \langle PS | \mathcal{P}_\uparrow   \psi_{(+)}(0) | n \rangle
    \right|^2 -
    \left|
      \langle PS | \mathcal{P}_\downarrow \psi_{(+)}(0) | n \rangle
    \right|^2
  \right\} .
  \label{good16}
  \sublabel{subgood16}
\end{eqnarray}
\end{subequations}
These expressions exhibit the probabilistic content of the leading-twist
polarised distributions $\DL{f}(x)$ and $\DT{f}(x)$: $\DL{f}(x)$ is the number
density of quarks with helicity $+$ minus the number density of quarks with
helicity $-$ (assuming the parent nucleon to have helicity $+$); $\DT{f}(x)$ is
the number density of quarks with transverse polarisation $\uparrow$ minus the
number density of quarks with transverse polarisation $\downarrow$ (assuming
the parent nucleon to have transverse polarisation $\uparrow$). It is important
to notice that $\DT{f}$ admits an interpretation in terms of probability
densities only in the transverse polarisation basis.

The three leading-twist quark distribution functions are contained in the
entries of the spin density matrix of quarks in the nucleon ($\lambda(x)$ is
the quark helicity density, $\Vec{s}_\perp(x)$ is the quark transverse spin
density):
\begin{eqnarray}
  \rho_{\lambda\lambda'} &=&
  \left(
    \begin{array}{cc}
      \rho_{++} & \rho_{+-}
\\
      \rho_{-+} & \rho_{--}
    \end{array}
  \right)
  \nonumber
\\
  &=&
  \frac12
  \left(
    \begin{array}{cc}
      1 + \lambda(x)     & s_x(x) - \I s_y(x) \\
      s_x(x) + \I s_y(x) & 1 - \lambda(x)
    \end{array}
  \right) .
  \label{good17}
\end{eqnarray}
Recalling the probabilistic interpretation of the spin density matrix elements
discussed in Sec.~\ref{densmat}, one finds that the spin components
$\Vec{s}_\perp,\lambda$ of the quark appearing in (\ref{good17}) are related to
the spin components $\Vec{S}_\perp,\lambda_N$ of the parent nucleon by
\begin{subequations}
\begin{eqnarray}
  \lambda_q(x) \, f(x)     &=& \lambda_N     \, \DL{f}(x) \, ,
  \label{good18}
\\
  \Vec{s}_\perp(x) \, f(x) &=& \Vec{S}_\perp \, \DT{f}(x) \, .
  \label{good19}
  \sublabel{subgood19}
\end{eqnarray}
\end{subequations}

\subsection{Vector, axial and tensor charges}
\label{charges}

If we integrate the correlation matrix $\Phi(k,P,S)$ over $k$, or equivalently
$\Phi(x)$ over $x$, we obtain a local matrix element (which we call $\Phi$,
with no arguments)
\begin{equation}
  \Phi_{ij} = \int \! \d^4k \, \Phi_{ij}(k, P, S) =
  \int \! \d{x} \, \Phi_{ij}(x) =
  \langle PS | \anti\psi_j(0) \psi_i(0) | PS \rangle \, ,
  \label{charge1}
\end{equation}
which, in view of (\ref{df9}), can be parametrised as
\begin{equation}
  \Phi = \half
  \left[
    g_V \, \slashed{P} +
    g_A \, \lambda_N  \, \gamma_5 \, \slashed{P} +
    g_T \, \gamma_5 \, \slashed{S}_\perp \, \slashed{P}
  \right] .
  \label{charge2}
\end{equation}
Here $g_V, g_A$ and $g_T$ are the vector, axial and tensor charge,
respectively. They are given by the following matrix elements, recall
(\ref{df5}--\ref{subdf7}):
\begin{subequations}
\begin{eqnarray}
  \langle PS | \anti\psi(0) \gamma^\mu \psi_i(0) | PS \rangle
  &=&
  2 g_V \, P^\mu \, ,
  \label{charge3}
\\
  \langle PS | \anti\psi(0) \gamma^\mu \gamma_5 \psi_i(0) | PS \rangle
  &=&
  2 g_A \, M S^\mu \, ,
  \label{charge4}
\\
  \langle PS | \anti\psi(0) \I \sigma^{\mu\nu} \gamma_5 \psi_i(0) | PS \rangle
  &=&
  2 g_T \, ( S^\mu P^\nu - S^\nu P^\mu ) \, .
  \label{charge5}
\end{eqnarray}
\end{subequations}
Warning: the tensor charge $g_T$ should not be confused with the twist-three
distribution function $g_T(x)$ encountered in Sec.~\ref{transvtarget}.

Integrating eqs.~(\ref{df19}--\ref{subdf21}) and using the symmetry relations
(\ref{adf8}--\ref{subadf10}) yields
\begin{subequations}
\begin{align}
  \int_{-1}^{+1} \d{x} \, f(x) &=
  \int_0^1 \d{x} \left[ f(x) - \anti{f}(x) \right]         = g_V \, ,
  \label{charge6}
\\
  \int_{-1}^{+1} \d{x} \, \DL{f}(x) &=
  \int_0^1 \d{x} \left[ \DL{f}(x) + \DL\anti{f}(x) \right] = g_A \, ,
  \label{charge7}
\\
  \int_{-1}^{+1} \d{x} \, \DT{f}(x) &=
  \int_0^1 \d{x} \left[ \DT{f}(x) - \DT\anti{f}(x) \right] = g_T \, .
  \label{charge8}
\end{align}
\end{subequations}
Note that $g_V$ is simply the valence number. As a consequence of the charge
conjugation properties of the field bilinears $\anti\psi\gamma^\mu\psi$,
$\anti\psi\gamma^\mu\gamma_5\psi$ and $\anti\psi\I\sigma^{\mu\nu}\gamma_5\psi$,
the vector and tensor charges are the first moments of flavour non-singlet
combinations (quarks minus antiquarks) whereas the axial charge is the first
moment of a flavour singlet combination (quarks plus antiquarks).

\subsection{Quark--nucleon helicity amplitudes}
\label{helamp}

\begin{figure}
  \centering
  \begin{picture}(370,85)(0,0)
    \SetWidth{2}
    \ArrowLine(20,30)(50,40)
    \ArrowLine(90,40)(120,30)
    \SetWidth{0.5}
    \ArrowLine(52,45)(52,65)
    \ArrowLine(88,65)(88,45)
    \GOval(70,40)(12,25)(0){0.5}
    \Text(25,22)[]{$+$}
    \Text(115,22)[]{$+$}
    \Text(47,65)[r]{$+$}
    \Text(93,65)[l]{$+$}
    \SetOffset(110,0)
    \SetWidth{2}
    \ArrowLine(20,30)(50,40)
    \ArrowLine(90,40)(120,30)
    \SetWidth{0.5}
    \ArrowLine(52,45)(52,65)
    \ArrowLine(88,65)(88,45)
    \GOval(70,40)(12,25)(0){0.5}
    \Text(25,22)[]{$+$}
    \Text(115,22)[]{$+$}
    \Text(47,65)[r]{$-$}
    \Text(93,65)[l]{$-$}
    \SetOffset(220,0)
    \SetWidth{2}
    \ArrowLine(20,30)(50,40)
    \ArrowLine(90,40)(120,30)
    \SetWidth{0.5}
    \ArrowLine(52,45)(52,65)
    \ArrowLine(88,65)(88,45)
    \GOval(70,40)(12,25)(0){0.5}
    \Text(25,22)[]{$+$}
    \Text(115,22)[]{$-$}
    \Text(47,65)[r]{$-$}
    \Text(93,65)[l]{$+$}
  \end{picture}
  \caption{The three quark--nucleon helicity amplitudes.}
  \label{ampl}
\end{figure}
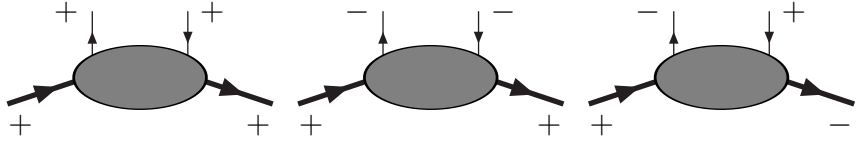

The \DIS hadronic tensor is related to forward virtual Compton scattering
amplitudes. Thus, leading-twist quark distribution functions can be expressed
in terms of quark--nucleon forward amplitudes. In the helicity basis these
amplitudes have the form $\mathcal{A}_{\Lambda\lambda,\Lambda'\lambda'}$, where
$\lambda,\lambda'$ ($\Lambda,\Lambda'$) are quark (nucleon) helicities. There
are in general 16 amplitudes. Imposing helicity conservation,
\begin{equation}
  \Lambda - \lambda' = \Lambda' - \lambda \, ,
  \quad \text{\ie,} \quad
  \Lambda + \lambda  = \Lambda' + \lambda' \, ,
  \label{forward1}
\end{equation}
only 6 amplitudes survive:
\begin{equation}
  \mathcal{A}_{++,++} \, , \;
  \mathcal{A}_{--,--} \, , \;
  \mathcal{A}_{+-,+-} \, , \;
  \mathcal{A}_{-+,-+} \, , \;
  \mathcal{A}_{+-,-+} \, , \;
  \mathcal{A}_{-+,+-} \, .
\end{equation}
Parity invariance implies
\begin{equation}
  \mathcal{A}_{ \Lambda  \lambda,  \Lambda'  \lambda'} =
  \mathcal{A}_{-\Lambda -\lambda, -\Lambda' -\lambda'} \, ,
  \label{forward2}
\end{equation}
and gives the following 3 constraints on the amplitudes:
\begin{eqnarray}
  \mathcal{A}_{++,++} &=& \mathcal{A}_{--,--} \, ,
  \nonumber
\\
  \mathcal{A}_{++,--} &=& \mathcal{A}_{--,++} \, ,
\\
  \mathcal{A}_{+-,-+} &=& \mathcal{A}_{-+,+-} \, .
  \nonumber
\end{eqnarray}
Time-reversal invariance,
\begin{equation}
  \mathcal{A}_{\Lambda  \lambda,  \Lambda' \lambda'} =
  \mathcal{A}_{\Lambda' \lambda', \Lambda  \lambda} \, ,
  \label{forward3}
\end{equation}
adds no further constraints. Hence, we are left with three independent
amplitudes (see Fig.~\ref{ampl})
\begin{equation}
  \mathcal{A}_{++,++} \, , \quad
  \mathcal{A}_{+-,+-} \, , \quad
  \mathcal{A}_{+-,-+} \, .
  \label{forward4}
\end{equation}

Two of the amplitudes in (\ref{forward4}), $\mathcal{A}_{++,++}$ and
$\mathcal{A}_{+-,+-}$, are diagonal in the helicity basis (the quark does not
flip its helicity: $\lambda=\lambda'$), the third, $\mathcal{A}_{+-,-+}$, is
off-diagonal (helicity flip: $\lambda=-\lambda'$). Using the optical theorem we
can relate these quark--nucleon helicity amplitudes to the three leading-twist
quark distribution functions, according to the scheme
\begin{subequations}
\begin{eqnarray}
  f(x) =
  & f_+(x) + f_-(x)
  & \sim \Im (\mathcal{A}_{++,++} + \mathcal{A}_{+-,+-}) \, ,
  \label{forward5}
\\
  \DL{f}(x) =
  & f_+(x) - f_-(x)
  & \sim \Im (\mathcal{A}_{++,++} - \mathcal{A}_{+-,+-}) \, ,
  \label{forward6}
\\
  \DT{f}(x) =
  & f_\uparrow(x) - f_\downarrow(x)
  & \sim \Im \mathcal{A}_{+-,-+} \, .
  \label{forward7}
  \sublabel{subforward7}
\end{eqnarray}
\end{subequations}
In a transversity basis (with $\uparrow$ directed along $y$)
\begin{equation}
  \begin{array}{lcr}
    | \uparrow\rangle &=& \frac1{\sqrtno2}
    \left[ |+\rangle + \I |-\rangle \strut \right] ,
  \\
    | \downarrow \rangle &=& \frac1{\sqrtno2}
    \left[ |+\rangle - \I |-\rangle \strut \right] ,
  \end{array}
\end{equation}
the transverse polarisation distributions $\DT{f}$ is related to a diagonal
amplitude
\begin{equation}
  \DT{f}(x) = f_\uparrow(x) - f_\downarrow(x) \sim
  \Im
  (
    \mathcal{A}_{\uparrow\uparrow,\uparrow\uparrow} -
    \mathcal{A}_{\uparrow\downarrow,\uparrow\downarrow}
  ) \, .
  \label{forward9}
\end{equation}

Reasoning in terms of parton--nucleon forward helicity amplitudes, it is easy
to understand why there is no such thing as leading-twist transverse
polarisation of gluons. A hypothetical $\DT{g}$ would imply an helicity flip
gluon--nucleon amplitude, which cannot exist owing to helicity conservation. In
fact, gluons have helicity $\pm1$ but the nucleon cannot undergo an helicity
change $\Delta\Lambda=\pm2$. Targets with higher spin may have an helicity-flip
gluon distribution.

If transverse momenta of quarks are not neglected, the situation becomes more
complicated and the number of independent helicity amplitudes increases. These
amplitudes combine to form six $\Vec{k}_\perp$-dependent distribution functions
(three of which reduce to $f(x)$, $\DL{f}(x)$ and $\DT{f}(x)$ when integrated
over $\Vec{k}_\perp$).

\subsection{The Soffer inequality}
\label{soffer}

\begin{figure}[htbp]
  \centering
  \includegraphics[width=9cm]{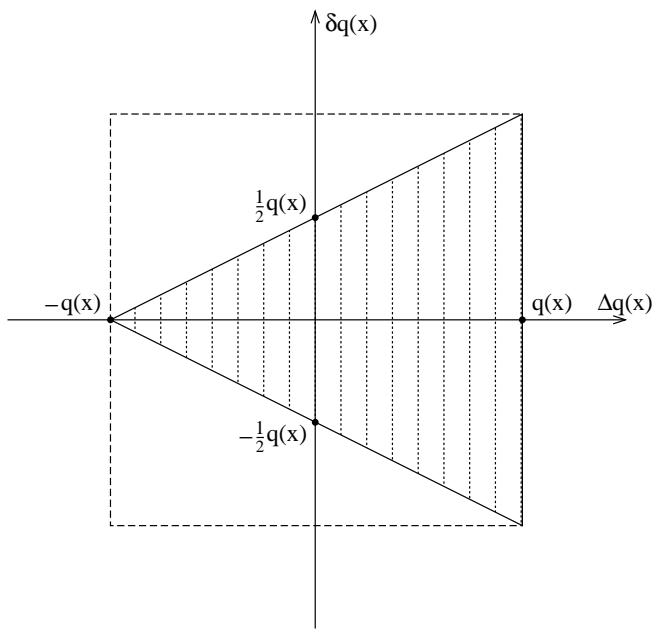}
  \caption{The Soffer bound on the leading-twist distributions
           \cite{Soffer:1995ww} (note that there $\DT{q}(x)$ was called
           $\delta{q}(x)$).}
  \label{fig_soffer}
\end{figure}

From the definitions of $f$, $\DL{f}$ and $\DT{f}$, that is,
$\DL{f}(x)=f_+(x)-f_-(x)$, $\DT{f}(x)=f_\uparrow(x)-f_\downarrow(x)$ and
$f(x)=f_+(x)+f_-(x)=f_\uparrow(x)+f_\downarrow(x)$, two bounds on $\DL{f}$ and
$\DT{f}$ immediately follow:
\begin{subequations}
\begin{eqnarray}
  | \DL{f}(x) | &\le& f(x) \, ,
  \label{soffer0.1}
\\
  | \DT{f}(x) | &\le& f(x) \, ,
  \label{soffer0.2}
  \sublabel{subsoffer0.2}
\end{eqnarray}
\end{subequations}
Similar inequalities are satisfied by the antiquark distributions. Another,
more subtle, bound, simultaneously involving $f$, $\DL{f}$ and $\DT{f}$, was
discovered by Soffer \cite{Soffer:1995ww}. It can be derived from the
expressions (\ref{forward5}--\ref{subforward7}) of the distribution functions
in terms of quark--nucleon forward amplitudes. Let us introduce the
quark--nucleon vertices $a_{\Lambda\lambda'}$:
\begin{center}
  \begin{picture}(130,90)(0,0)
    \SetOffset(35,0)
    \Text(5,40)[r]{$a_{\Lambda,\lambda'}\;\sim$}
    \SetWidth{2}
    \ArrowLine(25,20)(45,35)
    \SetWidth{0.5}
    \ArrowLine(50,70)(50,50)
    \Line(55,42)(75,42)
    \Line(55,46)(75,46)
    \Line(55,38)(75,38)
    \Line(55,34)(75,34)
    \GCirc(50,40){10}{0.5}
    \Text(20,20)[r]{$\Lambda$}
    \Text(45,70)[r]{$\lambda'$}
    \Text(80,40)[l]{$X$}
  \end{picture}
\end{center}
and rewrite eqs.~(\ref{forward5}--\ref{subforward7}) in the form
\begin{subequations}
\begin{eqnarray}
  f(x)
  &\sim&
  \Im (\mathcal{A}_{++,++} +\mathcal{A}_{+-,+-})
  \sim
  \sum_X (a_{++}^* a_{++} + a_{+-}^* a_{+-}) \, ,
  \label{soffer0.3}
\\
  \DL{f}(x)
  &\sim&
  \Im (\mathcal{A}_{++,++} -\mathcal{A}_{+-,+-})
  \sim
  \sum_X ( a_{++}^* a_{++}- a_{+-}^* a_{+-}) \, ,
  \label{soffer0.4}
\\
  \DT{f}(x)
  &\sim&
  \Im \mathcal{A}_{+-,-+}
  \sim
  \sum_X a_{--}^* a_{++} \, .
  \label{soffer0.5}
\end{eqnarray}
\end{subequations}
From
\begin{equation}
  \sum_X | a_{++} \pm a_{--} | ^2 \ge 0 \, ,
\end{equation}
using parity invariance, we obtain
\begin{equation}
  \sum_X a_{++}^* a_{++} \pm \sum_X a_{--}^* a_{++} \ge 0 \, ,
  \label{soffer1}
\end{equation}
that is
\begin{equation}
  f_+(x) \ge | \DT{f}(x) | \, ,
  \label{soffer2}
\end{equation}
which is equivalent to
\begin{equation}
  f(x) + \DL{f}(x) \ge 2 | \DT{f}(x) | \, .
  \label{soffer3}
\end{equation}
An analogous relation holds for the antiquark distributions.
Eq.~(\ref{soffer3}) is known as the Soffer inequality. It is an important
bound, which must be satisfied by the leading-twist distribution functions. The
reason it escaped attention until a relatively late discovery in
\cite{Soffer:1995ww} is that it involves three quantities that are not diagonal
in the same basis. Thus, to be derived, Soffer's inequality requires
consideration of probability amplitudes, not of probabilities themselves. The
constraint (\ref{soffer3}) is represented in Fig.~\ref{fig_soffer}.

We shall see in Sec.~\ref{sofferevol} that the Soffer bound, like the other two
-- more obvious -- inequalities (\ref{soffer0.1},~\ref{subsoffer0.2}), is
preserved by \QCD evolution, as it should be.

\subsection{Transverse motion of quarks}
\label{transvmot}

Let us now account for the transverse motion of quarks. This is necessary in
semi-inclusive \DIS, when one wants to study the $\Vec{P}_{h\perp}$
distribution of the produced hadron. Therefore, in this section we shall
prepare the field for later applications.

The quark momentum is now given by
\begin{equation}
  k^\mu \simeq xP^\mu + k_\perp^\mu \, ,
  \label{tm1}
\end{equation}
where we have retained $k_\perp^\mu$, which is zeroth order in $P^+$ and thus
suppressed by one power of $P^+$ with respect to the longitudinal momentum.

At leading twist, again, only the vector, axial and tensor terms in (\ref{df2})
appear and eqs.~(\ref{df2.2}, \ref{subdf2.3}, \ref{subdf2.5}) become
\begin{subequations}
\begin{eqnarray}
  \mathcal{V}^\mu &=&  \, A_1 \, P^\mu \, ,
  \label{tm2}
\\
  \mathcal{A}^\mu &=& \lambda_N \, A_2 \, P^\mu + \frac1{M} \,
  \widetilde{A}_1 \, \Vec{k}_\perp{\cdot}\Vec{S}_\perp \, P^\mu \, ,
  \label{tm3}
\\
  \mathcal{T}^{\mu\nu} &=&
  A_3 \, P_{\vphantom\perp}^{[\mu} \ST^{\nu]} \, +
  \frac{\lambda_N}{M} \,
  \widetilde{A}_2 \, P_{\vphantom\perp}^{[\mu} k_\perp^{\nu]} +
  \frac1{M^2} \,\widetilde{A}_3 \,
  \Vec{k}_\perp{\cdot}\Vec{S}_\perp \, P_{\vphantom\perp}^{[\mu} k_\perp^{\nu]} \, ,
  \label{tm4}
  \sublabel{subtm4}
\end{eqnarray}
\end{subequations}
where we have defined new real functions $\widetilde{A}_i(k^2,k{\cdot}P)$ (the tilde
signals the presence of $\Vec{k}_\perp$) and introduced powers of $M$, so that
all coefficients have the same dimension. The quark--quark correlation matrix
(\ref{df2}) then reads
\begin{eqnarray}
  \Phi(k, P, S) &=& \frac12
  \left\{ \Strut
    A_1 \, \slashed{P} +
    A_2 \, \lambda_N \, \gamma_5 \, \slashed{P} +
    A_3 \, \slashed{P} \, \gamma_5 \, \slashed{S}_\perp
  \right.
  \nonumber
\\
  && \qquad \null +
  \left.
    \frac1{M} \, \widetilde{A}_1 \, \Vec{k}_\perp{\cdot}\Vec{S}_\perp \,
    \gamma_5 \slashed{P}
    +
    \widetilde{A}_2 \, \frac{\lambda_N}{M} \,
    \slashed{P} \, \gamma_5 \, \slashed{k}_\perp
  \right.
  \nonumber
\\
  && \qquad \null +
  \left.
    \frac1{M^2} \, \widetilde{A}_3 \, \Vec{k}_\perp{\cdot}\Vec{S}_\perp \,
    \slashed{P} \, \gamma_5 \, \slashed{k}_\perp
  \right\}.
  \label{tm5}
\end{eqnarray}
We can project out the $A_i$'s and $\widetilde{A}_i$'s, as we did in
Sec.~\ref{leading}
\begin{subequations}
\begin{eqnarray}
  \frac1{2P^+} \, \Tr(\gamma^+ \Phi)
  &=&
  A_1 \, ,
  \label{tm6}
\\
  \frac1{2P^+} \, \Tr(\gamma^+ \gamma_5 \Phi) &=&
  \lambda_N \, A_2 +
  \frac1{M} \, \Vec{k}_\perp{\cdot}\Vec{S}_\perp \, \widetilde{A}_1 \, ,
  \label{tm7}
\\
  \frac1{2P^+} \, \Tr(\I \sigma^{i+} \gamma_5 \Phi) &=&
  \ST^i \, A_3 +
  \frac{\lambda_N}{M}\, k_\perp^i \, \widetilde{A}_2 +
  \frac1{M^2} \, \Vec{k}_\perp{\cdot}\Vec{S}_\perp \, k_\perp^i \, \widetilde{A}_3
  \, .
  \label{tm8}
  \sublabel{subtm8}
\end{eqnarray}
\end{subequations}
Let us rearrange the r.h.s.\ of the last expression in the following manner
\begin{eqnarray}
  \ST^i \, A_3 +
  \frac1{M^2} \, \Vec{k}_\perp{\cdot}\Vec{S}_\perp \, k_\perp^i \, \widetilde{A}_3
  &=&
  \ST^i
  \left( A_3 + \frac{\Vec{k}_\perp^2}{2M^2} \, \widetilde{A}_3 \right)
  \nonumber
\\
  && \null -
  \frac1{M^2}
  \left(
    k_\perp^i k_\perp^j + \frac12 \, \Vec{k}_\perp^2 \, g_\perp^{ij}
  \right) S_{\perp j} \, \widetilde{A}_3 \, . \hspace{2em}
  \label{tm9}
\end{eqnarray}
If we integrate eqs.~(\ref{tm6}--\ref{subtm8}) over $k$ with the constraint
$x=k^+/P^+$, the terms proportional to $\widetilde{A}_1$, $\widetilde{A}_2$ and
$\widetilde{A}_3$ in (\ref{tm7}, \ref{subtm8}) and (\ref{tm9}) vanish. We are
left with the three terms proportional to $A_1$, $A_2$ and to the combination
$A_3+(\Vec{k}_\perp^2/2M^2)\,\widetilde{A}_3$, which give, upon integration, the
three distribution functions $f(x)$, $\DL{f}(x)$ and $\DT{f}(x)$, respectively.
The only difference from the previous case of no quark transverse momentum is
that $\DT{f}(x)$ is now related to $A_3+(\Vec{k}_\perp^2/2M^2)\,\widetilde{A}_3$
and not to $A_3$ alone
\begin{equation}
  \DT{f}(x) \equiv
  \int \! \frac{\d^4k}{(2\pi)^4}
  \left( A_3 + \frac{\Vec{k}_\perp^2}{2M^2} \, \widetilde{A}_3 \right)
  \delta(x - k^+/P^+) \, .
  \label{tm10}
\end{equation}

If we do not integrate over $\Vec{k}_\perp$, we obtain six
$\Vec{k}_\perp$-dependent distribution functions. Three of them, which we call
$f(x,\Vec{k}_\perp^2)$, $\DL{f}(x,\Vec{k}_\perp^2)$ and
$\DT'f(x,\Vec{k}_\perp^2)$, are such that
$f(x)=\int\d^2\Vec{k}_{\perp}f(x,\Vec{k}_\perp^2)$, \etc. The other three are
completely new and are related to the terms of the correlation matrix
containing the $\widetilde{A}_i$'s. We shall adopt Mulders' terminology for
them \cite{Mulders:1996dh, Boer:1998nt}. Introducing the notation
\begin{eqnarray}
  \Phi^{[\Gamma]}
  &\equiv&
  \frac12 \int \! \frac{\d{k}^+ \, \d{k}^-}{(2\pi)^4} \,
  \Tr(\Gamma \, \Phi) \, \delta(k^+ - xP^+)
  \nonumber
\\
  &=&
  \int \! \frac{\d\xi^- \, \d^2\Vec\xi_\perp}{2(2\pi)^3}
  \e^{\I (x P^+ \xi^- - \Vec{k}_\perp{\cdot}\Vec\xi_\perp)}
  \langle PS
    | \anti\psi(0) \Gamma \psi(0, \xi^-, \Vec\xi_\perp)
  | PS \rangle \, , \qquad
  \label{tm10b}
\end{eqnarray}
we have
\begin{subequations}
\begin{eqnarray}
  \Phi^{[\gamma^+]}
  &=&
  \mathcal{P}_{q/N}(x,\Vec{k}_\perp) =
  f(x, \Vec{k}_\perp^2) \, ,
  \label{tm11}
\\
  \Phi^{[\gamma^+ \gamma_5]}
  &=&
  \mathcal{P}_{q/N}(x,\Vec{k}_\perp) \, \lambda (x,\Vec{k}_\perp)
  \nonumber
\\
  &=&
  \lambda_N \, \DL{f}(x,\Vec{k}_\perp^2) +
  \frac{\Vec{k}_\perp{\cdot}\Vec{S}_\perp}{M} \, g_{1T}(x, \Vec{k}_\perp^2) \, ,
  \label{tm12}
\\
  \Phi^{[\I \sigma^{i+} \gamma_5]}
  &=&
  \mathcal{P}_{q/N}(x, \Vec{k}_\perp) \, s_\perp^i (x, \Vec{k}_\perp)
  \nonumber
\\
  &=&
  \ST^i \, \DT'f(x,\Vec{k}_\perp^2) + \frac{\lambda_N}{M} \,
  k_\perp^i \, h_{1L}^\perp(x,\Vec{k}_\perp^2)
  \nonumber
\\
  && \null -
  \frac1{M^2}
  \left(
    k_\perp^i k_\perp^j + \frac12 \, \Vec{k}_\perp^2 \, g_\perp^{ij}
  \right)
  S_{\perp j} \, h_{1T}^\perp (x,\Vec{k}_\perp^2) \, ,
  \label{tm13}
  \sublabel{subtm13}
\end{eqnarray}
\end{subequations}
where $\mathcal{P}_{q/N}(x,\Vec{k}_\perp)$ is the probability of finding a
quark with longitudinal momentum fraction $x$ and transverse momentum
$\Vec{k}_\perp$, and $\lambda(x,\Vec{k}_\perp)$,
$\Vec{s}_\perp(x,\Vec{k}_\perp)$ are the quark helicity and transverse spin
densities, respectively. The spin density matrix of quarks now reads
\begin{equation}
  \rho_{\lambda\lambda'} = \frac12
  \left(
    \begin{array}{cc}
      1 + \lambda (x, \Vec{k}_\perp) &
      s_x(x, \Vec{k}_\perp) - \I s_y(x,\Vec{k}_\perp) \\
      s_x(x, \Vec{k}_\perp) + \I s_y(x,\Vec{k}_\perp) &
      1 - \lambda(x, \Vec{k}_\perp)
    \end{array}
  \right) ,
  \label{tm13b}
\end{equation}
and its entries incorporate the six distributions listed above, according to
eqs.~(\ref{tm11}--\ref{tm12}).

Let us now try to understand the partonic content of the
$\Vec{k}_\perp$-dependent distributions. If the target nucleon is unpolarised,
the only measurable quantity is $f(x,\Vec{k}_\perp^2)$, which coincides with
$\mathcal{P}_q(x,\Vec{k}_\perp)$, the number density of quarks with
longitudinal momentum fraction $x$ and transverse momentum squared
$\Vec{k}_\perp^2$.

If the target nucleon is transversely polarised, there is some probability of
finding the quarks transversely polarised along the same direction as the
nucleon, along a different direction, or longitudinally polarised. This variety
of situations is allowed by the presence of $\Vec{k}_\perp$. Integrating over
$\Vec{k}_\perp$, the transverse polarisation asymmetry of quarks along a
different direction with respect to the nucleon polarisation, and the
longitudinal polarisation asymmetry of quarks in a transversely polarised
nucleon disappear: only the case $\Vec{s}_\perp\parallel\Vec{S}_\perp$
survives.

Referring to Fig.~\ref{azimuth} for the geometry in the azimuthal plane and
using the following parametrisations for the vectors at hand (we assume full
polarisation of the nucleon):
\begin{eqnarray}
  \Vec{k}_\perp &=&
  ( | \Vec{k}_\perp | \, \cos\phi_k, \,
   -| \Vec{k}_\perp | \, \sin\phi_k) \, ,
  \label{tm14b}
\\
  \Vec{S}_\perp &=&
  ( \cos\phi_S, \, - \sin\phi_S) \, ,
  \label{tm14c}
\\
  \Vec{s}_\perp &=&
  ( | \Vec{s}_\perp | \, \cos\phi_s, \,
   -| \Vec{s}_\perp | \, \sin\phi_s) \, ,
  \label{tm14d}
\end{eqnarray}
we find for the $\Vec{k}_\perp$-dependent transverse polarisation distributions
of quarks in a transversely polarised nucleon (${\pm}$ denote, as usual,
longitudinal polarisation whereas $\uparrow\downarrow$ denote transverse
polarisation)
\begin{subequations}
\begin{eqnarray}
  \mathcal{P}_{q\uparrow/N\uparrow}(x,\Vec{k}_\perp) -
  \mathcal{P}_{q\downarrow/N\uparrow}(x,\Vec{k}_\perp)
  &=&
  \cos(\phi_S - \phi_s) \, \DT'f(x,\Vec{k}_\perp^2)
  \nonumber
\\
  && \hspace*{-5em} \null +
  \frac{\Vec{k}_\perp^2}{2M^2} \,
  \cos(2 \phi_k - \phi_S - \phi_s) \,
  h_{1T}^\perp(x,\Vec{k}_\perp^2) \, , \qquad
  \label{tm15}
\end{eqnarray}
and for the longitudinal polarisation distribution of quarks in a transversely
polarised nucleon
\begin{equation}
  \mathcal{P}_{q+/N\uparrow}(x,\Vec{k}_\perp) -
  \mathcal{P}_{q-/N\uparrow}(x,\Vec{k}_\perp) =
  \frac{| \Vec{k}_\perp |}{M} \, \cos(\phi_S - \phi_k) \,
  g_{1T}(x,\Vec{k}_\perp^2) \, .
  \label{tm16}
\end{equation}

\begin{figure}
  \centering
  \begin{picture}(280,220)(0,65)
    \LongArrow(20,160)(260,160)
    \LongArrow(140,280)(140,70)
    \SetWidth{1}
    \LongArrow(140,160)(230,220)
    \SetWidth{2}
    \LongArrow(140,160)(115,260)
    \SetWidth{2}
    \LongArrow(140,160)(65,220)
    \SetWidth{0.5}
    \LongArrowArc(140,160)(30,0,105)
    \LongArrowArc(140,160)(15,0,35)
    \LongArrowArc(140,160)(45,0,142)
    \Text(260,150)[t]{$x$}
    \Text(132,70)[r]{$y$}
    \Text(157,167)[l]{$\phi_k$}
    \Text(160,194)[]{$\phi_S$}
    \Text(150,215)[]{$\phi_s$}
    \Text(230,225)[b]{\large$\Vec{k}_\perp$}
    \Text(115,265)[b]{\large$\Vec{S}_\perp$}
    \Text(65,225)[b]{\large$\Vec{s}_\perp$}
  \end{picture}
  \caption{Our definition of the azimuthal angles in the plane orthogonal to
           the $\gamma^*N$ axis. The photon momentum, which is directed along
           the positive $z$ axis, points inwards. For our choice of the axes
           see Fig.~\ref{plane3}.}
  \label{azimuth}
\end{figure}
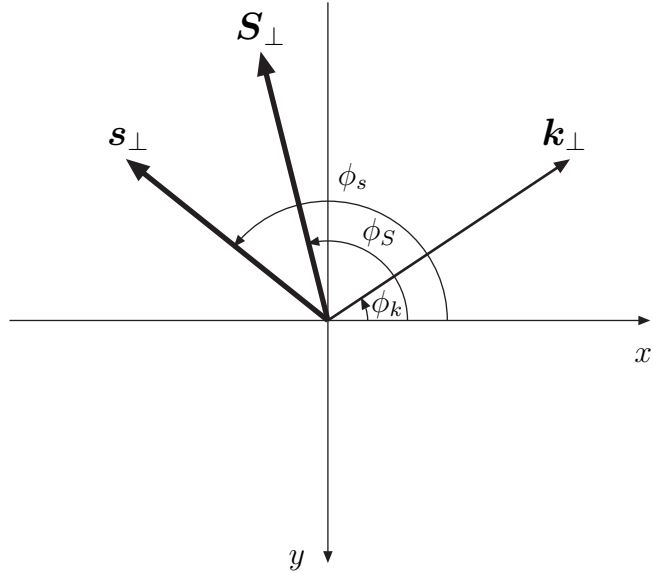

Due to transverse motion, quarks can also be transversely polarised in a
longitudinally polarised nucleon. Their polarisation asymmetry is
\begin{equation}
  \mathcal{P}_{q\uparrow/N+}(x,\Vec{k}_\perp) -
  \mathcal{P}_{q\downarrow/N+}(x,\Vec{k}_\perp)
  =
  \frac{| \Vec{k}_\perp |}{M} \,
  \cos(\phi_k - \phi_s) \, h_{1L}^\perp(x,\Vec{k}_\perp^2) \, .
  \label{tm17}
\end{equation}
\end{subequations}
As we shall see in Sec.~\ref{sidisasymm}, the $\Vec{k}_\perp$-dependent
distribution function $h_{1L}^\perp$ plays a r\^{o}le in the azimuthal asymmetries
of semi-inclusive leptoproduction.

\subsection{$T$-odd distributions}
\label{todd}

Relaxing the time-reversal invariance condition (\ref{df1d}) -- we postpone the
discussion of the physical relevance of this until the end of this subsection
-- two additional terms in the vector and tensor components of $\Phi$ arise
\begin{align}
  \mathcal{V}^\mu &= \dots + \frac1{M} \, A_1' \,
  \varepsilon^{\mu\nu\rho\sigma} \, P_\nu k_{\perp\rho} S_{\perp\sigma} \, ,
  \tag{\ref{df2.2}$''$}
  \label{todd1}
\\
  \mathcal{T}^{\mu\nu} &= \dots + \frac1{M} \, A_2' \,
  \varepsilon^{\mu\nu\rho\sigma} \, P_\rho k_{\perp\sigma} \, ,
  \tag{\ref{df2.5}$''$}
  \label{todd2}
\end{align}
which give rise to two $\Vec{k}_\perp$-dependent $T$-odd distribution
functions, $f_{1T}^\perp$ and $h_1^\perp$
\begin{subequations}
\begin{eqnarray}
  \Phi^{[\gamma^+]}
  &=& \dots -
  \frac{\varepsilon_\perp^{ij} k_{\perp i} S_{\perp j}}{M} \,
  f_{1T}^\perp(x,\Vec{k}_\perp^2) \, ,
  \label{todd3}
\\
  \Phi^{[ \I \sigma^{i +} \gamma_5]}
  &=& \dots -
  \frac{\varepsilon_\perp^{ij} k_{\perp j}}{M} \,
  h_{1}^\perp(x,\Vec{k}_\perp^2) \, .
  \label{todd4}
\end{eqnarray}
\end{subequations}

Let us see the partonic interpretation of the new distributions. The first of
them, $f_{1T}^\perp$, is related to the number density of unpolarised quarks in
a transversely polarised nucleon. More precisely, it is given by
\begin{subequations}
\begin{eqnarray}
  \mathcal{P}_{q/N \uparrow}   (x, \Vec{k}_\perp) -
  \mathcal{P}_{q/N \downarrow} (x, \Vec{k}_\perp)
  &=&
  \mathcal{P}_{q/N \uparrow} (x,   \Vec{k}_\perp) -
  \mathcal{P}_{q/N \uparrow} (x, - \Vec{k}_\perp)
  \nonumber
\\
  &=&
  -2 \frac{| \Vec{k}_\perp |}{M} \, \sin(\phi_k - \phi_S)\,
  f_{1T}^\perp(x, \Vec{k}_\perp^2) \, . \qquad
  \label{todd5}
\end{eqnarray}
The other $T$-odd distribution, $h_1^\perp$, measures quark transverse
polarisation in an unpolarised hadron \cite{Boer:1998nt} and is defined via
\begin{equation}
  \mathcal{P}_{q \uparrow/N}   (x, \Vec{k}_\perp) -
  \mathcal{P}_{q \downarrow/N} (x, \Vec{k}_\perp) =
  - \frac{| \Vec{k}_\perp |}{M} \, \sin(\phi_k - \phi_s)
  \, h_1^\perp(x, \Vec{k}_\perp^2) \, .
  \label{todd6}
\end{equation}
\end{subequations}
We shall encounter again these distributions in the analysis of hadron
production (Sec.~\ref{singlespintransv}). For later convenience we define two
quantities $\Delta_0^T{f}$ and $\DT^0{f}$, related to $f_{1T}^\perp$ and
$h_1^\perp$, respectively, by (for the notation see Sec.~\ref{notations})
\begin{subequations}
\begin{eqnarray}
  \Delta_0^T f(x, \Vec{k}_\perp^2)
  &\equiv&
  -2 \frac{| \Vec{k}_\perp |}{M} \, f_{1T}^\perp(x, \Vec{k}_\perp^2) \, ,
  \label{todd6b}
\\
  \DT^0 f(x, \Vec{k}_\perp^2) &\equiv&
  - \, \frac{| \Vec{k}_\perp |}{M} \, h_{1}^\perp(x, \Vec{k}_\perp^2) \, .
  \label{todd6c}
\end{eqnarray}
\end{subequations}

It is now time to comment on the physical meaning of the quantities we have
introduced in this section. One may legitimately wonder whether $T$-odd quark
distributions, such as $f_{1T}^\perp$ and $h_1^\perp$ that violate the
time-reversal condition (\ref{df1d}) make any sense at all. In order to justify
the existence of $T$-odd distribution functions, their proponents
\cite{Anselmino:1998yz} advocate initial-state effects, which prevent
implementation of time-reversal invariance by na\"{\i}vely imposing the condition
(\ref{df1d}). The idea, similar to that which leads to admitting $T$-odd
fragmentation functions as a result of final-state effects (see
Sec.~\ref{kappatransv}), is that the colliding particles interact strongly with
non-trivial relative phases. As a consequence, time reversal no longer implies
the constraint (\ref{df1d}).\footnote{Thus ``$T$-odd'' means that condition
(\ref{df1d}) is not satisfied, not that time reversal is violated.} If
hadronic interactions in the initial state are crucial to explain the existence
of $f_{1T}^\perp$ and $h_1^\perp$, these distributions should only be
observable in reactions involving two initial hadrons (Drell--Yan processes,
hadron production in proton--proton collisions \etc.). This mechanism is known
as the Sivers effect \cite{Sivers:1989cc, Sivers:1991fh}. Clearly, it should be
absent in leptoproduction.

A different way of looking at the $T$-odd distributions has been proposed in
\cite{Anselmino:1997jj, Drago:1998rd, Anselmino:2001xx}. By relying on a
general argument using time reversal, originally due to Wigner and recently
revisited by Weinberg \cite{Weinberg:1995mt-p100}, the authors of
\cite{Anselmino:2001xx} show that time reversal does not necessarily forbid
$f_{1T}^\perp$ and $h_1^\perp$. In particular, an explicit realisation of
Weinberg's mechanism, based on chiral Lagrangians, shows that $f_{1T}^\perp$
and $h_1^\perp$ may emerge from the time-reversal preserving chiral dynamics of
quarks in the nucleon, with no need for initial-state interaction effects. If
this idea is correct, the $T$-odd distributions should also be observable in
semi-inclusive leptoproduction. A conclusive statement on the matter will only
be made by experiments.

\subsection{Twist-three distributions}
\label{twist3}

At twist 3 the quark--quark correlation matrix, integrated over $k$, has the
structure \cite{Jaffe:1992ra}
\begin{equation}
  \Phi (x) = \dots + \frac{M}{2}
  \left\{
    e(x) + g_T(x) \, \gamma_5 \, \slashed{S}_\perp +
    \frac{\lambda_N}{2} \, h_L(x) \, \gamma_5
    \left[ \slashed{p}, \slashed{n} \right]
  \right\} ,
  \label{higher1}
\end{equation}
where the dots represent the twist-two contribution, eq.~(\ref{df23}), and $p$,
$n$ are the Sudakov vectors (see Appendix~\ref{sudakov}). Three more
distributions appear in (\ref{higher1}): $e(x)$, $g_T(x)$ and $h_L(x)$. We
already encountered $g_T(x)$, which is the twist-three partner of $\DL{f}(x)$:
\begin{subequations}
\begin{eqnarray}
  \int \! \frac{\d\tau}{2\pi} \, \e^{\I \tau x} \,
  \langle PS | \anti\psi(0) \gamma^\mu \gamma_5 \psi(\tau n) | PS \rangle
  &=&
  2 \lambda_N \, \DL{f}(x) \, p^\mu + 2 M \, g_T(x) \, \ST^\mu
  \nonumber
\\
  && \null  + \text{twist-4 terms} \, .
  \label{higher2}
\end{eqnarray}
Analogously, $h_L(x)$ is the twist-three partner of $\DT{f}(x)$ and appears in
the tensor term of the quark--quark correlation matrix:
\begin{eqnarray}
  \int \! \frac{\d\tau}{2\pi} \, \e^{\I \tau x} \,
  \langle PS | \anti\psi(0) \I\sigma^{\mu\nu} \gamma_5 \psi(\tau n) | PS \rangle
  &=&
  2 \DT{f}(x) \, p_{\vphantom\perp}^{[\mu} \ST^{\nu]} +
  2 M \, h_L(x) \,  p^{[\mu} n^{\nu]}
  \nonumber
\\
  && \null + \text{twist-4 terms.}
  \label{higher3}
\end{eqnarray}
The third distribution, $e(x)$, has no counterpart at leading twist. It appears
in the expansion of the scalar field bilinear:
\begin{equation}
  \int \! \frac{\d\tau}{2\pi} \, \e^{\I \tau x} \,
  \langle PS | \anti\psi(0) \psi(\tau n) | PS \rangle
  = 2 M \, e(x) + \; \text{twist-4 terms.}
  \label{higher5}
\end{equation}
\end{subequations}

The higher-twist distributions do not admit any probabilistic interpretation.
To see this, consider for instance $g_T(x)$. Upon separation of $\psi$ into
good and bad components, it turns out to be
\begin{eqnarray}
  g_T(x) &=&
  \frac{P^+}{M} \int \! \frac{\d\xi^-}{4\pi} \e^{\I xP^+ \xi^-}
  \langle PS |
  \psi_{(+)}^\dagger(0)
    \gamma^0 \gamma^1 \gamma_5
  \psi_{(-)}(0,\xi^-,\Vec{0}_\perp)
  \nonumber
\\
  && \hspace{9em} \null -
  \psi_{(-)}^\dagger(0)
    \gamma^0 \gamma^1 \gamma_5
  \psi_{(+)}(0,\xi^-,\Vec{0}_\perp)
  | PS \rangle \, . \qquad
  \label{higher6}
\end{eqnarray}
This distribution cannot be put into a form such as
(\ref{good15},~\ref{subgood16}). Thus $g_T$ cannot be regarded as a probability
density. Like all higher-twist distributions, it involves quark--quark--gluon
correlations and hence has no simple partonic meaning. It is precisely this
fact that renders $g_T(x)$ and the structure function that contains $g_T(x)$,
\ie, $g_2(x,Q^2)$, quite subtle and difficult to handle within the framework of
parton model.

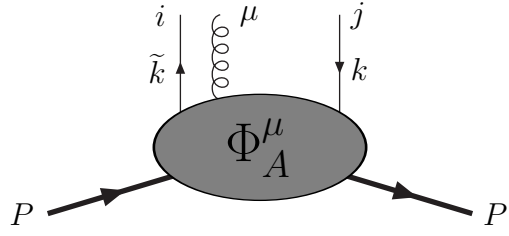
\begin{figure}
  \centering
  \begin{picture}(300,110)(0,0)
    \SetWidth{2}
    \ArrowLine(70,15)(120,30)
    \ArrowLine(180,30)(230,15)
    \SetWidth{0.5}
    \ArrowLine(120,50)(120,90)
    \ArrowLine(180,90)(180,50)
    \Gluon(135,58)(135,90){3}{4.5}
    \Text(115,90)[r]{$i$}
    \Text(185,90)[l]{$j$}
    \Text(143,90)[l]{$\mu$}
    \Text(65,15)[r]{$P$}
    \Text(235,15)[l]{$P$}
    \Text(115,70)[r]{$\widetilde{k}$}
    \Text(185,70)[l]{$k$}
    \GOval(150,40)(20,40)(0){0.5}
    \Text(150,40)[]{\LARGE$\Phi_A^\mu$}
  \end{picture}
  \caption{The quark--quark--gluon correlation matrix $\Phi_A^\mu$.}
  \label{phi2}
\end{figure}

It should be borne in mind that the twist-three distributions in
(\ref{higher1}) are, in a sense, ``effective'' quantities, which incorporate
various kinematical and dynamical effects that contribute to higher twist:
quark masses, intrinsic transverse motion and gluon interactions. It can be
shown \cite{Mulders:1996dh} that $e(x)$, $h_L(x)$ and $g_T(x)$ admit the
decomposition
\begin{subequations}
\begin{eqnarray}
  e(x) &=& \frac{m_q}{M} \, \frac{f(x)}{x} + \widetilde{e}(x) \, ,
  \label{higher6b}
\\
  h_L(x) &=& \frac{m_q}{M} \, \frac{\DL{f}(x)}{x} - \frac2{x} \,
  h_{1L}^{\perp (1)}(x) + \widetilde{h}_L(x) \, ,
  \label{higher6c}
\\
  g_T(x) &=& \frac{m_q}{M} \, \frac{\DT{f}(x)}{x} + \frac1{x} \,
  g_{1T}^{(1)}(x) + \widetilde{g}_T(x) \, ,
  \label{higher6d}
\end{eqnarray}
\end{subequations}
where we have introduced the weighted distributions
\begin{subequations}
\begin{eqnarray}
  h_{1L}^{\perp (1)}(x) &\equiv&
  \int \! \d^2\Vec{k}_\perp \, \frac{\Vec{k}_\perp^2}{2M^2} \,
  h_{1L}^\perp(x, \Vec{k}_\perp^2) \, ,
  \label{higher6db}
\\
  g_{1T}^{(1)}(x) &\equiv&
  \int \! \d^2\Vec{k}_\perp \, \frac{\Vec{k}_\perp^2}{2M^2} \,
  g_{1T}(x, \Vec{k}_\perp^2) \, .
  \label{higher6dc}
\end{eqnarray}
\end{subequations}
The three tilde functions $\widetilde{e}(x)$, $\widetilde{h}_L(x)$ and
$\widetilde{g}_T(x)$ are the genuine interaction-dependent twist-three parts of
the subleading distributions, arising from non-handbag diagrams like that of
Fig.~\ref{nonhand}. To understand the origin of such quantities, let us define
the quark correlation matrix with a gluon insertion (see Fig.~\ref{phi2})
\begin{eqnarray}
  \Phi_{Aij}^\mu(k, \widetilde{k}, P, S)
  &=&
  \int \! \d^4\xi \int \! \d^4 z \,
  \e^{\I\widetilde{k}{\cdot}\xi} \, \e^{\I(k - \widetilde{k}){\cdot}z}
  \nonumber
\\
  && \hspace{3em} \null \times
  \langle PS |
    \anti\phi_j(0) \, g A^\mu(z) \, \phi_i(\xi)
  | PS \rangle \, .
  \label{higher6e}
\end{eqnarray}
Note that in the diagram of Fig.~\ref{phi2} the momenta of the quarks on the
left and on the right of the unitarity cut are different. We call $x$ and $y$
the two momentum fractions, \ie,
\begin{equation}
  k = xP \, , \qquad \widetilde{k} = yP \, ,
  \label{higher6f}
\end{equation}
and integrate (\ref{higher6e}) over $k$ and $\widetilde{k}$ with the
constraints (\ref{higher6f})
\begin{eqnarray}
  \Phi_{Aij}^\mu(x,y)
  &=&
  \int\! \frac{\d^4k}{(2\pi)^4}
  \int\! \frac{\d^4 \widetilde{k}}{(2\pi)^4} \, \Phi_A^\mu(k, \widetilde{k}, P, S) \,
  \delta(x - k^+/P^+) \, \delta(y - \widetilde{k}^+/P^+)
  \nonumber
\\
  &=&
  \int\! \frac{\d\tau}{2\pi} \int\! \frac{\d\eta}{2\pi}
  \, \e^{\I \tau y} \, \e^{\I \eta(x - y)} \,
  \langle PS |
    \anti\phi_j(0) \, g A^\mu(\eta n ) \, \phi_i(\tau n)
  | PS \rangle \, , \hspace{4em}
  \label{higher6g}
\end{eqnarray}
where in the last equality we set $\tau=P^+\xi^-$ and $\eta=P^+z^-$, and $n$ is
the usual Sudakov vector. If a further integration over $y$ is performed, one
obtains a quark--quark--gluon correlation matrix where one of the quark fields
and the gluon field are evaluated at the same space--time point:
\begin{equation}
  \Phi_{Aij}^\mu(x) =
  \int \! \frac{\d\tau}{2\pi} \, \e^{\I \tau x}
  \langle PS |
    \anti\phi_j(0) \, g A^\mu(\tau n) \, \phi_i(\tau n)
  | PS \rangle \, .
  \label{higher6h}
\end{equation}

The matrix $\Phi_A^\mu(x,y)$ makes its appearance in the calculation of the
had\-ronic tensor at the twist-three level. It contains four multiparton
distributions $G_A$, $\widetilde{G}_A$, $H_A$ and $E_A$; and has the following
structure:
\begin{eqnarray}
  \Phi_A^\mu(x,y)
  &=&
  \frac{M}{2}
  \left\{
    \I \, G_A(x,y) \, \varepsilon_\perp^{\mu\nu} S_{\perp\nu} \slashed{P} +
    \widetilde{G}_A(x,y) \, \ST^\mu \gamma_5 \slashed{P}
  \right.
  \nonumber
\\
  && \qquad \null +
  \left.
    H_A(x,y) \, \lambda_N \gamma_5 \gamma_\perp^\mu \slashed{P} +
    E_A(x,y) \, \gamma_\perp^\mu \slashed{P}
  \right\} .
  \label{higher6i}
\end{eqnarray}
Time-reversal invariance implies that $G_A$, $\widetilde{G}_A$, $H_A$ and $E_A$
are real functions. By hermiticity $\widetilde{G}_A$ and $H_A$ are symmetric
whereas $G_A$ and $E_A$ are antisymmetric under interchange of $x$ and $y$.

It turns out that $\widetilde{g}_T(x), \widetilde{h}_L(x)$ and
$\widetilde{e}(x)$ are indeed related to $G_A$, $\widetilde{G}_A$, $H_A$ and
$E_A$, in particular to the integrals over $y$ of these functions. One finds,
in fact, \cite{Boer:1997bw}
\begin{subequations}
\begin{eqnarray}
  x \, \widetilde{g}_T(x) &=& \int \! \d{y}
    \left[ G_A(x,y) + \widetilde{G}_A(x,y) \right] ,
  \label{higher6j}
\\
  x \, \widetilde{h}_L(x) &=& 2 \int \! \d{y} \, H_A(x,y) \, ,
  \label{higher6k}
\\
  x \, \widetilde{e}(x)   &=& 2 \int \! \d{y} \, E_A(x,y) \, .
  \label{higher6l}
\end{eqnarray}
\end{subequations}

For future reference we give in conclusion the $T$-odd twist-three quark--quark
correlation matrix, which contains three more distribution functions
\cite{Boer:1998nt, Boer:1999mm}
\begin{equation}
  \left. \Phi (x) \right|_{T-\text{odd}} =
  \frac{M}{2}
  \left\{
    f_T(x) \, \varepsilon_\perp^{\mu\nu} S_{\perp\mu} \gamma_\nu -
    \I \lambda_N \, e_L(x) \, \gamma_5 \, +
    \frac{\I}{2} \, h(x) \left[ \slashed{p}, \slashed{n} \right]
  \right\} ,
  \label{higher7}
\end{equation}
We shall find these distributions again in Sec.~\ref{twist3drellyan}.

The quark (and antiquark) distribution functions at leading twist and twist~3
are collected in Table~\ref{table0}.
\begin{table}[htbp]
  \centering
  \caption{The quark distributions at twist 2 and 3. $S$ denotes the
           polarisation state of the parent hadron (0 indicates unpolarised).
           The asterisk indicates $T$-odd quantities.}
  \label{table0}
  \vspace{1ex}
  \begin{tabular}{cccc}
    \hline\hline
    \multicolumn{4}{c}{Quark distributions}
  \\
    \hline
    $S$ & 0 & L &  T \\
    \hline
    twist 2 & $f(x)$ & $\DL{f}(x)$ & $\DT{f}(x)$ \\
    \hline
    twist 3 & $e(x)$ & $h_L(x)$ & $g_T(x)$ \\
    (*) & $h(x)$ & $e_L(x)$ & $f_T(x)$ \\
    \hline\hline
  \end{tabular}
\end{table}

\subsection{Sum rules for the transversity distributions}
\label{sumrules}

A noteworthy relation between the twist-three distribution $h_L$ and $\DT{f}$,
arising from Lorentz covariance, is \cite{Boglione:2000jk}
\begin{equation}
  h_L(x) = \DT{f}(x) - \frac{\d}{\d{x}} \, h_{1L}^{\perp (1)}(x) \, .
  \label{sumrules1}
\end{equation}
where $h_{1L}^{\perp(1)}(x)$ has been defined in (\ref{higher6db}). Combining
(\ref{sumrules1}) with (\ref{higher6c}) and solving for $h_{1L}^{\perp(1)}$ we
obtain \cite{Jaffe:1992ra} (quark mass terms are neglected)
\begin{equation}
  h_L(x) =
  2x \int_x^1 \frac{\d{y}}{y^2} \, \DT{f}(y) +
  \widetilde{h}_L (x) -
  2x \int_x^1 \frac{\d{y}}{y^2} \, \widetilde{h}_L (y) \, .
  \label{sumrules3}
\end{equation}
On the other hand, solving for $h_L$ leads to
\begin{equation}
  x^3 \, \frac{\d}{\d{x}} \left( \frac{h_{1L}^{\perp (1)}}{x^2} \right) =
  x \, \DT{f} (x) - x \, \widetilde{h}_L(x) \, .
  \label{sumrules4}
\end{equation}

If the twist-three interaction dependent distribution $\widetilde{h}_L(x)$ is
set to zero one obtains from (\ref{sumrules3})
\begin{equation}
  h_L(x) = 2x \int_x^1 \frac{\d{y}}{y^2} \, \DT{f}(y) \, ,
  \label{sumrules5}
\end{equation}
and from (\ref{sumrules4}) and (\ref{sumrules1})
\begin{equation}
  h_{1L}^{\perp(1)}(x) =
  - \frac12 \, x h_L(x) =
  - x^2 \int_x^1 \frac{\d{y}}{y^2} \, \DT{f}(y) \, .
  \label{sumrules6}
\end{equation}
Equation (\ref{sumrules5}) is the transversity analogue of the \WW sum rule
\cite{Wandzura:1977qf} for the $g_1$ and $g_2$ structure functions, which reads
\begin{equation}
  g_2^\text{WW}(x,Q^2) = -g_1(x,Q^2) + \int_x^1 \frac{\d{y}}{y} \, g_1(y,Q^2) \, ,
  \label{sumrules7}
\end{equation}
where $g_2^\text{WW}$ is the twist-two part of $g_2$. In partonic terms, in
fact, the \WW sum rule can be rewritten as (see (\ref{pm38}))
\begin{equation}
  g_T (x) = \int_x^1 \frac{\d{y}}{y} \, \DL{f}(y) \, ,
  \label{sumrules8}
\end{equation}
which is analogous to (\ref{sumrules6}) and can be derived from
(\ref{higher6b}) and from a relation for $g_T(x)$ similar to (\ref{sumrules1}).

\section{Transversity distributions in quantum chromodynamics}
\label{transvqcd}

As well-known, the principal effect of \QCD on the na\"{\i}ve parton model is to
induce, via renormalisation, a (logarithmic) dependence on $Q^2$
\cite{Georgi:1974sr, Gross:1973ju, Gross:1974cs}, the energy scale at which the
distributions are defined or (in other words) the resolution with which they
are measured. The two techniques with which we shall exemplify the following
discussions of this dependence and of the general calculational framework are
the \RGE applied to the \OPE (providing a solid formal basis) and the
ladder-diagram summation approach \cite{Dokshitzer:1978hw, Craigie:1980fa}
(providing a physically more intuitive picture). The variation of the
distributions as functions of energy scale may be expressed in the form of the
standard \DGLAP so-called evolution equations.

Further consequences of higher-order \QCD are mixing and, beyond the \LL
approximation, eventual scheme ambiguity in the definitions of the various
quark and gluon distributions; \ie, the densities lose their precise meaning in
terms of real physical probability and require further conventional definition.
In this section we shall examine the $Q^2$ evolution of the transversity
distribution at \LO and \NLO. In particular, we shall compare its evolution
with that of both the unpolarised and helicity-dependent distributions. Such a
comparison is especially relevant to the so-called Soffer inequality
\cite{Soffer:1995ww}, which involves all three types of distribution. The
section closes with a detailed examination of the question of parton density
positivity and the generalised so-called positivity bounds (of which Soffer's
is then just one example).

\subsection{The renormalisation group equations}
\label{renormalisation}

In order to establish our conventions for the definition of operators and their
renormalisation \etc, it will be useful to briefly recall the \RGE as applied
to the \OPE in \QCD. Before doing so let us make two remarks related to the
problem of scheme dependence. Firstly, in order to lighten the notation, where
applicable and unless otherwise stated, all expressions will be understood to
refer to the so-called minimal modified subtraction $(\MSbar)$ scheme. A
further complication that arises in the case of polarisation at \NLO is the
extension of $\gamma_5$ to $d\not=4$ dimensions \cite{Akyeampong:1974vj,
Breitenlohner:1977hg, Chanowitz:1979zu}. An in-depth discussion of this problem
is beyond the scope of the present review and the interested reader is referred
to \cite{Kamal:1996as}, where it is also considered in the context of
transverse polarisation.

For a generic composite operator $\Oper$, the scale independent so-called bare
($\Oper^B$) and renormalised ($\Oper(\mu^2)$) operators are related via a
renormalisation constant $Z(\mu^2)$, where $\mu$ is then the renormalisation
scale:
\begin{equation}
  \Oper(\mu^2)
  =
  Z^{-1}(\mu^2) \, \Oper^{B} \, .
  \label{eq:qcd-op-ren}
\end{equation}
The scale dependence of $\Oper(\mu^2)$ is obtained by solving the \RGE, which
expressed in terms of the \QCD coupling constant, $\alpha_s=g^2/4\pi$, is
\begin{equation}
  \mu^2\frac{\partial\Oper(\mu^2)}{\partial\mu^2}
  + \gamma \left( \alpha_s(\mu^2) \right) \Oper(\mu^2)
  =
  0 \, ,
  \label{eq:qcd-1.2}
\end{equation}
where $\gamma(\alpha_s(\mu^2))$, the anomalous dimension for the operator
$\Oper(\mu^2)$, is defined by
\begin{equation}
  \gamma(\alpha_s(\mu^2))
  =
  \mu^2 \frac{\partial}{\partial\mu^2} \ln Z(\mu^2) \, .
  \label{eq:qcd-1.3}
\end{equation}
The formal solution is simply
\begin{equation}
  \Oper(Q^2)
  =
  \Oper(\mu^2) \,
  \exp
  \left[
    - \int_{\alpha_s(\mu^2)}^{\alpha_s(Q^2)} \!
      \d{\alpha_s} \frac{\gamma(\alpha_s)}{\beta(\alpha_s)}
  \right] ,
  \label{eq:qcd-RGEsoln}
\end{equation}
where $\beta(\alpha_s)$ is the \RGE function governing the renormalisation
scale dependence of the effective \QCD coupling constant $\alpha_s(\mu^2)$:
\begin{equation}
  \beta(\alpha_s)
  =
  \mu^2 \frac{\partial\alpha_s(\mu^2)}{\partial\mu^2} \, .
  \label{eq:qcd-RGEbeta}
\end{equation}
At \NLO the anomalous dimension $\gamma(\mu^2)$ and the \QCD $\beta$-function
$\beta(\alpha_s)$ can then be expanded perturbatively as
\begin{eqnarray}
  \gamma(\alpha_s)
  &=&
          \frac{\alpha_s}{2\pi}         \, \gamma^{(0)}
  + \left(\frac{\alpha_s}{2\pi}\right)^2 \, \gamma^{(1)}
  + \Ord(\alpha_s^3) \, ,
  \label{eq:qcd-nlo-gamma}
\\
  \beta(\alpha_s)
  &=&
  - \alpha_s
  \left[
          \frac{\alpha_s}{4\pi}         \, \betacft0
  + \left(\frac{\alpha_s}{4\pi}\right)^2 \, \betacft1
  + \Ord(\alpha_s^3)
  \right] .
  \label{eq:qcd-nlo-beta}
\end{eqnarray}
The first two coefficients of the  $\beta$-function are:
$\betacft0=\frac{11}3\CG-\frac43\TF=11-\frac23\Nf$ and
$\betacft1=\frac23(17\CG^2-10\CG\TF-6\CF\TF)=102-\frac{38}3\Nf$, where $\CG=\Nc$
and $\CF=(\Nc^2-1)/2\Nc$ are the usual Casimirs related respectively to the
gluon and the fermion representations of the colour symmetry group, $\SU(\Nc)$,
and $\TF=\half\Nf$, for active quark flavour number $\Nf$. This leads to the
following \NLO expression for the \QCD coupling constant:
\begin{equation}
  \frac{\alpha_s(Q^2)}{4\pi}
  \simeq
  \frac1{\betacft0 \ln Q^2/\Lambda^2}
  \left[
  1 - \frac{\betacft1}{\betacft0^2}
      \frac{\ln\ln Q^2/\Lambda^2}{\ln Q^2/\Lambda^2}
  \right] ,
  \label{eq:qcd-nlo-alpha}
\end{equation}
where $\Lambda$ is the \QCD scale parameter.

A generic observable derived from the operator $\Oper$ may be defined by
$f(Q^2)\sim\braket{PS}{\Oper(Q^2)}{PS}$. Inserting the above expansions into
(\ref{eq:qcd-RGEsoln}), the \NLO evolution equation for $f(Q^2)=\vev{\Oper}$ is
then obtained (note that the equations apply directly to $\Oper$ if it already
represents a physical observable):
\begin{align}
  \frac{f(Q^2)}{f(\mu^2)}
  &=
  \left(
    \frac{\alpha_s(Q^2)}{\alpha_s(\mu^2)}
  \right)^{-\frac{2\gamma^{(0)}}{\betacft0}}
  \left[
    \frac{\betacft0+\betacft1\alpha_s(  Q^2)/4\pi}
         {\betacft0+\betacft1\alpha_s(\mu^2)/4\pi}
  \right]^{-\left(\frac{4\gamma^{(1)}}{\betacft1}
                 -\frac{2\gamma^{(0)}}{\betacft0}
           \right)} ,
  \label{eq:qcd-nlo-n-evol-0}
\\
\intertext{which, to \NLO accuracy, may be expanded thus}
  &\simeq
  \left(
    \frac{\alpha_s(Q^2)}{\alpha_s(\mu^2)}
  \right)^{-\frac{2\gamma^{(0)}}{\betacft0}}
  \left[
    1 + \frac{\alpha_s(\mu^2)-\alpha_s(Q^2)}{\pi\betacft0}
    \left(
      \gamma^{(1)} - \frac{\betacft1}{2\betacft0} \, \gamma^{(0)}
    \right)
  \right] . \hspace{3em}
  \tag{\ref{eq:qcd-nlo-n-evol-0}$'$}
  \label{eq:qcd-nlo-n-evol}
\end{align}
In order to obtain physical hadronic cross-sections at \NLO level, the \NLO
contribution to $f$ has then to be combined with the \NLO contribution to the
relevant hard partonic cross-section; indeed, it is only this combination that
is fully scheme independent.

It turns out that the quantities typically measured experimentally (\ie,
cross-sections, \DIS structure functions or, more simply, quark distributions)
are related via Mellin-moment transforms to expectation values of composite
quark and gluon field operators. The definition we adopt for the Mellin
transform of structure functions, anomalous dimension \etc.\ is as
follows:\footnote{The definition with $n{-}1$ replaced by $n$ is also found in
the literature.}$\!^,$\footnote{We choose to write $n$ as an argument to avoid
confusion with the label indicating perturbation order.}
\begin{equation}
  f(n) = \int_0^1 \d{x} \, x^{n-1} f(x) \, .
  \label{eq:mellin}
\end{equation}

We may also define the \AP splitting function, $P(x)$, as precisely the
function of which the Mellin moments, eq.~(\ref{eq:mellin}), are just the
anomalous dimensions, $\gamma(n)$. Note that $P(x)$ may be expanded in powers
of the \QCD coupling constant in a manner analogous to $\gamma$ and therefore
also depends on $Q^2$. In $x$-space the evolution equations may be written in
the following schematic form:
\begin{equation}
  \frac{\d}{\d\ln Q^2} \, f(x,Q^2)
  =
  P(x,Q^2) \otimes f(x,Q^2) \, ,
  \label{eq:qcd-x-evol}
\end{equation}
where the symbol $\otimes$ stands for a convolution in $x$,
\begin{equation}
  g(x) \otimes f(x)
  =
  \int_x^1 \frac{\d{y}}{y} \, g\!\left(\frac{x}{y}\right) f(y) \, ,
  \label{eq:qcd-convolution}
\end{equation}
which becomes a simple product in Mellin-moment space. With these expressions
it is then possible to perform numerical evolution either via direct
integration of (\ref{eq:qcd-x-evol}) using suitable parametric forms to fit
data, or in the form of (\ref{eq:qcd-nlo-n-evol}) via inversion of the Mellin
moments.

The operators governing the twist-two\footnote{There are, of course, possible
higher-twist contributions too, but we shall ignore these here.} evolution of
moments of the $f_1$, $g_1$ and $h_1$ structure functions (in this section we
shall use $f_1$, $g_1$ and $h_1$ to generically indicate unpolarised, helicity
and transversity weighted parton densities respectively) are\footnote{All
composite operators appearing herein are to be considered implicitly
traceless.}
\begin{subequations}
\begin{eqnarray}
  \Oper^{f_1}(n)
  &=&
  \mathcal{S} \,
    \anti\psi
      \gamma^{\mu_1}
      \I\CovDer^{\mu_2} \dots \I\CovDer^{\mu_n}
    \psi \, ,
  \label{eq:qcd-f1-op}
\\
  \Oper^{g_1}(n)
  &=&
  \mathcal{S} \,
    \anti\psi
      \gamma_5 \gamma^{\mu_1}
      \I\CovDer^{\mu_2} \dots \I\CovDer^{\mu_n}
    \psi \, ,
  \label{eq:qcd-g1-op}
\\
  \Oper^{h_1}(n)
  &=&
  \mathcal{S}' \,
    \anti\psi
      \gamma_5 \gamma^{\mu_1} \gamma^{\mu_2}
      \I\CovDer^{\mu_3} \dots \I\CovDer^{\mu_n}
    \psi \, .
  \label{eq:qcd-h1-op}
\end{eqnarray}
\end{subequations}
where the symbol $\mathcal{S}$ conventionally indicates symmetrisation over the
indices $\mu_1,\mu_2,\dots\mu_n$ while the symbol $\mathcal{S}'$ indicates
simultaneous \emph{anti}symmetrisation over the indices $\mu_1$ and $\mu_2$ and
symmetrisation over the indices $\mu_2,\mu_3,\dots\mu_n$.

\subsection{QCD evolution at leading order}
\label{qcdleading}

The $Q^2$ evolution coefficients for $f_1$ and $g_1$ at \LO have long been known
while the \LO $Q^2$ evolution for $h_1$ was first specifically presented in
\cite{Artru:1990zv}. However, the first calculations of the one-loop anomalous
dimensions related to the operators governing the evolution of $\DT{q}(x,Q^2)$
date back, in fact, to early (though incomplete) work on the evolution of the
transverse-spin \DIS structure function $g_2$ \cite{Kodaira:1979ib}, which,
albeit in an indirect manner, involves the operators of interest here. Mention
(though again incomplete) may also be found in \cite{Antoniadis:1981dg}.
Following this, and with various approaches, the complete derivation of the
complex system of evolution equations for the twist-three operators governing
$g_2$ was presented \cite{Bukhvostov:1983te, Bukhvostov:1985rn,
Ratcliffe:1985mp}. Among the operators mixing with the leading contributions
one finds the following:
\begin{equation}
  \Oper(n)
  =
  \mathcal{S}'
  m \anti\psi
    \gamma_5 \gamma^{\mu_1} \gamma^{\mu_2}
    \I\CovDer^{\mu_3} \dots \I\CovDer^{\mu_n}
  \psi \, ,
\label{eq:qcd-qqma_m}
\end{equation}
where $m$ is the (current) quark mass. It is immediately apparent that this is
none other than the twist-two operator responsible for $\DT{q}(x)$, multiplied
here however by a quark mass and thereby rendered twist three---its evolution
is, of course, identical to the twist-two version.

For reasons already mentioned, see for example Eq.~(\ref{forward9}) and the
discussion following, calculation of the anomalous dimensions governing
$\DT{q}(x)$ turns out to be surprisingly simpler than for the other twist-two
structure functions, owing to its peculiar chiral-odd properties. Indeed, as we
have seen, the gluon field cannot contribute at \LO in the case of spin-half
hadrons as it would require helicity flip of two units in the corresponding
hadron--parton amplitude. Thus, in the case of baryons the evolution is of a
purely non-singlet type. Note that this is no longer the case for targets of
spin greater than one half and, as pointed out in \cite{Artru:1990zv}, a
separate contribution due to linear gluon polarisation is possible; we shall
also consider the situation for spin-1 mesons and/or indeed photons in what
follows.

\begin{figure}[htbp]
  \centering
  \begin{picture}( 300, 100)(   0,   0)
    \def\operbase{%
      \BCirc     (   0,  90){7}
      \Line      (   5,  85)(  -5,  95)
      \Line      (   5,  95)(  -5,  85)
      \Line      (  -5,  85)( -35,  15)
      \Line      (   5,  85)(  35,  15)
    }
    \SetOffset   (  30,   0)
    \operbase
    \Gluon       ( -20,  50)(  20,  50){2}{6}
    \Text        (   0,   0)[c]{(a)}
    \SetOffset   ( 150,   0)
    \operbase
    \GlueArc     ( -20,  50)(15,67,247){2}{6}
    \Text        (   0,   0)[c]{(b)}
    \SetOffset   ( 270,   0)
    \operbase
    \GlueArc     (  25,  78)(25,168.5,255){2}{5}
    \Text        (   0,   0)[c]{(c)}
  \end{picture}
  \caption{Example one-loop diagrams contributing to the $\Ord(\alpha_s)$
           anomalous dimensions of $\DT{q}$.}
  \label{fig:ope-ren-lo}
\end{figure}
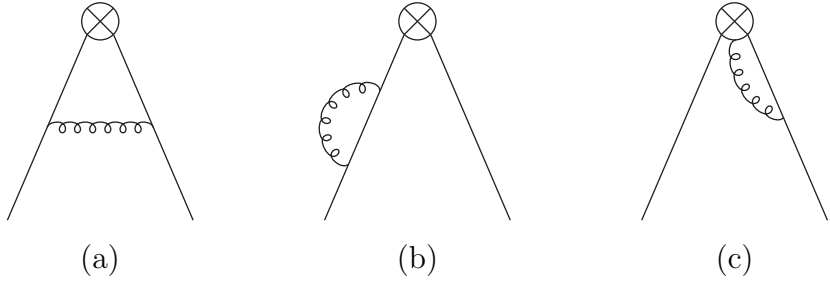

In dimensional regularisation ($d=4-\epsilon$ dimensions) calculation of the
anomalous dimensions requires evaluation of the $1/\epsilon$ poles in the
diagrams depicted in Fig.~\ref{fig:ope-ren-lo} (recall that at this order there
is no scheme dependence). Although not present in baryon scattering, the
linearly polarised gluon distribution, $\DT{g}$, can contribute to scattering
involving polarised spin-1 mesons. Thus, to complete the discussion of
leading-order evolution, we include here too the anomalous dimensions for this
density. For the four cases that are diagonal in parton type (spin-averaged and
helicity-weighted \cite{Altarelli:1977zs}; transversity and linear gluon
polarisation \cite{Artru:1990zv}) one finds in Mellin-moment space:
\begin{subequations}
\begin{eqnarray}
  \gamma_{qq}(n)
  &=&
  \CF \left[ \frac32 + \frac1{n(n+1)} - 2\sum_{j=1}^n\frac1j \right] ,
  \label{eq:qcd-qqf1lo}
\\[1ex]
  \DL\gamma_{qq}(n)
  &=&
  \gamma_{qq}(n) \, ,
  \label{eq:qcd-qqg1lo}
  \sublabel{subeq:qcd-qqg1lo}
\\[1ex]
  \DT\gamma_{qq}(n)
  &=&
  \CF \left[ \frac32 - 2\sum_{j=1}^n\frac1j \right]
  \nonumber
\\
  &=&
  \gamma_{qq}(n) - \CF \frac1{n(n+1)} \, ,
  \label{eq:qcd-qqh1lo}
\\[1ex]
  \DT\gamma_{gg}(n)
  &=&
  \CG \left[ \frac{11}{6n} - 2\sum_{j=1}^n\frac1j \right] - \frac23 \TF \, .
  \label{eq:qcd-ggh1lo}
\end{eqnarray}
\end{subequations}
The equality expressed in eq.~(\ref{eq:qcd-qqg1lo}) is a direct consequence of
fermion-helicity conservation by purely vector interactions in the limit of
negligible fermion mass.

The first point to stress is the commonly growing negative value, for
increasing $n$, indicative of the tendency of all the $x$-space distributions
to migrate towards $x=0$ with increasing $Q^2$. In other words, evolution has a
degrading effect on the densities. Secondly, in contrast to the behaviour of
both $q$ and $\DL{q}$, the anomalous dimensions governing $\DT{q}$ do
\emph{not} vanish for $n=0$ and hence there is no sum rule associated with the
tensor charge \cite{Jaffe:1992ra}. Moreover, comparison to $\DL{q}$ reveals
that $\DT\gamma_{qq}(n)<\DL\gamma_{qq}(n)$ for all $n$. This implies that for
(hypothetically) identical starting distributions (\ie,
$\DT{q}(x,Q_0^2)=\DL{q}(x,Q_0^2)$), $\DT{q}(x,Q^2)$ everywhere in $x$ will fall
more rapidly than $\DL{q}(x,Q^2)$ with increasing $Q^2$. We shall return in more
detail to this point in Sec.~\ref{sofferevol}.

For completeness, we also present the \AP splitting functions (\ie, the
$x$-space version of the anomalous dimensions)
$\gamma(n)=\int_0^1\d{x}\,x^{n-1}P(x)$, for the pure fermion sector:
\begin{subequations}
\begin{eqnarray}
  P_{qq}^{(0)}
  &=&
  \CF \left( \frac{1+x^2}{1-x} \right)_+ \, ,
  \label{eq:qcd-skf1lo}
\\[0.5ex]
  \DL{P}_{qq}^{(0)}
  &=&
  P_{qq}^{(0)} \, ,
  \label{eq:qcd-skg1lo}
\\[0.5ex]
  \DT{P}_{qq}^{(0)}
  &=&
  \CF \left[ \left( \frac{1+x^2}{1-x} \right)_+ - 1 + x \right]
  \nonumber
\\[0.5ex]
  &=&
  P_{qq}^{(0)}(z) - \CF (1-z) \, ,
  \label{eq:qcd-skh1lo}
\end{eqnarray}
\end{subequations}
where the ``plus'' regularisation prescription is defined in
Appendix~\ref{sec:mellin} in eq.~(\ref{eq:plus-def}) and we also have made use
of the identity given in eq.~(\ref{eq:plus-id}). Naturally, the plus
prescription is to be ignored when multiplying functions that vanish at $x=1$.
Once again, the inequality $\DT{P}_{qq}^{(0)}<\DL{P}_{qq}^{(0)}$ is manifest
for all $x<1$, indeed, one has
\begin{equation}
  \half
  \left[
         P_{qq}^{(0)}(x)
  + \DL{P}_{qq}^{(0)}(x)
  \right]
  - \DT{P}_{qq}^{(0)}(x)
  =
  \CF (1-x)
  \geq
  0 \, ,
  \label{eq:qcd-soffer-evol-lo}
\end{equation}

\begin{figure}[htbp]
  \centering
  \begin{picture} ( 300, 100)(   0,   0)
    \SetOffset      (  30,   0)
    \ArrowLine      ( -35,  55)( -35,  95)
    \ArrowLine      (  35,  95)(  35,  55)
    \ArrowLine      ( -35,  15)( -35,  55)
    \ArrowLine      (  35,  55)(  35,  15)
    \Gluon          ( -35,  55)(  35,  55){2}{6}
    \Text           (   0,   0)[c]{(a)}
    \SetOffset      ( 150,   0)
    \ArrowLine      ( -35,  55)( -35,  95)
    \ArrowLine      (  35,  95)(  35,  55)
    \Gluon          ( -35,  15)( -35,  55){2}{4}
    \ArrowLine      ( -37,  33)( -37,  37)
    \Gluon          (  35,  55)(  35,  15){2}{4}
    \ArrowLine      (  37,  37)(  37,  33)
    \ArrowLine      ( -35,  55)(  35,  55)
    \Text           (   0,   0)[c]{(b)}
    \SetOffset      ( 270,   0)
    \Gluon          ( -35,  55)( -35,  95){2}{4}
    \ArrowLine      ( -37,  73)( -37,  77)
    \Gluon          (  35,  95)(  35,  55){2}{4}
    \ArrowLine      (  37,  77)(  37,  73)
    \ArrowLine      ( -35,  15)( -35,  55)
    \ArrowLine      (  35,  55)(  35,  15)
    \ArrowLine      ( -35,  55)(  35,  55)
    \Text           (   0,   0)[c]{(c)}
  \end{picture}
  \caption{The \protect\OPI kernels contributing to the $\Ord(\alpha_s)$
           evolution of $\DT{q}$ in the axial gauge.}
  \label{fig:1PI-kernel}
\end{figure}
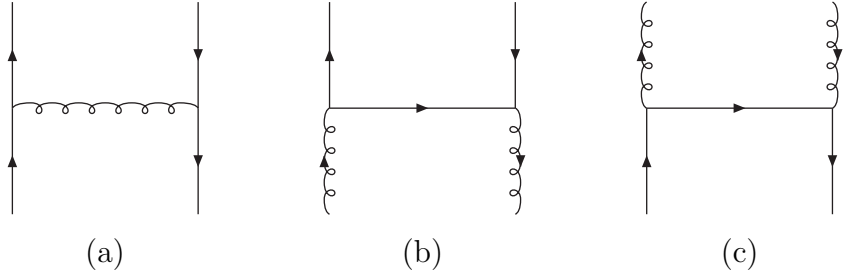

The non-mixing of the transversity distributions for quarks, $\DT{q}$, and
gluons, $\DT{g}$, is afforded a physical demonstration via the ladder-diagram
summation technique. In Fig.~\ref{fig:1PI-kernel} the general leading-order
\OPI kernels are displayed. If the four external lines are all quarks (\ie, a
gluon rung, see Fig.~\ref{fig:1PI-kernel}a), the kernel is clearly diagonal (in
parton type) and therefore contributes to the evolution of $\DT{q}$. For the
case in which one pair of external lines are quarks and the other gluons (\ie,
a quark rung, see Figs.~\ref{fig:1PI-kernel}b and c), helicity conservation
along the quark line in the chiral limit implies a vanishing contribution to
transversity evolution. Likewise, the known properties of four-body amplitudes,
namely t-channel helicity conservation, preclude any contribution that might
mix the evolution of $\DT{q}$ and $\DT{g}$.

The same reasoning clearly holds at higher orders since the only manner for
gluon and quark ladders to mix is via diagrams in which an incoming quark line
connects to its Hermitian-conjugate partner. Thus, quark-helicity conservation
in the chiral limit will always protect against such contributions.

Before continuing to \NLO, a comment is in order here on the recent debate in
the literature \cite{Meyer-Hermann:2000cy, Mukherjee:2001zx, Blumlein:2001ca}
regarding the calculation of the anomalous dimensions for $h_1$ and the
validity of certain approaches. The authors of \cite{Meyer-Hermann:2000cy}
attempted to calculate the anomalous dimensions relevant to $h_1$ exploiting a
method based on \cite{Ioffe:1995aa}. The motivation was that use of so-called
time-ordered or old-fashioned perturbation theory in the Weizs\"{a}cker--Williams
approximation \cite{vonWeizsacker:1934sx, Williams:1934ad} (as adopted in, \eg,
\cite{Artru:1990zv}) encounters a serious difficulty: it is only applicable to
the region $x<1$ while the end-point ($x=1$) contributions cannot be evaluated
directly. Where there is a conservation law (\eg, quark number), then appeal to
the resulting sum rule allows indirect extraction at this point (the common
$\delta$-function contribution). In the case of transversity no such conserved
quantum number exists and one might doubt the validity of such calculations.
Indeed, the claim in \cite{Meyer-Hermann:2000cy} was that direct calculation,
based on a dispersion relation approach, yielded a different result to that
reported in eq.~(\ref{eq:qcd-skh1lo}) above.

\emph{A priori}, from a purely theoretical point of view, such an apparent
discrepancy is hard to credit: were it real, it would imply precisely the type
of ambiguity to which the singularity structure of the theory on the light-cone
is supposedly immune. In \cite{Meyer-Hermann:2000cy} the anomalous dimensions
are calculated via the one-loop corrections to the Compton amplitude or classic
handbag diagram (\eg, see Fig.~\ref{handbag} with on-shell external quark
states and no lower hadronic blob). In order to mimic the required chiral
structure, one of the upper vertices is taken to be $\gamma_5\gamma^\mu$ and
the other $\gamma^\mu$ or $\unitop$ for $g_1$ or $h_1$ respectively. The
results for $g_1$ are in agreement with other approaches while the anomalous
dimensions for $h_1$ differ in the coefficient of the $\delta$-function
contribution. What immediately casts doubt on such findings is the fact that in
a physical gauge, as used for example in the ladder-diagram summation approach
\cite{Dokshitzer:1978hw, Craigie:1980fa} mentioned earlier, precisely all such
vertex corrections are in fact \emph{absent} (to this order in \QCD). Moreover,
it is just this property that gives rise, in that approach, to the universal
short-distance behaviour, \emph{independent} of the particular nature of the
vertices involved.

Various cross checks of these potentially disturbing findings have been
performed \cite{Mukherjee:2001zx, Blumlein:2001ca} with the conclusion that the
original calculations are after all correct. In particular, Bl\"{u}mlein
\cite{Blumlein:2001ca} has produced a very thorough appraisal of the situation.
Moreover, he has uncovered a fatal conceptual oversight: the scalar current is
\emph{not} conserved.\footnote{We are particularly grateful to Johannes
Bl\"{u}mlein for illuminating discussion on this point.} To appreciate the
relevance of this observation  let us briefly recall the salient points of the
\RGE approach (for more detail the reader is referred to
\cite{Blumlein:2001ca}).

A product of two currents (as in the peculiar Compton amplitude under
consideration) may be expanded as
\begin{eqnarray}
  j_1(z) j_2(0) = \sum_n C(n;z) \, \Oper(n;0) \, ,
\end{eqnarray}
where typically then $j_i=j_{V,A}$ (\ie, vector or axial vector currents) but
here a scalar current $j_S$ must be introduced. The \RGE for the Wilson
coefficients $C(n;z)$ is
\begin{eqnarray}
  \left[
    \mathcal{D} +
    \gamma_{j_1}(g) +
    \gamma_{j_2}(g) -
    \gamma_{\Oper}(n;g)
    \strut
  \right] C(n;z)
  = 0 \, ,
\end{eqnarray}
where $\gamma_{j_i}(g)$ and $\gamma_{\Oper}(n;g)$ are the anomalous dimensions
of the currents $j_i$ and the composite operators $\Oper(n)$ respectively, and
(neglecting quark masses) the RG operator is defined as
\begin{eqnarray}
  \mathcal{D} =
  \mu^2 \frac{\partial}{\partial\mu^2} +
  \beta(g) \, \frac{\partial}{\partial{g}} \, .
\end{eqnarray}
Thus, the \LL corrections to the Compton amplitude have coefficients
\begin{eqnarray}
  \gamma_C(n;g) =
  \gamma_{j_1}(g) +
  \gamma_{j_2}(g) -
  \gamma_{\Oper}(n;g) \, .
\end{eqnarray}
The point then is that while in the better-known spin-averaged and
helicity-weighted cases $\gamma_{j_i}=0$ for both currents (axial and/or
vector), the scalar current necessary for the transversity case is \emph{not}
conserved and $\gamma_{j_S}\not=0$. Thus, in contrast to the former, $\gamma_C$
and $-\gamma_{\Oper}$ do \emph{not} coincide in the calculation for $h_1$. The
discrepancy in \cite{Meyer-Hermann:2000cy} is due precisely to the neglect of
$\gamma_{j_S}$.

\subsection{QCD evolution at next-to-leading order}
\label{qcdnext}

Reliable \QCD analysis of the sort of data samples we have come to expect in
modern experiments requires full \NLO accuracy. For this it is necessary to
calculate both the anomalous dimensions to two-loop level and the constant
terms (\ie, the part independent of $\ln{Q^2}$) of the so-called coefficient
function (or hard-scattering process) at the one-loop level, together, of
course, with the two-loop $\beta$-function.

The two-loop anomalous-dimension calculation for $h_1$ has now been presented
in three papers: \cite{Hayashigaki:1997dn, Kumano:1997qp} using the MS scheme
in the Feynman gauge and \cite{Vogelsang:1998ak} using the $\MSbar$ scheme in
the light-cone gauge. These complement the earlier two-loop calculations for
the better-known twist-two structure functions: $f_1$ \cite{Floratos:1977au,
Floratos:1979ny, Gonzalez-Arroyo:1979df, Curci:1980uw, Furmanski:1980cm,
Floratos:1981hk, Floratos:1981hm} and $g_1$ \cite{Mertig:1996ny,
Vogelsang:1996vh, Vogelsang:1996im}. Such knowledge has been exploited in the
past for the phenomenological parametrisation of $f_1$ \cite{Gluck:1995uf,
Martin:1995ws, Lai:1997mg} and $g_1$ \cite{Gluck:1996yr, Gehrmann:1996ag,
Altarelli:1997nm} in order to perform global analyses of the experimental data;
and will certainly be of value when the time comes to analyse data on
transversity.

\begin{figure}[htbp]
  \centering
  \begin{picture} ( 300, 210)(   0,-110)
    \def\operbase{%
      \BCirc        (   0,  90){7}
      \Line         (   5,  85)(  -5,  95)
      \Line         (   5,  95)(  -5,  85)
      \Line         (  -5,  85)( -35,  15)
      \Line         (   5,  85)(  35,  15)
    }
    \SetOffset      (  30,   0)
    \operbase
    \Gluon          (-15.5,  60)(15.5,  60){2}{4}
    \Gluon          ( -24,  40)(  24,  40){2}{6}
    \Text           (   0,   5)[c]{(a)}
    \SetOffset      ( 150,   0)
    \operbase
    \Gluon          ( -20,  50)(  20,  50){2}{5}
    \GlueArc        ( -20,  50)(15,67,247){2}{5}
    \Text           (   0,   5)[c]{(b)}
    \SetOffset      ( 270,   0)
    \operbase
    \GlueArc        (  25,  78)(25,168.5,255){2}{4}
    \GlueArc        (  20,  50)(15,-67,113){2}{5}
    \Text           (   0,   5)[c]{(c)}
    \SetOffset      (  30,-110)
    \operbase
    \Gluon          (-15.5,  60)(  24,  40){2}{5}
    \Gluon          ( -24,  40)(15.5,  60){2}{5}
    \Text           (   0,   5)[c]{(d)}
    \SetOffset      ( 150,-110)
    \operbase
    \Gluon          (   0,  83)(   0,  42){2}{4}
    \Gluon          ( -24,  40)(  24,  40){2}{6}
    \Text           (   0,   5)[c]{(e)}
    \SetOffset      ( 270,-110)
    \operbase
    \Gluon          (-15.5,  60)(15.5,  60){2}{4}
    \GlueArc        (  35,  78)(35,172,248){2}{6}
    \Text           (   0,   5)[c]{(f)}
  \end{picture}
  \caption{Example two-loop diagrams contributing at $\Ord(\alpha_s^2)$ to the
           anomalous dimensions of $\DT{q}$.}
  \label{fig:ope-ren-nlo}
\end{figure}
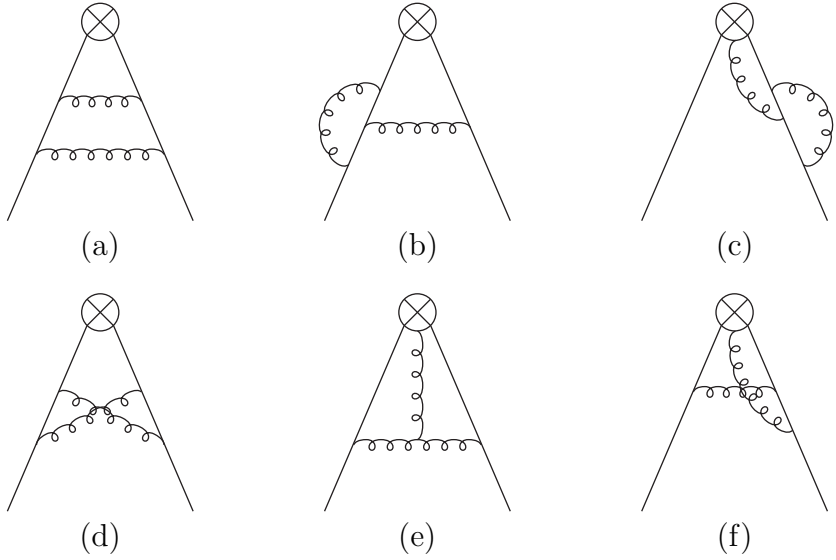

The situation at \NLO is still relatively simple, as compared to the
unpolarised or helicity-weighted cases. Examples of the relevant two-loop
diagrams are shown in Fig.~\ref{fig:ope-ren-nlo}. It remains impossible for the
gluon to contribute, for the reasons already given. The only complication is
the usual mixing, possible at this level, between quark and antiquark
distributions, for which quark helicity conservation poses no restriction since
the quark and antiquark lines do not connect directly to one another, see
Fig.~\ref{fig:ope-ren-nlo}d.

It is convenient to introduce the following combinations of quark transversity
distributions (the $\pm$ subscript is not to be confused with helicity):
\begin{subequations}
\begin{eqnarray}
  \DT{q}_{\pm}(n)
  &=&
  \DT{q}(n) \pm \DT\anti{q}(n) \, ,
  \label{eq:qcd-5.1}
\\
  \DT\widetilde{q}_+(n)
  &=&
  \DT{q}_+(n) - \DT{q}_+^\prime(n) \, ,
  \label{eq:qcd-5.2}
\\
  \DT\Sigma(n)
  &=&
  \sum_q \DT{q}_+(n) \, ,
  \label{eq:qcd-5.3}
\end{eqnarray}
\end{subequations}
where $q$ and $q'$ represent quarks of differing flavours. The specific
evolution equations may then be written as (\eg, see\cite{Ellis:1996nn})
\begin{subequations}
\begin{eqnarray}
  \frac{\d}{\d\ln Q^2} \DT{q}_-(n,Q^2)
  &=&
  \DT\gamma_{qq,-}(n,\alpha_s(Q^2)) \, \DT{q}_-(n,Q^2) \, ,
  \label{eq:qcd-evol-minus}
\\
  \frac{\d}{\d\ln Q^2} \DT\widetilde{q}_+(n,Q^2)
  &=&
  \DT\gamma_{qq,+}(n,\alpha_s(Q^2)) \, \DT\widetilde{q}_+(n,Q^2) \, ,
  \label{eq:qcd-evol-plus}
\\
  \frac{\d}{\d\ln Q^2} \DT\Sigma(n,Q^2)
  &=&
  \DT\gamma_{\Sigma\Sigma}(n,\alpha_s(Q^2)) \, \DT\Sigma(n,Q^2) \, ,
  \sublabel{subeq:qcd-evol-singlet}
\end{eqnarray}
\end{subequations}
Note that the first moment ($n=1$) in eq.~(\ref{eq:qcd-evol-minus}) corresponds
to evolution of the nucleon's tensor charge \cite{Jaffe:1991kp, Jaffe:1992ra,
Schmidt:1997vm}. The splitting functions $\DT\gamma_{qq,{\pm}}$ and
$\DT\gamma_{\Sigma\Sigma}$ have expansions in powers of the coupling constant
that take the following form:
\begin{equation}
  \DT\gamma_{ii}(n,\alpha_s)
  =
  \left( \frac{\alpha_s}{2\pi} \right)
  \DT\gamma_{qq}^{(0)}(n)
  +
  \left( \frac{\alpha_s}{2\pi} \right)^2
  \DT\gamma_{ii}^{(1)}(n) + \dots \, ,
  \label{eq:qcd-5.4}
\end{equation}
where $\{ii\}=\{qq,{\pm}\}$, $\{\Sigma\Sigma\}$ and we have taken into account the
fact that $\DT\gamma_{qq,+}$, $\DT\gamma_{qq,-}$ and $\DT\gamma_{\Sigma\Sigma}$
are all equal at \LO. It is convenient then to introduce\cite{Ellis:1996nn}
\begin{subequations}
\begin{eqnarray}
  \DT\gamma_{qq,{\pm}}^{(1)}(n)
  &=&
  \DT\gamma_{qq}^{(1)}(n) \pm \DT\gamma_{q\anti{q}}^{(1)}(n) \, ,
  \label{eq:qcd-gamma-pm}
\\
  \DT\gamma_{\Sigma\Sigma}^{(1)}(n)
  &=&
  \DT\gamma_{qq,+}^{(1)}(n) + \DT\gamma_{qq,PS}^{(1)}(n) \, .
  \label{eq:qcd-5.5}
\end{eqnarray}
\end{subequations}
Since it turns out that $\DT\gamma_{qq,PS}^{(1)}(n)=0$, the two evolution
eqs.~(\ref{eq:qcd-evol-plus},~\ref{subeq:qcd-evol-singlet}) may be replaced by
a single equation:
\begin{equation}
  \frac{\d}{\d\ln Q^2} \, \DT{q}_+(n,Q^2)
  =
  \DT\gamma_{qq,+}(n,\alpha_s(Q^2)) \, \DT{q}_+(n,Q^2) \, .
  \tag{\ref{eq:qcd-evol-plus}$'$}
  \label{eq:qcd-evol-simplified}
\end{equation}

The formal solution to eqs.~(\ref{eq:qcd-evol-minus}) and
(\ref{eq:qcd-evol-simplified}) is well-known (\eg, see \cite{Gluck:1982sc}) and
reads
\begin{eqnarray}
  \DT{q}_{\pm}(n,Q^2)
  &=&
  \left\{1 + \frac{\alpha_s(Q_0^2)
    - \alpha_s(Q^2)}{\pi\betacft0}
    \left[
      \DT\gamma_{qq,{\pm}}^{(1)}(n)
    - \frac{\betacft1}{2\betacft0} \DT\gamma_{qq}^{(0)}(n)
    \right]
  \right\}
  \nonumber
\\
  &&
  \times
  \left(
    \frac{\alpha_s(Q^2)}{\alpha_s (Q_0^2)}
  \right)^\frac{-2\DT\gamma_{qq}^{(0)}(n)}{\betacft0}
  \! \DT{q}_{\pm}(n,Q_0^2) \, ,
  \label{eq:qcd-5.6}
\end{eqnarray}
with the input distributions $\DT{q}_{\pm}(n,Q_0^2)$ given at the input scale $Q_0$.
Of course, the corresponding \LO expressions may be recovered from the above
expressions by setting the \NLO quantities, $\DT\gamma_{ij,{\pm}}^{(1)}$ and
$\betacft1$, to zero.

In the $\MSbar$ scheme the $\gamma^{(1)}(n)$ relevant to $h_1$ are as
follows:\footnote{Note that $\frac23S_1(n)$ in the second line of (A.8) in
\cite{Gonzalez-Arroyo:1979df} should read $\frac23S_3(n)$.}
\begin{eqnarray}
  \DT\gamma_{qq,\eta}^{(1)}(n)
  &=&
  \CF^2
  \left\{ \Strut
    \textfrac38
    + \textfrac2{n(n+1)} \delta_{\eta-}
    - 3 S_2(n)
    - 4 S_1(n)
    \left[
      S_2(n) - S'_2 \left( \textfrac{n}{2} \right)
    \right]
  \right.
  \nonumber
\\
  &&
  \hspace*{2em}
  \left. \Strut \null
  - 8 \widetilde{S}(n)
  + S'_3 \left( \textfrac{n}{2} \right)
  \right\}
  \nonumber
\\
  &+&
  \half \CF \Nc\left\{ \Strut
    \textfrac{17}{12} -
    \textfrac{2}{n(n+1)} \delta_{\eta-} -
    \textfrac{134}{9} S_1(n) +
    \textfrac{22}{3} S_2(n)
  \right.
  \nonumber
\\
  &&
  \hspace*{4em}
  \left. \Strut \null +
    4 S_1(n) \left[ 2 S_2(n) - S'_2 \left( \textfrac{n}{2} \right) \right] +
    8 \widetilde{S}(n) -
    S'_3 \left( \textfrac{n}{2} \right)
  \right\}
  \nonumber
\\
  &+&
  \textfrac23 \CF \TF
  \left\{ \Strut
    {-}\textfrac14 + \textfrac{10}{3}S_1(n) - 2S_2(n)
  \right\} ,
  \label{eq:qcd-nlo-gamma-h1}
\end{eqnarray}
where $\eta={\pm}$ and the $S$ functions are defined by
\begin{subequations}
\begin{eqnarray}
  \label{eq:qcd-s-sums}
  S_k(n)
  &=&
  \sum_{j=1}^n j^{-k},
\\
  S'_k \left( \textfrac{n}{2} \right)
  &=&
  2^k \kern-1em \sum_{j=2,\text{even}}^n j^{-k},
\\
  \widetilde{S}(n)
  &=&
  \sum_{j=1}^n (-1)^j S_1(j) j^{-2} \, .
\end{eqnarray}
\end{subequations}

\begin{figure}[htbp]
  \centering
  \includegraphics[width=\textwidth]{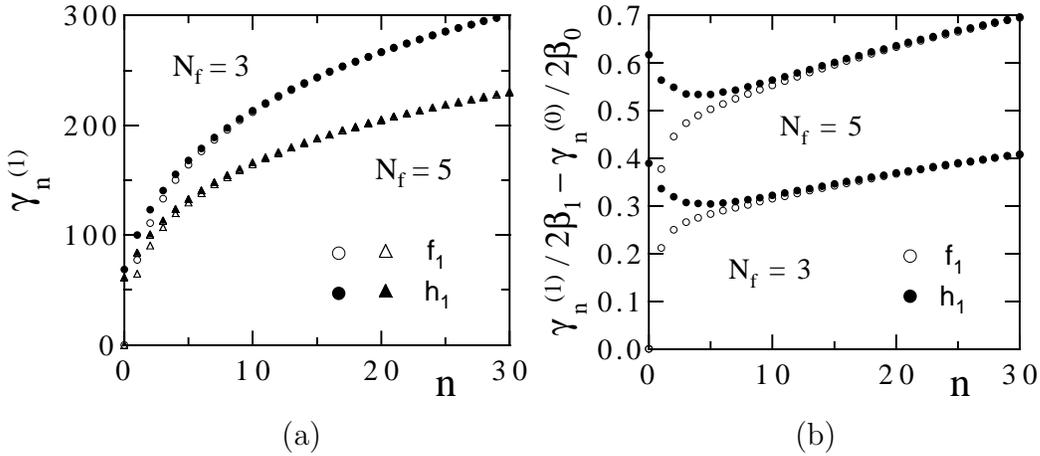}
  \hspace*{1cm}(a)\hspace{6.3cm}(b)
  \caption{Comparison between $f_1$ and $h_1$ of the variation with $n$ of
           (a) the two-loop anomalous dimensions $\gamma^{(1)}_n$ for $N_f=3$
           (circles) and 5 (triangles), and (b) the combination
           $\gamma^{(1)}(n)/2\beta_1-\gamma^{(0)}(n)/2\beta_0$ for $N_f=3$ and
           5; from~\cite{Hayashigaki:1997dn}.}
  \label{fig:HKK-fig23}
\end{figure}

\begin{figure}[htbp]
  \centering
  \includegraphics[width=\textwidth]{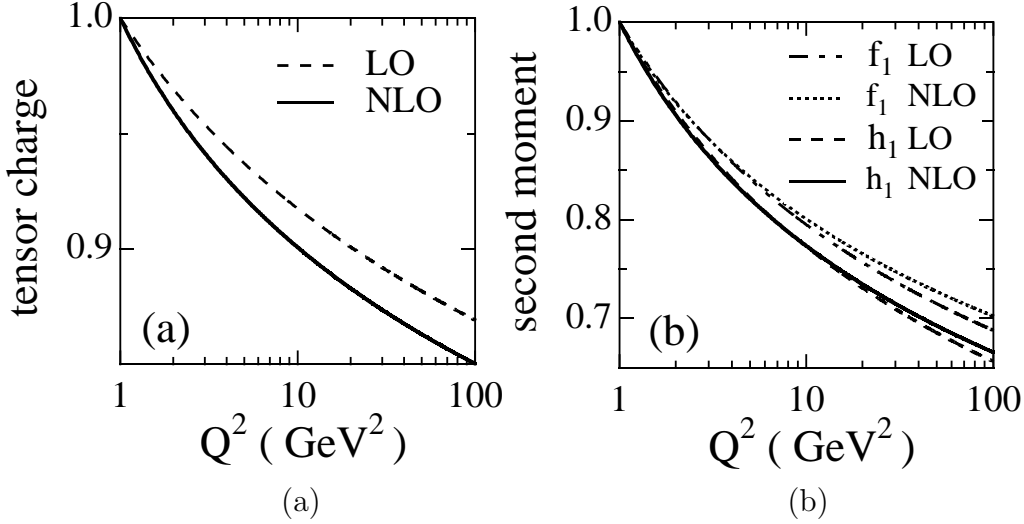}
  \hspace*{1.0cm}(a)\hspace{6.2cm}(b)
  \caption{The LO and NLO $Q^2$-evolution of (a) the tensor charge and (b) the
           second moments of $h_1(x,Q^2)$ and $f_1(x,Q^2)$ (both are normalised
           at $Q^2=1\GeV^2$), from~\cite{Hayashigaki:1997dn}.}
  \label{fig:HKK-fig4}
\end{figure}

In Fig.~\ref{fig:HKK-fig23} we show the $n$ dependence of the two-loop
anomalous dimensions (as presented in
\cite{Hayashigaki:1997dn}\footnote{According to the convention adopted for the
moments in \cite{Hayashigaki:1997dn}, $n=0$ there corresponds to $n=1$ in the
present report}). From the figure, one clearly sees that for $n$ small,
$\DT\gamma^{(1)}(n)$ is significantly larger than $\gamma^{(1)}(n)$ but, with
growing $n$, very quickly approaches $\gamma^{(1)}(n)$ while maintaining the
inequality $\DT\gamma^{(1)}(n)>\gamma^{(1)}(n)$. For the specific moments $n=1$
(corresponding to the tensor charge) and $n=2$, we display the $Q^2$ variation
in Fig.~\ref{fig:HKK-fig4}

To express the corresponding results in $x$ space it is convenient to introduce
the following definitions:\footnote{In order to avoid confusion with the
tensor-charge anomalous dimensions, the notation adopted here corresponding to
$\DDT{R}^{(0)}(x)$ is different than that commonly adopted.}
\begin{eqnarray}
  \DDT{R}^{(0)}(x)
  &=&
  \frac{2x}{(1-x)_+} \, ,
  \label{eq:qcd-delta-func}
\\
  S_2(x)
  &=&
  \int_{\frac{x}{1+x}}^{\frac1{1+x}} \frac{\d{z}}{z}
  \ln \left( \frac{1-z}{z} \right)
  \nonumber
\\
  &=&
  - 2\Li_2(-x) - 2\ln{x}\ln(1+x) + \textfrac12\ln^2x - \textfrac16\pi^2 \, ,
  \label{eq:qcd-s-function}
\end{eqnarray}
where $\Li_2(x)$ is the usual dilogarithm function. In the $\MSbar$ scheme,
defining
\begin{equation}
  \DT{P}_{qq,{\pm}}^{(1)}(x)
  =
  \DT{P}_{qq}^{(1)}(x) \pm \DT{P}_{q\anti{q}}^{(1)}(x) \, ,
  \label{eq:qcd-5.9}
\end{equation}
\cf eq.~(\ref{eq:qcd-gamma-pm}), one then has
\begin{subequations}
\begin{eqnarray}
  \DT{P}_{qq}^{(1)}(x)
  &=&
  \CF ^2
  \left\{ \Strut
    1 - x -
    \left[ \textfrac32 + 2\ln(1-x) \right] \ln{x} \, \DDT{R}^{(0)}(x)
  \right.
  \nonumber
\\
  &&
  \hspace*{5em}
  +
  \left. \Strut
    \left[ \textfrac38 - \half\pi^2 + 6\zeta(3) \right] \delta(1-x)
  \right\}
  \nonumber
\\
  &+&
  \half \CF \CG
  \left\{ \Strut
    - (1-x) +
    \left[
      \textfrac{67}{9} + \textfrac{11}{3}\ln{x} + \ln^2x - \textfrac{1}{3}\pi^2
    \right] \DDT{R}^{(0)}(x)
  \right.
  \nonumber
\\
  &&
  \hspace*{5em}
  +
  \left. \Strut
    \left[ \textfrac{17}{12} + \textfrac{11}{9} \pi^2 - 6\zeta(3) \right]
    \delta(1-x)
  \right\}
  \nonumber
\\
  &+&
  \textfrac23 \CF \TF
  \left\{ \Strut
    \left[ -\ln{x}- \textfrac53 \right] \DDT{R}^{(0)}(x)
  \right.
  \nonumber
\\
  &&
  \hspace*{5em}
  -
  \left. \Strut
    \left[ \textfrac14 + \textfrac{1}{3} \pi^2 \right] \delta(1-x)
  \right\} ,
  \label{eq:qcd-DTP1qq}
\\[1ex]
  \DT{P}_{q\anti{q}}^{(1)}(x)
  &=&
  \CF \left[ \CF - \half\CG \right]
  \left\{ \Strut
    - (1 -x) + 2 S_2(x) \, \DDT{R}^{(0)}(-x)
  \right\} ,
  \label{eq:qcd-DTP1qqbar}
  \sublabel{subeq:qcd-DTP1qqbar}
\end{eqnarray}
\end{subequations}
where $\zeta(3)\approx1.202057$ is the usual Riemann Zeta function. Note that
the plus prescription is to be ignored in $\DDT{R}^{(0)}(-x)$.

To complete this section we also report on the corresponding \NLO calculation
for linear (transverse) gluon polarisation \cite{Vogelsang:1998yd}. As already
noted, $\DT{g}$ is precluded in the case of spin-half hadrons -- it may,
however, be present in objects of spin one, such as the deuteron or indeed even
the photon\cite{Delduc:1980ef}.
\begin{eqnarray}
  \DT{P}_{gg}^{(1)}(x) &=&
  \CG^2 \left[
    \left(
      \textfrac{67}{18} +
      \textfrac12\ln^2x -
      2\ln{x}\ln(1-x) -
      \textfrac{1}{6}\pi^2
    \right) \DDT{R}^{(0)}(x)
    \right.
  \nonumber
\\
  && \hspace{1cm} \left. \null +
    \frac{1-x^3}{6x} + S_2(x) \DDT{R}^{(0)}(-x) +
    \left( \textfrac83 + 3\zeta(3) \right) \delta(1-x)
  \right]
  \nonumber
\\
  &+&
  \CG\TF
  \left[
   -\textfrac{10}{9}\DDT{R}^{(0)}(x) +
    \frac{1-x^3}{3x} -
    \textfrac43\delta(1-x)
  \right]
  \nonumber
\\
  &-&
  \CF\TF \left[ \frac{2(1-x^3)}{3x} + \delta(1-x) \right] \, .
  \label{eq:Pgg-nlo}
\end{eqnarray}
The corresponding expression in Mellin-moment space for the anomalous
dimensions is\footnote{We are very grateful to Werner Vogelsang for providing
us with the exact expression.}
\begin{eqnarray}
  \DT\gamma_{gg}^{(1)}(n)
  &=&
  \CG^2
  \left[
    \textfrac83 +
    \textfrac1{2(n-1)(n+2)} +
    S_1(n) \left( 2 S_2^{'}(\textfrac{n}2) - \textfrac{67}9 \right) +
    \textfrac12 S_3^{'}(\textfrac{n}2) - 4 \widetilde{S}(n)
  \right]
  \nonumber
\\
  &+&
  \CG \TF
  \left[
    -\textfrac43 + \textfrac{20}9 S_1(n) + \textfrac1{(n-1)(n+2)}
  \right]
  \nonumber
\\
  &-&
  \CF \TF \textfrac{n(n+1)}{(n-1)(n+2)} \, .
\end{eqnarray}
As noted in \cite{Vogelsang:1998yd} the result for the part $\sim\CF\TF$ in
(\ref{eq:Pgg-nlo}) was presented in \cite{Delduc:1980ef} for the region $x<1$
(corresponding to the two-loop splitting function for linearly polarised gluons
into \emph{photons}). However, the two calculations appear to be at variance:
the results of \cite{Delduc:1980ef} imply a small-$x$ behaviour of $\Ord(1/x^2)$
for the relevant splitting function, which would then be more singular than the
unpolarised case.

There are two aspects of the splitting function (\ref{eq:Pgg-nlo}) that warrant
particular comment. Firstly, the small-$x$ behaviour changes significantly on
going from \LO to \NLO. At \LO, the splitting function is $\Ord(x)$ for $x\to0$
whereas at \NLO there are $\Ord(1/x)$ terms (as in the unpolarised case): we
have
\begin{equation}
  \label{smallx1}
  \DT{P}_{gg}^{(1)}(x) \approx
  \frac1{6x} \left( \Nc^2 + 2\Nc\TF - 4\CF\TF \right) +
  \Ord(x) \qquad (x\to0) \, .
\end{equation}
Notice that all logarithmic terms ${\sim}x\ln^2x$ cancel in this limiting
region.

The second comment regards the so-called supersymmetric limit: namely
$\CF=\Nc=2\TF$ \cite{Bukhvostov:1985rn}, which was investigated for the
unpolarised and longitudinally polarised \NLO splitting functions in
\cite{Antoniadis:1981zv, Mertig:1996ny, Vogelsang:1996im, Vogelsang:1996vh},
for the time-like case in \cite{Stratmann:1997hn} and for the case of
transversity in \cite{Vogelsang:1998yd}. In the supersymmetric limit the \LO
splitting functions for quark transversity and for linearly polarised gluons
are equal\cite{Artru:1990zv, Delduc:1980ef}:
\begin{equation}
  \DT{P}_{qq}^{(0)}(x) =
  \Nc \left[ \frac{2x}{(1-x)_+} + \frac32 \delta(1-x) \right] =
  \DL{P}_{gg}^{(0)}(x) \, .
\end{equation}
Hence, we may consider linear polarisation of the gluon as the supersymmetric
partner to transversity (see also \cite{Bukhvostov:1985rn}). Indeed, as was
natural, we have already applied the terminology without distinction to
spin-half and spin-one.

In \cite{Vogelsang:1998yd} the check was performed that the supersymmetric
relation still holds at \NLO. To do so it is necessary to transform to a
regularisation scheme that respects supersymmetry, namely dimensional
reduction. As noted in \cite{Vogelsang:1998yd} the transformation is rendered
essentially trivial owing to the absence of $\Ord(\epsilon)$ terms in the
$d$-dimensional \LO splitting functions for transversity or linearly polarised
gluons at $x<1$; such terms are always absent in dimensional \emph{reduction}
but may be present in dimensional \emph{regularisation}. Thus, at \NLO the
results for the splitting functions for quark transversity -- see
eqs.~(\ref{eq:qcd-DTP1qq},\,\ref{subeq:qcd-DTP1qqbar}) -- and for linearly
polarised gluons, eq.~(\ref{eq:Pgg-nlo}), automatically coincide for $x<1$ with
their respective $\MSbar$ expressions in dimensional \emph{reduction}. These
expressions may therefore be immediately compared in the supersymmetric limit
and indeed for $\CF=\Nc=2\TF$
\begin{equation}
  \DT{P}_{qq,+}^{(1)}(x) \equiv \DT{P}_{gg}^{(1)}(x) \qquad (x<1) \, .
\end{equation}
Note, in addition, that the supersymmetric relation is trivially satisfied for
$x=1$; see~\cite{Vogelsang:1996im, Vogelsang:1996vh}, where the appropriate
factorisation-scheme transformation to dimensional reduction for $x=1$ is
given.

\subsection{Evolution of the transversity distributions}
\label{comparison}

The interest in the effects of evolution in the case of transversity is
two-fold: first, there is the obvious question of the relative magnitude of the
distributions at high energies given some low-energy starting point (\eg, a
non-perturbative model calculation, for a detailed discussion and examples see
Sec.~\ref{models}) and second is the problem raised by the Soffer inequality.
It is to the first that we now address our attention while we shall deal with
latter shortly.

Let us for the moment simply pose the question of the effect of \QCD evolution
\cite{Barone:1996un, Barone:1997fh, Barone:1997mj, Bourrely:1998bx,
Contogouris:1998tw, Hayashigaki:1997dn, Hirai:1997mm, Martin:1996sn,
Martin:1998rz, Martin:1999mg, Scopetta:1997mf, Vogelsang:1993jn,
Vogelsang:1998ak, Contogouris:1998iz, Hayashigaki:1997nw, Koike:1997sj,
Kumano:1997ng, Vogelsang:1998yd} on the overall magnitude of the transversity
densities that might be constructed at some low-energy scale. As already noted
above, there is no conservation rule associated with the tensor charge of the
nucleon (\cf the vector and axial-vector charges) and, indeed, the sign of the
anomalous dimensions at both \LO and \NLO is such that the first moment of
$h_1$ falls with increasing $Q^2$. Thus, one immediately deduces that the tensor
charge will eventually disappear in comparison to the vector and axial charges.
Such behaviour could have a dramatic impact on the feasibility of high-energy
measurement of $h_1$ and thus requires carefully study.

A analytic functional form for the \LO anomalous dimensions governing the
evolution of $h_1$ is
\begin{equation}
  \DT\gamma^{(0)}(n)
  =
  \textfrac43
  \left\{
    \textfrac32 - 2\left[ \strut \psi(n+1) + \gamma_\text{E} \right]
  \right\} ,
  \label{eq:qcd-func-form}
\end{equation}
where $\psi(z)=\frac{\d\ln\Gamma(z)}{\d{z}}$ is the digamma function and
$\gamma_\text{E}=0.5772157$ is the Euler--Mascheroni constant. Since
$\DT\gamma^{(0)}(1)=-\frac23$, the first moment of $h_1$ and the tensor
charges, $\dT{q}=\int_0^1\d{x}\,(\DT{q}-\DT\anti{q})$, decrease with $Q^2$ as
\begin{eqnarray}
  \dT{q}(Q^2)
  &=&
  \left[
    \frac{\alpha_s (Q^2)}{\alpha_s(Q_0^2)}
  \right]^{-2\DT\gamma_{qq}^{(0)}\!(1)/\betacft0} \,
  \dT{q}(Q_0^2)
  \nonumber
\\
  &=&
  \left[ \frac{\alpha_s(Q_0^2)}{\alpha_s(Q^2)} \right]^{-4/27} \,
  \dT{q}(Q_0^2) \, ,
  \label{eq:qcd-tens-chg-evol}
\end{eqnarray}
where, to obtain the second equality, we have set $\Nf=3$. Despite the
smallness of the exponent, $-4/27$, we shall see that the evolution of
$\DT{q}(x,Q^2)$ is rather different from that of the helicity distributions
$\DL{q}(x,Q^2)$, especially for small $x$. At \NLO this becomes
\begin{eqnarray}
  \dT{q}(Q^2)
  &=&
  \left[
    \frac{\alpha_s (Q^2)}{\alpha_s(Q_0^2)}
  \right]^{-2\DT\gamma_{qq}^{(0)}\!(1)/\beta_0}
  \nonumber
\\
  && \null \times
  \left[ 1 + \frac{\alpha_s(Q_0^2) - \alpha_s(Q^2)}{\pi\beta_0}
    \left(
      \DT\gamma_{qq,-}^{(1)}(1)
      - \frac{\beta_1}{2\beta_0}\,\DT\gamma_{qq}^{(0)}(1)
    \right)
  \right]
  \dT{q}(Q_0^2)
  \nonumber
\\
  &=&
  \left[
    \frac{\alpha_s (Q^2)}{\alpha_s(Q_0^2)}
  \right]^{\frac{4}{27}}
  \left[ \STRUT
    1 - \frac{337}{486\pi} \left( \alpha_s(Q_0^2) - \alpha_s(Q^2) \right)
  \right] ,
\end{eqnarray}
where in the second equality we have used
\begin{equation}
  \DT\gamma_{qq, -}^{(1)}(1)
  =
  \frac{19}{8} \, \CF^2 - \frac{257}{72} \, \CF \Nc + \frac{13}{18} \,
  \CF \TF = -\frac{181}{18} + \frac{13}{27} \, \Nf  \, ,
\end{equation}
and we have once again set $\Nf=3$.

Recall that the first moments of the $q\to{qg}$ polarised and unpolarised
splitting functions, vanish to all orders in perturbation theory and that the
$g\to{q}\anti{q}$ polarised anomalous dimension $\DL\gamma_{qg}(1)$ is zero at
\LO; thus, $\DL{q}(Q^2)$ is constant. This can be seen analytically by the
following argument \cite{Barone:1997fh} based on the double-log approximation.
The leading behaviour of the parton distributions for small $x$ is governed by
the rightmost singularity of their anomalous dimensions in Mellin-moment space.
From eq.~(\ref{eq:qcd-func-form}) we see that this singularity is located at
$n=-1$ for $\DT{q}$ at \LO. Expanding $\DT\gamma(n)$ around this point gives
\begin{equation}
  \DT\gamma(n) \sim \frac8{3(n+1)} + \Ord(1) \, .
  \label{eq:qcd-n-sing}
\end{equation}
Equivalently, in $x$-space, expanding the splitting function $\DT{P}$ in powers
of $x$ yields
\begin{equation}
  \DT{P}(x) \sim \frac83 x + \Ord(x^2) \, .
  \label{eq:qcd-x-sing}
\end{equation}
In contrast, the rightmost singularity for $\DL{q}$ in moment space is located
at $n=0$ and the splitting functions $\DL{P}_{qq}$ and $\DL{P}_{qg}$ behave as
constants as $x\to0$. Therefore, owing to \QCD evolution, $\DT{q}$ acquires and
an extra suppression factor of one power of $x$ with respect to $\DL{q}$ at
small $x$. We note too that at \NLO the rightmost singularity for $\DT{q}$ is
located at $n=0$, so that \NLO evolution renders the \DGLAP asymptotics for
$x\to0$ in the case of transversity compatible with Regge
theory\cite{Kirschner:1997jj}.

\begin{figure}[htbp]
  \centering
  \includegraphics[width=8cm]{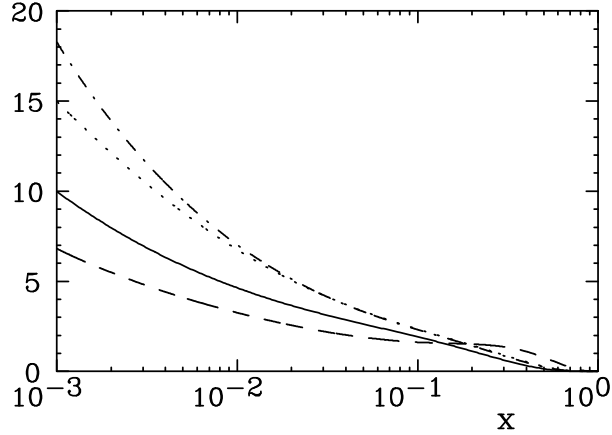}
  \caption{Evolution of the helicity and transversity distributions for the $u$
           flavour \cite{Barone:1997fh}. The dashed curve is the input
           $\DT{u}=\DL{u}$ at $Q_0^2=0.23\GeV^2$ taken from the GRV
           \cite{Gluck:1995yq} parametrisation. The solid (dotted) curve is
           $\DT{u}$ ($\DL{u}$) at $Q^2=25\GeV^2$. The dot--dashed curve is the
           result of the evolution of $\DT{u}$ at $Q^2=25\GeV^2$ driven by
           $P_{qq}$, \ie, with the term $\DT{P}$ in $P_h$ turned off.}
  \label{fig:h1evol}
\end{figure}

As mentioned earlier, this problem may be investigated numerically by
integrating the \DGLAP equations (\ref{eq:qcd-x-evol}) with suitable starting
input for $h_1$ and $g_1$. As a reasonable trial model one may assume the
various $\DT{q}$ and $\DL{q}$ to be equal at some small scale $Q_0^2$ and then
allow the two types of distributions to evolve separately, each according to
its own evolution equations. The input hypothesis
$\DT{q}(x,Q_0^2)=\DL{q}(x,Q_0^2)$ is suggested by various quark-model
calculations of $\DT{q}$ and $\DL{q}$ \cite{Jaffe:1992ra, Barone:1996un} (see
also Sec.~\ref{models} here), in which these two distributions are found to be
very similar at a scale $Q_0^2\lsim0.5\GeV^2$. For $\DL{q}(x,Q_0^2)$, we then use
the leading-order \GRV parametrisation \cite{Gluck:1995yq}, whose input scale
is $Q_0^2=0.23\GeV^2$. The result for the $u$-quark distributions is shown in
Fig.~\ref{fig:h1evol} (the situation is similar for the other flavours). The
dashed line is the input, the solid line and the dotted line are the results of
the evolution of $\DT{u}$ and $\DL{u}$, respectively, at $Q^2=25\GeV^2$. For
completeness, the evolution of $\DT{u}$ when driven only by $P_{qq}$ -- \ie,
with the $\DT{P}$ term turned off, see eq.~(\ref{eq:qcd-skh1lo}) -- is also
shown (dot--dashed line). The large difference in the evolution of $\DT{u}$
(solid curve) and $\DL{u}$ (dotted curve) at small $x$ is evident. Note also
the difference between the correct evolution of $\DT{u}$ and the evolution
driven purely by $P_{qq}$ (dot--dashed curve).

\begin{figure}[htbp]
  \centering
  \includegraphics[width=\textwidth]{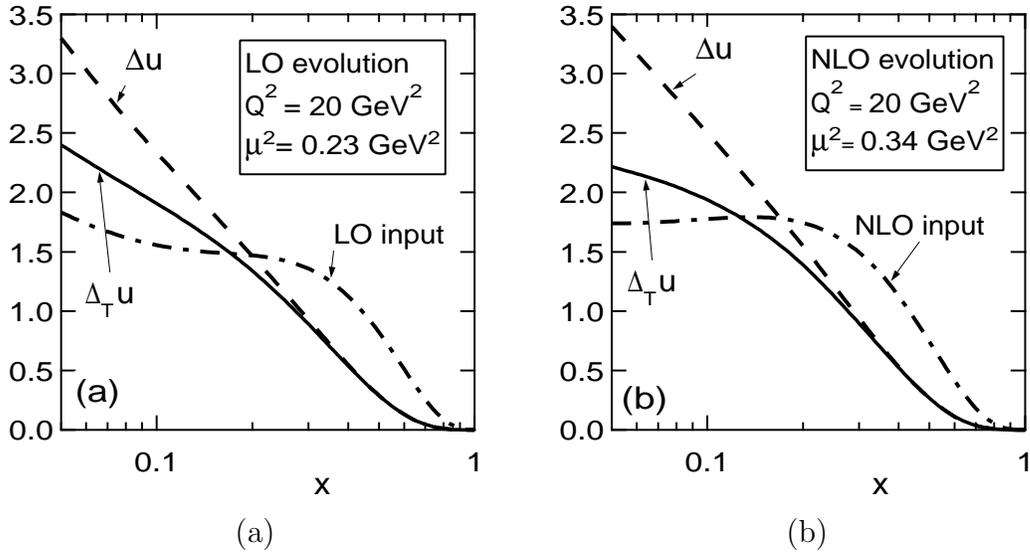}
  \hspace*{0.4cm}(a)\hspace{6.8cm}(b)
  \caption{Comparison of the $Q^2$-evolution of $\DT{u}(x,Q^2)$ and
           $\DL{u}(x,Q^2)$ at (a) LO and (b) NLO,
           from~\cite{Hayashigaki:1997dn}.}
  \label{fig:HKK-fig6}
\end{figure}

As a further comparison of the behaviour of $h_1$ and $g_1$, in
Figs.~\ref{fig:HKK-fig6} (a) and (b) we display the \LO and \NLO $Q^2$-evolution
of $\DT{u}$ and $\DL{u}$, starting respectively from the \LO and \NLO input
function for $\DL{u}$ given in \cite{Gluck:1996yr} for $Q^2=0.23$ and
$0.34\GeV^2$. Although \LO evolution leads to a significant divergence between
$\DT{u}$ and $\DL{u}$ at $Q^2=20\GeV^2$,\footnote{Such a difference was also
pointed out in \cite{Barone:1997mj}.} this tendency is strengthened by the \NLO
evolution, in particular, in the small-$x$ region. Although the evolution of
$\DL{u}$ shown in Fig.~\ref{fig:HKK-fig6} is affected by mixing with the gluon
distribution, the non-singlet quark distributions also show the same trend.

\begin{figure}[htbp]
  \centering
  \includegraphics[width=0.5\textwidth]{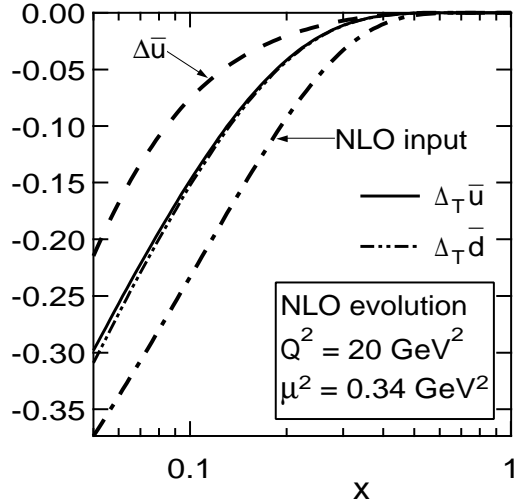}
  \caption{Comparison of the NLO $Q^2$-evolution of $\DT\bar{u}(x,Q^2)$,
           $\DT\bar{d}(x,Q^2)$ and $\DL\bar{u}(x,Q^2)$,
           from~\cite{Hayashigaki:1997dn}.}
  \label{fig:HKK-fig7}
\end{figure}

In Fig.~\ref{fig:HKK-fig7}, we compare the \NLO $Q^2$-evolution of
$\DT\anti{u}$, $\DT\anti{d}$ and $\DL\anti{u}$, starting from the same input
distribution function (the \NLO input function for the sea-quark distribution
to $g_1$ given in \cite{Gluck:1996yr}). The difference between $\DT\anti{u}$
and $\DL\anti{u}$ is again significant. Although the input sea-quark
distribution is taken to be flavour symmetric ($\DT\anti{u}=\DT\anti{d}$ at
$Q^2=0.34\GeV^2$), \NLO evolution violates this symmetry owing to the appearance
of $\DT{P}_{q\anti{q}}^{(1)}$ -- see (\ref{eq:qcd-DTP1qqbar}). However, this
effect is very small, as is evident from Fig.~\ref{fig:HKK-fig7} and discussed
in \cite{Martin:1998rz}.

\subsection{Evolution of the Soffer inequality and general positivity
  constraints}
\label{sofferevol}

Particular interest in the effects of evolution arises in connection with the
Soffer inequality \cite{Soffer:1995ww}, eq.~(\ref{soffer3}). It has been argued
\cite{Goldstein:1995ek} that this inequality, which was derived within a parton
model framework, may be spoilt by radiative corrections, much as the
Callan--Gross relation. Such an analogy, however, is somewhat misleading, since
the Soffer inequality is actually very similar to the more familiar positivity
bound $|\DL{q}(x)|\leq{q}(x)$. The \LO evolution of the inequality is governed
by eq.~(\ref{eq:qcd-soffer-evol-lo}) and hence it is not endangered, as pointed
out in \cite{Barone:1997fh}. At \NLO the situation is complicated by the
well-known problems of scheme dependence \etc.

Indeed, it is perhaps worth remarking that the entire question of positivity is
ill-defined beyond \LO, inasmuch as the parton distributions themselves as
physical quantities become ill-defined: \apriori there is no guarantee in a
given scheme that any form of positivity will survive higher-order corrections.
This observation may, of course, be turned on its head and used to impose
conditions on the scheme choice such that positivity will be guaranteed
\cite{Bourrely:1997nc}. At any rate, if this is possible then at the hadronic
level any natural positivity bounds should be respected, independently of the
regularisation scheme applied.

An instructive and rather general manner to examine the problem is to recast
the system of evolution equations into a form analogous to the Boltzmann
equation \cite{Bourrely:1998bx}. First of all, let us rewrite
eq.~(\ref{eq:qcd-x-evol}) in a slightly more suggestive form for the
non-singlet case:
\begin{equation}
  \frac{\d{q}(x,t)}{\d{t}}
  =
  \int_x^1 \frac{\d{y}}{y} \; P\!\left(\frac{x}{y},t\right) q(y,t) \, ,
  \label{eq:qcd-t-evol}
\end{equation}
where $t=\ln{Q^2}$. One may thus interpret the equation as describing the time,
$t$, evolution of densities, $f(x,t)$, in a one-dimensional $x$ space. The flow
is constrained to run from large to small $x$ owing to the ordering $x<y$ under
the integral. Such an interpretation facilitates dealing with the \IR
singularities present in the expressions for $P(x)$. Indeed, a key element is
provided by consideration of precisely the \IR
singularities\cite{Collins:1989wj, Durand:1987te}.

Let us now rewrite the plus regularisation in the following form:
\begin{equation}
  P_+(x,t)
  =
  P(x,t) - \delta(1-x) \int_0^1 \frac{\d{y}}{y} \; P(y,t) \, ,
  \label{eq:qcd-kin-plus}
\end{equation}
which then permits the evolution equations to be rewritten as
\begin{equation}
  \frac{\d{q}(x,t)}{\d{t}}
  =
  \int_x^1 \frac{\d{y}}{y} \; q(y,t) \, P\!\left( \frac{x}{y},t \right) -
  q(x,t) \int_0^1 \d{y} \; P(y,t) \, .
  \label{eq:qcd-kin-evol}
\end{equation}
Reading the second term as describing the flow of partons at the point $x$
\cite{Collins:1989wj}, the kinetic interpretation is immediate. It is useful to
render the analogy more direct by the change of variables $y\to{y/x}$ in the
second term, leading to the following more symmetric form:
\begin{eqnarray}
  \frac{\d{q}(x,t)}{\d{t}}
  =
  \int_x^1 \frac{\d{y}}{y} \; q(y,t) \, P\!\left( \frac{x}{y},t \right) -
  \int_0^x \frac{\d{y}}{x} \; q(x,t) \, P\!\left( \frac{y}{x\strut},t \right) .
  \label{eq:qcd-Boltzmann-1}
\end{eqnarray}
In this fashion the equation has been translated into a form analogous to the
Boltzmann equation: namely,
\begin{eqnarray}
  \frac{\d{q}(x,t)}{\d{t}}
  =
  \int_0^1 \d{y}
  \left[ \strut
    \sigma(y \to x;t) \, q(y,t)
  - \sigma(x \to y;t) \, q(x,t)
  \right] ,
  \label{eq:qcd-Boltzmann-2}
\end{eqnarray}
where the one-dimensional analogue of the Boltzmann ``scattering probability''
may be defined as
\begin{eqnarray}
  \sigma(y \to x;t)
  =
  \theta(y-x) \, \frac1{y} \, P\!\left( \frac{x}{y},t \right) .
  \label{eq:qcd-B-scatt-prob}
\end{eqnarray}
Cancellation of the \IR divergencies between contributions involving real and
virtual gluons is therefore seen to occur as a consequence of the continuity
condition on ``particle number''; \ie, the equality of flow in and out in the
neighbourhood of $y=x$ in both terms of eq.~(\ref{eq:qcd-Boltzmann-2}).

In the spin-averaged case the particle density (at some initialisation point)
is positive by definition. Now, the negative second term in
eq.~(\ref{eq:qcd-Boltzmann-2}) cannot change the sign of the distribution
because it is ``diagonal'' in $x$, \ie, it is proportional to the function at the
same point $x$. When the distribution is sufficiently close to zero, it stops
decreasing. This is true for both ``plus'' and $\delta(1-x)$ terms, for any value
of their coefficients (if positive, it only reinforces positivity of the
distribution).

Turning next to the spin-dependent case, for simplicity we consider first the
flavour non-singlet and allow the spin-dependent and spin-independent kernels
to be different, as they indeed are at \NLO. Rather than the usual helicity
sums and differences, it turns out to be convenient to cast the equations in
terms of definite parton helicities. Although such a form mixes contributions
of different helicities, the positivity properties emerge more clearly. We thus
have
\begin{subequations}
\begin{eqnarray}
  \label{eq:qcd-hel-fracs}
  \frac{\d{q}_+(x,t)}{\d{t}}
  =
    P_{++}(x,t) \otimes q_+(x,t)
  + P_{+-}(x,t) \otimes q_-(x,t) \, ,
\\
  \frac{\d{q}_-(x,t)}{\d{t}}
  =
    P_{+-}(x,t) \otimes q_+(x,t)
  + P_{++}(x,t) \otimes q_-(x,t) \, ,
\end{eqnarray}
\end{subequations}
where $P_{+\PM}(z,t)=\half[P(z,t)\pm\DL{P}(z,t)]$ are the evolution kernels for
helicity non-flip and flip respectively. For $x<y$, positivity of the initial
distributions, $q_{\pm}(x,t_0)\geq0$ or $|\DL{q}(x,t_0)|\leq{q}(x,t_0)$, is
preserved if both kernels $P_{+\PM}$ are positive, which is true if
\begin{eqnarray}
  |\DL{P}(z,t)| \leq P(z,t) \qquad (z<1) \, .
  \label{eq:qcd-hel-pos}
\end{eqnarray}
Terms that are singular at $z=1$ cannot alter positivity as they only appear in
the diagonal (in helicity) kernel, $P_{++}$; non-forward scattering is
completely \IR safe. Once again in the kinetic interpretation, the
distributions $q_+$ and $q_-$ stop decreasing on approaching zero.

To extend the proof to include the case in which there is quark--gluon mixing
is trivial---we need the full expressions for the evolutions of quark and gluon
distributions of each helicity:
\begin{subequations}
\begin{eqnarray}
  \label{eq:qcd-hel-evol}
  \frac{\d{q}_+(x,t)}{{\d{t}}}
  &=&
    P_{++}^{qq}(x,t) \otimes q_+(x,t)
  + P_{+-}^{qq}(x,t) \otimes q_-(x,t)
  \nonumber
\\
  &+&
    P_{++}^{qg}(x,t) \otimes g_+(x,t)
  + P_{+-}^{qg}(x,t) \otimes g_-(x,t) \, ,
\\[1ex]
  \frac{\d{q}_-(x,t)}{{\d{t}}}
  &=&
    P_{+-}^{qg}(x,t) \otimes q_+(x,t)
  + P_{++}^{qg}(x,t) \otimes q_-(x,t)
  \nonumber
\\
  &+&
    P_{+-}^{qg}(x,t) \otimes g_+(x,t)
  + P_{++}^{qg}(x,t) \otimes g_-(x,t) \, ,
\\[1ex]
  \frac{\d{g}_+(x,t)}{{\d{t}}}
  &=&
    P_{++}^{gq}(x,t) \otimes q_+(x,t)
  + P_{+-}^{gq}(x,t) \otimes q_-(x,t)
  \nonumber
\\
  &+&
    P_{++}^{gg}(x,t) \otimes g_+(x,t)
  + P_{+-}^{gg}(x,t) \otimes g_-(x,t) \, ,
\\[1ex]
  \frac{\d{g}_-(x,t)}{{\d{t}}}
  &=&
    P_{+-}^{gq}(x,t) \otimes q_+(x,t)
  + P_{++}^{gq}(x,t) \otimes q_-(x,t)
  \nonumber
\\
  &+&
    P_{+-}^{gg}(x,t) \otimes g_+(x,t)
  + P_{++}^{gg}(x,t) \otimes g_-(x,t) \, .
\end{eqnarray}
\end{subequations}
Since inequality (\ref{eq:qcd-hel-pos}) is clearly valid separately for each
type of parton\cite{Bourrely:1997nc},
\begin{eqnarray}
  |\DL{P}_{ij}(z,t)|
  \leq
  P_{ij}(z,t) \qquad (z<1,\quad i,j=q,g) \, ,
  \label{eq:qcd-deltap-ineq}
\end{eqnarray}
all the kernels appearing on the r.h.s.\ of this system, are positive. With
regard to the singular terms, they are again diagonal (in parton type here) and
hence cannot affect positivity. The validity of the equations at \LO is
guaranteed via their derivation, just as the (positive) helicity-dependent
kernels were in fact first calculated in \cite{Altarelli:1977zs}. At \NLO, the
situation is more complex\cite{Bourrely:1997nc}.

To conclude, the maintenance of positivity under $Q^2$ evolution has two
sources: (a) inequalities (\ref{eq:qcd-deltap-ineq}), leading to the increase
of distributions and (b) the kinetic interpretation of the decreasing terms.
For the latter, it is crucial that they are diagonal in $x$, helicity and also
parton type, which is a prerequisite for their \IR nature.

We now finally return to the Soffer inequality: in analogy with the previous
analysis, it is convenient to define the following ``super'' distributions
\begin{subequations}
\begin{eqnarray}
  \label{eq:qcd-super-q}
  Q_+(x) &=& q_+(x) + \DT{q}(x) \, ,
\\
  Q_-(x) &=& q_+(x) - \DT{q}(x) \, .
\end{eqnarray}
\end{subequations}
According to the Soffer inequality, both distributions are positive at some
scale (say $Q^2_0$) and the evolution equations for the non-singlet case take
the form (henceforth the argument $t$ will be suppressed)
\begin{subequations}
\begin{eqnarray}
  \label{eq:qcd-super-evol}
  \frac{\d{Q}_+(x)}{\d{t}}
  &=&
    P^Q_{++}(x) \otimes Q_+(x)
  + P^Q_{+-}(x) \otimes Q_-(x) \, ,
\\[1ex]
  \frac{\d{Q}_-(x)}{\d{t}}
  &=&
    P^Q_{+-}(x) \otimes Q_+(x)
  + P^Q_{++}(x) \otimes Q_-(x)) \, ,
\end{eqnarray}
\end{subequations}
where the ``super'' kernels at \LO are just
\begin{subequations}
\begin{eqnarray}
  \label{eq:qcd-super-kern}
  P^Q_{++}(z)
  &=&
  \half [P_{qq}^{(0)}(z) + P_h^{(0)}(z)]
  \nonumber
\\
  &=&
  \half \CF \left[ \frac{(1+z)^2}{(1-z)_+} + 3\delta(1-z) \right] ,
\\[1ex]
  P^Q_{+-}(z)
  &=&
  \half[P_{qq}^{(0)}(z) - P_h^{(0)}(z)]
  \nonumber
\\
  &=&
  \half \CF (1-z) \, .
\end{eqnarray}
\end{subequations}

One can easily see, that the inequalities analogous to
(\ref{eq:qcd-deltap-ineq}) are satisfied, so that both $P^Q_{++}(z)$ and
$P^Q_{+-}(z)$ are positive for $z<1$, while the singular term appears only in
the diagonal kernel. Thus, both requirements are fulfilled and the Soffer
inequality is maintained under \LO evolution. The extension to the singlet case
is trivial owing to the exclusion of gluon mixing. Therefore, only evolution of
quarks is affected, leading to the presence of the same extra terms on the
r.h.s., as in eqs.~(\ref{eq:qcd-hel-evol}):
\begin{subequations}
\begin{eqnarray}
  \label{eq:PQ1}
  \frac{\d{Q}_+(x)}{\d{t}}
  &=&
    P_{++}^Q   (x) \otimes Q_+(x)
  + P_{+-}^Q   (x) \otimes Q_-(x)
  \nonumber
\\
  &+&
    P_{+-}^{qG}(x) \otimes G_+(x)
  + P_{++}^{qG}(x) \otimes G_-(x) \, ,
\\[1ex]
  \frac{\d{Q}_-(x)}{\d{t}}
  &=&
    P_{+-}^Q   (x) \otimes Q_+(x)
  + P_{++}^Q   (x) \otimes Q_-(x)
  \nonumber
\\
  &+&
    P_{+-}^{qG}(x) \otimes G_+(x)
  + P_{++}^{qG}(x) \otimes G_-(x) \, ,
\end{eqnarray}
\end{subequations}
which are all positive and singularity free; this concludes the demonstration
that positivity is indeed preserved.

\section{Transversity in semi-inclusive leptoproduction}
\label{semiinclusive}

While it is usual to adopt \DIS as the defining process and point of reference
when discussing distribution functions, as repeatedly noted and explicitly
shown in Sec.~\ref{quarkdistr}, the case of transversity is somewhat special in
that it does not appear in \DIS. However, owing to the topology of the
contributing Feynman diagrams, transversity does play a r\^{o}le in semi-inclusive
\DIS, owing to the presence of \emph{two} hadrons: one in the initial state and
the other in the final state \cite{Artru:1990zv, Artru:1992jg, Cortes:1992ja,
Collins:1993kk, Jaffe:1993xb, Tangerman:1995hw, Anselmino:1997qx}. This process
is the subject of the present section.

\subsection{Definitions and kinematics}
\label{semidef}

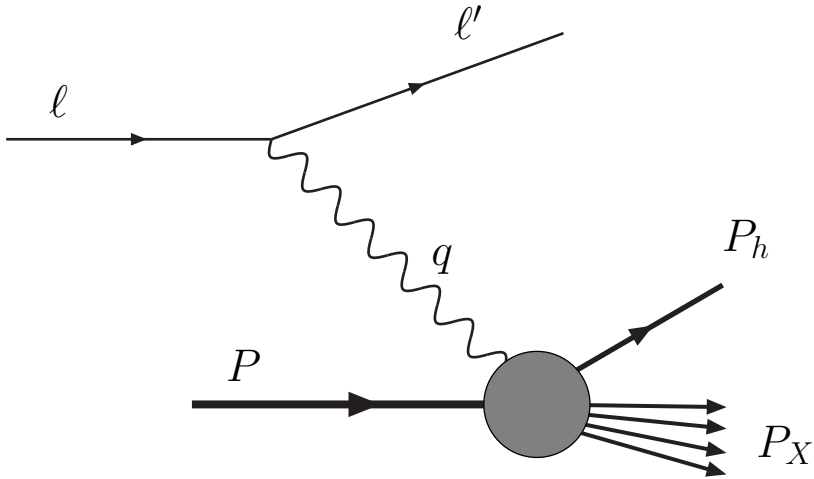
\begin{figure}[htbp]
  \centering
  \begin{picture}(340,175)(30,30)
    \SetWidth{1}
    \ArrowLine(50,150)(150,150)
    \Text(70,165)[]{\Large $\ell$}
    \ArrowLine(150,150)(260,190)
    \Text(225,195)[]{\Large $\ell'$}
    \Photon(150,150)(250,50){-5}{8}
    \Text(215,105)[]{\Large $q$}
    \SetWidth{3}
    \ArrowLine(120,50)(250,50)
    \Text(140,65)[]{\Large $P$}
    \SetWidth{2}
    \ArrowLine(260,60)(320,95)
    \SetWidth{1.5}
    \LongArrow(250,50)(320,49)
    \LongArrow(250,48)(320,41)
    \LongArrow(250,46)(320,32)
    \LongArrow(250,44)(320,24)
    \Text(345,35)[]{\Large $P_X$}
    \Text(330,113)[]{\Large $P_h$}
    \SetWidth{0.5}
    \GCirc(250,50){20}{0.5}
  \end{picture}
  \caption{Semi-inclusive deeply-inelastic scattering.}
  \label{sidis}
\end{figure}

Semi-inclusive -- or, to be more precise, single-particle inclusive --
leptoproduction (see Fig.~\ref{sidis}) is a \DIS reaction in which a hadron
$h$, produced in the current fragmentation region, is detected in the final
state (for the general formalism see \cite{Meng:1992da, Mulders:1996dh})
\begin{equation}
  l(\ell) \, + \, N(P) \to l'(\ell') \, + \, h(P_h) \, + \, X(P_X) \, .
  \label{sidis0}
\end{equation}
With a transversely polarised target, one can measure quark transverse
polarisation at leading twist either by looking at a possible asymmetry in the
$\Vec{P}_{h\perp}$ distribution of the produced hadron (the so-called Collins
effect \cite{Collins:1993kk, Collins:1994kq, Boer:1998nt, Mulders:1996dh}), or
by polarimetry of a transversely polarised final hadron (for instance, a
$\Lambda^0$ hyperon) \cite{Artru:1990zv, Artru:1991wq, Jaffe:1996wp,
Mulders:1996dh}. Transversity distributions also appear in the
$\Vec{P}_{h\perp}$-integrated cross-section at higher twist \cite{Jaffe:1993xb,
Mulders:1996dh}.

We define the invariants
\begin{equation}
  x = \frac{Q^2}{2P{\cdot}q} \, , \quad
  y = \frac{P{\cdot}q}{P{\cdot}\ell} \, , \quad
  z = \frac{P{\cdot}P_h}{P{\cdot}q} \, .
  \label{sidis7}
\end{equation}
We shall be interested in the limit where $Q^2\equiv-q^2,P{\cdot}q,P_h{\cdot}q$ and $P_h{\cdot}P$
become large while $x$ and $z$ remain finite.

\begin{figure}[htbp]
  \centering
  \begin{picture}(340,130)(0,20)
    \SetOffset(-80,-30)
    \Line(100,50)(300,50)
    \Line(250,100)(350,100) 
    \Line(100,50)(200,150)
    \Line(300,50)(400,150)
    \SetWidth{1}
    \Photon(200,100)(243,100){3}{3.5}
    \LongArrow(243,100)(250,100)
    \LongArrow(200,100)(180,115)
    \LongArrow(160,85)(200,100)
    \SetWidth{0.5}
    \Text(125,60)[l]{lepton plane}
    \Text(225,90)[]{$\Vec{q}$}
    \Text(160,78)[]{$\Vec\ell$}
    \Text(186,121)[l]{$\Vec\ell'$}
    \SetWidth{0.5}
    \Line(250,100)(270,180)
    \Line(350,100)(370,180)
    \Line(270,180)(370,180)
    \SetWidth{0.5}
    \Text(320,170)[L]{hadron plane}
    \Text(315,145)[L]{$\Vec{P}_H$}
    \Text(265,145)[L]{$\Vec{P}_{H\perp}$}
    \SetWidth{2}
    \LongArrow(250,100)(310,140)
    \LongArrow(250,100)(260,140)
    \SetWidth{0.5}
    \Line(200,150)(260,150)
    \Line(400,150)(365,150)
    \CArc(350,100)(20,45,75)
    \Text(368,130)[]{$\phi_h$}
    \LongArrow(380,100)(365,100)
    \LongArrow(380,100)(380,85)
    \LongArrow(380,100)(390,110)
    \Text(365,97)[t]{$z$}
    \Text(377,83)[r]{$y$}
    \Text(395,110)[l]{$x$}
  \end{picture}
  \caption{Lepton and hadron planes in semi-inclusive leptoproduction.}
  \label{plane2}
\end{figure}
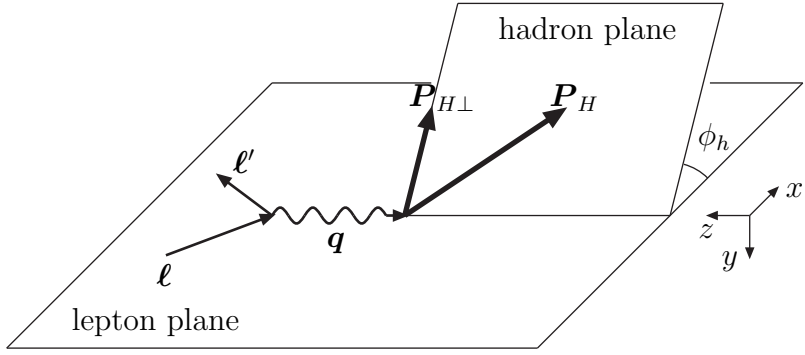

The geometry of the process is shown in Fig.~\ref{plane2}. The lepton
scattering plane is identified by $\Vec\ell$ and $\Vec\ell'$. The virtual
photon is taken to move along the $z$-axis. The three-momenta of the virtual
photon $\Vec{q}$ and of the produced hadron $\Vec{P}_h$ define a second plane,
which we call the hadron plane. The spin $S$ of the nucleon and the spin $S_h$
of the produced hadron satisfy $S^2=S_h^2=-1$ and $S{\cdot}P=S_h{\cdot}P_h=0$.

The cross-section for the reaction (\ref{sidis0}) is
\begin{eqnarray}
  \d\sigma &=&
  \frac1{4\ell{\cdot}P} \, \sum_{s_{l'}} \sum_X
  \int \! \frac{{\d^3} \Vec{P}_X}{(2\pi)^3 \, 2E_X}
  \nonumber
\\
  && \null \times
  (2\pi)^4 \, \delta^4(P + \ell - P_X - P_h - \ell') \,
  | \mathcal{M} |^2 \,
  \frac{{\d}^3 \Vec{\ell}'}{(2\pi)^3\,2E'} \,
  \frac{\d^3\Vec{P}_h}{(2\pi)^3\,2E_h} \, , \qquad
  \label{sidis1}
\end{eqnarray}
where we have summed over the spin $s_{l'}$ of the outgoing lepton. The squared
matrix element in (\ref{sidis1}) is
\begin{eqnarray}
  | \mathcal{M} |^2 &=&
  \frac{e^4}{q^4}
  \left[ \strut
    \anti{u}_{l'} (\ell', s_{l'}) \gamma_\mu u_l(\ell, s_l)
  \right]^*
  \left[ \strut
    \anti{u}_{l'} (\ell', s_{l'}) \gamma_\nu u_l(\ell, s_l)
  \right]
  \nonumber
\\
  && \null \times
  \langle X, P_h S_h | J^\mu (0) | PS \rangle^*
  \langle X, P_h S_h | J^\nu (0) | PS \rangle \, ,
  \label{sidis2}
\end{eqnarray}
Introducing the leptonic tensor
\begin{eqnarray}
  L_{\mu\nu}
  &=&
  \sum_{s_{l'}}
  \left[
    \strut \anti{u}_{l'} (\ell', s_{l'}) \gamma_\mu u_l(\ell, s_{l})
  \right]^*
  \left[ \strut
    \anti{u}_{l'} (\ell', s_{l'}) \gamma_\nu u_l(\ell, s_{l})
  \right]
  \nonumber
\\
  &=&
  2 (\ell_\mu \ell'_\nu + \ell_\nu \ell'_\mu - g_{\mu\nu} \, \ell{\cdot}\ell') +
  2\I \, \lambda_l \, \varepsilon_{\mu\nu\rho\sigma} \ell^\rho q^\sigma \, ,
  \label{sidis3}
\end{eqnarray}
and the hadronic tensor
\begin{eqnarray}
  W^{\mu\nu} &=&
  \frac1{(2\pi)^4} \, \sum_X
  \int \! \frac{{\d^3} \Vec{P}_X}{(2\pi)^3\,2E_X} \, (2\pi)^4  \,
  \delta^4(P + q - P_X - P_h )
  \nonumber
\\
  && \hspace{5em} \null \times
  \langle PS         | J^\mu (0) | X, P_h S_h \rangle
  \langle X, P_h S_h | J^\nu (0) | PS         \rangle \, ,
  \label{sidis4}
\end{eqnarray}
the cross-section becomes
\begin{equation}
  \d\sigma =
  \frac1{4\ell{\cdot}P} \, \frac{e^4}{Q^4} \, L_{\mu\nu}W^{\mu\nu} \, (2\pi)^4 \,
  \frac{{\d}^3 \Vec{\ell}'}{(2\pi)^3\,2E'} \,
  \frac{{\d}^3 \Vec{P}_h}{(2\pi)^3\,2E_h} \, ,
  \label{sidis5}
\end{equation}
In the target rest frame ($\ell{\cdot}P=ME$) one has
\begin{equation}
  2 E_h \, \frac{\d\sigma}{\d^3\Vec{P}_h \, \d{E}' \, \d\Omega}
  =
  \frac{\alpha_\text{em}^2}{2MQ^4} \, \frac{E'}{E} \,
  L_{\mu\nu} W^{\mu\nu} \, ,
  \label{sidis6}
\end{equation}
In terms of the invariants $x$, $y$ and $z$ eq.~(\ref{sidis6}) reads
\begin{equation}
  2 E_h \, \frac{\d\sigma}{\d^3\Vec{P}_h \, \d{x} \, \d{y} }
  =
  \frac{\pi\alpha_\text{em}^2y}{Q^4} \, L_{\mu\nu} W^{\mu\nu} \, .
  \label{sidis8}
\end{equation}
If we decompose the momentum $\Vec{P}_h$ of the produced hadron into a
longitudinal ($\Vec{P}_{h\parallel}$) and a transverse ($\Vec{P}_{h\perp}$)
component with respect to the $\gamma^*N$ axis and if $|\Vec{P}_{h\perp}|$ is
small compared to the energy $E_h$, then we can write approximately
\begin{equation}
  \frac{\d^3\Vec{P}_h}{2E_h} = \frac1{2z} \, \d{z} \, \d^2\Vec{P}_{h\perp} \, ,
\end{equation}
and re-express eq.~(\ref{sidis8}) as
\begin{equation}
  \frac{\d\sigma}{\d{x} \d{y} \d{z} \d^2\Vec{P}_{h\perp}} =
  \frac{\pi\alpha_\text{em}^2}{2Q^4} \, \frac{y}{z} \,
  L_{\mu\nu} W^{\mu\nu} \, .
  \label{sidis9}
\end{equation}

Instead of working in a $\gamma^*N$ collinear frame, it is often convenient to
work in a frame where the target nucleon and the produced hadron move
collinearly (the $hN$ collinear frame, see Appendix~\ref{hN}). In this frame
the virtual photon has a transverse momentum $\Vec{q}_T$, which is related to
$\Vec{P}_{h\perp}$, up to $1/Q^2$ corrections, by
$\Vec{q}_T\simeq-\Vec{P}_{h\perp}/z$. Thus eq.~(\ref{sidis9}) can be written as
\begin{equation}
  \frac{\d\sigma}{\d{x} \d{y} \d{z} \d^2\Vec{q}_{T}} =
  \frac{\pi\alpha_\text{em}^2}{2Q^4} \, yz \, L_{\mu\nu} W^{\mu\nu} \, .
  \label{sidis10}
\end{equation}

Let us now evaluate the leptonic tensor. In the $\gamma^*N$ collinear frame the
lepton momenta can be parametrised in terms of the Sudakov vectors $p$ and $n$
as
\begin{subequations}
\begin{eqnarray}
  \ell^\mu &=&
  \frac{x}{y} \, (1-y) \, p^\mu +
  \frac{Q^2}{2 x y} \, n^\mu +
  \ell_\perp^\mu \, ,
  \label{sidis20}
\\
  {\ell'}^\mu &=&
  \frac{x}{y} \, p^\mu +
  \frac{Q^2(1-y)}{2xy} \, n^\mu +
  \ell_\perp^\mu \, ,
  \label{sidis21a}
\end{eqnarray}
\end{subequations}
with $\Vec\ell_\perp^2=(\frac{1-y}{y^2})\,Q^2$. The symmetric part of the leptonic
tensor then becomes
\begin{subequations}
\begin{eqnarray}
  L^{\mu\nu\text{(S)}} &=&
  -\frac{Q^2}{y^2} \left[ 1 + (1-y)^2 \right] g_\perp^{\mu\nu} +
  \frac{4(1-y)}{y^2} \, t^\mu t^\nu
  \nonumber
\\
  && \null +
  \frac{2(2-y)}{y} \, (t^\mu \ell_\perp^\nu + t^\nu \ell_\perp^\mu ) +
  \frac{4Q^2\,(1-y)}{y^2}
  \left(
    \frac{\ell_\perp^\mu \ell_\perp^{\nu}}{\Vec\ell_\perp^2} +
    \frac12\,g_\perp^{\mu\nu}
  \right) , \hspace{4em}
  \label{sidis22}
\end{eqnarray}
where $t^\mu=2xp^\mu+q^\mu$; the antisymmetric part reads
\begin{equation}
  L^{\mu\nu (A)} =
  \lambda_l \,
  \varepsilon_{\mu\nu\rho\sigma}
  \left[
    \frac{Q^2 (2 -y)}{y} \, p^\rho n^\sigma
    +
    \frac{Q^2}{x} \, \ell_\perp^\rho n^\sigma -
    2x \, \ell_\perp^\rho p^\sigma
  \right] .
  \label{sidis23}
\end{equation}
\end{subequations}

At leading-twist level, only semi-inclusive \DIS processes with an unpolarised
lepton beam probe the transverse polarisation distributions of quarks
\cite{Mulders:1996dh}. Therefore, in what follows, we shall focus on this case
and take only the target nucleon (and, possibly, the outgoing hadron) to be
polarised. At twist 3 there are also semi-inclusive \DIS reactions with
polarised leptons, which allow extracting $\DT{f}$. For these higher-twist
processes we shall limit ourselves to presenting the cross-sections without
derivation.

\subsection{The parton model}
\label{sidisparton}

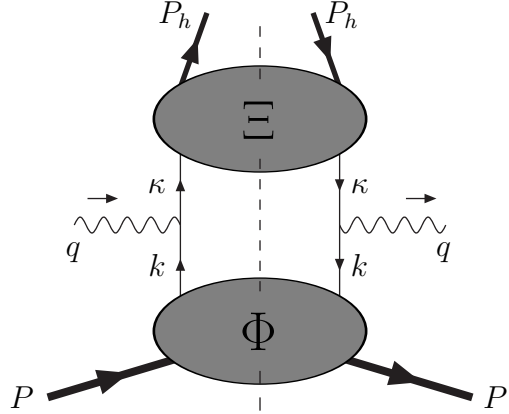
\begin{figure}[htbp]
  \centering
  \begin{picture}(300,160)(0,20)
    \SetWidth{3}
    \ArrowLine(70,25)(120,40)
    \ArrowLine(180,40)(230,25)
    \SetWidth{0.5}
    \ArrowLine(120,60)(120,90)
    \ArrowLine(180,90)(180,60)
    \ArrowLine(120,90)(120,120)
    \ArrowLine(180,120)(180,90)
    \Photon(80,90)(120,90){3}{4.5}
    \Photon(220,90)(180,90){3}{4.5}
    \GOval(150,50)(20,40)(0){0.5}
    \SetWidth{2}
    \ArrowLine(120,142)(130,170)
    \ArrowLine(170,170)(180,142)
    \SetWidth{0.5}
    \GOval(150,130)(20,40)(0){0.5}
    \LongArrow(85,100)(95,100)
    \LongArrow(205,100)(215,100)
    \Text(65,25)[r]{$P$}
    \Text(235,25)[l]{$P$}
    \Text(80,80)[]{$q$}
    \Text(220,80)[]{$q$}
    \Text(125,170)[r]{$P_h$}
    \Text(175,170)[l]{$P_h$}
    \Text(115,75)[r]{$k$}
    \Text(185,75)[l]{$k$}
    \Text(115,105)[r]{$\kappa$}
    \Text(185,105)[l]{$\kappa$}
    \Text(150,130)[]{\LARGE$\Xi$}
    \Text(150,50)[]{\LARGE$\Phi$}
    \DashLine(150,20)(150,35){4}
    \DashLine(150,65)(150,115){4}
    \DashLine(150,145)(150,165){4}
  \end{picture}
  \caption{Diagram contributing to semi-inclusive DIS at LO.}
  \label{handbag2}
\end{figure}

In the parton model the virtual photon strikes a quark (or antiquark), which
later fragments into a hadron $h$. The process is depicted in
Fig.~\ref{handbag2}. The relevant diagram is the handbag diagram with an upper
blob representing the fragmentation process.

Referring to Fig.~\ref{handbag2} for the notation, the hadronic tensor is given
by (for simplicity we consider only the quark contribution)
\begin{eqnarray}
  W^{\mu\nu} &=&
  \frac1{(2\pi)^4} \, \sum_a e_a^2 \, \sum_X
  \int \! \frac{{\d^3} \Vec{P}_X}{(2\pi)^3 \, 2E_X}
  \int \! \frac{\d^4k}{(2\pi)^4}
  \int \! \frac{\d^4\kappa}{(2\pi)^4}
  \nonumber
\\
  && \null \times
  (2\pi)^4 \, \delta^4(P - k - P_X) \;
  (2\pi)^4 \, \delta^4(k + q - \kappa) \;
  (2\pi)^4 \delta^4(\kappa - P_h - P_{X'})
  \nonumber
\\
  && \null \times
  \left[ \strut
    \anti\chi(\kappa; P_h,S_h) \gamma^\mu \phi(k; P, S)
  \right]^*
  \left[ \strut
    \anti\chi(\kappa; P_h,S_h) \gamma^\nu \phi(k; P, S)
  \right] ,
  \label{sidis10a}
\end{eqnarray}
where $\phi(k;P,S)$ and $\chi(\kappa;P_h,S_h)$ are matrix elements of the quark
field $\psi$, defined as
\begin{equation}
  \phi (k; P, S) = \langle X | \psi(0) | PS \rangle \, ,
  \label{sidis11}
\end{equation}
\begin{equation}
  \chi (\kappa; P_h, S_h) =
  \langle 0 | \psi(0) | P_h S_h, X \rangle \, .
  \label{sidis12}
\end{equation}
We now introduce the quark--quark correlation matrices
\begin{eqnarray}
  \Phi_{ij}(k; P, S) &=&
  \sum_X \int \! \frac{\d^3\Vec{P}_X}{(2\pi)^3\,2E_X} \;
  (2\pi)^4 \, \delta^4(P_X + k - P) \,
  \phi_i(k;P,S) \, \anti\phi_j(k;P,S)
  \nonumber
\\
  &=&
  \int \! \d^4\xi \, \e^{\I k{\cdot}\xi} \,
  \langle PS | \anti\psi_j(0) \psi_i (\xi) | PS \rangle ,
  \label{sidis13}
\end{eqnarray}
and
\begin{eqnarray}
  \Xi_{ij}(\kappa; P_h, S_h) &=&
  \sum_{X} \int \! \frac{\d^3\Vec{P}_{X}}{(2\pi)^3 \, 2E_X} \,
  (2\pi)^4 \, \delta^4(P_h + P_{X} - \kappa)
  \nonumber
\\
  && \hspace{2.5em} \null \times
  \chi_i(\kappa; P_h, S_h) \, \anti\chi_j(\kappa; P_h, S_h)
  \nonumber
\\
  &=& \sum_{X} \int \! \frac{\d^3\Vec{P}_{X}}{(2\pi)^3 \, 2E_X}
  \int \! \d^4 \xi \, \e^{\I \kappa{\cdot}\xi}
  \nonumber
\\
   && \hspace{2.5em} \null \times
  \langle 0 | \psi_i(\xi) | P_h S_h, X \rangle
  \langle P_h S_h, X | \anti\psi_j(0) | 0 \rangle \, ,
  \label{sidis14}
\end{eqnarray}
Here $\Phi$ is the matrix already encountered in inclusive \DIS, see
Secs.~\ref{parton} and \ref{qqcorr}, which incorporates the quark distribution
functions. $\Xi$ is a new quark--quark correlation matrix (sometimes called
decay matrix), which contains the fragmentation functions of quarks into a
hadron $h$. An average over colours is included in $\Xi$. Inserting
eqs.~(\ref{sidis13}, \ref{sidis14}) into (\ref{sidis10a}) yields
\begin{equation}
  W^{\mu\nu} =
  \sum_a e_a^2 \int \! \frac{\d^4k}{(2\pi)^4} \int \!
  \frac{\d^4\kappa}{(2\pi)^4} \, \delta^4(k + q - \kappa) \,
  \Tr [\Phi \, \gamma^\mu \, \Xi \gamma^\nu] \, .
  \label{sidis15}
\end{equation}

It is an assumption of the parton model that $k^2$, $k{\cdot}P$, $\kappa^2$ and
$\kappa{\cdot}P_h$ are much smaller than $Q^2$. Stated differently, when these
quantities become large, $\Phi$ and $\Xi$ are strongly suppressed. Let us work
in the $hN$ collinear frame (see Appendix~\ref{hN}), the photon momentum is
\begin{equation}
  q^\mu \simeq
  - xP^\mu + \frac1{z} \, P_h^\mu + q_T^\mu =
  \left( - xP^+, \, \frac1{z} P_h^-, \, \Vec{q}_T \right) .
  \label{sidis16}
\end{equation}
We recall that $\Vec{P}_{h\perp}\simeq-z\Vec{q}_T$. The quark momenta are
\begin{subequations}
\begin{eqnarray}
  k^\mu &\simeq&
  \alpha \, P^\mu + k_T^\mu = (\xi P^+ , 0, \Vec{0}_T)
  \label{sidis17b}
\\
  \kappa^\mu &\simeq&
  \beta \, P_h^\mu + \kappa_T^\mu = (0, P_h^-/\zeta, \Vec{0}_T) \, .
  \label{sidis17c}
\end{eqnarray}
\end{subequations}
Thus the delta function in (\ref{sidis15}) can be decomposed as
\begin{eqnarray}
  \delta^4(k + q - \kappa) &=&
  \delta(k^+ + q^+ - \kappa^+) \,
  \delta(k^- + q^- - \kappa^-) \,
  \delta^2(\Vec{k}_T + \Vec{q}_T - {\Vec\kappa}_T)
  \nonumber
\\
  &\simeq&
  \delta(k^+ - xP^+) \,
  \delta(k^- - P_h^- /z) \,
  \delta^2(\Vec{k}_T + \Vec{q}_T - {\Vec\kappa}_T) \, . \qquad
  \label{sidis18}
\end{eqnarray}
which implies $\alpha = x$ and $\beta = 1/z$, that is
\begin{subequations}
\begin{eqnarray}
  k^\mu &\simeq& xP^\mu + k_T^\mu
  \label{sidis19}
\\
  \kappa^\mu &\simeq& \frac1{z} P_h^\mu + \kappa_T^\mu \, .
  \label{sidis20a}
\end{eqnarray}
\end{subequations}
The hadronic tensor (\ref{sidis15}) then becomes
\begin{eqnarray}
  W^{\mu\nu} &=&
  \sum_a e_a^2
  \int \! \frac{\d{k}^+ \, \d{k}^- \, \d^2\Vec{k}_T}{(2\pi)^4}
  \int \! \frac{\d\kappa^+ \, \d\kappa^- \, \d^2\Vec\kappa_T}{(2\pi)^4}
  \nonumber
\\
  && \hspace{2.5em} \null \times
  \delta(k^+ - xP^+) \,
  \delta(k^- - P_h^- /z) \,
  \delta^2(\Vec{k}_T + \Vec{q}_T - {\Vec\kappa}_T)
  \nonumber
\\
  && \hspace{2.5em} \null \times
  \Tr [\Phi \, \gamma^\mu \, \Xi \, \gamma^\nu] \, .
  \label{sidis21b}
\end{eqnarray}
Exploiting the delta functions in the longitudinal momenta, we obtain
\begin{eqnarray}
  W^{\mu\nu}
  &=&
  \sum_a e_a^2
  \int \! \frac{\d{k}^- \, \d^2\Vec{k}_T}{(2\pi)^4}
  \int \! \frac{\d\kappa^+ \, \d^2\Vec\kappa_T}{(2\pi)^4}
  \nonumber
\\
  && \null \times
  \delta^2(\Vec{k}_T + \Vec{q}_T - \Vec\kappa_T)
  \, \Tr
  [ \Phi \, \gamma^\mu \, \Xi \gamma^\nu ]_{k^+=xP^+,\,\kappa^-=P_h^-/z} \, .
  \label{sidis22a}
\end{eqnarray}

To obtain the final form of $W^{\mu\nu}$, we must insert the explicit
expressions for $\Phi$ and $\Xi$ into (\ref{sidis22a}). The former has been
already discussed in Sec.~\ref{qqcorr}. In the following we shall concentrate
on the structure of $\Xi$.

\subsection{Systematics of fragmentation functions}
\label{systfrag}

The fragmentation functions are contained in the decay matrix $\Xi$, which we
rewrite here for convenience (from now on $\sum_X$ incorporates the integration
over $\Vec{P}_X$)
\begin{equation}
  \Xi_{ij}(\kappa; P_h, S_h) =
  \sum_{X} \int \! \d^4 \xi \, \e^{\I \kappa{\cdot}\xi}
  \langle 0 | \psi_i(\xi) | P_h S_h, X \rangle
  \langle P_h S_h, X | \anti\psi_j(0) | 0 \rangle \, .
  \label{frag1}
\end{equation}
We have omitted the path-ordered exponential
$\mathcal{L}=\mathcal{P}\,\exp\left(-{\I}g\int{\d}s_\mu\,A^\mu(s)\right)$,
needed to make (\ref{frag1}) gauge invariant, since in the $A^+=0$ gauge a
proper path may be chosen such that $\mathcal{L}=1$. Hereafter the formalism
will be similar to that developed in Sec.~\ref{qqcorr} for $\Phi$ and,
therefore, much detail will be suppressed.

The quark fragmentation functions are related to traces of the form
\begin{equation}
  \Tr[\Gamma \, \Xi] =
  \sum_{X} \int \! \d^4 \xi \, \e^{\I \kappa{\cdot}\xi}
  \Tr
  \langle 0 | \psi_i(\xi) | P_h S_h, X \rangle
  \langle P_h S_h, X | \anti\psi_j (0) \, \Gamma | 0 \rangle \, ,
  \label{frag1b}
\end{equation}
where $\Gamma$ is a Dirac matrix. $\Xi$ can be decomposed over a Dirac matrix
basis as
\begin{equation}
  \Xi(\kappa; P_h, S_h) = \frac12
  \left\{
    \mathcal{S} \, \one +
    \mathcal{V}_\mu \, \gamma^\mu +
    \mathcal{A}_\mu \gamma_5 \gamma^\mu +
    \I \mathcal{P}_5 \gamma_5 +
    \frac{\I}{2} \, \mathcal{T}_{\mu\nu} \, \sigma^{\mu\nu} \gamma_5
  \right\} ,
  \label{frag2}
\end{equation}
where the quantities $\mathcal{S}$, $\mathcal{V}^\mu$, $\mathcal{A}^\mu$,
$\mathcal{P}_5$, $\mathcal{T}^{\mu\nu}$, constructed with the momentum of the
fragmenting quark $\kappa^\mu$, the momentum of the produced hadron $P_h^\mu$
and its spin $S_h^\mu$, have the general form\footnote{We consider here
spin-$\frac{1}{2}$ (or spin-0) hadrons. For the production of spin-1 hadrons
see below, Sec.~\ref{spinone}.} \cite{Levelt:1994ac, Levelt:1994np,
Tangerman:1995hw, Mulders:1996dh, Boglione:1999pz}
\begin{subequations}
\begin{eqnarray}
  \mathcal{S} &=&
  \frac12 \, \Tr(\Xi) = \mathcal{C}_1
  \label{frag3}
\\
  \mathcal{V}^\mu &=&
  \frac12 \, \Tr(\gamma^\mu \, \Xi) =
  \mathcal{C}_2 \, P_h^\mu +
  \mathcal{C}_3 \, \kappa^\mu +
  \mathcal{C}_{10} \, \varepsilon^{\mu\nu\rho\sigma} S_{h\nu} P_{h\rho}
  \kappa_\sigma \, ,
  \label{frag4}
\\
  \mathcal{A}^\mu &=&
  \frac12 \, \Tr(\gamma^\mu \gamma_5 \, \Xi) =
  \mathcal{C}_4 \, S_h^\mu +
  \mathcal{C}_5 \, \kappa{\cdot}S_h \, P_h^\mu +
  \mathcal{C}_6 \, \kappa{\cdot}S_h \, \kappa^\mu
  \label{frag5}
\\
  \mathcal{P}_5 &=&
  \frac1{2\I} \, \Tr(\gamma_5 \, \Xi) =
  \mathcal{C}_{11} \, \kappa{\cdot}S_h
  \label{frag6}
\\
  \mathcal{T}^{\mu\nu} &=&
  \frac1{2\I} \, \Tr(\sigma^{\mu\nu} \gamma_5 \, \Xi) =
  \mathcal{C}_7 \, P_h^{[\mu}S_h^{\nu]} + \mathcal{C}_8 \,
  \kappa^{[\mu}S_h^{\nu ]}
  \nonumber
\\
  && \hspace{7.5em} \null +
  \mathcal{C}_9 \, \kappa{\cdot}S_h \, P_h^{[\mu}\kappa^{\nu]} +
  \mathcal{C}_{12} \, \varepsilon^{\mu\nu\rho\sigma} P_{h\rho}
  \kappa_\sigma \, .
  \label{frag7}
  \sublabel{subfrag7}
\end{eqnarray}
\end{subequations}
The quantities $\mathcal{C}_i=\mathcal{C}_i(\kappa^2,\kappa{\cdot}P_h)$ are real
functions of their arguments, owing to the hermiticity property of $\Xi$.

The presence of the terms with coefficients $\mathcal{C}_{10}$,
$\mathcal{C}_{11}$ and $\mathcal{C}_{12}$, which were forbidden in the
expansion of the $\Phi$ matrix by time-reversal invariance, is justified by the
fact that in the fragmentation case we cannot na\"{\i}vely impose a condition
similar to (\ref{df1d}), that is
\begin{equation}
  \Xi^*(\kappa; P_h, S_h) =
  \gamma_5 C
    \; \Xi (\widetilde\kappa; \widetilde{P}_h, \widetilde{S}_h) \;
  C^\dagger \gamma_5 \, .
  \label{frag7bis}
\end{equation}
In the derivation of (\ref{df1d}) the simple transformation property of the
nucleon state $|PS\rangle$ under $T$ is crucial. However, $\Xi$ contains the
states $|P_hS_h,X\rangle$ which are out-states with possible final-state
interactions between the hadron and the remnants. Under time reversal they do
not simply invert their momenta and spin but transform into in-states
\begin{equation}
  T \, | P_h S_h, X; \text{out} \rangle \propto
  | \widetilde{P}_h \widetilde{S}_h, \widetilde{X}; \text{in} \rangle \, .
  \label{frag7b}
\end{equation}
These may differ non-trivially from
$|\widetilde{P}_h\widetilde{S}_h,\widetilde{X};\text{out}\rangle$ owing to
final-state interactions, which can generate relative phases between the
various channels open in the $|\text{in}\rangle\to|\text{out}\rangle$
transition. Thus, the terms containing $\mathcal{C}_{10}$, $\mathcal{C}_{11}$
and $\mathcal{C}_{12}$ are allowed in principle. The fragmentation functions
related to these terms are called $T$-odd fragmentation functions
\cite{Levelt:1994np, Mulders:1996dh, Boglione:1999pz}. One of them, called
$H_1^\perp$, gives rise to the so-called Collins effect \cite{Collins:1993kk,
Mulders:1996dh, Boglione:1999pz}.

\begin{figure}[htbp]
  \centering
  \begin{picture}(340,100)(0,0)
    \SetWidth{3}
    \SetWidth{2}
    \Line(120,45)(140,75)
    \ArrowLine(140,75)(150,90)
    \ArrowLine(190,90)(220,45)
    \SetWidth{0.5}
    \ArrowLine(120,0)(120,30)
    \ArrowLine(220,30)(220,0)
    \Text(228,5)[l]{\large$\kappa$}
    \Text(112,5)[r]{\large$\kappa$}
    \Text(140,90)[r]{\large$P_h$}
    \Text(200,90)[l]{\large$P_h$}
    \GOval(170,40)(15,60)(0){0.8}
    \GOval(145,60)(18,6)(30){0.4}
    \DashLine(170,0)(170,95){8}
  \end{picture}
  \caption{A hypothetical mechanism giving rise to a $T$-odd fragmentation
           function.}
  \label{toddfrag}
\end{figure}
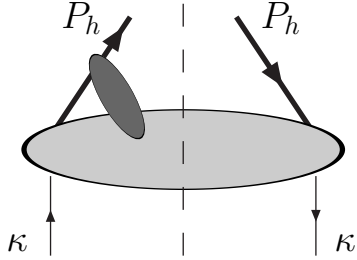

A generic mechanism giving rise to $T$-odd fragmentation functions is shown
diagrammatically in Fig.~\ref{toddfrag}. What is needed, in order to produce
such fragmentation functions, is an interference diagram in which the
final-state interaction (represented in figure by the dark blob) between the
produced hadron and the residual fragments cannot be reabsorbed into the
quark--hadron vertex \cite{Bianconi:1999cd}.

It has been argued in \cite{Jaffe:1998hf} that the relative phases between the
hadron and the $X$ system might actually cancel in the sum over $X$. This would
cause the $T$-odd distributions to disappear. Only experiments will settle the
question.

Working in a $hN$ collinear frame, the vectors (or pseudovectors) appearing in
(\ref{frag3}--\ref{subfrag7}) are
\begin{equation}
  P_h^\mu \, , \quad
  \kappa^\mu \simeq \frac1{z} \, P_h^\mu + \kappa_T \quad \text{and} \quad
  S_h^\mu \simeq \frac{\lambda_h}{M_h} \, P_h^\mu + S_{h T}^\mu \, ,
  \label{frag8}
\end{equation}
where we have to remember that the transverse components are suppressed by a
factor $1/P_h^-$ (that is, $1/Q$) compared to the longitudinal ones.

To start with, consider the case of collinear kinematics. If we ignore
$\Vec\kappa_T$, at leading twist (that is at order $\Ord(P_h^-)$) the terms
contributing to (\ref{frag2}) are
\begin{subequations}
\begin{eqnarray}
  \mathcal{V}^\mu &=&
  \frac12 \, \Tr(\gamma^\mu \, \Xi) =
  B_1 \, P_h^\mu \, ,
  \label{frag9}
\\
  \mathcal{A}^\mu &=& \frac12 \, \Tr(\gamma^\mu \gamma_5 \, \Xi) =
  \lambda_h \, B_2 \, P_h^\mu \, ,
  \label{frag10}
\\
  \mathcal{T}^{\mu\nu} &=&
  \frac1{2\I} \, \Tr(\sigma^{\mu\nu} \gamma_5 \, \Xi) =
  B_3 \, P_h^{[\mu} S_{hT}^{\nu]} \, ,
  \label{frag11}
  \sublabel{subfrag11}
\end{eqnarray}
\end{subequations}
where we introduced the functions $B_i(\kappa^2,\kappa{\cdot}P_h)$. The decay matrix
then reads
\begin{equation}
  \Xi(\kappa; P_h, S_h) =
  \half
  \left\{B_1 \,
    \slashed{P}_h +
    \lambda_h \, B_2 \, \gamma_5 \, \slashed{P}_h +
    B_3 \, \slashed{P}_h \, \gamma_5 \slashed{S}_{hT}
  \right\}.
  \label{frag12}
\end{equation}
Recalling that $P_h$ only has a $P_h^-$ component,
eqs.~(\ref{frag9}--\ref{subfrag11}) become
\begin{subequations}
\begin{eqnarray}
  \frac1{2P_h^-} \, \Tr(\gamma^- \, \Xi) &=&
  B_1 \, ,
  \label{frag13}
\\
  \frac1{2P_h^-} \, \Tr(\gamma^- \gamma_5 \, \Xi) &=&
  \lambda_h \, B_2 \, ,
  \label{frag14}
\\
  \frac1{2P_h^-} \, \Tr(\I \sigma^{i-} \gamma_5 \, \Xi) &=&
  S_{h T}^i \, B_3 \, \, .
  \label{frag15}
\end{eqnarray}
\end{subequations}
The three leading-twist fragmentation functions: the unpolarised fragmentation
function $D_q(x)$, the longitudinally polarised fragmentation function
$\DL{D}_q(x)$, and the transversely polarised fragmentation function
$\DT{D}_q(x)$, are obtained by integrating $B_1$, $B_2$ and $B_3$,
respectively, over $\kappa$, with the constraint $1/z=\kappa^-/P_{h}^-$. For
instance
\begin{eqnarray}
  D(z) &=&
  \frac{z}{2} \int \! \frac{\d^4\kappa}{(2\pi)^4} \,
  B_1(\kappa^2, \kappa{\cdot}P_h) \, \delta(1/z - \kappa^-/P_h^-)
  \nonumber
\\
  &=&
  \frac{z}{4} \int \! \frac{\d^4\kappa}{(2\pi)^4} \,
  \Tr(\gamma^- \Xi) \, \delta(\kappa^- - P_h^-/z)
  \nonumber
\\
  &=&
  \frac{z}{4} \, \sum_X \int \! \frac{\d\xi^+}{2\pi} \,
  \e^{\I P_h^- \xi^+/z} \,
  \nonumber
\\
   && \hspace{2.5em} \null \times
  \langle 0 | \psi(\xi^+, 0, \Vec{0}_\perp) | P_h S_h, X \rangle
  \langle P_h S_h, X | \anti\psi(0) \gamma^- | 0 \rangle \, . \qquad
  \label{frag16}
\end{eqnarray}
The normalisation of $D(z)$ is such that
\begin{equation}
  \sum_h \sum_{S_h} \int \! \d{z} \, z \, D(z) = 1 \, ,
  \label{frag17}
\end{equation}
where $\sum_h$ is a sum over all produced hadrons. Hence, $D(z)$ is the number
density of hadrons of type $h$ with longitudinal momentum fraction $z$ in the
fragmenting quark.

Analogously, we have for $\DL{D}(z)$ (with $\lambda_h=1$)
\begin{eqnarray}
  \DL{D}(z) &=&
  \frac{z}{2} \int \! \frac{\d^4\kappa}{(2\pi)^4} \,
  B_2(\kappa^2, \kappa{\cdot}P_h) \, \delta(1/z - \kappa^-/P_h^-)
  \nonumber
\\
  &=&
  \frac{z}{4} \, \sum_X \int \! \frac{\d\xi^+}{2\pi} \, \e^{\I P_h^- \xi^+/z}
  \nonumber
\\
  && \hspace{1em} \null \times
  \langle 0 | \psi(\xi^+, 0, \Vec{0}_\perp) | P_h S_h, X \rangle
  \langle P_h S_h, X | \anti\psi(0) \gamma^- \gamma_5 | 0 \rangle \, , \qquad
  \label{frag18}
\end{eqnarray}
and for $\DT{D}(z)$ (with $S_{hT}^i=(1,0)$ for definiteness)
\begin{eqnarray}
  \DT{D}(z) &=&
  \frac{z}{2} \int \! \frac{\d^4\kappa}{(2\pi)^4} \,
  B_3(\kappa^2, \kappa{\cdot}P_h) \, \delta(1/z - \kappa^-/P_h^-)
  \nonumber
\\
  &=& \frac{z}{4} \, \sum_X \int \! \frac{\d\xi^+}{2\pi} \,
  \e^{\I P_h^- \xi^+/z} \,
  \nonumber
\\
  && \hspace{1em} \null \times
  \langle 0 | \psi(\xi^+, 0, \Vec{0}_\perp) | P_h S_h, X \rangle
  \langle P_h S_h, X | \anti\psi(0) \I \sigma^{1-} \gamma_5 | 0 \rangle \, .
  \qquad
  \label{frag19}
\end{eqnarray}
Note that $\DT{D}(z)$ is the fragmentation function analogous to the transverse
polarisation distribution function $\DT{f}(x)$. In the literature $\DT{D}$ is
often called $H_1(z)$ \cite{Mulders:1996dh}.

Introducing the $\kappa$-integrated matrix
\begin{equation}
  \Xi(z) =
  \frac{z}{2} \int \! \frac{\d^4\kappa}{(2\pi)^4} \;
  \Xi(\kappa; P_h, S_h) \, \delta(1/z - \kappa^-/P_h^-) \, ,
  \label{frag20}
\end{equation}
the leading-twist structure of the fragmentation process is summarised in the
expression of $\Xi(z)$, which is
\begin{equation}
  \Xi(z) = \half
  \left\{
    D(z) \, \slashed{P}_h +
    \lambda_h \, \DL{D}(z) \, \gamma_5 \slashed{P}_h +
    \DT{D}(z) \, \slashed{P}_h \gamma_5 \slashed{S}_{hT}
  \right\} .
  \label{frag21}
\end{equation}

The probabilistic interpretation of $D(z)$, $\DL{D}(z)$ and $\DT{D}(z)$ is
analogous to that of the corresponding distribution functions (see
Sec.~\ref{probab}). If we denote by $\mathcal{N}_{h/q}(z)$ the probability of
finding a hadron with longitudinal momentum fraction $z$ inside a quark $q$,
then we have (using $\pm$ to label longitudinal polarisation states and
$\uparrow\downarrow$ to label transverse polarisation states)
\begin{subequations}
\begin{eqnarray}
  D(z) &=&
  \mathcal{N}_{h/q}(z) \, ,
  \label{frag22}
\\
  \DL{D}(z) &=&
  \mathcal{N}_{h/q+}(z) - \mathcal{N}_{h/q-}(z) \, ,
  \label{frag23}
\\
  \DT{D}(z) &=&
  \mathcal{N}_{h/q\uparrow}(z) - \mathcal{N}_{h/q\downarrow}(z) \, .
  \label{frag24}
\end{eqnarray}
\end{subequations}

\subsection{$\Vec\kappa_T$-dependent fragmentation functions}
\label{kappatransv}

\begin{figure}[htbp]
  \centering
  \begin{picture}(305,160)(0,25)
    \Photon(40,150)(146,150){3}{9.5}
    \LongArrow(146,150)(150,150)
    \ArrowLine(215,148)(198,148)
    \ArrowLine(182,148)(150,148)
    \ArrowLine(150,152)(190,152)
    \SetWidth{2}
    \ArrowLine(260,148)(220,148)
    \ArrowLine(190,152)(220,177)
    \SetWidth{0.5}
    \GOval(220,148)(5,10)(90){0.5}
    \GOval(190,152)(5,10)(90){0.5}
    \Text(150,130)[]{(a)}
    \Text(265,148)[l]{$\Vec{P}$}
    \Text(225,177)[l]{$\Vec{P}_h$}
    \Text(95,160)[]{$\Vec{q}$}
    \SetOffset(0,20)
    \DashLine(40,50)(260,50){5}
    \Photon(40,75)(146,51){3}{9.5}
    \LongArrow(146,51)(150,50)
    \ArrowLine(220,50)(150,50)
    \ArrowLine(150,50)(190,25)
    \SetWidth{2}
    \ArrowLine(260,50)(220,50)
    \ArrowLine(190,25)(230,25)
    \SetWidth{0.5}
    \GOval(220,50)(5,10)(90){0.5}
    \GOval(190,25)(5,10)(90){0.5}
    \Text(150,10)[]{(b)}
    \Text(95,75)[]{$\Vec{q}$}
    \Text(265,50)[l]{$\Vec{P}$}
    \Text(235,25)[l]{$\Vec{P}_h$}
  \end{picture}
  \caption{Kinematics in (a) the $\gamma^*N$ frame and (b) the $hN$ frame.}
  \label{kappat}
\end{figure}
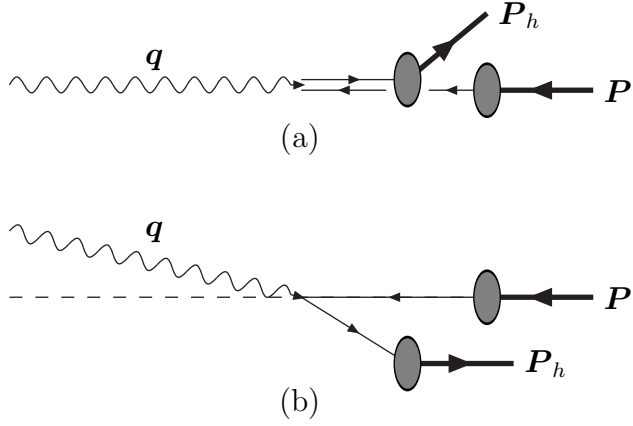

In the collinear case ($\Vec{k}_{T}=\Vec\kappa_T=0$) the produced hadron is
constrained to have zero transverse momentum
($\Vec{P}_{h\perp}=-z\Vec{q}_{T}=0$). Therefore, in order to investigate its
$\Vec{P}_{h\perp}$ distribution within the parton model, one has to account for
the transverse motion of quarks (in \QCD transverse momenta of quarks emerge at
\NLO owing to gluon emission). The kinematics in the $\gamma^*N$ and $hN$
frames is depicted in Fig.~\ref{kappat} (for simplicity the case of no
transverse motion of quarks inside the target is illustrated).

Reintroducing $\Vec\kappa_T$, we have at leading twist \cite{Levelt:1994ac,
Levelt:1994np, Tangerman:1995hw, Mulders:1996dh, Boglione:1999pz}
\begin{subequations}
\begin{eqnarray}
  \mathcal{V}^\mu &=&
  \frac12 \, \Tr(\gamma^\mu \, \Xi) =
  B_1 \, P_h^\mu +
  \frac1{M_h} \, B'_1 \, \varepsilon^{\mu\nu\rho\sigma} \,
  P_{h\nu} \kappa_{T\rho} S_{h T \sigma} \, ,
  \label{kappaT01}
\\
  \mathcal{A}^\mu &=&
  \frac12 \, \Tr(\gamma^\mu \gamma_5 \, \Xi) =
  \lambda_h \, B_2 \, P_h^\mu + \frac1{M_h} \, \widetilde{B}_1 \,
  \Vec\kappa_{T}{\cdot}\Vec{S}_{h T} \, P_h^\mu \, ,
  \label{kappaT02}
\\
  \mathcal{T}^{\mu\nu} &=&
  \frac1{2\I} \, \Tr(\sigma^{\mu\nu} \gamma_5 \, \Xi) =
  B_3 \, P_h^{[\mu} S_{h T}^{\nu ]} \, +
  \frac{\lambda_h}{M_h} \, \widetilde{B}_2 \, P_h^{[\mu} \kappa_{T}^{\nu]}
  \null +
  \frac1{M_h^2} \, \widetilde{B}_3 \, \Vec\kappa_{T}{\cdot}\Vec{S}_{hT} \,
  P_h^{[\mu} \kappa_{T}^{\nu]}
  \nonumber
\\
  && \hspace{8em} \null +
  \frac1{M_h} \, B'_2 \, \varepsilon^{\mu\nu\rho\sigma} \,
  P_{h \rho} \kappa_{T \sigma} \, ,
  \label{kappaT03}
  \sublabel{subkappaT03}
\end{eqnarray}
\end{subequations}
where we have introduced new functions $\widetilde{B}_i(\kappa^2,\kappa{\cdot}P_h)$
(the tilde signals the presence of $\Vec\kappa_{T}$),
$B'_i(\kappa^2,\kappa{\cdot}P_h)$ (the prime labels the $T$-odd terms) and inserted
powers of $M_h$ so that all coefficients have the same dimension. Contracting
eqs.~(\ref{kappaT01}--\ref{subkappaT03}) with $P_\mu$ yields
\begin{subequations}
\begin{eqnarray}
  \frac1{2P_h^-} \, \Tr(\gamma^- \Xi) &=&
  B_1 +
  \frac1{M_h} \, B'_1 \, \varepsilon_{T}^{ij} \kappa_{T i} S_{h T j} \, ,
  \label{kappaT1}
\\
  \frac1{2P^-} \, \Tr(\gamma^- \gamma_5 \Xi) &=&
  \lambda_h \, B_2 +
  \frac1{M_h} \, \Vec\kappa_{T}{\cdot}\Vec{S}_{hT} \, \widetilde{B}_1,
  \label{kappaT2}
\\
  \frac1{2P^-} \, \Tr(\I \sigma^{i -} \gamma_5 \Xi) &=&
  \left(
    B_3 + \frac{\Vec\kappa_{T}^2}{2M_h^2} \, \widetilde{B}_3
  \right) S_{h T}^i +
  \frac{\lambda_h}{M_h} \, \widetilde{B}_2 \, \kappa_T^i
  \nonumber
\\
  && \hspace{-4em} \null -
  \frac1{M_h^2} \, \widetilde{B}_3
  \left(
    \kappa_{T}^i \kappa_{T}^j + \frac12 \, \Vec\kappa_{T}^2 \, g_\perp^{ij}
  \right) S_{h T j}
  +
  \frac1{M_h} \, B'_2 \, \varepsilon_T^{ij} \kappa_{T j} \, . \qquad
  \label{kappaT3}
  \sublabel{subkappaT3}
\end{eqnarray}
\end{subequations}
The eight $\Vec\kappa_T$-dependent fragmentation functions are obtained from
the $B$ coefficients as follows
\begin{equation}
  D(z, {\Vec\kappa'_T}^2) = \frac1{2z}
  \int \! \frac{\d\kappa^+ \d\kappa^-}{(2\pi)^4} \,
  B_1(\kappa^2, \kappa{\cdot}P_h) \, \delta(1/z - \kappa^-/P_h^- ) \, ,
  \label{kappaT4}
\end{equation}
\etc, where $\Vec\kappa'_T\equiv-z\Vec\kappa_T$ is the transverse momentum of
the hadron $h$ with respect to the fragmenting quark, see eq.~(\ref{sidis20a}).
If the transverse motion of quarks inside the target is ignored, then
$\Vec\kappa'_T$ coincides with $\Vec{P}_{h\perp}$.

Defining the integrated trace
\begin{eqnarray}
  \Xi^{[\Gamma]} &\equiv&
  \frac1{4z} \int \! \frac{\d\kappa^+ \d\kappa^-}{(2\pi)^4} \,
  \Tr(\Gamma \Xi) \, \delta(\kappa^- - P_h^-/z)
  \nonumber
\\
  &=&
  \frac1{4z} \, \sum_X \int \! \frac{\d\xi^+ \d^2\Vec\xi_T}{(2\pi)^3} \,
  \e^{\I (P_h^- \xi^+/z - \Vec\kappa_T{\cdot}\Vec\xi_T)}
  \nonumber
\\
  && \hspace{2.5em} \null \times
  \Tr
  \langle 0 | \psi(\xi^+, 0, \Vec\xi_T) | P_h S_h, X \rangle
  \langle P_h S_h, X | \anti\psi(0) \Gamma | 0 \rangle \, ,
  \label{kappaT5}
\end{eqnarray}
we obtain from (\ref{kappaT1}--\ref{subkappaT3})
\begin{subequations}
\begin{eqnarray}
  \Xi^{[\gamma^-]} &=&
  \mathcal{N}_{h/q}(z, \Vec\kappa'_T)
  \nonumber
\\
  &=&
  D(z, {\Vec\kappa'_T}^2) +
  \frac1{M_h} \, \varepsilon_{T}^{ij}
  \kappa_{Ti} S_{hTj} \, D_{1T}^\perp(z, {\Vec\kappa'_T}^2) \, ,
  \label{kappaT6}
\\
  \Xi^{[\gamma^- \gamma_5]}
  &=&
  \mathcal{N}_{h/q}(z, \Vec\kappa'_T) \, \lambda'(z, \Vec\kappa'_T)
  \nonumber
\\
  &=&
  \lambda_h \, \DL{D}(z, {\Vec\kappa'_T}^2) +
  \frac1{M_h} \, \Vec\kappa_{T}{\cdot}\Vec{S}_{h T} \,
  G_{1T}(z, {\Vec\kappa'_T}^2) \, ,
  \label{kappaT7}
\\
  \Xi^{[\I \sigma^{i-} \gamma_5]}
  &=&
  \mathcal{N}_{h/q}(z, \Vec\kappa'_T) \, {s'}_\perp^i(z, \Vec\kappa'_T)
  \nonumber
\\
  &=&  S_{h T}^i \, \Delta'_T D(z, {\Vec\kappa'_T}^2) +
  \frac{\lambda_h}{M_h} \, \kappa_T^i H_{1L}^\perp(z, {\Vec\kappa'_T}^2)
  \nonumber
\\
  && \null -
  \frac1{M_h^2}
  \left(
    \kappa_{T}^i \kappa_{T}^j +
    \frac12 \, \Vec\kappa_{T}^2 \, g_\perp^{ij}
  \right) S_{h T j} \, H_{1 T}^\perp(z, {\Vec\kappa'_T}^2)
  \nonumber
\\
  && \null +
  \frac1{M_h} \, \varepsilon_T^{ij}
  \kappa_{T j} \, H_1^\perp(z, {\Vec\kappa'_T}^2) \, ,
  \label{kappaT8}
  \sublabel{subkappaT8}
\end{eqnarray}
\end{subequations}
where $\Vec{s}'=(\Vec{s}'_\perp,\lambda')$ is the spin of the quark and
$\mathcal{N}_{h/q}(z,\Vec\kappa'_T)$ is the probability of finding a hadron
with longitudinal momentum fraction $z$ and transverse momentum
$\Vec\kappa'_T=-z\Vec\kappa_T$, with respect to the quark momentum, inside a
quark~$q$.

In (\ref{kappaT6}--\ref{subkappaT8}) we have adopted a more traditional
notation for the three fragmentation functions, $D$, $\DL{D}$ and $\DT{D}$,
that survive upon integration over $\Vec\kappa_T$ whereas we have resorted to
Mulders' terminology \cite{Mulders:1996dh} for the other, less familiar,
fragmentation functions, $D_{1T}^\perp$, $G_{1T}$, $H_{1L}^\perp$,
$H_{1T}^\perp$ and $H_{1}^\perp$ (note that in Mulders' scheme $D$, $\DL{D}$
and $\Delta'_T D$ are called $D_1$, $G_{1L}$ and $H_{1T}$, respectively, and
$D_1$, $G_{1}$ and $H_{1}$, once integrated over $\Vec\kappa_T$). The
integrated fragmentation functions $D(z)$ and $\DL{D}(z)$, are obtained from
$D(z,{\Vec\kappa'_T}^2)$ and $\DL{D}(z,{\Vec\kappa'_T}^2)$, via
\begin{subequations}
\begin{align}
       D(z) &= \int \! \d^2\Vec\kappa'_T \,      D(z,{\Vec\kappa'_T}^2) \, ,
  \label{kappaT9}
\\
  \DL{D}(z) &= \int \! \d^2\Vec\kappa'_T \, \DL{D}(z,{\Vec\kappa'_T}^2) \, ,
  \label{kappaT9a}
\intertext{whereas $\DT{D}(z)$ is given by}
  \DT{D}(z) &= \int \! \d^2\Vec\kappa'_T
  \left(
    \Delta'_T D(z,{\Vec\kappa'_T}^2) +
    \frac{{\Vec\kappa'_T}^2}{2M_h^2} \, H_{1T}^\perp(z,{\Vec\kappa'_T}^2)
  \right) .
  \label{kappaT10}
\end{align}
\end{subequations}

Among the unintegrated fragmentation functions, the $T$-odd quantity
$H_1^\perp(z,{\Vec\kappa'_T}^2)$ plays an important r\^{o}le in the phenomenology of
transversity as it is related to the Collins effect, \ie, the observation of
azimuthal asymmetries in single-inclusive production of unpolarised hadrons at
leading twist. In partonic terms, $H_1^\perp$ is defined -- see
eq.~(\ref{todd6}) for the corresponding distribution function $h_1^\perp$ --
via
\begin{equation}
  \mathcal{N}_{h/q \uparrow}  (z,\Vec\kappa'_{T}) -
  \mathcal{N}_{h/q \downarrow}(z,\Vec\kappa'_{T}) =
  \frac{|\Vec\kappa_{T}|}{M_h} \, \sin(\phi_\kappa - \phi_{s'}) \;
  H_1^\perp(z,{\Vec\kappa'_T}^2) \, ,
  \label{kappaT11}
\end{equation}
where $\phi_\kappa$ and $\phi_{s'}$ are the azimuthal angles of the quark
momentum and polarisation, respectively, defined in a plane perpendicular to
$\Vec{P}_h$. The angular factor in (\ref{kappaT11}), that is (recall that
$\Vec{P}_h$ is directed along $-z$)
\begin{equation}
  \sin(\phi_\kappa - \phi_{s'}) =
  \frac{(\Vec\kappa \vprod \Vec{P}_h){\cdot}\Vec{s}'}
       {|\Vec\kappa \vprod \Vec{P}_h| \, |\Vec{s}'|} \, ,
  \label{kappaT11b}
\end{equation}
is related to the so-called Collins angle (see Sec.~\ref{sidisasymm}), as we
now show. First of all, note that on neglecting $\Ord(1/Q)$ effects, azimuthal
angles in the plane perpendicular to the $hN$ axis coincide with the azimuthal
angles defined in the plane perpendicular to the $\gamma^*N$ axis. Then, if we
ignore the intrinsic motion of quarks inside the target, we have
$\Vec\kappa_T=-\Vec{P}_{h\perp}/z$ and
\begin{equation}
  \phi_\kappa = \phi_h - \pi \, .
  \label{kappaT11c}
\end{equation}
The angle in (\ref{kappaT11b}) is therefore
\begin{equation}
  \phi_\kappa - \phi_{s'} = \phi_h - \phi_{s'} - \pi
  = - \Phi_C - \pi \, ,
  \label{kappaT11d}
\end{equation}
so that
\begin{equation}
  \sin(\phi_{\kappa} - \phi_{s'}) = \sin\Phi_C \, ,
  \label{kappaT11e}
\end{equation}
where $\Phi_C$, the azimuthal angle between the spin vector of the fragmenting
quark and the momentum of the produced hadron, is what is known as the Collins
angle \cite{Collins:1993kk}.

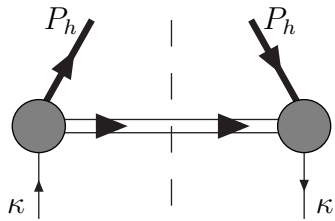
\begin{figure}[htbp]
  \centering
  \begin{picture}(200,85)(0,10)
    \SetWidth{0.5}
    \ArrowLine(50,15)(50,40)
    \ArrowLine(150,40)(150,15)
    \Line(50,47.5)(150,47.5)
    \Line(50,52.5)(150,52.5)
    \SetWidth{2.5}
    \ArrowLine(52,58)(70,90)
    \ArrowLine(130,90)(148,58)
    \SetWidth{3}
    \LongArrow(80,50)(82,50)
    \LongArrow(126,50)(128,50)
    \SetWidth{0.5}
    \DashLine(100,20)(100,90){10}
    \GCirc(50,50){10}{0.5}
    \GCirc(150,50){10}{0.5}
    \Text(65,90)[r]{$P_h$}
    \Text(135,90)[l]{$P_h$}
    \Text(45,20)[r]{$\kappa$}
    \Text(155,20)[l]{$\kappa$}
  \end{picture}
  \caption{Toy model for fragmentation.}
  \label{fragmod}
\end{figure}

Just to show how the $T$-odd fragmentation function $H_1^\perp$ may arise from
non-trivial final-state interactions, as discussed in Sec.~\ref{systfrag}, let
us consider a toy model \cite{Bianconi:1999cd} (see Fig.~\ref{fragmod}) that
provides a simple example of the mechanism symbolically presented in
Fig.~\ref{toddfrag}. Thus, we assume that the quark, with momentum $\kappa$ and
mass $m$, fragments into an unpolarised hadron, leaving a remnant which is a
point-like scalar diquark. The fragmentation function $H_1^\perp$ is contained
in the tensor component of the matrix $\Xi$
\begin{equation}
  \Xi(\kappa, P_h) = \frac12
  \left\{
    \dots +
    \frac{\I}{2} \, \mathcal{T}_{\mu\nu} \sigma^{\mu\nu} \gamma_5
  \right\} ,
  \label{kappaT12}
\end{equation}
where -- see eq.~(\ref{kappaT03}),
\begin{equation}
  \mathcal{T}_{\mu\nu} =
  \dots +
  \frac1{M_h} \, B'_2 \, \varepsilon_{\mu\nu\rho\sigma} \,
  P_h^\rho \kappa^\sigma \, ,
  \label{kappaT13}
\end{equation}
so that, using $\gamma_5\sigma^{\mu\nu}=
\half\I\,\varepsilon^{\mu\nu\alpha\beta}\,\sigma_{\alpha\beta}$,
\begin{equation}
  \Xi(\kappa, P_h) = \frac12
  \left\{
    \dots \, + \frac1{M_h} \, B'_2 \, \sigma^{\mu\nu} P_{h\mu} \kappa_\nu
  \right\} .
  \label{kappaT14}
\end{equation}
If we describe the hadron $h$ by a plane wave
\begin{equation}
  \psi_h (x) \sim u(P_h) \, \e^{\I \, P_h{\cdot}x} \, ,
  \label{kappaT15}
\end{equation}
it is easy to show that the fragmentation matrix $\Xi$ is
\begin{eqnarray}
  \Xi(\kappa, P_h) &\sim& \frac{- \I}{\slashed\kappa - m} \, u(P_h) \,
  \anti{u}(P_h) \, \frac{\I}{\slashed\kappa - m}
  \nonumber
\\
  & \sim &
  \frac{\slashed\kappa + m}{\kappa^2 - m^2}
  \left( \slashed{P}_h + M_h \right)
  \frac{\slashed\kappa + m}{\kappa^2 - m^2} \, ,
  \label{kappaT16}
\end{eqnarray}
where we have omitted inessential factors.

We cannot extract a factor proportional to $\sigma^{\mu\nu}P_{h\mu}\kappa_\nu$
(hence, producing $H_1^\perp$) from (\ref{kappaT16}).

Let us now suppose that a residual interaction of $h$ with the intermediate
state generates a phase in the hadron wave function. If, for instance, in
(\ref{kappaT16}) we make the replacement (assuming only two fragmentation
channels)
\begin{equation}
  u(P_h) \to u(P_h) + \e^{\I \chi} \, \slashed\kappa \, u(P_h) \, ,
  \label{kappaT17}
\end{equation}
by a little algebra one can show that a term of the type (\ref{kappaT14})
emerges in $\Xi$, with
\begin{equation}
  B'_2 \sim \frac{M_h}{\kappa^2 - m^2} \, \sin\chi \, .
  \label{kappaT18}
\end{equation}
Therefore, if the interference between the fragmentation channels produces a
non-zero phase $\chi$, $T$-odd contributions may appear. The proliferation of
channels, however, might lead, as suggested in \cite{Jaffe:1998hf}, to the
vanishing of such phases and of the resulting $T$-odd fragmentation functions.

Another microscopic mechanism that may give rise to a $T$-odd fragmentation
function has been recently investigated in \cite{Bacchetta:2001di}. Using a
simple pseudoscalar coupling between pions and quarks to model the
fragmentation process, these authors show that the inclusion of one-loop
self-energy and vertex  corrections generates a non-vanishing $H_1^{\perp}$.

\subsection{Cross-sections and asymmetries in semi-inclusive leptoproduction}
\label{sidisasymm}

We shall now calculate the trace in (\ref{sidis22a}).

At leading twist, as already mentioned, transverse polarisation distributions
are probed by unpolarised lepton beams. In this case, the leptonic tensor is
symmetric and couples to the symmetric part of $W^{\mu\nu}$, that is
\begin{eqnarray}
  W^{\mu\nu (S)} &=&
  \frac12 \, \sum_a e_a^2
  \int \! \frac{\d{k}^- \, \d^2\Vec{k}_T}{(2\pi)^4}
  \int \! \frac{\d\kappa^+ \, \d^2\Vec\kappa_T }{(2\pi)^4}
  \nonumber
\\
  && \hspace{1em} \null \times
  \delta^2(\Vec{k}_T + \Vec{q}_T - \Vec\kappa_T) \,
  \Tr
  \left[
    \Phi \, \gamma^{\{\mu} \, \Xi \, \gamma^{\nu\}}
  \right]_{k^+=xP^+,\,\kappa^-=P_h^-/z} \, . \qquad
  \label{sicross1}
\end{eqnarray}
Using the Fierz identity we can decompose the trace in (\ref{sicross1}) as
\begin{eqnarray}
  \Tr \left[ \Phi \, \gamma^{\{\mu} \, \Xi \, \gamma^{\nu\}} \right] &=&
  \half
  \left\{ \strut
  \Tr \left[ \strut \Phi \right] \Tr \left[ \Xi \right] +
  \Tr \left[ \strut \I \, \Phi \, \gamma_5 \right]
  \Tr \left[ \strut \I \, \Xi  \, \gamma_5 \right]
  \right.
  \nonumber
\\
  && \null -
  \Tr \left[ \strut \Phi \, \gamma^\alpha \right]
  \Tr \left[ \strut \Xi  \, \gamma_\alpha \right] -
  \Tr \left[ \strut \Phi \, \gamma^\alpha \, \gamma_5 \right]
  \Tr \left[ \strut \Xi  \, \gamma_\alpha \, \gamma_5 \right]
  \nonumber
\\
  && \left. \null +
  \half \,
  \Tr \left[ \strut \I \, \Phi \, \sigma^{\alpha\beta} \, \gamma_5 \right]
  \Tr \left[ \strut \I \, \Xi  \, \sigma_{\alpha\beta} \, \gamma_5 \right]
  \right\} \, g^{\mu\nu}
  \nonumber
\\
  && \null +
  \half \,
  \Tr \left[ \Phi \, \gamma^{\{\mu} \right]
  \Tr \left[ \Xi  \, \gamma^{\nu\}} \right] +
  \half \,
  \Tr \left[ \Phi \, \gamma^{\{\mu} \, \gamma_5 \right]
  \Tr \left[ \Xi  \, \gamma^{\nu\}} \, \gamma_5 \right]
  \nonumber
\\
  && \null +
  \half \,
  \Tr \left[ \I \, \Phi \, \sigma^{\alpha\{\mu} \, \gamma_5 \right]
  \Tr \left[ \I \, \Xi  \, \sigma^{\nu\}}_{\alpha} \, \gamma_5 \right] .
  \label{sicross2}
\end{eqnarray}

If we insert eqs.~(\ref{tm2}--\ref{subtm4}) and
(\ref{kappaT01}--\ref{subkappaT03}) into (\ref{sicross1}) and integrate over
$k^-$ and $\kappa^+$ making use of eqs.~(\ref{tm11}--\ref{subtm13}) and
(\ref{kappaT6}--\ref{subkappaT8}), after some algebra we obtain
\cite{Levelt:1994ac, Levelt:1994np, Tangerman:1995hw, Mulders:1996dh}
\begin{eqnarray}
  W^{\mu\nu (S)} &=&
  2 \sum_a e_a^2 \, z \int \! \d^2\Vec\kappa_T \int \! \d^2\Vec{k}_T \,
  \delta^2(\Vec{k}_T +  \Vec{q}_T - \Vec\kappa_T) \,
  \nonumber
\\
  &\times&
  \left\{
  - g_T^{\mu\nu}
  \left[
    f(x, \Vec{k}_T^2) \, D(z, {\Vec\kappa'_T}^2) +
    \frac1{M_h} \, \varepsilon_T^{\rho\sigma} \kappa_{T\rho} S_{h T \sigma} \,
    f(x, \Vec{k}_T^2) \, D_{1T}^\perp(z, {\Vec\kappa'_T}^2)
  \right]
  \STRUT \right.
  \nonumber
\\
  && \null -
  \left[
    S_T^{\{\mu} S_{h T}^{\nu \}} + \Vec{S}_T{\cdot}\Vec{S}_{hT} \, g_T^{\mu\nu}
  \right]
  \Delta'_T f(x, \Vec{k}_T^2) \, \Delta'_T D(z, {\Vec\kappa'_T}^2)
  \nonumber
\\
  && \null -
  \frac{\left[S_T^{\{\mu} \kappa_{T}^{\nu \}} +
          \Vec{S}_T{\cdot}\Vec\kappa_T \, g_T^{\mu\nu}
        \right] \Vec\kappa_T{\cdot}\Vec{S}_{hT}}{M_h^2} \,
  \DT{f}(x, \Vec{k}_T^2) \, H_{1T}^\perp(z, {\Vec\kappa'_T}^2)
  \nonumber
\\
  && \null -
  \frac{S_T^{\{\mu} \varepsilon_T^{\nu\}\rho} \kappa_{T\rho} +
  \kappa_T^{\{\mu} \varepsilon_T^{\nu\}\rho} S_{T\rho}}{2M_h} \, \Delta'_T
  f(x, \Vec{k}_T^2) \, H_{1}^\perp(z, {\Vec\kappa'_T}^2) + \dots
  \left. \STRUT \right\} .
  \label{sicross4}
\end{eqnarray}
In (\ref{sicross4}) we have considered only unpolarised and transversely
polarised terms, and we have omitted the $\Vec{k}_T$-dependent contributions
(in the following we shall assume that transverse motion of quarks inside the
target can be neglected).

Neglecting higher-twist (\ie, $\Ord(1/Q)$) contributions, the transverse ($T$)
vectors and tensors appearing in (\ref{sicross4}) coincide with the
corresponding perpendicular ($\perp$) vectors and tensors. The contraction of
$W^{\mu\nu(S)}$ with the leptonic tensor $L_{\mu\nu}^{(S)}$ is performed by
means of the identities  \cite{Mulders:1996dh}
\begin{subequations}
\begin{align}
  g_\perp^{\mu\nu} \, L_{\mu\nu}^{(S)} &= -
  \frac{2 Q^2}{y^2} \left[ 1 + (1-y)^2 \right] ,
  \label{sicross5}
\\
  \left[
    a_\perp^{\{\mu} b_\perp^{\nu\}} +
    \Vec{a}_\perp{\cdot}\Vec{b}_\perp \, g_\perp^{\mu\nu}
  \right] L_{\mu\nu}^{(S)}
  &= \frac{4Q^2(1-y)}{y^2} \, | \Vec{a}_\perp | | \Vec{b}_\perp |
  \cos(\phi_a + \phi_b) \, ,
  \label{sicross6}
\\
  \half
  \left[
    a_\perp^{\{\mu} \varepsilon_\perp^{\nu\}\rho} b_{\perp\rho} +
    b_\perp^{\{\mu} \varepsilon_\perp^{\nu\}\rho} a_{\perp\rho}
  \right] L_{\mu\nu}^{(S)}
  &=
  -
  \frac{4Q^2(1-y)}{y^2} \,
  | \Vec{a}_\perp | | \Vec{b}_\perp | \sin(\phi_a + \phi_b) \, ,
  \label{sicross7}
  \sublabel{subsicross7}
\end{align}
\end{subequations}
where $\phi_a$ and $\phi_b$ are the azimuthal angles in the plane perpendicular
to the photon--nucleon axis. Combining eq.~(\ref{sicross4}) with
eqs.~(\ref{sicross5}--\ref{subsicross7}) leads quite straight-forwardly to the
parton-model formul{\ae} for the cross-sections. To obtain the leading-order \QCD
expressions, one must simply insert the $Q^2$ dependence into the distribution
and fragmentation functions.

\subsubsection{Integrated cross-sections}

Consider, first of all, the cross-sections integrated over $\Vec{P}_{h\perp}$.
In this case, the $\Vec{k}_T$ and $\Vec\kappa_T$ integrals decouple and can be
performed, yielding the integrated distribution and fragmentation functions.
Hence we obtain
\begin{eqnarray}
  \frac{\d\sigma}{\d{x} \d{y} \d{z}} &=&
  \frac{4\pi\alpha_\text{em}^2s}{Q^4} \, \sum_a e_a^2 \, x
  \left\{
    \frac12  \left[ 1 + (1-y)^2 \right] f_a(x) \, D_a(z)
  \right.
  \nonumber
\\
  &-&
  \left. \Strut
    (1-y) \, | \Vec{S}_\perp | \, | \Vec{S}_{h\perp} | \,
    \cos(\phi_S + \phi_{S_{h}}) \, \DT{f}_a(x) \, \DT{D}_a(z)
  \right\} . \qquad
  \label{sicross8}
\end{eqnarray}
As one can see, at leading twist, the transversity distributions are probed
only when both the target and the produced hadron are transversely polarised.

From (\ref{sicross8}) we can extract the transverse polarisation
$\Vec{\mathcal{P}}_h$ of the detected hadron, defined so that (`unp' =
unpolarised)
\begin{equation}
  \d\sigma =
  \d\sigma_\text{unp} \, (1 + \Vec{\mathcal{P}}_h {\cdot}\Vec{S}_{h}) \, .
  \label{sicross8b}
\end{equation}
If we denote by $\mathcal{P}_{hy}^\uparrow$ the transverse polarisation of $h$
along $y$, when the target nucleon is polarised along $y$ ($\uparrow$), and by
$\mathcal{P}_{hx}^{\to}$ the transverse polarisation of $h$ along $x$, when the
target nucleon is polarised along $x$ ($\to$), we find
\begin{equation}
  \mathcal{P}_{hy}^\uparrow = -
  \mathcal{P}_{hx}^{\to} =
  \frac{2(1-y)}{1 + (1-y)^2} \,
  \frac{\sum_a e_a^2 \, \DT{f}_a (x) \, \DT{D}_a (z)}
       {\sum_a e_a^2 \,      f_a (x) \,      D_a (z)} \, .
  \label{sicross10}
\end{equation}

If the hadron $h$ is not transversely polarised, or -- \emph{a fortiori} -- is
spinless, the leading-twist $\Vec{P}_{h\perp}$-integrated cross-section does
not contain $\DT{f}$. In this case, in order to probe the transversity
distributions, one has to observe the $\Vec{P}_{h\perp}$ distributions, or
consider higher-twist contributions (Sec.~\ref{sidistwist3}). In the next
section we shall discuss the former possibility.

\subsubsection{Azimuthal asymmetries}
\label{azimasym}

We now study the (leading-twist) $\Vec{P}_{h\perp}$ distributions in
semi-inclusive \DIS and the resulting azimuthal asymmetries. We shall assume
that the detected hadron is spinless, or that its polarisation is not observed.
For simplicity, we also neglect (at the beginning, at least) the transverse
motion of quarks inside the target. Thus (\ref{sicross4}) simplifies as follows
(recall that only the unpolarised and the transversely polarised terms are
considered)
\begin{eqnarray}
  W^{\mu\nu (S)} &=&
  2 \sum_a e_a^2 \, z \int \! \d^2\Vec\kappa_{T} \,
  \delta^2(\Vec\kappa_{T} + \Vec{P}_{h\perp}/z)
  \nonumber
\\
  && \hspace{1em} \times
  \left\{ \Strut \right.
    - g_\perp^{\mu\nu} \, f(x) \, D(z, {{\Vec\kappa}'_{T}}^2)
  \nonumber
\\
  && \hspace{3em} -
  \frac{\left(
          S_T^{\{\mu} \varepsilon_T^{\nu\}\rho} \kappa_{T\rho} +
          \kappa_T^{\{\mu} \varepsilon_T^{\nu\}\rho} S_{T\rho}
        \right)}
       {2M_h} \, \DT{f}(x) \, H_{1}^\perp(z, {\Vec\kappa'_T}^2)
  \nonumber
\\
  && \hspace{3em} + \dots \left. \Strut \right\} .
  \label{sicross11}
\end{eqnarray}
Contracting $W^{\mu\nu(S)}$ with the leptonic tensor (\ref{sidis22}) and
inserting the result into (\ref{sidis9}) gives the cross-section
\begin{eqnarray}
  \frac{\d\sigma}{\d{x} \d{y} \d{z} \d^2\Vec{P}_{h\perp}} &=&
  \frac{4\pi\alpha_\text{em}^2s}{Q^4} \, \sum_a e_a^2 \, x
  \left\{
    \frac12 \left[ 1 + ( 1 - y)^2 \right]
    f_a(x) \, D_a(z, \Vec{P}_{h\perp}^2)
  \right.
  \nonumber
\\
  && \hspace{7em} \null +
  (1- y) \, \frac{| \Vec{P}_{h\perp} |}{z M_h} \,
  | \Vec{S}_\perp | \, \sin(\phi_S + \phi_h)
  \nonumber
\\
  && \hspace{8em} \null \times
  \left. \Strut
    \DT{f}_a(x) \, H_{1a}^\perp(z, \Vec{P}_{h\perp}^2)
  \right\} .
  \label{sicross12}
\end{eqnarray}
From this we obtain the transverse single-spin asymmetry
\begin{eqnarray}
  A_T^h &\equiv&
  \frac{\d\sigma(\Vec{S}_\perp) - \d\sigma(-\Vec{S}_\perp)}
       {\d\sigma(\Vec{S}_\perp) + \d\sigma(-\Vec{S}_\perp)}
  \nonumber
\\
  &=&
  \frac{2 (1-y)}{1 + (1-y)^2} \,
  \frac{\sum_a e_a^2 \, \DT{f}_a(x) \, \DT^0 D_a(z, \Vec{P}_{h\perp}^2)}
       {\sum_a e_a^2 \,      f_a(x) \,       D_a(z, \Vec{P}_{h\perp}^2)} \,
  | \Vec{S}_\perp | \sin(\phi_S + \phi_h) \, . \hspace{3em}
  \label{sicross13}
\end{eqnarray}
Here we have defined the $T$-odd fragmentation function
$\DT^0D(z,\Vec{P}_{h\perp}^2)$ as -- see (\ref{kappaT11})
\begin{equation}
  \DT^0 D(z, \Vec{P}_{h\perp}^2)
  =
  \frac{| \Vec{P}_{h\perp} |}{z M_h} \, H_1^\perp(z, \Vec{P}_{h\perp}^2) \, .
  \label{sicross14}
\end{equation}
Note that our $\DT^0D$ is related to $\Delta^ND$ of \cite{Anselmino:1999pw} by
$\DT^0D=\Delta^ND/2$ (our notation is explained in Sec.~\ref{notations}).

The existence of an azimuthal asymmetry in transversely polarised
leptoproduction of spinless hadrons at leading twist, which depends on the
$T$-odd fragmentation function $H_1^\perp$ and arises from final-state
interaction effects, was predicted by Collins \cite{Collins:1993kk} and is now
known as the Collins effect.

\begin{figure}[htbp]
  \centering
  \begin{picture}(280,220)(0,65)
    \LongArrow(20,160)(260,160)
    \LongArrow(140,280)(140,70)
    \SetWidth{2}
    \LongArrow(140,160)(230,220)
    \LongArrow(140,160)(180,230)
    \LongArrow(140,160)(108,216)
    \DashLine(180,230)(196,258){4}
    \LongArrow(196,258)(200,265)
    \SetWidth{0.5}
    \LongArrowArcn(140,160)(40,90,35)
    \LongArrowArc(140,160)(40,90,118)
    \Text(260,150)[t]{$x$}
    \Text(132,70)[r]{$y$}
    \Text(155,210)[]{$\Phi_C$}
    \Text(230,225)[b]{\large$\Vec{P}_{h\perp}$}
    \Text(200,270)[b]{\large$\Vec{S}_\perp$}
    \Text(108,222)[b]{\large$\Vec{s}'_\perp$}
    \Text(185,230)[l]{\large$\Vec{s}_\perp$}
  \end{picture}
  \caption{The transverse spin vectors and the transverse momentum of the
           outgoing hadron in the plane perpendicular to the $\gamma^*N$ axis.
           $\Phi_C$ is the Collins angle.}
  \label{azimuth2}
\end{figure}
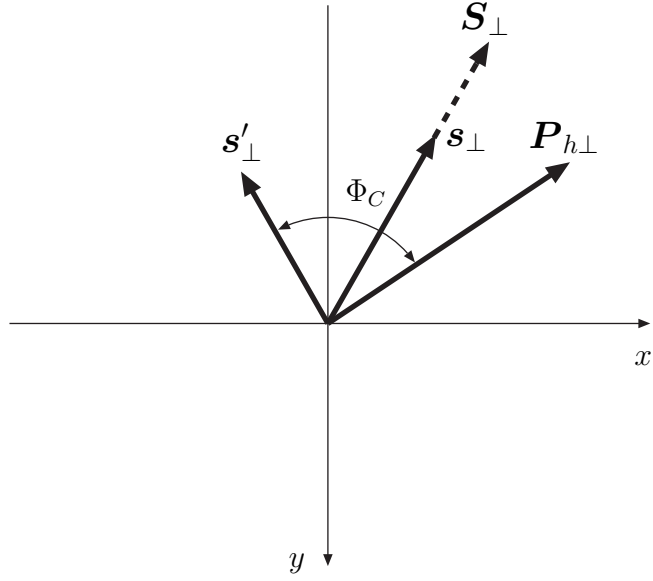

The Collins angle $\Phi_C$ was originally defined in \cite{Collins:1993kk} as
the angle between the transverse spin vector of the fragmenting quark and the
transverse momentum of the outgoing hadron, \ie,
\begin{equation}
  \Phi_C = \phi_{s'} - \phi_h \, .
  \label{sicross15}
\end{equation}
Thus, one has
\begin{equation}
  \sin\Phi_C =
  \frac{( \Vec{q} \vprod \Vec{P}_h){\cdot}\Vec{s}'}
       {| \Vec{q} \vprod \Vec{P}_h | \, | \Vec{s}' |} \, .
  \label{sicross16}
\end{equation}
Since, as dictated by QED (see Sec.~\ref{factsidis}), the directions of the
final and initial quark spins are related to each other by (see
Fig.~\ref{azimuth2})
\begin{equation}
  \phi_{s'} = \pi - \phi_s \, ,
  \label{sicross17}
\end{equation}
(\ref{sicross16}) becomes $\Phi_C=\pi-\phi_s-\phi_h$. Ignoring the transverse
motion of quarks in the target, the initial quark spin is parallel to the
target spin (\ie, $\phi_s=\phi_S$) and $\Phi_C$ can finally be expressed in
terms of measurable angles as
\begin{equation}
  \Phi_C = \pi - \phi_S - \phi_h \, .
  \label{sicross18}
\end{equation}

If the transverse motion of quarks in the target is taken into account the
cross-sections become more complicated. We limit ourselves to a brief overview
of them.  Let us start from the unpolarised cross-section, which reads
\begin{equation}
  \frac{\d\sigma_\text{unp}}{\d{x} \d{y} \d{z} \d^2\Vec{P}_{h\perp}} =
  \frac{4\pi\alpha_\text{em}^2s}{Q^4} \,
  \sum_a \frac{e_a^2}{2} \, x \left[ 1 + ( 1 -  y)^2 \right]
  I[ f_a \, D_a ] \, ,
  \label{sicross19}
\end{equation}
where we have introduced the integral $I$, defined as \cite{Mulders:1996dh}
\begin{eqnarray}
  I[ f \, D ] (x,z) &\equiv&
  \int \! \d^2\Vec{k}_T \, \d^2\Vec\kappa_T \,
  \delta^2(\Vec{k}_T + \Vec{q}_T - \Vec\kappa_T) \, f(x, \Vec{k}_T^2) \,
  D(z, {\Vec\kappa'_T}^2)
  \nonumber
\\
  &=&
  \int \! \d^2\Vec{k}_T \, \, f(x, \Vec{k}_T^2) \,
  D(z, | \Vec{P}_{h\perp} - z\Vec{k}_T |^2) \, .
  \label{sicross20}
\end{eqnarray}
The cross-section for a transversely polarised target takes the form
\begin{eqnarray}
  \frac{\d\sigma(\Vec{S}_\perp)}{\d{x} \d{y} \d{z} \d^2\Vec{P}_{h\perp}} &=&
  \frac{4\pi\alpha_\text{em}^2s}{Q^4} \, | \Vec{S}_\perp | \,
  \sum_a e_a^2 \, x(1-y)
  \nonumber
\\
  && \null \times
  I
  \left[
    \frac{\hat{\Vec{h}}{\cdot}\Vec\kappa_\perp}{M_h} \, \DT{f}_a \, H_{1a}^\perp
  \right] \sin(\phi_S + \phi_h) + \dots \, , \qquad
  \label{sicross21}
\end{eqnarray}
where $\hat{\Vec{h}}\equiv\Vec{P}_{h\perp}/|\Vec{P}_{h\perp}|$ and a term
giving rise to a $\sin(3\phi_h-\phi_S)$ asymmetry, but not involving $\DT{f}$,
has been omitted. As we shall see in Sec.~\ref{pionlepto} there are presently
some data on semi-inclusive \DIS off nucleons polarised along the scattering
axis, that are of a certain interest for the study of transversity. It is
therefore convenient, in view of the phenomenological analysis of those
measurements,  to give also the unintegrated cross-section for a longitudinally
polarised target, which, although not containing $\DT{f}$, depends on the
Collins fragmentation function $H_1^\perp$, a crucial ingredient in the
phenomenology of transversity. One finds
\begin{eqnarray}
  \frac{\d\sigma(\lambda_N)}{\d{x} \d{y} \d{z} \d^2\Vec{P}_{h\perp}} &=& -
  \frac{4\pi\alpha_\text{em}^2s}{Q^4} \, \lambda_N \sum_a e_a^2 \, x(1-y)
  \nonumber
\\
  &\times&
  I\left[
    \frac{2 \,
    (\hat{\Vec{h}}{\cdot}\Vec\kappa_\perp) \, (\hat{\Vec{h}}{\cdot}\Vec{k}_\perp) -
    \Vec\kappa_\perp{\cdot}\Vec{k}_\perp}{M M_h} \,
    h_{1La}^\perp \, H_{1a}^\perp
  \right] \sin(2 \phi_h) \, . \hspace{4em}
  \label{sicross22}
\end{eqnarray}
Note the characteristic $\sin(2\phi_h)$ dependence of (\ref{sicross21}) and the
appearance of the $\Vec{k}_\perp$-dependent distribution function
$h_{1L}^\perp$.

One can factorise the $x$ and $z$ dependence in the above expressions by
properly weighting the cross-sections with some function that depends on the
azimuthal angles \cite{Kotzinian:1997wt, Boer:1998nt}. This procedure also
singles out the different contributions to the cross-section for a given spin
configuration of the target (and of the incoming lepton). To see how it works
let us consider the  case of a transversely polarised target. We redefine the
azimuthal angles so that the orientation of the lepton plane is given by a
generic angle $\phi_{\ell}$ in the transverse space. Equation~(\ref{sicross21})
then becomes
\begin{eqnarray}
  \frac{\d\sigma(\Vec{S}_\perp)}{\d{x} \d{y} \d{z}
  \d\phi_{\ell} \d^2\Vec{P}_{h\perp}} &=&
  \frac{2 \alpha_\text{em}^2s}{Q^4} \, | \Vec{S}_\perp | \,
  \sum_a e_a^2 \, x(1-y)
  \nonumber
\\
  && \null \times
  I
  \left[
    \frac{\hat{\Vec{h}}{\cdot}\Vec\kappa_\perp}{M_h} \, \DT{f}_a \, H_{1a}^\perp
  \right]
  \sin(\phi_S + \phi_h - 2 \phi_{\ell})
  \nonumber
\\
  && \null + \sin( 3\phi_h - \phi_S - 2\phi_{\ell} ) \;\; \text{term}
  \label{sicross21b}
\end{eqnarray}
The weighted cross-section that projects the first term of (\ref{sicross21b})
out, and leads to a factorised expression in $x$ and $z$, is
\begin{eqnarray}
  & & \int \d\phi_{\ell} \, \d^2 \Vec{P}_{h\perp} \,
  \frac{\vert \Vec{P}_{h\perp}
  \vert}{M_h z} \, \sin(\phi_S + \phi_h - 2 \phi_\ell) \,
  \frac{\d\sigma(\Vec{S}_\perp)}{\d{x} \d{y} \d{z}
  \d\phi_{\ell} \d^2\Vec{P}_{h\perp}}
  \nonumber
\\
  && \null \hspace{0.8cm} =
  \frac{4\pi\alpha_\text{em}^2s}{Q^4} \, | \Vec{S}_\perp | \,
  \sum_a e_a^2 \, x(1-y)
  \DT{f}_a(x) \, H_{1a}^{\perp (1)}(z) \,,
  \label{sicross21c}
\end{eqnarray}
where the weighted fragmentation function  $H_1^{\perp(1)}$ is defined as
\begin{equation}
  H_1^{\perp(1)}(z) =
  z^2 \int \! \d^2\Vec\kappa_T
  \left( \frac{\Vec\kappa_T^2}{2M_h^2} \right) H_1^\perp(z, z^2\Vec\kappa_T^2) \, .
  \label{pion26}
\end{equation}

For a more complete discussion of transversely polarised semi-inclusive
leptoproduction, with or without intrinsic quark motion, we refer the reader to
the vast literature on the subject \cite{Levelt:1994ac, Tangerman:1995hw,
Mulders:1996dh, Kotzinian:1995dv, Kotzinian:1996cz, Jaffe:1996wp,
Kotzinian:1997wt, Mulders:1998kh, Boer:1998nt, Mulders:1999nt, Boer:1999uu,
Boglione:2000jk}. In Sec.~\ref{transvlepnuc} we shall present some predictions
and some preliminary experimental results on $A_T^h$.

\subsection{Semi-inclusive leptoproduction at twist three}
\label{sidistwist3}

Let us now see how transversity distributions appear at the higher-twist level.
We shall consider only twist-three contributions and limit ourselves to quoting
the main results without derivation (which may be found in
\cite{Mulders:1996dh}).

If the lepton beam is unpolarised, the cross-section for leptoproduction of
unpolarised (or spinless) hadrons with a transversely polarised target is
\begin{eqnarray}
  \frac{\d\sigma(\Vec{S}_\perp)}{\d{x} \d{y} \d{z}} &=&
  \frac{4\pi\alpha_\text{em}^2s}{Q^4} \, | \Vec{S}_\perp | \, \frac{M}{Q} \,
  \sum_a e_a^2 \, 2(2-y)\sqrt{1-y}
  \nonumber
\\
  & \times &
  \left\{
    \frac{M_h}{M} \, \sin\phi_S \, x \, \DT{f}_a(x) \,
    \frac{\widetilde{H}_a (z)}{z}
  \right.
  \nonumber
\\
  &-&
  \left.
    \lambda_h \, \cos\phi_S
    \left[
      x^2 \, g_T^a (x) \, \DL{D}_a(z) +
      \frac{M_h}{M} \, x \, \DT{f}_a(x) \, \frac{\widetilde{H}_L^a(z)}{z}
    \right]
  \right\},  \qquad
  \label{twist31}
\end{eqnarray}
where the factor $M/Q$ signals that (\ref{twist31}) is a twist-three quantity.
Adding (\ref{twist31}) to the transverse component of (\ref{sicross8}) gives
the complete $\Vec{P}_{h\perp}$-inte\-grated cross-section of semi-inclusive
DIS off a transversely polarised target up to twist 3.  Note that in
(\ref{twist31}) the leading-twist transversity distributions $\DT{f}(x)$ are
coupled to the twist-three fragmentation functions $\widetilde{H}(z)$ and
$\widetilde{H}_L(z)$, while the leading-twist helicity fragmentation function
$\DL{D}(z)$ is coupled to the twist-three distribution $g_T(x)$.
$\widetilde{H}(z)$ is a $T$-odd fragmentation function.

At twist 3, the transversity distributions also contribute to the scattering of
a longitudinally polarised lepton beam. The corresponding cross-section is
\begin{eqnarray}
  \frac{\d\sigma(\lambda_l, \Vec{S}_\perp)}{\d{x} \d{y} \d{z}} &=&
  -\frac{4\pi\alpha_\text{em}^2s}{Q^4} \, \lambda_l \,
  | \Vec{S}_\perp | \, \frac{M}{Q} \, \sum_a e_a^2 \, 2y\sqrt{1-y}
  \nonumber
\\
  &&  \null \times
  \left\{
    \cos\phi_S
    \left[
      x^2 \, g_T^a (x) \, D_a(z) +
      \frac{M_h}{M} \, x \, \DT{f}_a(x) \, \frac{\widetilde{E}_a (z)}{z}
    \right]
  \right.
  \nonumber
\\
  && \qquad \null +
  \left.
    \lambda_h \, \sin\phi_S
    \left[
      \frac{M_h}{M} \, x \, \DT{f}(x) \, \frac{\widetilde{E}_L^a (z)}{z}
    \right]
  \right\} .
  \label{twist32}
\end{eqnarray}
Here, again, the leading-twist transversity distributions $\DT{f}(x)$ are
coupled to the twist-three fragmentation functions $\widetilde{E}(z)$ and
$\widetilde{E}_L(z)$, and the leading-twist unpolarised fragmentation function
$D(z)$ is coupled to the twist-three distribution $g_T(x)$.
$\widetilde{E}_L(z)$ is a $T$-odd fragmentation function.

Up to order $\Ord(1/Q)$, there are no other observables in semi-inclusive
leptoproduction involving transversity distributions. The twist-two and
twist-three contributions to semi-inclusive leptoproduction involving the
transversity distributions $\DT{f}$ are collected in Tables~\ref{table1} and
\ref{table2}.

\begin{table}[htbp]
  \centering
  \caption{The contributions to the $\Vec{P}_{h\perp}$-integrated cross-section
           involving the transversity distributions. $T$, $L$ and $0$ denote
           transverse, longitudinal and no polarisation, respectively. The
           asterisk indicates $T$-odd observables.}
  \label{table1}
  \vspace{1ex}
  \begin{tabular}{cccc|l}
   \hline
   \multicolumn{5}{c}{Cross-section integrated over $\Vec{P}_{h\perp}$}
  \\ \hline\hline & $\ell$ & $N$ & $h$ & \multicolumn{1}{|c}{observable}
  \\ \hline
   twist 2 & 0 & T & T & $\DT{f}(x) \, \DT{D}(z)$
  \\ \hline
   twist 3 & 0 & T & 0 & $\DT{f}(x) \, \widetilde{H}(z)$ (*) \\
           & 0 & T & L & $\DT{f}(x) \, \widetilde{H}_L(z)$ \\
           & L & T & 0 & $\DT{f}(x) \, \widetilde{E}(z)$ \\
           & L & T & L & $\DT{f}(x) \, \widetilde{E}_L(z)$ (*)
  \\ \hline
  \end{tabular}
\end{table}

\begin{table}[htbp]
  \centering
  \caption{Contributions to the $\Vec{P}_{h\perp}$ distributions involving
           transversity. The produced hadron $h$ is taken to be unpolarised.
           The notation is as in Table~\ref{table1}.}
  \label{table2}
  \vspace{1ex}
  \begin{tabular}{ccc|l}
   \hline
   \multicolumn{4}{c}{$\Vec{P}_{h\perp}$ distribution ($h$ unpolarised)}
  \\ \hline\hline
   & $\ell$ & $N$ & \multicolumn{1}{|c}{observable}
  \\ \hline\hline
   twist 2 & 0 & T & $\DT{f} \otimes H_1^\perp$ (*)
  \\ \hline\hline
   twist 3 & 0 & T & $\DT{f} \otimes \widetilde{H}$ (*) \\
           & 0 & T & $\DT{f} \otimes H_1^\perp$ (*) \\
           & L & T & $\DT{f} \otimes \widetilde{E}$
  \\ \hline
  \end{tabular}
\end{table}

\subsection{Factorisation in semi-inclusive leptoproduction}
\label{factsidis}

It is instructive to use a different approach, based on \QCD factorisation, to
rederive the results on semi-inclusive \DIS presented in Sec.~\ref{sidisasymm}.
We start by considering the collinear case, that is ignoring the transverse
motion of quarks both in the target and in the produced hadron. In this case a
factorisation theorem is known to hold. This theorem was originally
demonstrated for the production of unpolarised particles \cite{Collins:1984ju,
Collins:1985ue, Bodwin:1984hc, Collins:1988ig, Collins:1988gx} and then also
shown to apply if the detected particles are polarised \cite{Collins:1993xw}.
In contrast, when the transverse motion of quarks is taken into account,
factorisation is not proven and can only be regarded as a reasonable
assumption.

\subsubsection{Collinear case}
\label{collsidis}

The \QCD factorisation theorem states that the cross-section for semi-inclusive
\DIS can be written, to all orders of perturbation theory, as
\begin{eqnarray}
  \d\sigma &=&
  \sum_{a b} \sum_{\lambda\lambda'\eta\eta'} \int \! \d\xi \, \d\zeta \,
  f_a(\xi, \mu) \, \rho_{\lambda'\lambda}
  \nonumber
\\
  && \hspace{5em} \null \times
  \d\hat\sigma_{\lambda\lambda'\eta\eta'}(x/\xi, Q/\mu, \alpha_s(\mu)) \;
  \mathcal{D}_{h/b}^{\eta'\eta} (\zeta, \mu) \, ,
  \label{sidisfac1}
\end{eqnarray}
where $\sum_{ab}$ is a sum over initial ($a$) and final ($b$) partons,
$\rho_{\lambda\lambda'}$ is the spin density matrix of parton $a$ in the
nucleon, and $\d\hat\sigma$ is the perturbatively calculable cross-section of
the hard subprocesses that contribute to the reaction. In (\ref{sidisfac1})
$\xi$ is the fraction of the proton momentum carried by the parton $a$, $\zeta$
is the fraction of the momentum of parton $b$ carried by the produced hadron,
and $\mu$ is the factorisation scale. Lastly, $\mathcal{D}_{h/b}(z)$ is the
fragmentation matrix of parton $b$ into the hadron $h$
\begin{equation}
  \mathcal{D}_{h/b} =
  \left(
    \begin{array}{cc}
      \mathcal{D}_{h/b}^{++} & \mathcal{D}_{h/b}^{+-}
    \\
      \mathcal{D}_{h/b}^{-+} & \mathcal{D}_{h/b}^{--}
    \end{array}
  \right) .
  \label{sidisfac15}
\end{equation}
It is defined in such a manner that
\begin{equation}
  \half \, \sum_{\eta} \mathcal{D}_{h/b}^{\eta \eta}(z) =
  \half
  \left[ \mathcal{D}_{h/b}^{++}(z) + \mathcal{D}_{h/b}^{--}(z) \right] =
  D_{h/b}(z) \, ,
  \label{sidisfac16}
\end{equation}
where $D_{h/b}(z)$ is the usual unpolarised fragmentation function, that is the
probability of finding a hadron $h$ with longitudinal momentum fraction $z$
inside a parton $b$. The difference of the diagonal elements of
$\mathcal{D}_{h/b}(z)$ gives the longitudinal polarisation fragmentation
function
\begin{subequations}
\begin{align}
  \half
  \left[ \mathcal{D}_{h/b}^{++}(z) - \mathcal{D}_{h/b}^{--}(z) \right] &=
  \lambda_h \, \DL{D}_{h/b}(z) \, ,
  \label{sidisfac16b}
\\
\intertext{whereas the off-diagonal elements are related to transverse
  polarisation}
  \half
  \left[ \mathcal{D}_{h/b}^{+-}(z) + \mathcal{D}_{h/b}^{-+}(z) \right] &=
  S_{hx} \, \DT{D}_{h/b}(z) \, ,
  \label{sidisfac16c}
\\
  \textfrac{\I}{2}
  \left[ \mathcal{D}_{h/b}^{+-}(z) - \mathcal{D}_{h/b}^{-+}(z) \right] &=
  S_{hy} \, \DT{D}_{h/b}(z) \, .
  \label{sidisfac16d}
\end{align}
\end{subequations}
Note that $\mathcal{D}_{h/b}$ is normalised such that for an unpolarised hadron
it reduces to the unit matrix.

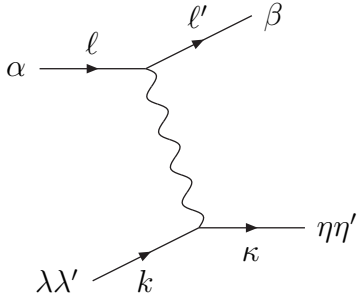
\begin{figure}[htbp]
  \centering
  \begin{picture}(160,100)(0,20)
    \ArrowLine(50,20)(90,40)
    \ArrowLine(90,40)(130,40)
    \Photon(70,100)(90,40){3}{4.5}
    \ArrowLine(30,100)(70,100)
    \ArrowLine(70,100)(110,120)
    \Text(25,100)[r]{$\alpha$}
    \Text(115,120)[l]{$\beta$}
    \Text(45,20)[r]{$\lambda \lambda'$}
    \Text(135,40)[l]{$\eta \eta'$}
    \Text(50,110)[]{$\ell$}
    \Text(90,120)[]{$\ell'$}
    \Text(70,20)[]{$k$}
    \Text(110,30)[]{$\kappa$}
  \end{picture}
  \caption{Lepton--quark (--antiquark) scattering.}
  \label{lq}
\end{figure}

At lowest order the only elementary process contributing to $\d\hat\sigma$ is
$l\,q\,(\anti{q})\to{l}\,q\,(\anti{q})$ (see Fig.~\ref{lq}). Thus, the sum
$\sum_{ab}$ runs only over quarks and antiquarks, and $a=b$.
Eq.~(\ref{sidisfac1}) then becomes (omitting energy scales)
\begin{eqnarray}
  E' E_h \, \frac{\d\sigma}{\d^3\Vec{\ell}' \, \d^3\Vec{P}_h} &=&
  \sum_a \sum_{\lambda\lambda'\eta\eta'} \int \! \d\xi \,
  \frac{\d\zeta}{\zeta^2} \, f_a(\xi) \, \rho_{\lambda'\lambda}
  \nonumber
\\
  && \hspace{4em} \null \times E' E_\kappa
  \left(
    \frac{\d\hat\sigma}{\d^3\Vec{\ell}' \, \d^3\Vec{\kappa}}
  \right)_{\lambda\lambda'\eta\eta'} \,
     \mathcal{D}_{h/a}^{\eta'\eta} (\zeta) \, . \qquad
  \label{sidisfac3}
\end{eqnarray}
The elementary cross-section in (\ref{sidisfac3}) is ($\hat{s}=xs$ is the
centre-of-mass energy squared of the partonic scattering, with the hat
labelling quantities defined at the subprocess level)
\begin{eqnarray}
  E' E_{\kappa}
  \left(
    \frac{\d\hat\sigma}{\d^3\Vec{\ell}' \, \d^3\Vec{\kappa}}
  \right)_{\lambda\lambda'\eta\eta'} &=&
  \frac1{32 \pi^2 \hat{s}} \,
  \frac12 \sum_{\alpha\beta}
  \mathcal{M}_{\lambda\alpha\eta\beta} \,
  \mathcal{M}^{*}_{\lambda'\alpha\eta'\beta} \,
  \delta^4(\ell + k - \ell' - \kappa)
  \nonumber
\\
  &=&
  \frac1{2\pi}
  \left( \frac{\d\hat\sigma}{\d{y}} \right)_{\lambda\lambda'\eta\eta'} \,
  \delta^4(\ell + k - \ell' - \kappa) \, ,
  \label{sidisfac4}
\end{eqnarray}
where
\begin{equation}
  \left( \frac{\d\hat\sigma}{\d{y}} \right)_{\lambda\lambda'\eta\eta'} =
  \frac1{16\pi\hat{s}} \, \frac12 \, \sum_{\alpha\beta}
  \mathcal{M}_{\lambda\alpha\eta\beta} \,
  \mathcal{M}^{*}_{\lambda'\alpha\eta'\beta} \, ,
  \label{sidisfac5}
\end{equation}
with the sum being performed over the helicities of the incoming and outgoing
leptons.

Working in the $hN$ collinear frame, where the photon momentum is
$q^\mu\simeq-xP^\mu+\frac1{z}\,P_h^\mu+q_T^\mu$, that is, in light-cone
components, $q^\mu\simeq(-xP^+,\frac1{z}\,P_h^-,\Vec{q}_T)$, the
energy--momentum conservation delta function may be written as
\begin{eqnarray}
  \delta^4(\ell + k - \ell' - \kappa) &=&
  \delta^4(q + k - \kappa)
  \nonumber
\\
  &\simeq&
  \delta(q^+ + k^+) \, \delta(q^- - \kappa^-) \, \delta^2(\Vec{q}_T)
  \nonumber
\\
  &=&
  \frac{2xz}{Q^2} \, \delta(\xi - x) \, \delta(\zeta - z) \,
  \delta^2(\Vec{q}_T) \, .
  \label{sidisfac6}
\end{eqnarray}

The integrations over $\xi$ and $\zeta$ in (\ref{sidisfac3}) can now be
performed and the cross-section for semi-inclusive \DIS (expressed in terms of
the invariants $x,y,z$ and of the transverse momentum of the outgoing hadron
$\Vec{P}_{h\perp}$) reads
\begin{equation}
  \frac{\d\sigma}{\d{x} \d{y} \d{z} \d^2\Vec{P}_{h\perp}} =
  \sum_a \sum_{\lambda\lambda'\eta\eta'}
  f_a(\xi) \, \rho_{\lambda'\lambda}
  \left( \frac{\d\hat\sigma}{\d{y}} \right)_{\lambda\lambda'\eta\eta'} \,
  \mathcal{D}_{h/a}^{\eta'\eta}(z) \, \delta^2(\Vec{P}_{h\perp}) \, .
  \label{sidisfac7}
\end{equation}
Note the $\delta^2(\Vec{P}_{h\perp})$ factor coming from the kinematics of the
hard subprocess at lowest order. Integrating the cross-section over the hadron
transverse momentum we obtain
\begin{equation}
  \frac{\d\sigma}{\d{x} \d{y} \d{z}} =
  \sum_a \sum_{\lambda\lambda'\eta\eta'}
  f_a(\xi) \, \rho_{\lambda \lambda'}
  \left( \frac{\d\hat\sigma}{\d{y}} \right)_{\lambda\lambda'\eta\eta'} \,
  \mathcal{D}_{h/a}^{\eta \eta'}(z) \, .
  \label{sidisfac8}
\end{equation}

Let us now look at the helicity structure of the $lq$ scattering process. By
helicity conservation, the only non-vanishing scattering amplitudes are
($y=-\hat{t}/\hat{s}=\half(1-\cos\theta)$)
\begin{subequations}
\begin{eqnarray}
  \mathcal{M}_{++++} &=& \mathcal{M}_{----} =
  4 \I \, e^2 e_a \, \frac1{\cos\theta} =
  2\I \, e^2 e_a \, \frac1{y} \, ,
   \label{sidisfac9a}
\\
  \mathcal{M}_{+-+-} &=& \mathcal{M}_{-+-+} =
  2\I \, e^2 e_a \, \frac{1 + \cos\theta}{1 - \cos\theta} =
  2\I \, e^2 e_a \, \frac{1-y}{y}\, ,
  \label{sidisfac9b}
\end{eqnarray}
\end{subequations}
where $\theta$ is the scattering angle in the $lq$ centre-of-mass frame. The
elementary cross-sections contributing to (\ref{sidisfac8}) are
\begin{subequations}
\begin{eqnarray}
  \left( \frac{\d\hat\sigma}{\d{y}} \right)_{++++} =
  \left( \frac{\d\hat\sigma}{\d{y}} \right)_{----} &=&
  \frac1{16\pi\hat{s}} \, \frac12
  \left( | \mathcal{M}_{++++} |^2 +  | \mathcal{M}_{+-+-} |^2 \right)
  \nonumber
\\
  &=&
  \frac{4\pi\alpha^2xs}{Q^4} \, \frac{e_a^2}{2}
  \left[ 1 + (1-y)^2 \right],
  \label{sidisfac10}
\\
  \left( \frac{\d\hat\sigma}{\d{y}} \right)_{+-+-} =
  \left( \frac{\d\hat\sigma}{\d{y}} \right)_{-+-+} &=&
  \frac1{16\pi\hat{s}} \,
  \Re \mathcal{M}_{++++} \, \mathcal{M}^*_{+-+-}
  \nonumber
\\
  &=&
  \frac{4\pi\alpha^2xs}{Q^4} \, e_a^2 \, (1-y) \, ,
  \label{sidisfac11}
  \sublabel{subsidisfac11}
\end{eqnarray}
\end{subequations}
and the cross-section (\ref{sidisfac8}) then reads
\begin{eqnarray}
  \frac{\d\sigma}{\d{x} \d{y} \d{z}} &=&
  \sum_a f_a(x)
  \left\{
    \left( \frac{\d\hat\sigma}{\d{y}} \right)_{++++}
    \left(
      \rho_{++} \, \mathcal{D}_{h/a}^{++} + \rho_{--} \, \mathcal{D}_{h/a}^{--}
    \right)
  \right.
  \nonumber
\\
  && \hspace{5em} \null +
  \left.
    \left( \frac{\d\hat\sigma}{\d{y}} \right)_{+-+-}
    \left(
      \rho_{+-} \, \mathcal{D}_{h/a}^{+-} + \rho_{-+} \, \mathcal{D}_{h/a}^{-+}
    \right)
  \right\} . \qquad
  \label{sidisfac12}
\end{eqnarray}

Inserting (\ref{sidisfac10},~\ref{subsidisfac11}) into (\ref{sidisfac12}) and
using (\ref{good17}--\ref{good19}) and (\ref{sidisfac16}--\ref{sidisfac16d}),
we obtain
\begin{eqnarray}
  \frac{\d\sigma}{\d{x} \d{y} \d{z}}
  &=&
  \frac{4\pi\alpha_\text{em}^2s}{Q^4} \, \sum_a e_a^2 \, x
  \nonumber
\\
  && \hspace{-4em} \null \times
  \left\{ \Strut \right.
  \frac12 \left[ 1 + ( 1 - y)^2 \right]
  \left[
    f_a(x) \, D_{h/a}(z) +
    \lambda_N \lambda_h \, \DL{f}_a(x) \, \DL{D}_{h/a}(z)
  \right]
  \nonumber
\\
  && \hspace{-2.5em} \null +
  (1-y) \, | \Vec{S}_\perp | \, | \Vec{S}_{h\perp} | \,
  \cos(\phi_S + \phi_{S_h}) \, \DT{f}_a(x) \, \DT{D}_{h/a}(z)
  \left. \Strut \right\} . \qquad
  \label{sidisfac23}
\end{eqnarray}
which coincides with the result already obtained in Sec.~\ref{sidisasymm}.

In the light of the present derivation of (\ref{sidisfac23}), we understand the
origin of the $y$-dependent factor in (\ref{sicross10}) and (\ref{sicross13}).
This factor,
\begin{equation}
  \hat{a}_T \equiv
  \frac{\d\hat\sigma_{+-+-}}{\d\hat\sigma_{++++}} =
  \frac{2(1-y)}{1+(1-y)^2} \, .
  \label{sidisfac24}
\end{equation}
is a spin transfer coefficient, \ie, the transverse polarisation of the final
quark generated by an initial transversely polarised quark in the $lq\to{l}q$
process. To see this, let us call $H_{\eta\eta'}$ the quantity
\begin{equation}
  H_{\eta \eta'} \equiv \rho_{\lambda'\lambda}
  \left( \frac{\d\hat\sigma}{\d{y}} \right)_{\lambda\lambda'\eta\eta'} \, ,
  \label{sidisfac24b}
\end{equation}
and introduce the spin density matrix of the final quark, defined via
\begin{equation}
  H_{\eta \eta'} = H_\text{unp} \, \rho'_{\eta \eta'} \, ,
  \label{sidisfac24c}
\end{equation}
where
\begin{equation}
  H_\text{unp} = H_{++} + H_{--} = (\d\hat\sigma)_{++++} \, .
  \label{sidisfac24d}
\end{equation}
We find explicitly
\begin{equation}
  \rho' =
  \left(
    \begin{array}{cc}
      \rho_{++} & \hat{a}_T \, \rho_{-+}
    \\
      \hat{a}_T \, \rho_{+-} & \rho_{--}
    \end{array}
  \right) ,
  \label{sidisfac24e}
\end{equation}
and, recalling that the final quark travels along $-z$, we finally obtain for
its spin vector $\Vec{s}'$
\begin{equation}
  s'_x = -\hat{a}_T \, s_x \, , \qquad
  s'_y =  \hat{a}_T \, s_y \, .
  \label{sidisfac25}
\end{equation}
Thus, the initial and final quark spin directions are specular with respect to
the $y$ axis. The factor $\hat{a}_T$ is also known as the \emph{depolarisation
factor}. It decreases with $y$, being unity at $y=0$ and zero at $y =1$.

\subsubsection{Non-collinear case}
\label{noncollsidis}

If quarks are allowed to have transverse momenta, \QCD factorisation is no
longer a proven property, but only an assumption. In this case we write, in
analogy with (\ref{sidisfac3}),
\begin{eqnarray}
  E' E_h \, \frac{\d\sigma}{\d^3\Vec{\ell}' \, \d^3\Vec{P}_h} &=&
  \sum_a \sum_{\lambda\lambda'\eta\eta'} \int \! \d\xi
  \int \! \frac{\d\zeta}{\zeta^2}
  \int \! \d^2\Vec{k}_T \int \! \d^2\Vec\kappa_T' \,
  \mathcal{P}_a(\xi, \Vec{k}_T) \, \rho_{\lambda'\lambda}
  \nonumber
\\
  && \hspace{2.5em} \null \times
  E' E_\kappa
  \left(
    \frac{\d\hat\sigma}{\d^3\Vec{\ell}' \, \d^3\Vec{\kappa}}
  \right)_{\lambda\lambda'\eta\eta'} \,
  \mathcal{D}_{h/a}^{\eta'\eta} (\zeta, \Vec\kappa'_T) \, ,
  \label{noncoll1}
\end{eqnarray}
where $\mathcal{P}_a(\xi,\Vec{k}_T)$ is the probability of finding a quark $a$
with momentum fraction $x$ and transverse momentum $\Vec{k}_T$ inside the
target nucleon, and $\mathcal{D}_{h/a}(\zeta,\Vec\kappa'_T)$ is the
fragmentation matrix of quark $a$ into the hadron $h$, having transverse
momentum $\Vec\kappa'_T\equiv=-z\Vec\kappa_T$ with respect to the quark
momentum. Evaluating eq.~(\ref{noncoll1}), as we did with
eq.~(\ref{sidisfac3}), we obtain the cross-section in terms of the invariants
$x,y,z$ and of $\Vec{P}_{h\perp}^2$
\begin{eqnarray}
  \frac{\d\sigma}{\d{x} \d{y} \d{z} \d^2\Vec{P}_{h\perp}} &=&
  \sum_a \sum_{\lambda\lambda'\eta\eta'}
  \int \! \d^2\Vec{k}_T \int \! \d^2\Vec\kappa'_T \, \mathcal{P}_a(\xi, \Vec{k}_T) \,
  \rho_{\lambda'\lambda} \,
  \nonumber
\\
  &\times &
  \left( \frac{\d\hat{\sigma}}{\d{y}} \right)_{\lambda\lambda'\eta\eta'} \,
  \mathcal{D}_{h/a}^{\eta'\eta}(z, \Vec\kappa'_T) \,
  \delta^2(z \Vec{k}_T - \Vec\kappa'_T - \Vec{P}_{h\perp}) \, .
  \label{noncoll2}
\end{eqnarray}
Inserting the elementary cross-sections (\ref{sidisfac10},~\ref{subsidisfac11})
in (\ref{noncoll2}) and writing explicitly the sum over the helicities, we
obtain
\begin{eqnarray}
  \frac{\d\sigma}{\d{x} \d{y} \d{z} \d^2\Vec{P}_{h\perp}} &=&
  \frac{4\pi\alpha_\text{em}^2s}{Q^4} \,
  \sum_a e_a^2 \, x \int \! \d^2\Vec{k}_T \int \! \d^2\Vec\kappa'_T \,
  \mathcal{P}_a(x, \Vec{k}_T)
  \nonumber
\\
  && \hspace{2.5em} \null \times
  \left\{
    \frac12 \left[ 1 + (1-y)^2 \right]
    \left[
      \rho_{++} \, \mathcal{D}_{h/a}^{++} +
      \rho_{--} \, \mathcal{D}_{h/a}^{--}
    \right]
  \right.
  \nonumber
\\
  && \hspace{5em} \null +
  \left. \Strut
    (1-y)
    \left[
      \rho_{+-} \, \mathcal{D}_{h/a}^{+-} +
      \rho_{-+} \, \mathcal{D}_{h/a}^{-+}
    \right]
  \right\}
  \nonumber
\\
  && \hspace{2.5em} \null \times
  \delta^2(z\Vec{k}_T - \Vec\kappa'_T - \Vec{P}_{h\perp}) \, .
  \label{noncoll3}
\end{eqnarray}

Let us suppose now that the hadron $h$ is unpolarised. Using the correspondence
(\ref{good17}) between the spin density matrix elements and the spin of the
initial quark, and the analogous relations for the fragmentation matrix
obtained from (\ref{kappaT6}--\ref{subkappaT8}), that is
\begin{subequations}
\begin{eqnarray}
   \frac12 \, ( \mathcal{D}_{h/a}^{++} + \mathcal{D}_{h/a}^{--} ) &=&
   D(z, {\Vec\kappa'_T}^2) + \frac1{M_h} \, \varepsilon_{T}^{ij}
  \kappa_{T i} S_{hTj} \, D_{1T}^\perp(z, {\Vec\kappa'_T}^2) \, ,
  \label{noncoll3b}
\\
  \frac12 \, (\mathcal{D}_{h/a}^{++} - \mathcal{D}_{h/a}^{--}) &=&
  \lambda_h \, \DL{D}(z, {\Vec\kappa'_T}^2) + \frac1{M_h} \,
  \Vec\kappa_{T}{\cdot}\Vec{S}_{h T} \, G_{1T}(z, {\Vec\kappa'_T}^2) \, , \qquad
  \label{noncoll3c}
\\
  \frac12 \, (\mathcal{D}_{h/a}^{+-} + \mathcal{D}_{h/a}^{-+} ) &=&
  S_{hT}^1 \, \Delta'_T D(z, {\Vec\kappa'_T}^2) +
  \frac{\lambda_h}{M_h}\, \kappa_T^1 H_{1L}^\perp(z, {\Vec\kappa'_T}^2)
  \nonumber
\\
  && \null -
  \frac1{M_h^2}
  \left(
    \kappa_{T}^1 \kappa_{T}^j + \frac12 \, \Vec\kappa_{T}^2 \, g_\perp^{1j}
  \right)
  S_{hTj} \, H_{1T}^\perp(z, {\Vec\kappa'_T}^2)
  \nonumber
\\
  && \null +
  \frac1{M_h} \, \varepsilon_T^{1j} \kappa_{Tj} \,
  H_1^\perp(z, {\Vec\kappa'_T}^2) \, ,
  \label{noncoll3d}
\\
  -\frac1{2\I} \, (\mathcal{D}_{h/a}^{+-} - \mathcal{D}_{h/a}^{-+} ) &=&
  S_{hT}^2 \, \Delta'_T D(z, {\Vec\kappa'_T}^2) +
  \frac{\lambda_h}{M_h}\, \kappa_T^2 H_{1L}^\perp(z, {\Vec\kappa'_T}^2)
  \nonumber
\\
  &&  \null -
  \frac1{M_h^2}
  \left(
    \kappa_{T}^1 \kappa_{T}^j + \frac12 \, \Vec\kappa_{T}^2 \, g_\perp^{2j}
  \right)
  S_{hTj} \, H_{1T}^\perp(z, {\Vec\kappa'_T}^2)
  \nonumber
\\
  && \null +
  \frac1{M_h} \, \varepsilon_T^{2j} \kappa_{Tj} \,
  H_1^\perp(z, {\Vec\kappa'_T}^2) \, ,
  \label{noncoll3e}
  \sublabel{subnoncoll3e}
\end{eqnarray}
\end{subequations}
the transverse polarisation contribution to the cross-section turns out to be
\begin{eqnarray}
  \frac{\d\sigma(\Vec{S}_\perp)}{\d{x} \d{y} \d{z} \d^2\Vec{P}_{h\perp} } &=& -
  \frac{4\pi\alpha_\text{em}^2s}{Q^4} \, \sum_a e_a^2 \, x(1-y)
  \int \! \d^2\Vec{k}_T \int \! \d^2\Vec\kappa'_T \, \mathcal{P}_a(x, \Vec{k}_T)
  \nonumber
\\
  && \null \times
  \frac1{M_h} \, (s_x \kappa_{Ty} + s_y \kappa_{Tx}) \,
  H_{1 a}^\perp(z, {\Vec\kappa'}_T^2)
  \nonumber
\\
  && \null \times
  \delta^2(z \Vec{k}_T - \Vec\kappa'_T - \Vec{P}_{h\perp}) \, .
  \label{noncoll4}
\end{eqnarray}

If, for simplicity, we neglect the transverse momentum of the quarks inside the
target, then $\Vec{s}_\perp\,\mathcal{P}_a(x)=\Vec{S}_\perp\,\DT{f}_a(x)$. The
integration over $\Vec\kappa'_T$ can be performed giving the constraint
$\Vec\kappa'_T\equiv-z\Vec\kappa_T=\Vec{P}_{h\perp}$, and eq.~(\ref{noncoll4})
becomes, with our convention for the axes and azimuthal angles
\begin{eqnarray}
  \frac{\d\sigma(\Vec{S}_\perp)}
       {\d{x} \, \d{y} \, \d{z} \, \d^2\Vec{P}_{h\perp}} &=&
  \frac{4\pi\alpha_\text{em}^2s}{Q^4} \, | \Vec{S}_\perp | \,
  \sum_a e_a^2 \, x(1-y) \, \DT{f}_a(x)
   \nonumber
\\
  && \hspace{2.5em} \null \times
  \frac{| \Vec{P}_{h\perp} |}{z M_h} \, H_{1 a}^\perp(z, \Vec{P}_{h\perp}^2) \,
  \sin(\phi_{\kappa} + \phi_S) \, . \hspace{2em}
  \label{noncoll5}
\end{eqnarray}
Since $\Vec\kappa_T=-\Vec{P}_{h\perp}/z$, we have
\begin{equation}
  \phi_\kappa + \phi_S = \phi_h - \pi + \phi_S = \Phi_C - \pi \, ,
  \label{noncoll6}
\end{equation}
and (\ref{noncoll5}) reduces to the transverse polarisation term of
(\ref{sicross12}).

\subsection{Two-hadron leptoproduction}
\label{twoparticle}

\begin{figure}[htbp]
  \centering
  \begin{picture}(360,175)(30,25)
    \SetWidth{1}
    \ArrowLine(50,150)(150,150)
    \Text(70,165)[]{\Large$\ell$}
    \ArrowLine(150,150)(260,190)
    \Text(225,195)[]{\Large$\ell'$}
    \Photon(150,150)(250,50){-5}{8}
    \Text(215,107)[]{\Large$q$}
    \SetWidth{3}
    \ArrowLine(120,50)(250,50)
    \Text(140,65)[]{\Large$P$}
    \SetWidth{2}
    \ArrowLine(260,55)(320,85)
    \ArrowLine(260,60)(310,100)
    \SetWidth{1.5}
    \LongArrow(250,50)(320,49)
    \LongArrow(250,48)(320,41)
    \LongArrow(250,46)(318,33)
    \LongArrow(250,44)(316,25)
    \Text(330,35)[l]{\Large$P_X$}
    \Text(330,85)[l]{\Large$P_2$}
    \Text(310,108)[b]{\Large$P_1$}
    \SetWidth{0.5}
    \GCirc(250,50){20}{0.5}
  \end{picture}
  \caption{Two-particle leptoproduction.}
  \label{fig:sidis2}
\end{figure}
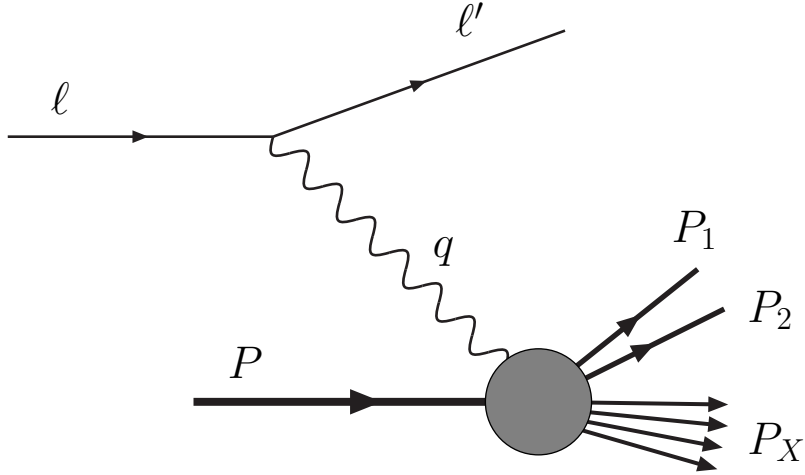

Another partially inclusive \DIS reaction that may provide important
information on transversity is two-particle leptoproduction (see
Fig.~\ref{fig:sidis2}):
\begin{equation}
  l (\ell) \, + \, N (P)
  \to
  l' (\ell') \, + \, h_1 (P_1) \, + \, h_2 (P_2) \, + \, X (P_X) \, .
  \label{twopart1}
\end{equation}
with the target transversely polarised. In this reaction two hadrons (for
instance, two pions) are detected in the final state.

Two-hadron leptoproduction has been proposed and studied by various authors
\cite{Collins:1994kq, Jaffe:1998hf, Bianconi:1999cd, Bianconi:1999uc} as a
process that can probe the transverse polarisation distributions of the
nucleon, coupled to some interference fragmentation functions. The idea is to
look at angular correlations of the form $(\Vec{P}_1\vprod\Vec{P}_2){\cdot}\Vec{s}'$,
where $\Vec{P}_1$ and $\Vec{P}_2$ are the momenta of the two produced hadrons
and $\Vec{s}'_\perp$ is the transverse spin vector of the fragmenting quark.
These correlations are not forbidden by time-reversal invariance owing to
final-state interactions between the two hadrons. To our knowledge, the first
authors who suggested resonance interference as a way to produce non-diagonal
fragmentation matrices of quarks were Cea \etal. \cite{Cea:1988gw} in their
attempt to explain the observed transverse polarisation of $\Lambda^0$ hyperons
produced in $pN$ interactions \cite{Bunce:1976yb}.

Hereafter we shall consider an unpolarised lepton beam and unpolarised hadrons
in the final state. The cross-section for the reaction (\ref{twopart1}) reads,
\cf eq.~(\ref{sidis5})
\begin{equation}
  \d\sigma =
  \frac1{4\ell{\cdot}P} \,
  \frac{e^4}{Q^4} \, L_{\mu\nu} W^{\mu\nu} \, (2\pi)^4 \,
  \frac{\d^3\Vec{\ell}'}{(2\pi)^3 \, 2E'} \,
  \frac{\d^3\Vec{P}_1}{(2\pi)^3 \, 2E_1} \,
  \frac{\d^3\Vec{P}_2}{(2\pi)^3 \, 2E_2} \, ,
  \label{twopart2}
\end{equation}
where $L_{\mu\nu}$ is the usual leptonic tensor, eq.~(\ref{sidis3}), and
$W^{\mu\nu}$ is the hadronic tensor
\begin{eqnarray}
  W^{\mu\nu} &=&
  \frac1{(2\pi)^4} \, \sum_X \int \!
  \frac{{\d^3}P_X}{(2\pi)^3 \, 2E_X} \, (2\pi)^4  \,
  \delta^4(P + q - P_X - P_1 - P_2)
  \nonumber
\\
  && \hspace{2.5em} \null \times
  \langle PS | J^\mu (0) | X, P_1 P_2 \rangle
  \langle X, P_1 P_2 | J^\nu (0) | PS \rangle \, ,
  \label{twopart3}
\end{eqnarray}

Following \cite{Bianconi:1999cd}, we introduce the combinations
\begin{equation}
  P_h \equiv P_1 + P_2, \quad R  \equiv \half \, (P_1 - P_2) \, ,
  \label{twopart1a}
\end{equation}
and the invariants
\begin{align}
    z_1 &= \frac{P{\cdot}P_1}{P{\cdot}q} \, , &
    z_2 &= \frac{P{\cdot}P_2}{P{\cdot}q} \, ,
\\
    z = z_1 + z_2 &= \frac{P{\cdot}P_h}{P{\cdot}q} \, , &
    \xi &= \frac{z_1}{z} = \frac{P{\cdot}P_1}{P{\cdot}P_h} = 1 - \frac{z_2}{z} \, ,
  \label{twopart4}
\end{align}
in terms of which the cross-section becomes
\begin{equation}
  \frac{\d\sigma}
       {\d{x} \, \d{y} \, \d{z} \, \d\xi \,
        \d^2\Vec{P}_{h\perp} \, \d^2\Vec{R}_\perp} =
  \frac{\pi\alpha_\text{em}^2}{4(2\pi)^3 \, Q^4} \,
  \frac{y}{\xi(1-\xi) \, z} \, L_{\mu\nu} W^{\mu\nu} \, .
  \label{twopart5}
\end{equation}
Using
\begin{equation}
  \d^2\Vec{R}_\perp =
  \half \, \d\Vec{R}_\perp^2 \, \d\phi_R =
  \half \, \xi(1-\xi) \, \d{M}_h^2 \, \d\phi_R \, ,
  \label{twopart6}
\end{equation}
where $M_h^2=P_h^2=(P_1+P_2)^2$ is the invariant-mass squared of the two hadrons
and $\phi_R$ is the azimuthal angle of $\Vec{R}$ in the plane perpendicular to
the $\gamma^*N$ axis, the cross-section can then be re-expressed as
\begin{equation}
  \frac{\d\sigma}
  {\d{x} \, \d{y} \, \d{z} \, \d\xi \,
   \d^2\Vec{P}_{h\perp} \, \d{M}_h^2 \, \d\phi_R}
  =
  \frac{\pi\alpha_\text{em}^2y}{2(2\pi)^3Q^4z} \, L_{\mu\nu} W^{\mu\nu} \, .
  \label{twopart7}
\end{equation}

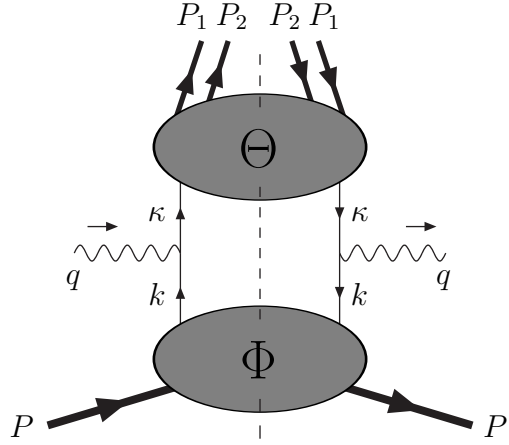
\begin{figure}[htbp]
  \centering
  \begin{picture}(300,170)(0,20)
    \SetWidth{3}
    \ArrowLine(70,25)(120,40)
    \ArrowLine(180,40)(230,25)
    \SetWidth{0.5}
    \ArrowLine(120,60)(120,90)
    \ArrowLine(180,90)(180,60)
    \ArrowLine(120,90)(120,120)
    \ArrowLine(180,120)(180,90)
    \Photon(80,90)(120,90){3}{4.5}
    \Photon(220,90)(180,90){3}{4.5}
    \GOval(150,50)(20,40)(0){0.5}
    \SetWidth{2}
    \ArrowLine(118,140)(128,170)
    \ArrowLine(130,145)(138,170)
    \ArrowLine(162,170)(170,145)
    \ArrowLine(172,170)(182,140)
    \SetWidth{0.5}
    \GOval(150,130)(20,40)(0){0.5}
    \LongArrow(85,100)(95,100)
    \LongArrow(205,100)(215,100)
    \Text(65,25)[r]{$P$}
    \Text(235,25)[l]{$P$}
    \Text(80,80)[]{$q$}
    \Text(220,80)[]{$q$}
    \Text(125,175)[b]{$P_1$}
    \Text(140,175)[b]{$P_2$}
    \Text(160,175)[b]{$P_2$}
    \Text(175,175)[b]{$P_1$}
    \Text(115,75)[r]{$k$}
    \Text(185,75)[l]{$k$}
    \Text(115,105)[r]{$\kappa$}
    \Text(185,105)[l]{$\kappa$}
    \Text(150,50)[]{\LARGE$\Phi$}
    \Text(150,130)[]{\LARGE$\Theta$}
    \DashLine(150,20)(150,35){4}
    \DashLine(150,65)(150,115){4}
    \DashLine(150,145)(150,165){4}
  \end{picture}
  \caption{Diagram contributing to two-hadron leptoproduction at lowest order.}
  \label{handbag3}
\end{figure}

In the parton model (see Fig.~\ref{handbag3}) the hadronic tensor has a form
similar to that of the single-particle case
\begin{eqnarray}
  W^{\mu\nu} &=&
  \sum_q e_q^2 \int \! \frac{\d{k}^- \, \d^2\Vec{k}_T}{(2\pi)^4}
  \int \! \frac{\d\kappa^+ \, \d^2\Vec\kappa_T }{(2\pi)^4}
  \nonumber
\\
  && \null \times
  \delta^2(\Vec{k}_T + \Vec{q}_T - {\Vec\kappa}_T) \,
  \Tr [\Phi \, \gamma^\mu \, \Theta \, \gamma^\nu
      ]_{k^+=xP^+,\,\kappa^-=P_h^-/z} \, ,
  \label{twopart8}
\end{eqnarray}
except that there now appears a decay matrix for the production of a pair of
hadrons
\begin{equation}
  \Theta_{ij}(\kappa; P_1, P_2) =
  \sum_{X} \int \! \d^4 \zeta \, \e^{\I \kappa{\cdot}\zeta}
  \langle 0 | \psi_i(\zeta) | P_1 P_2, X \rangle
  \langle P_1 P_2, X | \anti\psi_j (0) | 0 \rangle \, .
  \label{twopart9}
\end{equation}

Working in a frame where $P$ and $P_h$ are collinear (transverse vectors in
this frame are denoted, as usual, by a $T$ subscript), the matrix
(\ref{twopart9}) can be decomposed as was (\ref{frag2}). At leading twist the
contributing terms are (remember that hadrons $h_1$ and $h_2$ are unpolarised)
\begin{subequations}
\begin{eqnarray}
  \mathcal{V}^\mu &=&
  \half \, \Tr(\gamma^\mu \, \Theta) = \, \mathcal{B}_1 \, P_h^\mu \, ,
  \label{twopart10}
\\
  \mathcal{A}^\mu &=&
  \frac12 \, \Tr(\gamma^\mu \gamma_5 \, \Theta) =
  \frac1{M_1 \, M_2} \, \mathcal{B}'_1 \, \varepsilon^{\mu\nu\rho\sigma} \,
  P_{h\nu} R_\rho \kappa_{T\sigma} \, ,
  \label{twopart11}
\\
  \mathcal{T}^{\mu\nu} &=&
  \frac1{2\I} \, \Tr(\sigma^{\mu\nu} \, \gamma_5 \, \Theta) =
  \frac1{M_1 + M_2}
  \left[ \strut
    \mathcal{B}'_2 \, \varepsilon^{\mu\nu\rho\sigma} P_{h\rho} \kappa_{T\sigma}
  \right.
  \nonumber
\\
  && \hspace{13em}
  \left. \strut \null +
    \mathcal{B}'_3 \, \varepsilon^{\mu\nu\rho\sigma} P_{h\rho} R_\sigma
  \right] ,
  \label{twopart12}
  \sublabel{subtwopart12}
\end{eqnarray}
\end{subequations}
where $M_1$ and $M_2$ are the masses of $h_1$ and $h_2$, respectively. In
(\ref{twopart12}) $\mathcal{B}_i$ and $\mathcal{B}'_i$ are functions of the
invariants constructed with $\kappa$, $P$, $P_h$ and $R$. The prime labels the
so-called $T$-odd terms (but one should bear in mind that $T$-invariance is
\emph{not} actually broken).

Contracting eqs.~(\ref{twopart10}--\ref{subtwopart12}) with $P_\mu$ results in
\begin{subequations}
\begin{eqnarray}
  \frac1{2P_h^-} \, \Tr(\gamma^- \, \Theta) &=& \mathcal{B}_1 \, ,
  \label{twopart13}
\\
  \frac1{2P_h^-} \, \Tr(\gamma^- \gamma_5 \, \Theta) &=&
  \frac1{M_1 M_2} \, \mathcal{B}'_1 \,
  \varepsilon_T^{ij} R_{Ti} \kappa_{Tj} \, ,
  \label{twopart14}
\\
  \frac1{2P_h^-} \, \Tr(\I \sigma^{i-} \gamma_5 \, \Theta) &=&
  \frac1{M_1 + M_2}
  \left[ \strut
    \mathcal{B}'_2 \, \varepsilon_T^{ij} \kappa_{Tj} +
    \mathcal{B}'_3 \, \varepsilon_T^{ij} R_{Tj}
  \right] . \qquad
  \label{twopart15}
  \sublabel{subtwopart15}
\end{eqnarray}
\end{subequations}
Introducing the integrated trace
\begin{eqnarray}
  \Theta^{[\Gamma]} &=&
  \frac1{4z} \int \! \frac{\d\kappa^ + \d\kappa^-}{(2\pi)^4} \,
  \Tr(\Gamma \, \Theta) \,
  \delta \! \left( \kappa^- - \frac1{z}  \, P_h^- \right)
  \nonumber
\\
  &=&
  \frac1{4z} \, \sum_X \int \! \frac{\d\zeta^+ \d^2\Vec\zeta_T}{(2\pi)^3} \,
  \e^{\I (P_h^- \zeta^+/z - \Vec\kappa_T{\cdot}\Vec\zeta_T)} \,
  \nonumber
\\
  && \hspace{3em} \null \times
  \Tr
  \langle 0 | \psi(\zeta^+, 0, \Vec{0}_\perp) | P_1 P_2, X \rangle
  \langle P_1 P_2, X | \anti\psi(0) \Gamma | 0 \rangle \, , \qquad
  \label{twopart16}
\end{eqnarray}
we can rewrite eqs.~(\ref{twopart13}--\ref{subtwopart15}) as
\cite{Bianconi:1999cd}
\begin{subequations}
\begin{eqnarray}
  \Theta^{[\gamma^-]} &=&
  \mathcal{N}_{h_1h_2/q}(z, \xi, {\Vec\kappa'_T}, \Vec{R}_T)
  \nonumber
\\
  &=& D(z, \xi, {\Vec\kappa'}_T^2, \Vec{R}_T^2, \Vec\kappa'_T{\cdot}\Vec{R}_T ) \, ,
  \label{twopart17}
\\
  \Theta^{[\gamma^- \gamma_5]} &=&
  \mathcal{N}_{h_1h_2/q}(z, \xi, \Vec\kappa'_T, \Vec{R}_T) \, \lambda'_q
  \nonumber
\\
  &=&
  \frac1{M_1 M_2} \, \varepsilon_T^{ij} R_{Ti} \kappa_{Tj}
  G_1^\perp(z, \xi, {\Vec\kappa'}_T^2, \Vec{R}_T^2, \Vec\kappa'_T{\cdot}\Vec{R}_T ), \,
  \label{twopart18}
\\
  \Theta^{[\I \sigma^{i-} \gamma_5]} &=&
  \mathcal{N}_{h_1h_2/q}(z, \xi, \Vec\kappa'_T, \Vec{R}_T) \, {s'_\perp}^i
  \nonumber
\\
  &=&
  \frac1{M_1 + M_2}
  \left[
    \varepsilon_T^{ij} \kappa_{Tj} \,
    H_1^\perp(z, \xi, {\Vec\kappa'}_T^2, \Vec{R}_T^2, {\Vec\kappa'}_T{\cdot}\Vec{R}_T)
  \strut \right.
  \nonumber
\\
  && \hspace{5em}
  \left. \strut \null +
    \varepsilon_T^{ij} R_{Tj} \,
    \widetilde{H}_1^\perp
      (z, \xi, {\Vec\kappa'}_T^2, \Vec{R}_T^2, {\Vec\kappa'}_T{\cdot}\Vec{R}_T)
  \right] , \qquad
  \label{twopart19}
\end{eqnarray}
\end{subequations}
where $\mathcal{N}_{h_1h_2/q}(z,\xi,\Vec\kappa'_T,\Vec{R}_T)$ is the
probability for a quark $q$ to produce two hadrons $h_1,h_2$.

In eq.~(\ref{twopart19}), $D$, $G_1^\perp$, $H_1^\perp$ and
$\widetilde{H}_1^\perp$ are interference fragmentation functions of quarks into
a pair of unpolarised hadrons. In particular, $H_1^\perp$ and
$\widetilde{H}_1^\perp$ are related to quark transverse polarisation in the
target. $H_1^\perp$ has an analogue in the case of single-hadron production
(where it has been denoted by the same symbol), while $\widetilde{H}_1^\perp$
is a genuinely new function. It is important to notice that
$\widetilde{H}_1^\perp$ is the only fragmentation function, besides $D$, that
survives when the quark transverse momentum is integrated over.

The symmetric part $W^{\mu\nu(S)}$ of the hadronic tensor, the component
contributing to the cross-section when the lepton beam is unpolarised (as in
our case), is given by (with the same notation as in Sec.~\ref{sidisasymm} and
retaining only the unpolarised and the transverse polarisation terms)
\begin{eqnarray}
  W^{\mu\nu (S)} &=&
  2 \sum_a e_a^2 \, z \int \! \d^2\Vec\kappa_T \int \! \d^2\Vec{k}_T \,
  \delta^2(\Vec{k}_T +  \Vec{q}_T - \Vec\kappa_T) \,
  \nonumber
\\
  && \null \times
  \left\{ \Strut
    -g_T^{\mu\nu} \, f(x,\Vec{k}_T^2) \, D(z,{\Vec\kappa'_T}^2)
  \right.
  \nonumber
\\
  && \hspace{2em} \null -
  \frac{S_T^{\{\mu} \varepsilon_T^{\nu\}\rho} \kappa_{T\rho} +
        \kappa_T^{\{\mu} \varepsilon_T^{\nu\}\rho} S_{T\rho}}
       {2(M_1 + M_2)}
  \nonumber
\\
  && \hspace{3em} \null \times
  \Delta'_T f(x, \Vec{k}_T^2) \,
  H_{1}^\perp(z, \xi, {\Vec\kappa'_T}^2, \Vec{R}_T^2, \Vec\kappa_T'{\cdot}\Vec{R}_T)
  \nonumber
\\
  && \hspace{2em} \null -
  \frac{S_T^{\{\mu} \varepsilon_T^{\nu\}\rho} R_{T\rho} +
        R_T^{\{\mu} \varepsilon_T^{\nu\}\rho} S_{T\rho}}
       {2(M_1 + M_2)}
  \nonumber
\\
  && \hspace{3em} \left. \null \times
    \Delta'_T f(x, \Vec{k}_T^2) \,
    \widetilde{H}_{1}^\perp
      (z, \xi, {\Vec\kappa'_T}^2, \Vec{R}_T^2, \Vec\kappa_T'{\cdot}\Vec{R}_T)
    + \dots
  \Strut \right\} . \hspace{4em}
  \label{twopart20}
\end{eqnarray}

Let us now neglect the intrinsic motion of quarks inside the target. This
implies that $\Vec\kappa_T=-\Vec{P}_{h\perp}/z$. Contracting $W^{\mu\nu (S)}$
with the leptonic tensor $L_{\mu\nu}^{(S)}$ by means of the relations
(\ref{sicross5}--\ref{subsicross7}) and integrating over $\Vec{P}_{h\perp}$, we
obtain the cross-section (limited to the unpolarised and transverse
polarisation contributions)
\begin{eqnarray}
  \frac{\d\sigma}
       {\d{x} \, \d{y} \, \d{z} \, \d\xi \, \d{M}_h^2 \, \d\phi_R} &=&
  \frac{4\pi\alpha_\text{em}^2s}{(2\pi)^3\,Q^4} \, \sum_a e_a^2 \, x
  \nonumber
\\
  && \hspace{-9em} \null \times
  \left\{ \STRUT
    \frac12 \, [1+(1-y)^2] \, f_a(x) \, D_a(z,\xi,M_h^2)
  \right.
  \nonumber
\\
  && \hspace{-7.5em} \left. \null +
    (1- y) \, \frac{| \Vec{S}_\perp | \, | \Vec{R}_\perp |}{M_1 + M_2} \,
    \sin(\phi_S + \phi_R) \,
    \DT{f}_a(x) \, \widetilde{H}_{1a}^\perp(z, \xi, \Vec{R}_\perp^2)
  \Strut \right\} . \hspace{4em}
  \label{twopart21}
\end{eqnarray}
The fragmentation functions appearing here are integrated over
$\Vec{P}_{h\perp}^2$.

We define now the interference fragmentation function $\DT{I}(z,\xi,M_h^2)$ as
\begin{eqnarray}
  \DT{I}(z, \xi, M_h^2) &=&
  \frac{| \Vec{R}_\perp |}{M_1 + M_2} \, \widetilde{H}_1^\perp(z, \xi, M_h^2)
  \nonumber
\\
  &\propto&
  \mathcal{N}_{h_1 h_2/q \uparrow}   (z, \xi, \Vec{R}_\perp) -
  \mathcal{N}_{h_1 h_2/q \downarrow} (z, \xi, \Vec{R}_\perp) \, ,
  \label{twopart22}
\end{eqnarray}
where, we recall,
\begin{equation}
  \Vec{R}_\perp^2 = \xi(1-\xi) \, M_h^2 - (1 - \xi) \, M_1^2 - \xi \, M_2^2 \, .
  \label{twopart23}
\end{equation}
Integrating (\ref{twopart21}) over $\xi$, we finally obtain
\begin{eqnarray}
  \frac{\d\sigma}{\d{x} \, \d{y} \, \d{z} \, \d{M}_h^2 \, \d\phi_R} &=&
  \frac{4\pi\alpha_\text{em}^2s}{(2\pi)^3 \, Q^4} \, \sum_a e_a^2 \, x
  \left\{
    \frac12 \, [ 1 + ( 1 - y)^2 ] \,  f_a(x) \, D_a(z, M_h^2)
  \right.
  \nonumber
\\
  && \hspace{-4em} \null +
  \left.
    (1- y) \, | \Vec{S}_\perp | \, \sin(\phi_S + \phi_R)
    \DT{f}_a(x) \, \DT{I}_a(z, M_h^2)
  \Strut \right\} .
  \label{twopart24}
\end{eqnarray}
From (\ref{twopart24}) we obtain the transverse single-spin asymmetry
\begin{eqnarray}
  A_T^{h_1 h_2} &\equiv&
  \frac{\d\sigma(\Vec{S}_\perp) - \d\sigma(- \Vec{S}_\perp)}
       {\d\sigma(\Vec{S}_\perp) + \d\sigma(- \Vec{S}_\perp)} =
  \frac{2(1-y)}{1+(1-y)^2}
  \nonumber
\\
  && \null \times
  \frac{\sum_a e_a^2 \, \DT{f}_a(x) \, \DT{I}_a(z, M_h^2)}
       {\sum_a e_a^2 \, f_a(x) \, D_a(z, M_h^2)} \,
  | \Vec{S}_\perp | \, \sin(\phi_S + \phi_R) \, ,
  \label{twopart25}
\end{eqnarray}
which probes the transversity distributions along with the interference
fragmentation function $\DT{I}$.

We can introduce, into two-hadron leptoproduction, the analogue of the Collins
angle $\Phi_C$ of single-hadron leptoproduction, which we call $\Phi'_C$. We
define $\Phi'_C$ as the angle between the final quark transverse spin
$\Vec{s}'_\perp$ and $\Vec{R}_\perp$, \ie,
\begin{equation}
  \Phi'_C \equiv \phi_{s'} - \phi_R \, .
  \label{twopart26}
\end{equation}
We have
\begin{equation}
  \sin\Phi'_C \equiv
  \frac{(\Vec{P}_h \vprod \Vec{R}){\cdot}\Vec{s}'}
       {|\Vec{P}_h \vprod \Vec{R}| \, |\Vec{s}'|} =
  \frac{(\Vec{P}_2 \vprod \Vec{P}_1){\cdot}\Vec{s}'}
       {|\Vec{P}_2 \vprod \Vec{P}_1| \, |\Vec{s}'|} \, .
  \label{twopart27}
\end{equation}
Since $\phi_{s'}=\pi-\phi_s$, where $\phi_s$ is the azimuthal angle of the
initial quark transverse spin, we can also write
\begin{equation}
  \Phi'_C = \pi - \phi_s - \phi_R \, .
  \label{twopart28}
\end{equation}
If the initial quark has no transverse momentum with respect to the nucleon,
then $\phi_s=\phi_S$ and $\Phi'_C$ is given, in terms of measurable angles, by
\begin{equation}
  \Phi'_C = \pi - \phi_S - \phi_R \, .
  \tag{\ref{twopart28}$'$}
  \label{twopart29}
\end{equation}

In the language of \QCD factorisation the cross-section for two-hadron
leptoproduction is written as
\begin{eqnarray}
  \frac{\d\sigma}{\d{x} \, \d{y} \, \d{z} \, \d{M}_h^2 \, \d\phi_R} &=&
  \sum_a \sum_{\lambda\lambda'\eta\eta'} f_a(x) \, \rho_{\lambda'\lambda}
  \left( \frac{\d\sigma}{\d{y}} \right)_{\lambda\lambda'\eta\eta'}
  \nonumber
\\
  && \hspace{5em} \null \times
  \left[
    \frac{\d\mathcal{D}(z, M_h^2, \phi_R)}{\d{M}_h^2 \, \d\phi_R}
  \right]_{\eta'\eta} \, .
  \label{twopart30}
\end{eqnarray}
What we have found above is that the fragmentation matrix
$\d\mathcal{D}/\d{M}_h^2\,\d\phi_R$ factorises into $z$- and $M_h^2$-dependent
fragmentation functions and certain angular coefficients. For the case at hand,
the angular dependence is given by the factor $\sin(\phi_S+\phi_R)$ in
(\ref{twopart24}).

\begin{figure}[htbp]
  \centering
  \begin{picture}(300,185)(0,20)
    \SetWidth{3}
    \ArrowLine(70,25)(120,40)
    \ArrowLine(180,40)(230,25)
    \SetWidth{0.5}
    \ArrowLine(120,60)(120,90)
    \ArrowLine(180,90)(180,60)
    \ArrowLine(120,90)(120,120)
    \ArrowLine(180,120)(180,90)
    \Photon(80,90)(120,90){3}{4.5}
    \Photon(220,90)(180,90){3}{4.5}
    \GOval(150,50)(20,40)(0){0.5}
    \SetWidth{2}
    \DashArrowLine(120,142)(125,170){3}
    \DashArrowLine(175,170)(180,142){3}
    \SetWidth{1.5}
    \ArrowLine(125,170)(115,190)
    \ArrowLine(125,170)(140,190)
    \ArrowLine(160,190)(175,170)
    \ArrowLine(185,190)(175,170)
    \SetWidth{0.5}
    \GCirc(125,170){3}{0.5}
    \GCirc(175,170){3}{0.5}
    \GOval(150,130)(20,40)(0){0.5}
    \Text(65,25)[r]{$N$}
    \Text(235,25)[l]{$N$}
    \Text(115,155)[r]{$h,h'$}
    \Text(185,155)[l]{$h,h'$}
    \Text(115,195)[b]{$h_1$}
    \Text(140,195)[b]{$h_2$}
    \Text(160,195)[b]{$h_2$}
    \Text(185,195)[b]{$h_1$}
    \Text(75,90)[r]{$\gamma^*$}
    \Text(225,90)[l]{$\gamma^*$}
    \DashLine(150,20)(150,170){4}
  \end{picture}
  \caption{Leptoproduction of two hadrons $h_1$ and $h_2$ via resonance
           ($h,h'$) formation.}
  \label{handbag4}
\end{figure}
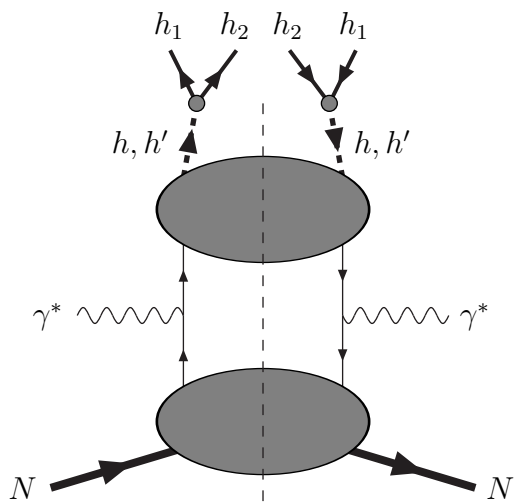

An explicit mechanism giving rise to an interference fragmentation function
like $\DT{I}$ has been suggested by Jaffe, Jin and Tang \cite{Jaffe:1998hf,
Jaffe:1998pv} (a similar mechanism was considered earlier in a different but
related context \cite{Cea:1988gw}). The process considered in
\cite{Jaffe:1998hf, Jaffe:1998pv} and shown diagrammatically in
Fig.~(\ref{handbag4}) is the production of a $\pi^+\pi^-$ pair, via formation
of a $\sigma$ ($I=0$, $L=0$) and $\rho$ ($I=1$, $L=1$) resonance. The
single-spin asymmetry then arises from interference between the $s$- and
$p$-wave of the pion system. Similar processes are ${\pi}K$ production near to
the $K^*$ resonance, and $K\anti{K}$ production near to the $\phi$. In all
these cases two mesons, $h_1$ and $h_2$, are generated from the decay of two
resonances $h\;(L=0)$ and $h'\;(L=1)$. The final state can be written as a
superposition of two resonant states with different relative phases
\begin{equation}
  | h_1 \, h_2,  \, X \rangle =
  \e^{\I\delta_0} \, | h,  \, X \rangle +
  \e^{\I\delta_1} \, | h', \, X \rangle \, .
  \label{twopart31}
\end{equation}
The interference between the two resonances is proportional to
$\sin(\delta_0-\delta_1)$. The values of $\delta_0$ and $\delta_1$ depend on
the invariant mass $M_h$ of the two-meson system.

It turns out that the interference fragmentation function $\DT{I}$ has the
following structure
\begin{equation}
  \DT{I}(z, M_h^2) \sim
  \sin\delta_0 \, \sin\delta_1 \, \sin(\delta_0 - \delta_1) \,
  \DT\hat{I}(z, M_h^2) \, ,
  \label{twopart32}
\end{equation}
where the phase factor $\sin\delta_0\,\sin\delta_1\,\sin(\delta_0-\delta_1)$
depends on $M_h^2$. The maximal value that this factor can attain is
$3\sqrtno{3}/8$.

The two-hadron spin-averaged fragmentation function $D_{h_1h_2}(z,M_h^2)$ is the
superposition of the unpolarised fragmentation functions of the two resonances
weighted by their phases
\begin{equation}
  D_{h_1 h_2}(z, M_h^2) =
  \sin^2\delta_0 \; D_{h}(z) +
  \sin^2\delta_1 \; D_{h'}(z) \, .
  \label{twopart33}
\end{equation}

The resulting single-spin asymmetry is then
\begin{eqnarray}
  A_T^{h_1 h_2} &\equiv&
  \frac{\d\sigma(\Vec{S}_\perp) - \d\sigma(-\Vec{S}_\perp)}
       {\d\sigma(\Vec{S}_\perp) + \d\sigma(-\Vec{S}_\perp)}
  \nonumber
\\
  &\propto&
  \frac{2(1-y)}{1+(1-y)^2} \,
  | \Vec{S}_\perp | \, \sin\delta_0 \, \sin\delta_1 \,
  \sin(\delta_0 - \delta_1) \, \sin(\phi_S + \phi_R)
  \nonumber
\\
  && \null \times \,
  \frac{\sum_a e_a^2 \, \DT{f}_a(x) \, \DT\hat{I}_a(z, M_h^2)}
       {\sum_a e_a^2 \, f_a(x) \,
         [ \sin^2\delta_0 \, \hat{D}_{h/a}(z) +
           \sin^2\delta_1 \, \hat{D}_{h'/a} (z)
         ]} \, .
  \label{twopart34}
\end{eqnarray}
We remark that the angle $\phi$ defined in \cite{Jaffe:1998hf} corresponds to
our $\phi_S-\phi_R-\pi/2$.

\begin{figure}[htbp]
  \centering
  \includegraphics[width=8cm]{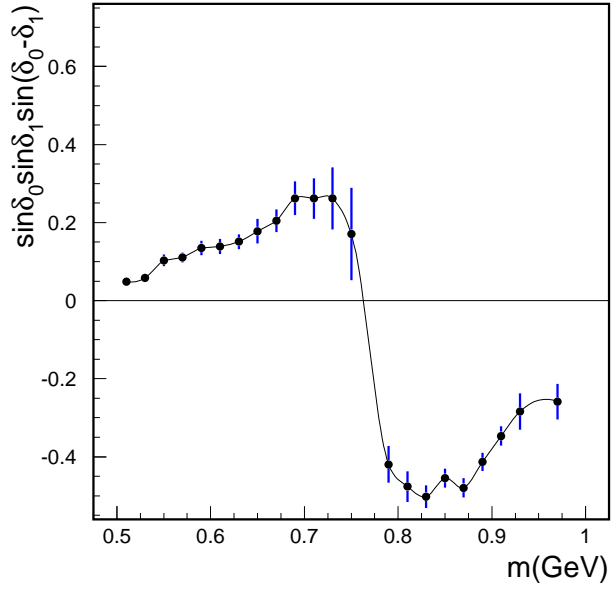}
  \caption{The factor $\sin\delta_0\,\sin\delta_1\,\sin(\delta_0-\delta_1)$
           obtained from $\pi\pi$ phase shifts (figure
           from~\cite{Jaffe:1998hf}).}
  \label{jjt}
\end{figure}

In the case of two-pion production, $\delta_0$ and $\delta_1$ can be obtained
from the data on $\pi\pi$ phase shifts \cite{Estabrooks:1974vu}. The factor
$\sin\delta_0\,\sin\delta_1\,\sin(\delta_0-\delta_1)$ is shown in
Fig.~\ref{jjt}. It is interesting to observe that the experimental value of
this quantity reaches 75\% of its theoretical maximum.

\subsection{Leptoproduction of spin-1 hadrons}
\label{spinone}

As first suggested by Ji \cite{Ji:1994vw} (see also \cite{Anselmino:1996vq})
the transversity distribution can be also probed in  leptoproduction of vector
mesons (\eg, $\rho, K^*, \phi$). The fragmentation process into spin-1 hadrons
has been fully analysed, from a formal viewpoint, in \cite{Bacchetta:2000jk,
Bacchetta:2001rb}. The polarisation state of a spin-1 particle is described by
a spin vector $\Vec{S}$ and by a rank-2 spin tensor $T^{ij}$. The latter
contains five parameters, usually called $S_{LL}$, $S_{LT}^x$, $S_{LT}^y$,
$S_{TT}^{xy}$ and $S_{TT}^{xx}$ \cite{Bacchetta:2000jk}. The transversity
distribution $\DT{f}$ emerges when an unpolarised beam strikes a transversely
polarised target. The cross-section in this case is \cite{Bacchetta:2000jk} (we
retain only the terms containing $\DT{f}$ and we use the notation of
Sec.~\ref{azimasym})
\begin{eqnarray}
& &  \frac{\d\sigma(\Vec{S}_\perp)}{\d{x} \d{y} \d{z} \d^2\Vec{P}_{h\perp}} =
  \frac{4\pi\alpha_\text{em}^2s}{Q^4} \, | \Vec{S}_\perp | \,
  \sum_a e_a^2 \, x(1-y)
  \nonumber
\\
  && \null \hspace{0.5cm} \times \left \{ \Strut
  \vert S_{LT} \vert \,
  \sin (\phi_{LT} + \phi_{S})\, I [\DT{f}_a \, H_{1LT}^a ]
\right. \nonumber \\
  && \null \hspace{0.5cm} + \vert S_{TT} \vert \,
  \sin (2 \phi_{TT} + \phi_S - \phi_h) \,
  I \left [\frac{\hat{\Vec{h}} \cdot \Vec\kappa_\perp}{M_h}
  \, \DT{f}_a \, H_{1TT}^a \right ]
\nonumber \\
  && \null \hspace{0.5cm}
  + S_{LL} \, \sin(\phi_S + \phi_h) \,
  I \left [\frac{\hat{\Vec{h}} \cdot \Vec\kappa_\perp}{M_h}
  \, \DT{f}_a \, H_{1LL}^{\perp a} \right ]
\nonumber \\
  && \null \hspace{0.5cm} + \vert S_{LT} \vert \,
  \sin(\phi_{LT} - \phi_S - 2 \phi_h)
  \, I \left [
  \frac{2 \, (\hat{\Vec{h}} \cdot \Vec\kappa_\perp)^2
  - \Vec\kappa_\perp^2}{2 M_h^2} \, \DT{f}_a
  \, H_{1 LT}^{\perp a} \right ]
\nonumber \\
  & & \null \hspace{0.5cm} - \vert S_{TT} \vert \,
  \sin (2 \phi_{TT} - \phi_S - 3 \phi_h)
\nonumber \\
  & & \null \hspace{1cm} \times \left.
  I \left [
    \frac{(\hat{\Vec{h}} \cdot \Vec\kappa_\perp) \,
           [ 2 \, (\hat{\Vec{h}} \cdot \Vec\kappa_\perp)^2
           - 3 \Vec\kappa_\perp^2 / 2]}{M_h^3} \,
    \DT{f}_a \, H_{1 TT}^{\perp a}
  \right ] \Strut \right \} + \ldots \, .
  \label{spinone1}
\end{eqnarray}
For simplicity, we have omitted the subscript $h$ in the tensor spin parameters
(which are understood to pertain to the produced hadron). The azimuthal angles
$\phi_{LT}$ and $\phi_{TT}$ are defined by
\begin{equation}
  \tan \phi_{LT} = \frac{S_{LT}^y}{S_{LT}^x} \; ,
  \quad
  \tan \phi_{TT} = \frac{S_{TT}^{xy}}{S_{TT}^{xx}} \, ,
  \label{spinone2}
\end{equation}
and
\begin{equation}
  \vert S_{LT} \vert = \sqrt{(S_{LT}^x)^2 + (S_{LT}^y)^2} \; ,
  \quad
  \vert S_{TT} \vert = \sqrt{(S_{TT}^{xx})^2 + (S_{TT}^{xy})^2} \, .
  \label{spinone3}
\end{equation}

Note that $\DT{f}$ couples in (\ref{spinone1}) to five different fragmentation
functions: $H_{1LT}$, $H_{1TT}$, $H_{1LL}^\perp$, $H_{1LT}^\perp$,
$H_{1TT}^\perp$. All these functions are $T$-odd. If we integrate the
cross-section over $\Vec{P}_{h\perp}$, only one term survives, namely
\begin{eqnarray}
  \frac{\d\sigma(\Vec{S}_\perp)}{\d{x} \d{y} \d{z}} &=&
  \frac{4\pi\alpha_\text{em}^2s}{Q^4} \, | \Vec{S}_\perp | \,
\vert S_{LT} \vert \sin (\phi_{LT} + \phi_S) \nonumber \\
& & \null \times
  \sum_a e_a^2 \, x(1-y) \, \DT{f}_a (x) \, H_{1LT}^a (z)\,.
\label{spinone4}
\end{eqnarray}
The fragmentation function appearing here, $H_{1LT}$, is called
$\hat{h}_{\bar1}$ by Ji \cite{Ji:1994vw}. It is a $T$-odd and chirally-odd
function which can be measured at leading twist and without considering
intrinsic transverse momenta. Probing the transversity by (\ref{spinone4})
requires polarimetry on the produced meson. For a self-analysing particle this
can be done by studying the angular distribution of its decay products (\eg,
$\rho^0\to\pi^+\pi^-$). Thus, the vector-meson fragmentation function $H_{1LT}$
represents a specific contribution to two-particle production near the vector
meson mass.

\subsection{Transversity in exclusive leptoproduction processes}

\begin{figure}[htbp]
  \centering
    \begin{picture}(395,130)(0,0)
    \SetOffset(-40,0)
    \SetWidth{2}
    \ArrowLine(80,25)(120,35)
    \ArrowLine(180,35)(220,25)
    \SetWidth{0.5}
    \ArrowLine(120,50)(120,80)
    \ArrowLine(180,80)(180,50)
    \GOval(150,40)(15,40)(0){0.5}
    \Text(110,75)[]{$-$}
    \Text(190,75)[]{$+$}
    \Text(75,20)[r]{\large$P$}
    \Text(225,20)[l]{\large$P'$}
    \Text(150,0)[]{(a)}
    \SetOffset(150,0)
    \SetWidth{2}
    \ArrowLine(80,25)(120,35)
    \ArrowLine(180,35)(220,25)
    \SetWidth{0.5}
    \Line(120,50)(140,80)
    \Line(160,80)(180,50)
    \Photon(87,108)(150,90){3}{4}
    \Line(150,91)(213,109)
    \Line(150,89)(213,107)
    \SetWidth{2}
    \ArrowLine(185,100)(192,102)
    \SetWidth{0.5}
    \GOval(150,40)(15,40)(0){0.5}
    \GCirc(150,90){15}{1}
    \Text(115,70)[]{$-$}
    \Text(185,70)[]{$+$}
    \Text(210,115)[b]{\large$V$}
    \Text(85,115)[b]{\large$\gamma^*$}
    \Text(150,0)[]{(b)}
  \end{picture}
  \caption{The off-diagonal transversity distribution (a) and its contribution
           to exclusive vector meson production (b).}
  \label{off}
\end{figure}
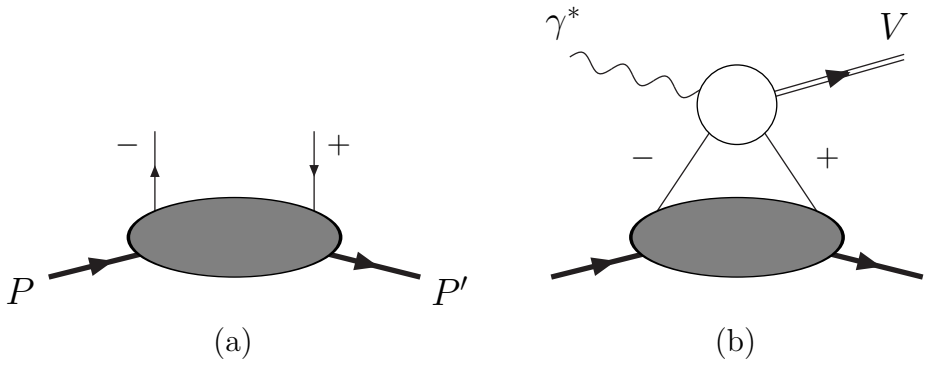

Let us now consider the possibility of observing the transversity distributions
in exclusive leptoproduction processes. Collins, Frankfurt and Strikman
\cite{Collins:1997fb}  remarked that the exclusive production of a transversely
polarised vector meson in \DIS, that is $lp\to{l}Vp$, involves the chirally-odd
off-diagonal parton distributions in the proton. These distributions (also
called ``skewed'' or ``off-forward'' distributions) depend on two variables $x$ and
$x'$ since the incoming and outgoing proton states have different momenta $P$
and $P'$, with $(P'-P)^2=t$ (the reader may consult \cite{Ji:1998pc,
Radyushkin:1997ki, Martin:1998wy} on skewed distributions). For instance, the
off-forward transversity distribution (represented in Fig.~\ref{off}a) contains
a matrix element of the form
$\langle{PS}|\anti\psi(0)\gamma^+\gamma_\perp\gamma_5\psi(\xi^-)|P'S\rangle$.
At low $x$ the difference between $x$ and $x'$ is small and the off-diagonal
distributions are completely determined by the corresponding diagonal ones.

In \cite{Mankiewicz:1998uy} it was shown that the chirally-odd contribution to
vector-meson production (see Fig.~\ref{off}b) is actually zero at \LO in
$\alpha_s$. This result was later extended in \cite{Diehl:1998pd}, where it was
observed that the vanishing of the chirally-odd contribution is due to angular
momentum and chirality conservation in the hard scattering and hence holds at
leading twist to all orders in the strong coupling. Thus, the (off-diagonal)
transversity distributions cannot be probed in exclusive vector-meson
leptoproduction.

\section{Transversity in hadronic reactions}
\label{transvhad}

The second class of reactions probing quark transversity is hadron scattering
with at least one of the two colliding particles in a transverse polarisation
state. We shall first consider the case where both initial hadrons are
transversely polarised. In particular, Drell--Yan production with two
transversely polarised hadrons turns out to be the most favourable reaction for
studying the transversity distributions. Indeed, the pioneering work of Ralston
and Soper \cite{Ralston:1979ys} and Pire and Ralston \cite{Pire:1983tv}
concentrated precisely on this process. We shall then see how transversity may
emerge even when only one of the colliding hadrons is transversely polarised.
This possibility, however, is more uncertain.

\subsection{Double-spin transverse asymmetries}
\label{doublespin}

When both hadrons are transversely polarised, the typical observables are
double-spin transverse asymmetries of the form
\begin{equation}
  A_{TT} =
  \frac{\d\sigma(\Vec{S}_T, \Vec{S}_T) - \d\sigma(\Vec{S}_T, -\Vec{S}_T)}
       {\d\sigma(\Vec{S}_T, \Vec{S}_T) + \d\sigma(\Vec{S}_T, -\Vec{S}_T)} \, ,
  \label{att}
\end{equation}
Since there is no gluon transversity distribution for spin-half hadrons,
transversely polarised $pp$ reactions which are dominated at the partonic level
by $qg$ or $gg$ scattering are expected to yield a very small $A_{TT}$
\cite{Ji:1992ev, Jaffe:1996ik}. Thus, direct-photon production (with
lowest-order subprocesses $gq\to{q}\gamma$ and $q\anti{q}\to{\gamma}g$),
heavy-quark production ($q\anti{q}\to{Q}\anti{Q}$ and $gg\to{Q}\anti{Q}$), and
two-gluon-jet production ($gg\to{gg}$ and $q\anti{q}\to{gg}$) do not seem to be
promising reactions to detect quark transverse polarisation.

The only good candidate process for measuring transversity in doubly polarised
$pp$ (or $p\anti{p}$) collisions is Drell--Yan lepton pair production
\cite{Ralston:1979ys, Cortes:1991ka, Jaffe:1992ra}. We shall see that at lowest
order $A_{TT}^\text{DY}$ contains combinations of the products
\begin{displaymath}
  \DT{f}(x_A) \; \DT\anti{f}(x_B) \, .
\end{displaymath}
The advantage of studying quark transverse polarisation via Drell--Yan is
twofold: \emph{i}) transversity distributions appear at leading-twist level;
\emph{ii}) the cross-section contains no unknown quantities, besides the
transversity distributions themselves. This renders theoretical predictions
relatively easier, with respect to other reactions.

\subsection{The Drell--Yan process}
\label{drellyan}

Drell--Yan lepton-pair production is the process
\begin{equation}
  A \, (P_1) \, + \, B\, (P_2)
  \to
  l^+ \, (\ell) \, + \, l^- \, (\ell') \, + \, X \, ,
  \label{dy1}
\end{equation}
where $A$ and $B$ are protons or antiprotons and $X$ is the undetected hadronic
system. The centre-of-mass energy squared of this reaction is
$s=(P_1+P_2)^2\simeq2\,P_1{\cdot}P_2$ (in the following, the hadron masses $M_1$ and
$M_2$ will be systematically neglected, unless otherwise stated). The lepton
pair originates from a virtual photon (or from a $Z^0$) with four-momentum
$q=\ell+\ell'$. Note that, in contrast to \DIS, $q$ is a time-like vector:
$Q^2=q^2>0$. This is also the invariant mass of the lepton pair. We shall
consider the deeply inelastic limit where $Q^2,s\to\infty$, while the ratio
$\tau=Q^2/s$ is fixed and finite.

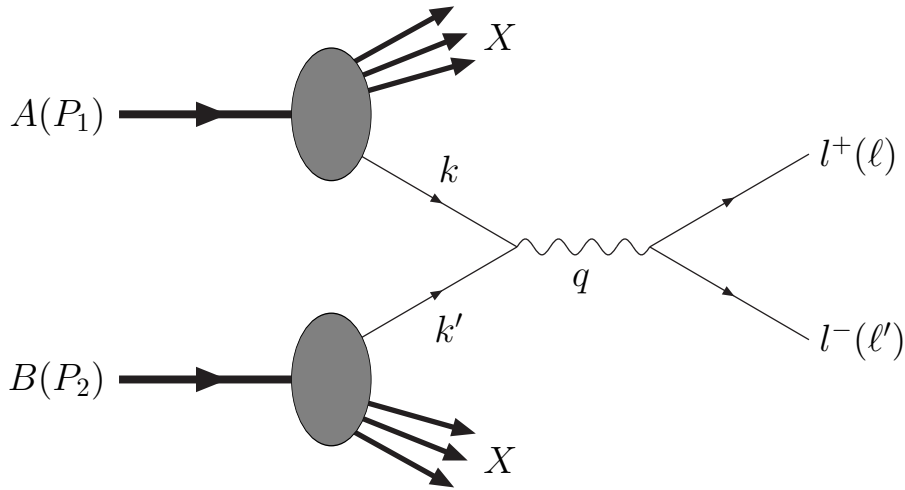
\begin{figure}[htbp]
  \centering
  \begin{picture}(380,190)(0,50)
    \SetOffset(10,0)
    \SetWidth{3}
    \ArrowLine(50,100)(120,100)
    \ArrowLine(50,200)(120,200)
    \SetWidth{0.5}
    \ArrowLine(140,115)(200,150)
    \ArrowLine(140,185)(200,150)
    \Photon(200,150)(250,150){3}{4}
    \ArrowLine(250,150)(310,115)
    \ArrowLine(250,150)(310,185)
    \SetWidth{2}
    \LongArrow(130,205)(183,220)
    \LongArrow(130,210)(180,230)
    \LongArrow(130,215)(175,240)
    \LongArrow(130,95)(183,80)
    \LongArrow(130,90)(180,70)
    \LongArrow(130,85)(175,60)
    \SetWidth{0.5}
    \GOval(130,100)(25,15)(0){0.5}
    \GOval(130,200)(25,15)(0){0.5}
    \Text(45,200)[r]{\large$A(P_1)$}
    \Text(45,100)[r]{\large$B(P_2)$}
    \Text(315,115)[l]{\large$l^-(\ell')$}
    \Text(315,185)[l]{\large$l^+(\ell)$}
    \Text(225,142)[t]{\large$q$}
    \Text(188,230)[l]{\large$X$}
    \Text(188,70)[l]{\large$X$}
    \Text(175,180)[]{\large$k$}
    \Text(175,120)[]{\large$k'$}
  \end{picture}
  \caption{Drell--Yan dilepton production.}
  \label{dy}
\end{figure}

The \DY cross-section is
\begin{equation}
  \d\sigma =
  \frac{8\pi^2\alpha_\text{em}^2}{s\,Q^4} \,
  2 L_{\mu\nu} W^{\mu\nu} \, (2\pi)^4 \,
  \frac{\d^3\Vec{\ell}}{(2\pi)^3 \, 2E} \,
  \frac{\d^3\Vec{\ell}'}{(2\pi)^3 \, 2E'} \, ,
  \label{dy9}
\end{equation}
where the leptonic tensor, neglecting lepton masses and ignoring their
polarisation, is given by
\begin{equation}
  L_{\mu\nu} = 2
  \left(
    \ell_\mu \ell'_\nu + \ell_\nu \ell'_\mu - \frac{Q^2}{2} \, g_{\mu\nu}
  \right) ,
  \label{dy8}
\end{equation}
and the hadronic tensor is defined as
\begin{eqnarray}
  W^{\mu\nu} &=&
  \frac1{(2\pi)^4} \, \sum_X \int \! \frac{\d^3\Vec{P}_X}{(2\pi)^3 \, 2E_X} \,
  (2\pi)^4  \, \delta^4(P_1 + P_2 - q - P_X )
  \nonumber
\\
  && \null \times
  \langle P_1 S_1, P_2 S_2 | J^\mu (0) | X \rangle
  \langle X | J^\nu (0) | P_1 S_1, P_2 S_2 \rangle
  \nonumber
\\
  &=&
  \frac1{(2\pi)^4} \int \! \d^4\xi \, \e^{\I q{\cdot}\xi} \,
  \langle P_1 S_1, P_2 S_2 | J^\mu(z) J^\nu(0) | P_1 S_1, P_2 S_2 \rangle \, .
  \label{dy7}
\end{eqnarray}
The phase space in (\ref{dy9}) can be rewritten as
\begin{equation}
  \frac{\d^3\Vec{\ell}}{(2\pi)^3\,2E} \,
  \frac{\d^3\Vec{\ell}'}{(2\pi)^3\,2E'} =
  \frac{\d\Omega \, \d^4 q}{8(2\pi)^6} \, ,
  \label{dy10}
\end{equation}
where $\Omega$ is the solid angle identifying the direction of the leptons in
their rest frame. Using (\ref{dy10}) the \DY cross-section takes the form
\begin{equation}
  \frac{\d\sigma}{\d^4q \, \d\Omega} =
  \frac{\alpha_\text{em}^2}{2s\,Q^4} \, L_{\mu\nu} W^{\mu\nu} \, .
  \label{dy11}
\end{equation}

We define now the two invariants
\begin{equation}
  x_1 = \frac{Q^2}{2P_1{\cdot}q} \, , \qquad
  x_2 = \frac{Q^2}{2P_2{\cdot}q} \, .
  \label{dy12}
\end{equation}
In the parton model $x_1$ and $x_2$ will be interpreted as the fractions of the
longitudinal momenta of the hadrons $A$ and $B$ carried by the quark and the
antiquark which annihilate into the virtual photon.

In a frame where the two colliding hadrons are collinear ($A$ is taken to move
in the positive $z$ direction), the photon momentum can be parametrised as
\begin{equation}
  q^\mu =
  \frac{Q^2}{x_2 s} \, P_1^\mu +
  \frac{Q^2}{x_1 s} \, P_2^\mu + q_T^\mu \, .
  \label{dy13}
\end{equation}
Neglecting terms of order $\Ord(1/Q^2)$, one finds $Q^2/x_1x_2s\simeq1$, that is
$\tau=Q^2/s=x_1x_2$, and therefore
\begin{equation}
  q^\mu =
  x_1 \, P_1^\mu +
  x_2 \, P_2^\mu +
  q_T^\mu =
  \left[ x_1 \, P_1^+, x_2 \, P_2^-, \Vec{q}_T \strut \right] .
  \label{dy14}
\end{equation}
Note that
\begin{equation}
  x_1 \simeq \frac{P_2{\cdot}q}{P_1{\cdot}P_2} \, , \qquad
  x_2 \simeq \frac{P_1{\cdot}q}{P_1{\cdot}P_2} \, .
  \label{dy15}
\end{equation}
In terms of $x_1$, $x_2$ and $\Vec{q}_T$ the \DY cross-section reads
\begin{equation}
  \frac{\d\sigma}{\d{x}_1 \, \d{x}_2 \, \d^2\Vec{q}_T \, \d\Omega} =
  \frac{\alpha_\text{em}^2}{4Q^4} \, L_{\mu\nu} W^{\mu\nu} \, .
  \label{dy16}
\end{equation}

\begin{figure}[htbp]
  \centering
  \begin{picture}(340,140)(80,45)
    \Line(100,50)(300,50)
    \Line(150,100)(350,100) 
    \Line(100,50)(150,100)
    \Line(300,50)(350,100)
    \Text(125,60)[l]{lepton plane}
    \Line(150,100)(165,180)
    \Line(350,100)(365,180)
    \Line(165,180)(365,180)
    \Line(350,100)(400,150)
    \Line(400,150)(360,150)
    \DashLine(360,150)(200,150){8}
    \DashLine(150,100)(200,150){8}
    \CArc(350,100)(30,45,80)
    \Text(370,135)[]{$\phi$}
    \LongArrow(350,100)(370,100)
    \LongArrow(350,100)(353,116)
    \Text(370,95)[t]{$\hat{z}$}
    \Text(348,116)[r]{$\hat{x}$}
    \SetWidth{2}
    \LongArrow(195,110)(248,100)
    \LongArrow(310,115)(252,100)
    \SetWidth{0.5}
    \Text(192,110)[r]{$\Vec{P}_1$}
    \Text(305,120)[b]{$\Vec{P}_2$}
    \SetWidth{1}
    \LongArrow(250,100)(280,80)
    \DashLine(250,100)(223,118){5}
    \LongArrow(223,118)(220,120)
    \SetWidth{0.5}
    \CArc(250,100)(20,-30,0)
    \Text(280,92)[]{$\theta$}
    \Text(280,75)[t]{$\Vec\ell$}
    \Text(220,125)[b]{$\Vec\ell'$}
  \end{picture}
  \caption{The geometry of Drell--Yan production in the rest frame of the
           lepton pair.}
  \label{dyplane}
\end{figure}
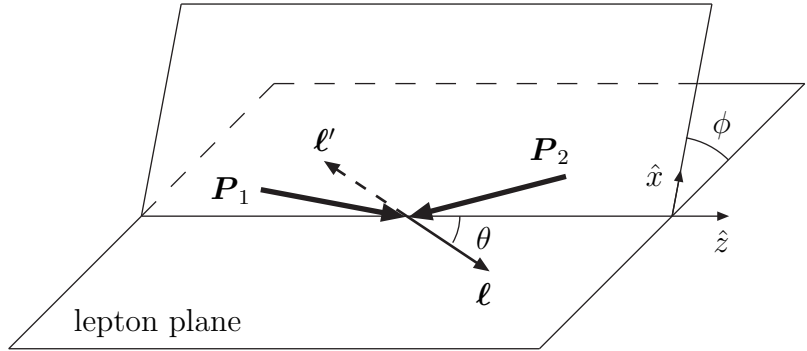

It is customary \cite{Ralston:1979ys, Tangerman:1995eh} to introduce three
vectors $Z^\mu$, $X^\mu$ and $Y^\mu$ defined as
\begin{subequations}
\begin{eqnarray}
  Z^\mu &=&
  \frac{P_2{\cdot}q}{P_1{\cdot}P_2} \, P_1^\mu - \frac{P_1{\cdot}q}{P_1{\cdot}P_2} \, P_2^\mu \, ,
  \label{dy17}
\\
  X^\mu &=& - \frac1{P_1{\cdot}P_2}
  \left[
    P_2{\cdot}Z \, \widetilde{P}_1^\mu -
    P_1{\cdot}Z \, \widetilde{P}_2^\mu
  \right] ,
  \label{dy18}
\\
  Y^\mu &=&
  \frac1{P_1{\cdot}P_2} \, \varepsilon^{\mu\nu\rho\sigma} \,
  P_{1\nu} P_{2\rho} q_\sigma \, .
  \label{dy19}
\end{eqnarray}
\end{subequations}
where $\widetilde{P}_{1,2}^\mu=P_{1,2}^\mu-(P_{1,2}{\cdot}q /q^2)\,q^\mu$. These
vectors are mutually orthogonal and orthogonal to $q^\mu$, and satisfy
\begin{equation}
  Z^2 \simeq - Q^2 \, , \qquad X^2 \simeq Y^2 \simeq - \Vec{q}_T^2 \, .
  \label{dy20}
\end{equation}
Thus, they form a set of spacelike axes and have only spatial components in the
dilepton rest frame. Using (\ref{dy14}), $Z^\mu$, $X^\mu$ and $Y^\mu$ can be
expressed as
\begin{equation}
  Z^\mu = x_1 \, P_1^\mu - x_2 \, P_2^\mu \, , \qquad
  X^\mu = q_T^\mu \, , \qquad
  Y^\mu = \varepsilon_T^{\sigma\mu} \, q_{T\sigma} \, ,
  \label{dy21}
\end{equation}
where
\begin{equation}
  \varepsilon_T^{\sigma\mu} \equiv
  \frac1{P_1{\cdot}P_2} \, \varepsilon^{\sigma\mu\nu\rho} P_{1\nu} P_{2\rho} \, .
\end{equation}
In terms of the unit vectors
\begin{equation}
  \hat{x}^\mu = \frac{X^\mu}{\sqrt{-X^2}} \, , \quad
  \hat{y}^\mu = \frac{Y^\mu}{\sqrt{-Y^2}} \, , \quad
  \hat{z}^\mu = \frac{Z^\mu}{\sqrt{-Z^2}} \, , \quad
  \hat{q}^\mu = \frac{q^\mu}{\sqrt{Q^2}} \, ,
  \label{dy22}
\end{equation}
the lepton momenta can be expanded as
\begin{subequations}
\begin{eqnarray}
  \ell^\mu &=&
  \half \, q^\mu +
  \half \, Q \, (\sin\theta \, \cos\phi \, \hat{x}^\mu +
                 \sin\theta \, \sin\phi \, \hat{y}^\mu +
                 \cos\theta \, \hat{z}^\mu) \, ,
  \label{dy23}
\\
  {\ell'}^\mu &=&
  \half \, q^\mu -
  \half \, Q \, (\sin\theta \, \cos\phi \, \hat{x}^\mu +
                 \sin\theta \, \sin\phi \, \hat{y}^\mu +
                 \cos\theta \, \hat{z}^\mu) \, .
  \label{dy24}
\end{eqnarray}
\end{subequations}
The geometry of the process in the dilepton rest frame is shown in
Fig.~\ref{dyplane}.

The leptonic tensor then reads
\begin{eqnarray}
  L^{\mu\nu} &=&
  - \half Q^2
  \left[
    (1+\cos^2\theta) \, g_\perp^{\mu\nu} -
    2\sin^2\theta \, \hat{z}^\mu \hat\zeta^\nu
  \right.
  \nonumber
\\
  && \hspace{2.5em} \null +
    2\sin^2\theta \, \cos2\phi \,
    ( \hat{x}^\mu \hat{x}^\nu + \half g_\perp^{\mu\nu} ) +
    \sin^2\theta \, \sin2\phi \; \hat{x}^{\{\mu} \hat{y}^{\nu\}}
  \nonumber
\\
  && \hspace{2.5em} \null
  \left. +
    \sin2\theta \, \cos\phi \; \hat{z}^{\{\mu} \hat{x}^{\nu\}} +
    \sin2\theta \, \sin\phi \; \hat{z}^{\{\mu} \hat{y}^{\nu\}}
  \right] ,
  \label{dy25}
\end{eqnarray}
where
\begin{equation}
  g_\perp^{\mu\nu} =
  g^{\mu\nu} - \hat{q}^\mu \hat{q}^\nu  + \hat{z}^\mu \hat{z}^\nu.
  \label{dy26}
\end{equation}

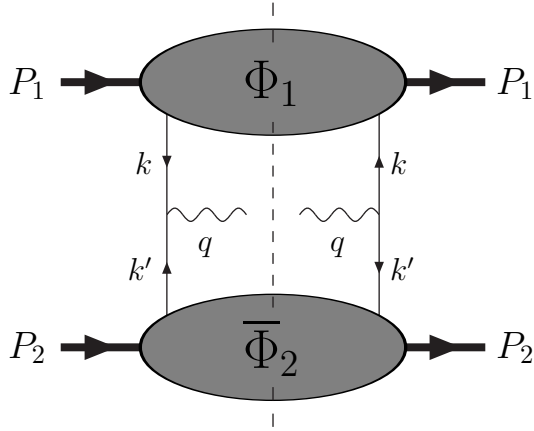
\begin{figure}[htbp]
  \centering
  \begin{picture}(300,170)(0,15)
    \SetWidth{3}
    \ArrowLine(70,50)(100,50)
    \ArrowLine(200,50)(230,50)
    \ArrowLine(70,150)(100,150)
    \ArrowLine(200,150)(230,150)
    \SetWidth{0.5}
    \ArrowLine(110,60)(110,100)
    \ArrowLine(190,100)(190,60)
    \ArrowLine(110,140)(110,100)
    \ArrowLine(190,100)(190,140)
    \Photon(110,100)(140,100){2.5}{2.5}
    \Photon(160,100)(190,100){2.5}{2.5}
    \GOval(150,50)(20,50)(0){0.5}
    \GOval(150,150)(20,50)(0){0.5}
    \Text(65,50)[r]{\large$P_2$}
    \Text(65,150)[r]{\large$P_1$}
    \Text(235,50)[l]{\large$P_2$}
    \Text(235,150)[l]{\large$P_1$}
    \Text(125,92)[t]{$q$}
    \Text(175,92)[t]{$q$}
    \Text(105,80)[r]{$k'$}
    \Text(195,80)[l]{$k'$}
    \Text(105,120)[r]{$k$}
    \Text(195,120)[l]{$k$}
    \Text(150,150)[]{\LARGE$\Phi_1$}
    \Text(150,50)[]{\LARGE$\overline{\Phi}_2$}
    \DashLine(150,20)(150,35){4}
    \DashLine(150,65)(150,135){4}
    \DashLine(150,165)(150,180){4}
  \end{picture}
  \caption{The parton-model diagram for the Drell--Yan hadronic tensor.}
  \label{handbag5}
\end{figure}

In the parton model, calling $k$ and $k'$ the momenta of the quark (or
antiquark) coming from hadron $A$ and $B$ respectively, the hadronic tensor is
(see Fig.~\ref{handbag5})
\begin{equation}
  W^{\mu\nu} =
  \frac1{3} \, \sum_a e_a^2 \int \! \frac{\d^4k}{(2\pi)^4}
  \int \! \frac{\d^4 k'}{(2\pi)^4} \, \delta^4(k + k' - q) \,
  \Tr [\Phi_1 \, \gamma^\mu \, \anti\Phi_2 \gamma^\nu].
  \label{dy27}
\end{equation}
Here $\Phi_1$ is the quark correlation matrix for hadron $A$, eq.~(\ref{df1}),
$\anti\Phi_2$ is the antiquark correlation matrix for hadron $B$,
eq.~(\ref{adf1}), and the factor $1/3$ has been added since in $\Phi_1$ and
$\anti\Phi_2$ summations over colours are implicit. It is understood that, in
order to obtain the complete expression of the hadronic tensor, one must add to
(\ref{dy27}) a term with $\Phi_1$ replaced by $\Phi_2$ and $\anti\Phi_2$
replaced by $\anti\Phi_1$, which accounts for the case where a quark is
extracted from $B$ and an antiquark is extracted from $A$. In the following
formul{\ae} we shall denote this term symbolically by $[1\leftrightarrow2]$.

Hereafter the quark transverse motion (which is discussed at length in
\cite{Tangerman:1995eh}) will be ignored and only the ordinary collinear
configuration will be considered.

We now evaluate the hadronic tensor in a frame where $A$ and $B$ move
collinearly, with large longitudinal momentum. Setting $k\simeq\xi_1\,P_1$,
$k'\simeq\xi_2\,P_2$ and using (\ref{dy14}), the delta function in (\ref{dy27})
gives
\begin{eqnarray}
  \delta^4(k + k' - \kappa)
  &=&
  \delta(k^+ + {k'}^+ - q^+) \,
  \delta(k^- + {k'}^- - q^-) \,
  \delta^2(\Vec{q}_T)
  \nonumber
\\
  &\simeq&
  \delta(\xi_1 P_1^+ - x_1 P_1^+) \,
  \delta(\xi_2 P_2^- - x_2 P_2^-) \,
  \delta^2(\Vec{q}_T)
  \nonumber
\\
  &=&
  \frac1{P_1^+ \, P_2^-} \,
  \delta(\xi_1 - x_1) \,
  \delta(\xi_2 - x_2) \,
  \delta^2(\Vec{q}_T) \, .
  \label{dy28}
\end{eqnarray}
The hadronic tensor then becomes
\begin{eqnarray}
  W^{\mu\nu}
  &=&
  \frac1{3} \, \sum_a e_a^2
  \int \! \frac{\d{k}^-  \, \d^2\Vec{k}_{T} }{(2\pi)^4}
  \int \! \frac{\d{k'}^+ \, \d^2\Vec{k}_{T}'}{(2\pi)^4} \,
  \delta^2(\Vec{q}_T)
  \nonumber
\\
  &&
  \hspace{2.5em} \null
  \times
  \Tr
  [
    \Phi_1 \, \gamma^\mu \, \anti\Phi_2 \gamma^\nu
  ]_{k^+=x_1P_1^+, \,{k'}^-=x_2 P_2^-}
  +
  [1 \leftrightarrow 2] \, .
  \label{dy29}
\end{eqnarray}

Since the leptonic tensor is symmetric, only the symmetric part of $W^{\mu\nu}$
contributes to the cross-section. For
$\Tr[\Phi_1\gamma^{\{\mu}\anti\Phi_2\gamma^{\nu\}}]$ we resort to the Fierz
decomposition (\ref{sicross2}), with the replacements $\Phi\to\Phi_1$,
$\Xi\to\anti\Phi_2$. Using (\ref{df16}--\ref{subdf18}) and
(\ref{adf4}--\ref{adf6}), we then obtain (the spins of the two hadrons are
$\Vec{S}_1=(\Vec{S}_{1T},\lambda_1)$ and $\Vec{S}_2=(\Vec{S}_{2T},\lambda_2)$)
\begin{eqnarray}
  \int \! \d^2\Vec{q}_T \, W^{\mu\nu}
  &=&
  \frac1{3} \sum_a e_a^2
  \left\{
    -g_T^{\mu\nu}
    \left[
      f_a(x_1) \, \anti{f}_a(x_2) +
      \lambda_1 \lambda_2 \, \DL{f}_a(x_1) \, \DL\anti{f}_a(x_2)
    \right]
  \right.
  \nonumber
\\
  && \null -
  \left.
    \left[
      S_{1T}^{\{\mu} S_{2T}^{\nu \}} +
      \Vec{S}_{1T}{\cdot}\Vec{S}_{2T} \, g_T^{\mu\nu}
    \right] \DT{f}_a(x_1) \, \DT\anti{f}_a(x_2)
  \right\}
  \nonumber
\\
  && \null + [1\leftrightarrow2] \, .
  \label{dy30}
\end{eqnarray}

In contracting the leptonic and the hadronic tensors, it is convenient to pass
from the $AB$ collinear frame to the $\gamma^*A$ collinear frame. We recall
that, at leading twist, the transverse ($T$) vectors approximately coincide
with the vectors perpendicular to the photon direction (denoted by a subscript
$\perp$): $\Vec{S}_{1T}\simeq\Vec{S}_{1\perp}$,
$\Vec{S}_{2T}\simeq\Vec{S}_{2\perp}$, $g_T^{\mu\nu}\simeq{}g_\perp^{\mu\nu}$.
Therefore, the contraction $L_{\mu\nu}\,W^{\mu\nu}$ can be performed by means
of the following identities
\begin{subequations}
\begin{eqnarray}
  - g_\perp^{\mu\nu} L_{\mu\nu} &=& Q^2 (1+\cos^2\theta),
  \label{dy31}
\\
  \left[
    S_{1\perp}^{\{\mu} S_{2\perp}^{\nu\}} +
    \Vec{S}_{1\perp}{\cdot}\Vec{S}_{2\perp} \, g_\perp^{\mu\nu}
  \right] L_{\mu\nu} &=&
  \nonumber
\\
  && \hspace{-8em} \null -
  Q^2 \sin^2\theta | \Vec{S}_{1\perp} | | \Vec{S}_{2\perp} |
  \cos(2\phi - \phi_{S_1} - \phi_{S_2}) \, ,
  \label{dy32}
\end{eqnarray}
\end{subequations}
where $\theta$ is the polar angle of the lepton pair in the dilepton rest frame
and $\phi_{S_1}$ ($\phi_{S_2}$) is the azimuthal angle of $\Vec{S}_{1\perp}$
($\Vec{S}_{2\perp}$), measured with respect to the lepton plane. For the
Drell--Yan cross-section we finally obtain
\begin{eqnarray}
  \frac{\d\sigma}{\d\Omega \d{x}_1 \d{x}_2} &=&
  \frac{\alpha_\text{em}^2}{4Q^2} \, \sum_a \, \frac{e_a^2}{3}
  \left\{
    \left[ \strut f_a(x_1) \, \anti{f}_a(x_2) \right.
  \right.
  \nonumber
\\
  && \null +
  \left.
    \lambda_1 \, \lambda_2 \, \DL{f}_a (x_1) \, \DL\anti{f}_a(x_2)
  \strut \right] (1 + \cos^2\theta)
  \nonumber
\\
  && \null +
  \left.
    | \Vec{S}_{1\perp} | \,
    | \Vec{S}_{2\perp} | \,
    \cos(2\phi - \phi_{S_1} - \phi_{S_2}) \,
    \DT{f}_a(x_1) \, \DT\anti{f}_a(x_2)\, \sin^2\theta
  \strut \right\}
  \nonumber
\\
  && \null + [1\leftrightarrow2] \, .
  \label{dy33}
\end{eqnarray}
From this we derive the parton-model expression for the double transverse
asymmetry:
\begin{eqnarray}
  A_{TT}^\text{DY} &=&
  \frac{\d\sigma(\Vec{S}_{1\perp}, \Vec{S}_{2\perp}) -
        \d\sigma(\Vec{S}_{1\perp},-\Vec{S}_{2\perp})}
       {\d\sigma(\Vec{S}_{1\perp}, \Vec{S}_{2\perp}) +
        \d\sigma(\Vec{S}_{1\perp},-\Vec{S}_{2\perp})}
  \nonumber
\\
  &=&
  | \Vec{S}_{1\perp} |
  | \Vec{S}_{2\perp} | \, \frac{\sin^2\theta \,
  \cos(2\phi - \phi_{S_1} - \phi_{S_2})}{1 + \cos^2\theta}
  \nonumber
\\
  && \null \times
  \frac{\sum_a e_a^2 \, \DT{f}_a(x_1) \DT\anti{f}_a(x_2) + [1\leftrightarrow2]}
       {\sum_a e_a^2 \,      f_a(x_1)    \anti{f}_a(x_2) + [1\leftrightarrow2]}
  \, ,
  \label{dy34}
\end{eqnarray}
and we see that a measurement of $A_{TT}^\text{DY}$ directly provides the
product of quark and antiquark transverse polarisation distributions
$\DT{f}(x_1)\,\DT{f}(x_2)$, with no mixing with other unknown quantities. Thus,
the Drell--Yan process seems to be, at least in principle, a very good reaction
to probe transversity. Note that in leading-order \QCD, eq.~(\ref{dy34}) is
still valid, with $Q^2$ dependent distribution functions, namely
\begin{eqnarray}
  A_{TT}^\text{DY} &=&
  | \Vec{S}_{1\perp} |
  | \Vec{S}_{2 \perp} | \,
  \frac{\sin^2\theta \, \cos(2 \phi - \phi_{S_1} - \phi_{S_2})}
       {1 + \cos^2\theta}
  \nonumber
\\
  && \null \times
  \frac{\sum_a e_a^2\,\DT{f}_a(x_1,Q^2)\DT\anti{f}_a(x_2,Q^2) + [1\leftrightarrow2]}
       {\sum_a e_a^2\,     f_a(x_1,Q^2)   \anti{f}_a(x_2,Q^2) + [1\leftrightarrow2]}
  \, .
  \label{dy35}
\end{eqnarray}
Here $\DT{f}(x,Q^2)$ are the transversity distributions evolved at \LO. In
Sec.~\ref{transvhh} we shall see some predictions for $A_{TT}^\text{DY}$.

\subsubsection{$Z^0$-mediated Drell--Yan process}
\label{Z0drellyan}

If Drell--Yan dilepton production is mediated by the exchange of a $Z^0$ boson,
the vertex $e_i\,\gamma^\mu$, where $e_i$ is the electric charge of particle
$i$ (quark or lepton), is replaced by $(V_i+A_i\,\gamma_5)\,\gamma^\mu$, where
the vector and axial-vector couplings are
\begin{equation}
\begin{split}
  V_i &= T_3^i - 2 e_i \, \sin^2\vartheta_W \, ,
\\
  A_i &= T_3^i \, .
  \label{dyZ1}
\end{split}
\end{equation}
The weak isospin $T_3^i$ is $+\half$ for $i=u$ and $-\half$ for $i=l^-,d,s$.

The resulting double transverse asymmetry has a form similar to (\ref{dy34}),
with the necessary changes in the couplings. Omitting the interference
contributions, it reads
\begin{eqnarray}
  A_{TT}^{\text{DY}, Z}
  &=&
  | \Vec{S}_{1\perp} | | \Vec{S}_{2\perp} | \,
  \frac{\sin^2\theta \, \cos(2\phi - \phi_{S_1} - \phi_{S_2})}
       {1 + \cos^2\theta}
  \nonumber
\\
  && \null \times
  \frac{\sum_a (V_a^2 - A_a^2) \,
        \DT{f}_a(x_1) \DT\anti{f}_a(x_2) + [1\leftrightarrow2]}
       {\sum_a (V_a^2 + A_a^2) \,
        f_a(x_1) \anti{f}_a(x_2) + [1\leftrightarrow2]} \, .
  \label{dy35a}
\end{eqnarray}

\subsection{Factorisation in Drell--Yan processes}
\label{dyfactorisation}

With a view to extending the results previously obtained to \NLO in \QCD, we
now rederive them in the framework of \QCD factorisation \cite{Collins:1993xw}.
The Drell--Yan cross-section is written in a factorised form as (hereafter we
omit the exchanged term $[1\leftrightarrow2]$)
\begin{eqnarray}
  \d\sigma &=&
  \sum_a \sum_{\alpha\alpha'\beta\beta'} \int \! \d\xi_1 \int \! \d\xi_2 \;
  \rho_{\alpha'\alpha}^{(1)} \, f_a(\xi_1, \mu^2) \,
  \rho_{\beta'\beta}^{(2)} \, \anti{f}_a(\xi_2, \mu^2)
  \nonumber
\\
  && \hspace{8em} \null \times
  \left[
    \d\hat\sigma(Q^2/\mu^2, \alpha_s(\mu^2))
  \right]_{\alpha\alpha'\beta\beta'} \, ,
  \label{dyfac0}
\end{eqnarray}
where $\xi_1$ and $\xi_2$ are the momentum fractions of the quark (from hadron
$A$) and antiquark (from $B$), $\rho^{(1)}$ and $\rho^{(2)}$ are the quark and
antiquark spin density matrices, and
$(\d\hat\sigma)_{\alpha\alpha'\beta\beta'}$ is the cross-section matrix (in the
quark and antiquark spin space) of the elementary subprocesses. As usual, $\mu$
denotes the factorisation scale.

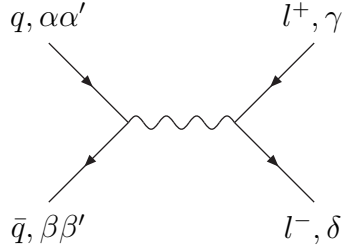
\begin{figure}[htbp]
  \centering
  \begin{picture}(200,90)(0,15)
    \ArrowLine(80,60)(50,30)
    \ArrowLine(50,90)(80,60)
    \Photon(80,60)(120,60){2.5}{3.5}
    \ArrowLine(120,60)(150,30)
    \ArrowLine(150,90)(120,60)
    \Text(50,100)[]{$q,\alpha\alpha'$}
    \Text(50,20)[]{$\bar{q},\beta\beta'$}
    \Text(150,20)[]{$l^-,\delta$}
    \Text(150,100)[]{$l^+,\gamma$}
  \end{picture}
  \caption{The $q\bar{q}\to{}l^+l^-$ process contributing to Drell--Yan
           production at LO.}
  \label{dyqqbar}
\end{figure}

At \LO $\d\hat\sigma$ incorporates a delta function of energy--momentum
conservation, namely $\delta^4(k+k'-q)$, which sets $\xi_1=x_1$, $\xi_2=x_2$
and $\Vec{q}_T=0$. Thus eq.~(\ref{dyfac0}) becomes (omitting the scales)
\begin{equation}
  \frac{\d\sigma}{\d\Omega \d{x}_1 \d{x}_2} =
  \sum_a \sum_{\alpha\alpha'\beta\beta'}
  \rho_{\alpha'\alpha}^{(1)} \, \rho_{\beta'\beta}^{(2)}
  \left( \frac{\d\hat\sigma}{\d\Omega} \right)_{\alpha\alpha'\beta\beta'}
  f_a (\xi_1) \, \anti{f}_a(\xi_2) \, ,
  \label{dyfac1}
\end{equation}
where the only subprocess is $q\anti{q}\to{l^+l^-}$ (see Fig.~\ref{dyqqbar})
and its cross-section is
\begin{equation}
  \left(\frac{\d\hat\sigma}{\d\Omega} \right)_{\alpha\alpha'\beta\beta'} =
  \frac1{64\pi^2\hat{s}} \, \sum_{\gamma\delta}
  \mathcal{M}^*_{\alpha\beta\gamma\delta} \,
  \mathcal{M}_{\alpha'\beta'\gamma\delta} \, .
  \label{dyfac2}
\end{equation}

The contributing scattering amplitudes are
\begin{equation}
  \mathcal{M}_{++++} = \mathcal{M}_{----} \, , \qquad
  \mathcal{M}_{++--} = \mathcal{M}_{--++} \, ,
  \label{dyfac3}
\end{equation}
and the cross-section (\ref{dyfac1}) reduces to
\begin{eqnarray}
  \frac{\d\sigma }{\d\Omega \d{x}_1 \d{x}_2} &=&
  \sum_a
  \left\{
    \left(
      \rho_{++}^{(1)} \rho_{++}^{(2)} +
      \rho_{--}^{(1)} \rho_{--}^{(2)}
    \right)
    \left( \frac{\d\hat\sigma}{\d\Omega} \right)_{++++}
  \right.
  \nonumber
\\
  && \null +
  \left.
    \left(
      \rho_{+-}^{(1)} \rho_{+-}^{(2)} +
      \rho_{-+}^{(1)} \rho_{-+}^{(2)}
    \right)
    \left( \frac{\d\hat\sigma}{\d\Omega} \right)_{+-+-}
  \right\} f_a(x_1) \, \anti{f}_a(x_2) \, . \qquad
  \label{dyfac4}
\end{eqnarray}
Here the spin density matrix elements are, for the quark
\begin{subequations}
\begin{equation}
  \begin{array}{rcl@{\qquad}rcl}
    \rho_{++}^{(1)} &=& \half (1 + \lambda) \, , &
    \rho_{--}^{(1)} &=& \half (1 - \lambda) \, ,
  \\
    \rho_{+-}^{(1)} &=& \half (s_{x} - \I \, s_{y}) \, , &
    \rho_{-+}^{(1)} &=& \half (s_{x} + \I \, s_{y}) \, ,
  \end{array}
\end{equation}
and for the antiquark
\begin{equation}
  \begin{array}{rcl@{\qquad}rcl}
    \rho_{++}^{(2)} &=& \half (1 + \anti\lambda) \, , &
    \rho_{--}^{(2)} &=& \half (1 - \anti\lambda) \, ,
  \\
    \rho_{+-}^{(2)} &=& \half (\anti{s}_{x} - \I \, \anti{s}_{y}) \, , &
    \rho_{-+}^{(2)} &=& \half (\anti{s}_{x} + \I \, \anti{s}_{y}) \, ,
  \end{array}
\end{equation}
\end{subequations}
so that (\ref{dyfac4}) becomes
\begin{eqnarray}
  \frac{\d\sigma}{\d\Omega \d{x}_1 \d{x}_2} &=&
  \frac12 \sum_a
  \left\{
    (1 + \lambda \, \anti\lambda)
    \left( \frac{\d\hat\sigma}{\d\Omega} \right)_{++++}
  \right.
  \nonumber
\\
  && \hspace{2.5em} \null +
  \left.
    (s_{x} \, \anti{s}_{x} - s_{y} \, \anti{s}_{y})
    \left( \frac{\d\hat\sigma}{\d\Omega} \right)_{+-+-}
  \right\} \, f_a(x_1) \, \anti{f}_a(x_2) \, . \qquad
  \label{dyfac5}
\end{eqnarray}

At \LO, the scattering amplitudes for the $q\anti{q}$ annihilation process are
\begin{subequations}
\begin{eqnarray}
  \mathcal{M}_{++++} = \mathcal{M}_{----} &=& e^2 \, e_a \, (1-\cos\theta) \, ,
  \label{dyfac6a}
\\
  \mathcal{M}_{++--} = \mathcal{M}_{--++} &=& e^2 \, e_a \, (1+\cos\theta) \, ,
  \label{dyfac6b}
\end{eqnarray}
\end{subequations}
and the elementary cross-sections then read
\begin{subequations}
\begin{eqnarray}
  \left( \frac{\d\hat\sigma}{\d\Omega} \right)_{++++}  =
  \left( \frac{\d\hat\sigma}{\d\Omega} \right)_{----} &=&
  \frac1{64\pi^2\hat{s}}
  \left( | \mathcal{M}_{++++} |^2 + | \mathcal{M}_{++--} |^2 \right)
  \nonumber
\\
  &=&
  \frac{\alpha_\text{em}^2 \, e_a^2}{2\hat{s}} \, \frac1{3}
  \left( 1 + \cos^2\theta \right) ,
  \label{dyfac7}
\\
  \left( \frac{\d\hat\sigma}{\d\Omega} \right)_{+-+-} =
  \left( \frac{\d\hat\sigma}{\d\Omega} \right)_{-+-+} &=&
  \frac1{64\pi^2\hat{s}} \,
  2 \Re \left( \mathcal{M}^*_{++++} \, \mathcal{M}_{++--} \right)
  \nonumber
\\
  &=&
  \frac{\alpha_\text{em}^2 \, e_a^2}{2\hat{s}} \, \frac1{3} \, \sin^2\theta \, .
  \label{dyfac8}
  \sublabel{subdyfac8}
\end{eqnarray}
\end{subequations}
Inserting (\ref{dyfac7},~\ref{subdyfac8}) into (\ref{dyfac5}) we obtain
\begin{eqnarray}
  \frac{\d\sigma}{\d\Omega \d{x}_1 \d{x}_2} &=&
  \frac{\alpha_\text{em}^2}{4Q^2} \, \sum_a \, \frac{e_a^2}{3}
  \left\{
    ( 1 + \lambda \anti\lambda ) \, (1 + \cos^2\theta)
  \right.
  \nonumber
\\
  && \null +
  \left.
    (s_x \, \anti{s}_x - s_y \, \anti{s}_y) \, \sin^2\theta
  \right\} \,  f_a(x_1) \, \anti{f}_a(x_2) \, .
  \label{dyfac9}
\end{eqnarray}
Using
\begin{subequations}
\begin{eqnarray}
  \lambda \, f_a(x_1) &=& \lambda_1 \, \DL{f}_a(x_1) \, , \qquad
  \Vec{s}_\perp \, f_a(x_1) = \Vec{S}_{1\perp} \, \DT{f}_a(x_1) \, ,
  \label{dyfac10}
\\
  \anti\lambda \, \anti{f}_a(x_2) &=&
  \lambda_2 \, \DL\anti{f}_a(x_2) \, , \qquad
  \anti{\Vec{s}}_\perp \, \anti{f}_a(x_2) =
  \Vec{S}_{2\perp} \, \DT\anti{f}_a(x_2)
  \, ,
  \label{dyfac10b}
\end{eqnarray}
\end{subequations}
we obtain
\begin{eqnarray}
  \frac{\d\sigma}{\d\Omega \d{x}_1 \d{x}_2} &=&
  \frac{\alpha_\text{em}^2}{4Q^2} \, \sum_a \, \frac{e_a^2}{3}
  \left\{ \left[ f_a(x_1) \, \anti{f}_a(x_2) \Strut \right. \right.
  \nonumber
\\
  && \hspace{6em} \left. \null + \Strut
  \lambda_1 \, \lambda_2 \, \DL{f}_a(x_1) \, \DL\anti{f}_a(x_2) \right]
  (1 + \cos^2\theta)
  \nonumber
\\[1ex]
  && \hspace{6em} \null +
  | \Vec{S}_{1\perp} | \,
  | \Vec{S}_{2\perp} | \, \cos( 2\phi - \phi_{S_1} - \phi_{S_2} )
  \nonumber
\\[1ex]
  && \hspace{7em} \null \times
  \left. \Strut
    \DT{f}_a(x_1) \, \DT\anti{f}_a(x_2) \, \sin^2\theta
  \right\} ,
  \label{dyfac11}
\end{eqnarray}
which is what we obtained in Sec.~\ref{drellyan} in a different manner (see
eq.~(\ref{dy33})). Note that the angular factor appearing in $A_{TT}^\text{DY}$
-- eq.~(\ref{dy34}) -- is the elementary double-spin transverse asymmetry of
the $q\anti{q}$ scattering process, namely
\begin{eqnarray}
  \hat{a}_{TT} &\equiv&
  \frac{\d\hat\sigma(\Vec{s}_\perp, \anti{\Vec{s}}_\perp) -
        \d\hat\sigma(\Vec{s}_\perp,-\anti{\Vec{s}}_\perp)}
       {\d\hat\sigma(\Vec{s}_\perp, \anti{\Vec{s}}_\perp) +
        \d\hat\sigma(\Vec{s}_\perp,-\anti{\Vec{s}}_\perp)}
  \nonumber
\\
  &=&
  \frac{\d\hat\sigma_{+-+-}}{\d\hat\sigma_{++++}} \,
  (s_x \anti{s}_x - s_y \anti{s}_y) =
  \frac{\sin^2\theta}{1+\cos^2\theta} \, \cos(2\phi - \phi_s - \phi_{\anti{s}})
  \, . \qquad
  \label{dyfac11bis}
\end{eqnarray}

The Drell--Yan cross-section is most often expressed as a function of the
rapidity of the virtual photon $y$ defined as
\begin{equation}
  y \equiv \frac12 \, \ln \frac{q^+}{q^-} =
  \frac12 \, \ln \frac{x_1}{x_2} \, .
  \label{dyfac11b}
\end{equation}
In the lepton c.m.\ frame $y=\half(1+\cos\theta)$. From
$x_1x_2\equiv\tau=Q^2/s$, we obtain
\begin{equation}
  x_1 = \sqrt\tau \, \e^y    \, , \qquad
  x_2 = \sqrt\tau \, \e^{-y} \, ,
  \label{dyfac11c}
\end{equation}
and (\ref{dyfac11}) becomes ($\d{y}\,\d{Q^2}=s\,\d{x_1}\,\d{x_2}$)
\begin{eqnarray}
  \frac{\d\sigma}{\d\Omega \d{y} \d{Q}^2} &=&
  \frac{\alpha_\text{em}^2}{4Q^2s} \, \sum_a \, \frac{e_a^2}{3}
  \left\{ \Strut
    \left[ f_a(x_1) \, \anti{f}_a(x_2) \right.
  \right.
  \nonumber
\\
  && \hspace{6em} \null +
  \left.
    \lambda_1 \, \lambda_2 \, \DL{f}_a (x_1) \, \DL\anti{f}_a(x_2)
  \right] (1 + \cos^2\theta)
  \nonumber
\\[0.5ex]
  && \hspace{6em} \null +
  | \Vec{S}_{1\perp} | \,
  | \Vec{S}_{2\perp} | \, \cos(2\phi - \phi_{S_1} - \phi_{S_2} )
  \nonumber
\\
  && \hspace{7em} \null \times
  \left. \Strut
    \DT{f}_a (x_1) \, \DT\anti{f}_a(x_2)\, \sin^2\theta
  \right\} .
  \label{dyfac12}
\end{eqnarray}
If we integrate over $\cos\theta$, we obtain
\begin{eqnarray}
  \frac{\d\sigma}{\d{y} \d{Q}^2 \d\phi}
  &=&
  \frac{2\alpha_\text{em}^2}{9Q^2s} \, \sum_a \, e_a^2
  \left\{ \Strut
    \left[
      f_a(x_1) \, \anti{f}_a(x_2) +
      \lambda_1 \, \lambda_2 \, \DL{f}_a(x_1) \, \DL\anti{f}_a(x_2)
    \right]
  \right.
  \nonumber
\\
  && \hspace{-2em} \null +
  \left. \Strut
    \half \, | \Vec{S}_{1\perp} | \, | \Vec{S}_{2 \perp} | \,
    \cos(2\phi - \phi_{S_1} - \phi_{S_2}) \, \DT{f}_a(x_1) \, \DT\anti{f}_a(x_2)
  \right\} . \hspace{4em}
  \label{dyfac13}
\end{eqnarray}

\begin{figure}[htbp]
  \centering
  \begin{picture}(360,85)(0,10)
    \ArrowLine(80,60)(65,45)
    \ArrowLine(65,45)(50,30)
    \ArrowLine(50,90)(65,75)
    \ArrowLine(65,75)(80,60)
    \Photon(80,60)(110,60){2.5}{3}
    \GlueArc(80,60)(20,135,225){3}{4}
    \Text(80,15)[t]{(a)}
    \ArrowLine(170,60)(140,30)
    \ArrowLine(140,90)(170,60)
    \Photon(170,60)(200,60){2.5}{3}
    \GlueArc(155,75)(10,-45,135){3}{4}
    \Text(170,15)[t]{(b)}
    \ArrowLine(230,90)(270,90)
    \ArrowLine(270,90)(270,30)
    \ArrowLine(270,30)(230,30)
    \Photon(270,90)(310,90){2.5}{3}
    \Gluon(270,30)(310,30){3}{5}
    \Text(270,15)[t]{(c)}
  \end{picture}
  \caption{Elementary processes contributing to the transverse Drell--Yan
           cross-section at NLO: (a, b) virtual-gluon corrections and (c)
           real-gluon emission.}
  \label{dyqqg2}
\end{figure}
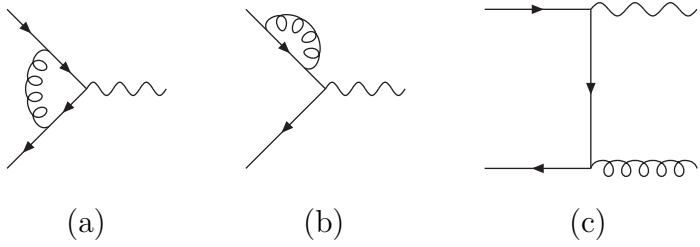

Let us now extend the previous results to \NLO. Here we are interested in the
transverse polarisation contribution to the Drell--Yan cross-section, which can
be written as (reintroducing the scales)
\begin{eqnarray}
  \frac{\d\sigma_T}{\d{y} \, \d{Q}^2 \, \d\phi}
  &=&
  \sum_a e_a^2 \int \! \d\xi_1 \int \! \d\xi_2 \;
  \frac{\d\hat\sigma_T(Q^2/\mu^2, \alpha_s)}{\d{y} \, \d{Q}^2 \, \d\phi}
  \nonumber
\\
  && \hspace{5em} \null \times
  \DT{f}_a(\xi_1, \mu^2) \, \DT{f}_a(\xi_2, \mu^2) \, .
  \label{dyfac14}
\end{eqnarray}
We have seen that at LO, \ie, $\Ord(\alpha_s^0)$, the elementary cross-section
is
\begin{equation}
  \text{LO}: \quad
  \frac{\d\hat\sigma_T}{\d{y} \, \d{Q}^2 \, \d\phi} =
  \frac{\alpha_\text{em}^2}{9Q^2 s} \,
  \cos(2\phi - \phi_{s} - \phi_{\anti{s}}) \,
  \delta(\xi_1 - x_1) \, \delta(\xi_2 - x_2) \, ,
  \label{dyfac15}
\end{equation}
where $\phi_s(\phi_{\anti{s}})$ is the azimuthal angle of the quark (antiquark)
spin, with respect to the lepton plane. Integrating over $y$ we obtain
\begin{equation}
  \text{LO}: \quad
  \frac{\d\hat\sigma_T}{\d{Q}^2 \, \d\phi}
  =
  \frac{\alpha_\text{em}^2}{9Q^2\hat{s}} \,
  \cos(2\phi - \phi_{s} - \phi_{\anti{s}}) \, \delta(1-z) \, ,
  \label{dyfac16}
\end{equation}
with $z\equiv\tau/\xi_1\xi_2=Q^2/\xi_1\xi_2s$.

At \NLO, \ie, $\Ord(\alpha_s)$, the subprocesses contributing to Drell--Yan
production are those shown in Fig.~\ref{dyqqg2}: virtual-gluon corrections and
real-gluon emission. The \NLO cross-section
$\d\hat\sigma_T/\d{y}\,\d{Q}^2\,\d\phi$ exhibits ultraviolet singularities
(arising from loop integrations), infrared singularities (due to soft gluons),
and collinear singularities (when the gluon is emitted parallel to the quark or
the antiquark). Summing virtual and real diagrams, only the collinear
divergences survive. Working in $d=4-2\epsilon$ they are of the type
$1/\epsilon$. These singularities are subtracted and absorbed in the definition
of the parton distributions.

The \NLO elementary cross-sections have been computed by several authors with
different methods \cite{Vogelsang:1993jn, Contogouris:1994ws, Kamal:1996as,
Vogelsang:1998ak}.\footnote{We recall that \NLO corrections to unpolarised
Drell--Yan were presented in \cite{Altarelli:1979ub, Harada:1979bj}, and to
longitudinally polarised Drell--Yan in \cite{Ratcliffe:1983yj}.} Vogelsang and
Weber \cite{Vogelsang:1993jn} were the first to perform this calculation using
massive gluons to regularise the divergences. Soon after the authors of
\cite{Contogouris:1994ws} presented a calculation based on dimensional
reduction. The result was then translated into dimensional regularisation in
\cite{Kamal:1996as}. As a check of the expression given in \cite{Kamal:1996as},
Vogelsang \cite{Vogelsang:1998ak} has shown how to exploit the earlier result
obtained in \cite{Vogelsang:1993jn}. From the detailed structure of the
collinear singularities for both dimensional and off-shell regularisation, it
is straight-forward to translate results from one scheme to another.

The expression for $\d\hat\sigma_T/\d{y}\,{\d}Q^2\,\d\phi$ is rather cumbersome
and we do not repeat it here (instead, we refer the reader to the original
papers). The $y$-integrated cross-section is more legible and reads, in the
$\MSbar$ scheme \cite{Vogelsang:1998ak}
\begin{eqnarray}
  \text{NLO}: \quad \frac{\d\hat\sigma_T}{\d{Q}^2 \, \d\phi}
  &=&
  \frac{\alpha_\text{em}^2}{9Q^2\hat{s}} \,
  \cos(2\phi - \phi_s - \phi_{\anti{s}})
  \nonumber
\\
  && \null \times \CF
  \left\{
    8z \left[ \frac{\ln(1-z)}{1-z} \right]_+ -
    \frac{4z \, \ln z}{1-z} -
    \frac{6z \, \ln^2z}{1-z}
  \right.
  \nonumber
\\
  && \hspace{3em} \null +
  \left.
    4(1-z) + \left( \frac{2\pi^2}{3} - 8 \right) \delta(1-z)
  \right\} . \qquad
  \label{dyfac18}
\end{eqnarray}
The quantity in curly brackets is the \NLO Wilson coefficient
$\DT{C}^{(1)}_\text{DY}$ for Drell--Yan. If we call $\DT\widetilde{C}$ the
quantity to be added to the Wilson coefficient in order to pass from the scheme
of \cite{Kamal:1996as} to $\MSbar$, the result (\ref{dyfac18}) coincides with
that of \cite{Kamal:1996as} for the choice $\DT\widetilde{C}=-\delta(1-z)$. On
the other hand, the expression for $\DT\widetilde{C}$ claimed in
\cite{Kamal:1996as} as providing the translation to the $\MSbar$ scheme (in
dimensional regularisation) is $\DT\widetilde{C}=-\delta(1-z)+2(1-z)$. In
\cite{Vogelsang:1998ak} it is noted that the reason for this difference lies in
the discrepancy between the calculation presented there and that of
\cite{Kamal:1996as} for the $(4-2\epsilon)$-dimensional \LO splitting function,
where extra $\Ord(\epsilon)$ terms were found. The correctness of the result in
\cite{Vogelsang:1998ak} for this quantity is supported by the observation that
the $d$-dimensional $2\to3$ squared matrix element for the process
$q\anti{q}\to\mu^+\mu^-g$ (with transversely polarised incoming (anti)quarks)
given there factorises into the product of the $d$-dimensional $2\to2$ squared
matrix element for $q\anti{q}\to\mu^+\mu^-$ multiplied by the splitting
function $\DT{P}_{qq}^{(0)}$ in $d=4-2\epsilon$ dimensions, when the collinear
limit of the gluon aligning parallel to one of the incoming quarks is correctly
performed. It is thus claimed that the result of \cite{Kamal:1996as} for the
\NLO transversely polarised Drell--Yan cross-section (in dimensional
regularisation) corresponds to a different (non-$\MSbar$) factorisation scheme.

Finally, a first step towards a \NNLO calculation of the transversely polarised
Drell--Yan cross-section was taken in \cite{Chang:1997ns}.

\subsubsection{Twist-three contributions to the Drell--Yan process}
\label{twist3drellyan}

At twist 3, transversity distributions are also probed in Drell--Yan processes
when one of the two hadrons is transversely polarised and the other is
longitudinally polarised. In this case, ignoring subtleties related to quark
masses and transverse motion (so that $h_L(x)=\widetilde{h}_L(x)$ and
$g_T(x)=\widetilde{g}_T(x)$, see Sec.~\ref{twist3}), the cross-section is
\cite{Jaffe:1992ra, Tangerman:1995eh, Boer:1997bw} (the transversely polarised
hadron is $A$)
\begin{eqnarray}
  \frac{\d\sigma}{\d\Omega \d{x}_1 \d{x}_2} &=&
  \frac{\alpha_\text{em}^2}{4Q^2} \,
  \sum_a \, \frac{e_a^2}{3}
  \left\{ \STRUT
    \dots \, +
    | \Vec{S}_{1\perp} | \, \lambda_2 \, \sin2\theta \, \cos(\phi-\phi_{S_1})
  \right.
  \nonumber
\\
  && \hspace{-4em} \null \times
  \left.
    \left[
      \frac{2M_1}{Q} \, x_1 \, g_T^a(x_1) \, \DL\anti{f}_a(x_2) +
      \frac{2M_2}{Q} \, x_2 \, \DT{f}_a(x_1) \, \anti{h}_L^a(x_2)
    \right]
  \right\} , \qquad
  \label{dytwist31}
\end{eqnarray}
where the dots denote the leading-twist contributions presented in
eq.~(\ref{dy33}). The transversity distribution of quarks in hadron $A$ is
coupled to the twist-three antiquark distribution $\anti{h}_L$. The
longitudinal--transverse asymmetry resulting from (\ref{dytwist31}) is (we
assume the masses of the two hadrons to be equal, \ie, $M_1=M_2\equiv{M}$)
\begin{eqnarray}
  A_{LT}^\text{DY} &=&
  | \Vec{S}_{1\perp} | \, \lambda_2 \,
  \frac{2\sin2\theta \, \cos(\phi - \phi_{S_1})}
       {1 + \cos^2\theta} \, \frac{M}{Q}
  \nonumber
\\
  && \null \times
  \frac{\sum_a e_a^2
        \left[
          x_1 \, g_T^a(x_1)\, \DL\anti{f}_a(x_2) +
          x_2 \, \DT{f}_a(x_1) \, \anti{h}_L^a(x_2)
        \right]}
       {\sum_a e_a^2 \, f_a(x_1)
   \anti{f}_a(x_2)} \, .
  \label{dytwist32}
\end{eqnarray}

Let us now consider the case where one of the two hadrons is unpolarised while
the other is transversely polarised. Time-reversal invariance implies the
absence of single-spin asymmetries, even at twist 3. Such asymmetries might
arise as a result of initial-state interactions that generate $T$-odd
distribution functions. If such a mechanism occurs, relaxing the na\"{\i}ve
time-reversal invariance condition (see Sec.~\ref{todd}), the Drell--Yan
cross-section acquires extra terms and (\ref{dytwist31}) becomes
\begin{eqnarray}
  \frac{\d\sigma}{\d\Omega \d{x}_1 \d{x}_2} &=&
  \frac{\alpha_\text{em}^2}{4Q^2} \, \sum_a \, \frac{e_a^2}{3}
  \left\{ \STRUT
    \dots + | \Vec{S}_{1\perp} | \, \sin2\theta \, \sin(\phi-\phi_{S_1})
  \right.
  \nonumber
\\
  && \hspace{-2em}
  \left. \null \times
    \left[
      \frac{2M_1}{Q} \, x_1 \, \widetilde{f}_T^a(x_1) \, \anti{f}_a(x_2) +
      \frac{2M_2}{Q} \, x_2 \, \DT{f}_a(x_1) \, \widetilde{\anti{h}}_a(x_2)
    \right]
  \right\} . \hspace{2em}
  \label{dytwist33}
\end{eqnarray}
Here $\widetilde{f}_T(x)$ and $\widetilde{\anti{h}}(x)$ are the twist-three
$T$-odd distribution functions introduced in Sec.~\ref{todd}. From
(\ref{dytwist33}) we obtain the single-spin asymmetry
\begin{eqnarray}
  A_{T}^\text{DY} &=&
  | \Vec{S}_{1\perp} | \,
  \frac{2\sin2\theta \, \sin(\phi-\phi_{S_1})}{1+\cos^2\theta} \, \frac{M}{Q}
  \nonumber
\\
  && \null \times
  \frac{\sum_a e_a^2
        \left[
          x_1 \, f_T^a(x_1)\, \anti{f}_a(x_2) +
          x_2 \, \DT{f}_a(x_1) \, \anti{h}_a(x_2)
        \right]}
       {\sum_a e_a^2 \, f_a(x_1) \anti{f}_a(x_2)} \, .
  \label{dytwist34}
\end{eqnarray}

The existence of $T$-odd distribution functions has also been advocated by Boer
\cite{Boer:1999mm} to explain, at leading-twist level, an anomalously large
$\cos2\phi$ term in the unpolarised Drell--Yan cross-section
\cite{Falciano:1986wk, Guanziroli:1988rp, Conway:1989fs}, which cannot be
accounted for by \LO or \NLO \QCD \cite{Brandenburg:1993cj} (it can however be
attributed to higher-twist effects \cite{Berger:1979du, Berger:1980xz,
Brandenburg:1994wf}). Boer has shown that, on introducing initial-state $T$-odd
effects, the unpolarised Drell--Yan cross-section indeed acquires a $\cos2\phi$
contribution involving the product
$h_1^\perp(x_1,\Vec{k}_\perp^2)\,\anti{h}_1^\perp(x_2,{\Vec{k}'_\perp}^2)$. If
hadron $A$ is transversely polarised, the same mechanism generates a
$\sin(\phi+\phi_{S_1})$ term, which depends on
$\DT{f}(x_1,\Vec{k}_\perp^2)\,\anti{h}_1^\perp(x_2,{\Vec{k}'_\perp}^2)$.

It must be stressed once again that the mechanism based on initial-state
interactions is highly hypothetical, if not at all unlikely. However, it was
shown in \cite{Hammon:1996pw, Boer:1997bw} that single-spin asymmetries might
arise in Drell--Yan processes owing to the so-called gluonic poles in
twist-three multiparton correlation functions \cite{Qiu:1991pp, Qiu:1992wg,
Qiu:1998ia}. Let us briefly address this issue (for a general discussion of
higher twists in hadron scattering see \cite{Efremov:1983eb, Ratcliffe:1985mp,
Qiu:1991xx, Qiu:1991xy, Ratcliffe:1998pq}).

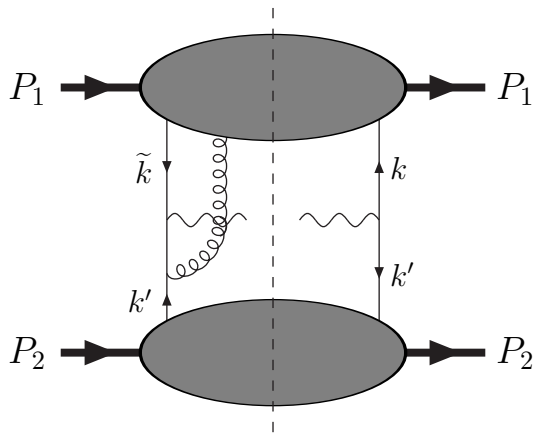
\begin{figure}[htbp]
  \centering
  \begin{picture}(300,170)(0,15)
    \SetWidth{3}
    \ArrowLine(70,50)(100,50)
    \ArrowLine(200,50)(230,50)
    \ArrowLine(70,150)(100,150)
    \ArrowLine(200,150)(230,150)
    \SetWidth{0.5}
    \ArrowLine(110,60)(110,80)
    \Line(110,80)(110,100)
    \ArrowLine(190,100)(190,60)
    \ArrowLine(110,140)(110,100)
    \ArrowLine(190,100)(190,140)
    \Photon(110,100)(140,100){2.5}{2.5}
    \Photon(160,100)(190,100){2.5}{2.5}
    \Gluon(130,135)(130,94){2.5}{6.8}
    \GlueArc(110,100)(20,270,363){2.5}{5.7}
    \GOval(150,50)(20,50)(0){0.5}
    \GOval(150,150)(20,50)(0){0.5}
    \Text(65,50)[r]{\large$P_2$}
    \Text(65,150)[r]{\large$P_1$}
    \Text(235,50)[l]{\large$P_2$}
    \Text(235,150)[l]{\large$P_1$}
    \Text(105,70)[r]{$k'$}
    \Text(195,80)[l]{$k'$}
    \Text(105,120)[r]{$\widetilde{k}$}
    \Text(195,120)[l]{$k$}
    \DashLine(150,20)(150,180){4}
  \end{picture}
  \caption{One of the diagrams contributing to the Drell--Yan cross-section at
           twist~3.}
  \label{handbag6}
\end{figure}

At twist 3 the Drell--Yan process is governed by diagrams such as that in
Fig.~\ref{handbag6}. The hadronic tensor is then (we drop the subscripts 1 and
2 from the quark correlation matrices for simplicity)
\begin{eqnarray}
  W^{\mu\nu} &=&
  \frac1{3} \, \sum_a e_a^2
  \left\{ \SSTRUT
    \int \! \frac{\d^4k}{(2\pi)^4} \int \! \frac{\d^4k'}{(2\pi)^4} \,
    \delta(k + k' - q) \,
    \Tr \left[ \Phi \, \gamma^\mu \, \overline\Phi \, \gamma^\nu \right]
  \right.
  \nonumber
\\
  && \hspace{2.5em} \null -
  \int \! \frac{\d^4k}{(2\pi)^4} \int \! \frac{\d^4k'}{(2\pi)^4}
  \int \! \frac{\d^4\widetilde{k}}{(2\pi)^4} \,
  \delta(k + k' - q)
  \nonumber
\\
  && \hspace{4em} \null \left. \times
  \Tr
  \left[
    \gamma_\sigma \,
    \frac{\slashed{\widetilde{k}} - \slashed{q}}
         {(\widetilde{k} - q)^2 + \I\epsilon} \,
      \gamma^\nu \, \Phi_{A}^\sigma \, \gamma^\mu \, \overline\Phi
  \right]
  + \dots \right\} ,
  \label{dytwist35}
\end{eqnarray}
where we have retained only one of the twist-three contributions, and
$\Phi_{A}^\sigma$ is the quark--quark--gluon correlation matrix defined in
(\ref{higher6e}). Neglecting $1/Q^2$ terms, the quark propagator in
(\ref{dytwist35}) gives ($\widetilde{k}=y_1P_1$)
\begin{equation}
  \frac{\slashed{\widetilde{k}} - \slashed{q}}{(\widetilde{k} - q)^2 + \I\epsilon} \to
  -\frac{\gamma^-}{2x_2P_2^-} \, \frac{x_1 - y_1}{x_1 - y_1 + \I\epsilon} \, .
  \label{dytwist36}
\end{equation}

Let us now introduce another quark--quark--gluon correlator
\begin{eqnarray}
  \Phi_{Fij}^\mu(x,y) &=&
  \int \! \frac{\d\tau}{2\pi}
  \int \! \frac{\d\eta}{2\pi} \,
  \e^{\I \tau y} \, \e^{\I \eta(x - y)}
  \nonumber
\\
  && \hspace{5em} \null \times
  \langle PS |
    \anti\phi_j(0) \, F^{+\mu}(\eta n) \, \phi_i(\tau n)
  | PS \rangle \, ,
  \label{dytwist310}
\end{eqnarray}
which can be parametrised as, see the analogous decomposition of
$\Phi_A^\mu(x,y)$ eq.~(\ref{higher6i}),
\begin{eqnarray}
  \Phi_F^\mu(x,y) &=&
  \frac{M}{2}
  \left\{
    \I \, G_F(x,y) \, \varepsilon_\perp^{\mu\nu} S_{\perp\nu} \slashed{P} +
    \widetilde{G}_F(x,y) \, \ST^\mu \gamma_5 \slashed{P}
  \right.
  \nonumber
\\
  && \hspace{2.5em} \null +
  \left.
    H_F(x,y) \, \lambda_N \gamma_5 \gamma_\perp^\mu \slashed{P} +
    E_F(x,y) \, \gamma_\perp^\mu \slashed{P}
  \right\} .
  \label{dytwist311}
\end{eqnarray}

In the $A^+=0$ gauge one has $F^{+\mu}=\partial^+A_\perp^\mu$ and by partial
integration one finds the following relation between $\Phi_A^\mu(x,y)$ and
$\Phi_F^\mu(x,y)$
\begin{equation}
  (x - y) \, \Phi_{A}^\mu(x,y) = - \I \, \Phi_F^\mu(x,y) \, .
  \label{dytwist312}
\end{equation}
Thus, if some projection of $\Phi_{F}^\mu(x,x)$ is non-zero, the corresponding
projection of $\Phi_{A}^\mu(x,x)$ must have a pole (the ``gluonic pole'').

From (\ref{dytwist36}) and (\ref{dytwist312}), we see that the trace in the
twist-three term of (\ref{dytwist35}) contains the quantity ($\PV$ stands for
principal value)
\begin{eqnarray}
  \frac{x_1 - y_1}{x_1 - y_1 + \I\epsilon} \, \Phi_{A}^\sigma(x_1, y_1) &=&
  \frac{-\I}{x_1 - y_1 + \I\epsilon} \, \Phi_{F}^\sigma(x_1, y_1)
  \nonumber
\\
  &=& \PV \frac{- \I}{x - y} \, \Phi_F^\mu(x,y)
  \nonumber
\\
  && \null -
  \pi \, \delta( x_1 - y_1) \, \Phi_{F}^\sigma(x_1, x_1) \, .
  \label{dytwist313}
\end{eqnarray}
Keeping the real term in (\ref{dytwist313}) one ultimately finds that the
Drell--Yan cross-section with one transversely polarised hadron and one
unpolarised hadron involves, at twist 3, the multiparton distributions
$G_F(x_1, x_1)$ and $E_F(x_1, x_1)$ (the former is proportional to the
distribution $T(x_1, x_1)$ introduced in \cite{Qiu:1992wg, Hammon:1996pw}). The
single-spin asymmetry is then expressed as
\begin{eqnarray}
  A_{T}^\text{DY}
  &\propto&
  | \Vec{S}_{1\perp} | \,
  \frac{2\sin2\theta \, \sin(\phi - \phi_{S_1})}{1+\cos^2\theta} \, \frac{M}{Q}
  \nonumber
\\
  && \null \times
  \frac{\sum_a e_a^2 \,
        [ G_F^a(x_1, x_1)\, \anti{f}_a(x_2) +
          \DT{f}_a(x_1) \, E_F^a(x_2, x_2)
        ]}
       {\sum_a e_a^2 \, f_a(x_1) \anti{f}_a(x_2)} \, .
  \label{dytwist314}
\end{eqnarray}

To establish a connection between (\ref{dytwist314}) and (\ref{dytwist34}), let
us now invert $F^{+\mu}=\partial^+A_\perp^\mu$, hence obtaining
\begin{eqnarray}
  \Phi_A^\mu(x,y) &=&
  \frac12 \, \delta(x-y)
  \left[ \Phi^\mu_{A(\infty)}(x) + \Phi^\mu_{A(-\infty)}(x) \right]
  \nonumber
\\
  && \null +
  \PV \frac{-\I}{x - y} \, \Phi_F^\mu(x,y) \, .
  \label{dytwist321}
\end{eqnarray}
If we impose antisymmetric boundary conditions \cite{Boer:1997bw}, \ie,
\begin{equation}
  \Phi^\mu_{A( \infty)}(x) = -
  \Phi^\mu_{A(-\infty)}(x) \, ,
\end{equation}
then (\ref{dytwist321}) reduces to
\begin{equation}
  \Phi_A^\mu(x,y) = \PV \frac{- \I}{x - y} \, \Phi_F^\mu(x,y) \, .
  \label{dytwist322}
\end{equation}
and (\ref{dytwist313}) becomes (``eff'' stands for ``effective'')
\begin{eqnarray}
  \Phi_{A}^{\sigma \text{eff}} &\equiv&
  \frac{x_1 - y_1}{x_1 - y_1 + \I\epsilon} \, \Phi_{A}^\sigma(x_1, y_1)
  \nonumber
\\
  &=&
  \Phi_{A}^\sigma(x_1, y_1) - \pi \, \delta( x_1 - y_1) \,
  \Phi_{F}^\sigma(x_1, x_1) \, .
  \label{dytwist323}
\end{eqnarray}
Now, the important observation is that $\Phi_A^\sigma$ and $\Phi_F^\sigma$ have
opposite behaviour with respect to time reversal and hence
$\Phi_A^{\sigma\text{eff}}$ has no definite behaviour under this
transformation. Consequently, the $T$-even functions of $\Phi_{F}^\sigma$ can
be identified with $T$-odd functions in the effective correlation matrix
$\Phi_A^{\sigma\text{eff}}$. This mechanism gives rise to two effective $T$-odd
distributions $\widetilde{f}_T^\text{eff}(x)$ and
$\widetilde{h}^\text{eff}(x)$, which are related to the multiparton
distribution functions by \cite{Boer:1997bw} (omitting some factors)
\begin{subequations}
\begin{eqnarray}
  \widetilde{f}_T^\text{eff}(x) &\sim&
  \int \! \d{y} \, \Im G_A^\text{eff}(x,y) \sim G_F(x,x) \, ,
  \label{dytwist324}
\\
  \widetilde{h}^\text{eff}(x) &\sim&
  \int \! \d{y} \, \Im E_A^\text{eff}(x,y) \sim E_F(x,x) \, .
  \label{dytwist325}
\end{eqnarray}
\end{subequations}
In the light of this correspondence one can see that eq.~(\ref{dytwist34}),
based on $T$-odd distributions, and eq.~(\ref{dytwist314}), based on
multiparton distributions, translate into each other. Thus, at least in the
case in which the $T$-odd functions appear at twist 3, there is an explanation
for them in terms of quark--gluon interactions, with no need for initial-state
effects.

In conclusion, let us summarise the various contributions of transversity to
the Drell--Yan cross-section in Table~\ref{table3}.
\begin{table}[htbp]
  \centering
  \caption{Contributions to the Drell--Yan cross-section involving
           transversity. The asterisk denotes $T$-odd terms.}
  \label{table3}
  \vspace{1ex}
  \begin{tabular}{ccc|l}
    \hline
    \multicolumn{4}{c}{Drell--Yan cross-section}
  \\ \hline\hline
            & $A$ & $B$ & \multicolumn{1}{|c}{observable}
  \\ \hline\hline
    twist 2 &  T  &  T  & $\DT{f}(x_1) \DT\anti{f}(x_2)$ \\
    \hline\hline
    twist 3 &  T  &  0  & $\DT{f}(x_1) \, \anti{h}(x_2)$ (*) \\
            &  T  &  L  & $\DT{f}(x_1) \, \anti{h}_L(x_2)$
  \\ \hline
  \end{tabular}
  \vspace{1cm}
\end{table}

\subsection{Single-spin transverse asymmetries}
\label{singlespintransv}

Consider now inclusive hadron production, $A+B\to{h+X}$. If only one of the
initial-state hadrons is transversely polarised and the final-state hadron is
spinless (or its polarisation is unobserved), what is measured is the
single-spin asymmetry
\begin{equation}
  A_{T}^h =
  \frac{\d\sigma(\Vec{S}_T) - \d\sigma(-\Vec{S}_T)}
       {\d\sigma(\Vec{S}_T) + \d\sigma(-\Vec{S}_T)} \, .
  \label{at}
\end{equation}
As we shall see, single-spin asymmetries are expected to vanish in
leading-twist in perturbative \QCD (this observation is originally due to Kane,
Pumplin and Repko \cite{Kane:1978nd}). They can arise, however, as a
consequence of quark transverse-motion effects \cite{Sivers:1989cc,
Collins:1993kk, Collins:1994kq} and/or higher-twist contributions
\cite{Efremov:1981sh, Efremov:1984ip, Qiu:1991pp, Qiu:1992wg}. In the former
case, one probes the following quantities related to transversity:
\begin{itemize}
\item
Distribution functions: $\DT{f}(x)$ (transversely polarised quarks in a
transversely polarised hadron), $f_{1T}^\perp(x,\Vec{k}_T^2)$ (unpolarised
quarks in a transversely polarised hadron), $h_{1}^\perp(x,\Vec{k}_T^2)$
(transversely polarised quarks in an unpolarised hadron).
\item
Fragmentation functions: $H_1^\perp(z,\Vec\kappa_T^2)$ (transversely polarised
quarks fragmenting into an unpolarised hadron), $D_{1T}^\perp(z,\Vec\kappa_T^2)$
(unpolarised quarks fragmenting into a transversely polarised hadron).
\end{itemize}

The twist-three single-spin asymmetries involving the transversity
distributions contain, besides the familiar unpolarised quantities, the
quark--gluon correlation function $E_F(x,y)$ of the incoming unpolarised hadron
and a twist-three fragmentation function of the outgoing hadron (see below,
Sec.~\ref{singlespintwist3}).

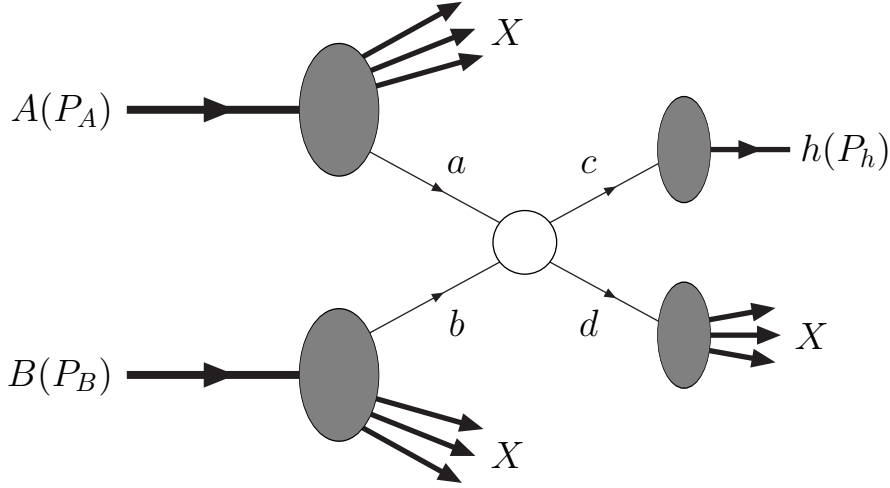
\begin{figure}[htbp]
  \centering
  \begin{picture}(380,190)(-20,55)
    \SetWidth{3}
    \ArrowLine(50,100)(120,100)
    \ArrowLine(50,200)(120,200)
    \SetWidth{0.5}
    \ArrowLine(140,115)(195,145)
    \ArrowLine(140,185)(195,155)
    \ArrowLine(205,145)(260,115)
    \ArrowLine(205,155)(260,185)
    \GCirc(200,150){12}{1}
    \SetWidth{2}
    \LongArrow(130,205)(183,220)
    \LongArrow(130,210)(180,230)
    \LongArrow(130,215)(175,240)
    \LongArrow(130,95)(183,80)
    \LongArrow(130,90)(180,70)
    \LongArrow(130,85)(175,60)
    \ArrowLine(270,185)(300,185)
    \LongArrow(265,115)(295,115)
    \LongArrow(265,120)(293,125)
    \LongArrow(265,110)(293,105)
    \SetWidth{0.5}
    \GOval(260,115)(20,10)(0){0.5}
    \GOval(260,185)(20,10)(0){0.5}
    \GOval(130,100)(25,15)(0){0.5}
    \GOval(130,200)(25,15)(0){0.5}
    \Text(45,200)[r]{\large$A(P_A)$}
    \Text(45,100)[r]{\large$B(P_B)$}
    \Text(188,230)[l]{\large$X$}
    \Text(188,70)[l]{\large$X$}
    \Text(175,180)[]{\large$a$}
    \Text(175,120)[]{\large$b$}
    \Text(225,120)[]{\large$d$}
    \Text(225,180)[]{\large$c$}
    \Text(303,115)[l]{\large$X$}
    \Text(305,185)[l]{\large$h(P_h)$}
  \end{picture}
  \caption{Hadron--hadron scattering with inclusive production of a particle
           $h$.}
  \label{hh}
\end{figure}

Let us now enter into some detail. We consider the following reaction (see
Fig.~\ref{hh}):
\begin{equation}
  A^\uparrow(P_A) \, + \, B(P_B) \, \to \, h(P_h) \, + \, X \, ,
  \label{hh1}
\end{equation}
where $A$ is transversely polarised and the hadron $h$ is produced with a large
transverse momentum $\Vec{P}_{hT}$, so that perturbative \QCD is applicable. In
typical experiments $A$ and $B$ are protons while $h$ is a pion.

The cross-section for (\ref{hh1}) is usually expressed as a function of
$\Vec{P}_{hT}^2$ and of the Feynman variable
\begin{equation}
  x_F \equiv \frac{2P_{hL}}{\sqrt{s}} = \frac{t-u}{s} \, ,
  \label{hh2}
\end{equation}
where $P_{hL}$ is the longitudinal momentum of $h$, and $s,t,u$ are the
Mandelstam variables
\begin{equation}
  s = (P_A + P_B)^2 \, , \qquad
  t = (P_A - P_h)^2 \, , \qquad
  u = (P_B - P_h)^2 \, .
  \label{hh3}
\end{equation}

The elementary processes at lowest order in \QCD are two-body partonic
processes
\begin{equation}
  a(k_a)\, + \, b(k_b)\, \to \, c(k_c) \, + \, d(k_d) \, .
  \label{hh4}
\end{equation}
In the collinear case we set
\begin{equation}
  k_a = x_a \, P_A \, , \qquad
  k_b = x_b \, P_B \, , \qquad
  k_c = \frac1{z} \, P_h \, .
  \label{hh5}
\end{equation}
and the partonic Mandelstam invariants are
\begin{subequations}
\begin{eqnarray}
  \hat{s} &=& (k_a + k_b)^2 \simeq x_a x_b s \, ,
  \label{hh6a}
\\
  \hat{t} &=& (k_a - k_c)^2 \simeq \frac{x_a t}{z} \, ,
  \label{hh6b}
\\
  \hat{u} &=& (k_b - k_c)^2 \simeq \frac{x_b u}{z} \, .
  \label{hh6c}
\end{eqnarray}
\end{subequations}
Thus the condition $\hat{s}+\hat{t}+\hat{u}=0$ implies
\begin{equation}
  z = - \frac{x_a t - x_b u}{x_a x_b s} \, .
  \label{hh7}
\end{equation}

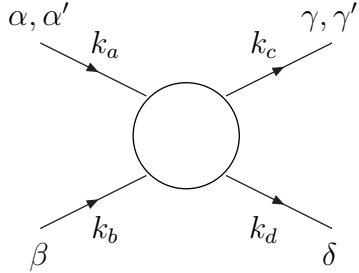
\begin{figure}[htbp]
  \centering
  \begin{picture}(200,95)(100,105)
    \SetWidth{0.5}
    \ArrowLine(145,115)(185,135)
    \ArrowLine(145,185)(185,165)
    \ArrowLine(215,135)(255,115)
    \ArrowLine(215,165)(255,185)
    \GCirc(200,150){20}{1}
    \Text(145,190)[b]{$\alpha,\alpha'$}
    \Text(255,190)[b]{$\gamma,\gamma'$}
    \Text(145,110)[t]{$\beta$}
    \Text(255,110)[t]{$\delta$}
    \Text(170,120)[t]{$k_b$}
    \Text(170,180)[b]{$k_a$}
    \Text(230,180)[b]{$k_c$}
    \Text(230,120)[t]{$k_d$}
  \end{picture}
  \caption{Elementary processes contributing to hadron--hadron scattering.}
  \label{hhelem2}
\end{figure}

According to the \QCD factorisation theorem the differential cross-section for
the reaction (\ref{hh1}) can formally be written as
\begin{equation}
  \d\sigma =
  \sum_{abc} \sum_{\alpha\alpha'\gamma\gamma'} \rho^a_{\alpha'\alpha} \,
  f_a(x_a) \otimes f_b(x_b) \otimes
  \d\hat\sigma_{\alpha\alpha'\gamma\gamma'} \otimes
  \mathcal{D}_{h/c}^{\gamma' \gamma}(z) \, .
  \label{hh8}
\end{equation}
Here $f_a$ ($f_b$) is the distribution of parton $a$ ($b$) inside the hadron
$A$ ($B$), $\rho^a_{\alpha\alpha'}$ is the spin density matrix of parton $a$,
$\mathcal{D}_{h/c}^{\gamma\gamma'}$ is the fragmentation matrix of parton $c$
into hadron $h$, and $\d\hat\sigma/\d\hat{t}$ is the elementary cross-section:
\begin{equation}
  \left(\frac{\d\hat\sigma}{\d\hat{t}} \right)_{\alpha\alpha'\gamma\gamma'} =
  \frac1{16\pi\hat{s}^2} \, \frac12 \, \sum_{\beta\delta}
  \mathcal{M}_{\alpha\beta\gamma\delta}
  \mathcal{M}^*_{\alpha'\beta\gamma'\delta} \, ,
  \label{hh9}
\end{equation}
where $\mathcal{M}_{\alpha\beta\gamma\delta}$ is the scattering amplitude for
the elementary partonic process (see Fig.~\ref{hhelem2}).

If the produced hadron is unpolarised, or spinless, as will always be the case
hereafter, only the diagonal elements of $\mathcal{D}_{h/c}^{\gamma\gamma'}$
are non-vanishing, \ie,
$\mathcal{D}_{h/c}^{\gamma\gamma'}=\delta_{\gamma\gamma'}\,D_{h/c}$, where
$D_{h/c}$ is the unpolarised fragmentation function. Together with helicity
conservation in the partonic subprocess, this implies $\alpha=\alpha'$.
Therefore, in (\ref{hh9}) there is no dependence on the spin of hadron $A$ and
all single-spin asymmetries are zero.

To escape such a conclusion we must consider either the intrinsic transverse
motion of quarks, or higher-twist effects.

\subsubsection{Transverse motion of quarks and single-spin asymmetries}
\label{transvmotsinglespin}

Let us first of all see how the transverse motion of quarks can generate
single-spin asymmetries. This can happen in three different ways:
\begin{enumerate}
\item
Intrinsic $\Vec\kappa_T$  in hadron $h$ implies that
$\mathcal{D}_{h/c}^{\gamma\gamma'}$ is not necessarily diagonal (owing to
$T$-odd effects at level of fragmentation functions).
\item
Intrinsic $\Vec{k}_T$ in hadron $A$ implies that $f_a(x_a)$ in (\ref{hh8})
should be replaced by the probability density $\mathcal{P}_a(x_a,\Vec{k}_T)$,
which may depend on the spin of hadron $A$ (again, owing to $T$-odd effects but
at the level of distribution functions).
\item
Intrinsic $\Vec{k}'_T$ in hadron $B$ implies that $f_b(x_b)$ in (\ref{hh8})
should be replaced by $\mathcal{P}_b(x_b,\Vec{k}'_T)$. The transverse spin of
parton $b$ inside the unpolarised hadron $B$ may then couple to the transverse
spin of parton $a$ inside $A$ (this too is a $T$-odd effect at the level of
distribution functions).
\end{enumerate}
Effect 1 is the Collins effect \cite{Collins:1993kk}, effect 2 is the Sivers
effect \cite{Sivers:1989cc}, and effect 3 is the effect studied by Boer
\cite{Boer:1999mm} in the context of Drell--Yan processes
(Sec.~\ref{twist3drellyan}). We stress that all these intrinsic $\Vec\kappa_T$,
$\Vec{k}_T$, or $\Vec{k}'_T$ effects are $T$-odd.

When the intrinsic transverse motion of quarks is taken into account, the \QCD
factorisation theorem is not proven. We assume, however, its validity and write
a factorisation formula similar to (\ref{hh8}), that is explicitly
\begin{eqnarray}
  E_h \, \frac{\d\sigma}{\d^3\Vec{P}_h} &=&
  \sum_{abc} \sum_{\alpha\alpha'\beta\beta'\gamma\gamma'}
  \int \! \d{x}_a \int \! \d{x}_b
  \int \! \d^2\Vec{k}_T \int \! \d^2\Vec{k}'_T \int \! \d^2\Vec\kappa_T \,
  \frac1{\pi z}
  \nonumber
\\
  && \hspace{-4em} \null \times
  \mathcal{P}_a(x_a, \Vec{k}_T) \, \rho^a_{\alpha'\alpha} \,
  \mathcal{P}_b(x_b, \Vec{k}'_T) \, \rho^b_{\beta'\beta}
  \left(
    \frac{\d\hat\sigma}{\d\hat{t}}
  \right)_{\alpha\alpha'\beta\beta'\gamma\gamma'}
  \mathcal{D}_{h/c}^{\gamma'\gamma}(z, \Vec\kappa_T) \, , \qquad
  \label{hh10}
\end{eqnarray}
where
\begin{equation}
  \left(
    \frac{\d\hat\sigma}{\d\hat{t}}
  \right)_{\alpha\alpha'\beta\beta'\gamma\gamma'}
  =
  \frac1{16\pi\hat{s}^2} \, \sum_{\beta\delta}
  \mathcal{M}_{\alpha\beta\gamma\delta}
  \mathcal{M}^*_{\alpha'\beta\gamma'\delta} \, .
  \label{hh11}
\end{equation}

To start with, let us consider the Collins mechanism for single-spin
asymmetries \cite{Collins:1993kk, Anselmino:1999pw}. We take into account the
intrinsic transverse motion of quarks inside the produced hadron $h$ (which is
responsible for the effect), and we neglect the transverse momenta of all other
quarks. Thus eq.~(\ref{hh10}) becomes
\begin{eqnarray}
  E_h \, \frac{\d\sigma}{\d^3\Vec{P}_h} &=&
  \sum_{abc} \sum_{\alpha\alpha'\gamma\gamma'}
  \int \! \d{x}_a \int \! \d{x}_b \int \! \d^2\Vec\kappa_T \, \frac1{\pi z}
  \nonumber
\\
  && \null \times
  f_a(x_a) \, \rho^a_{\alpha'\alpha} \, f_b(x_b)
  \left(
    \frac{\d\hat\sigma}{\d\hat{t}}
  \right)_{\alpha\alpha'\gamma\gamma'}
  \mathcal{D}_{h/c}^{\gamma'\gamma}(z,\Vec\kappa_T) \, ,
  \label{hh12}
\end{eqnarray}
and the elementary cross-sections are given by (\ref{hh9}), with $\Vec\kappa_T$
retained. We are interested in transverse spin asymmetries
$\d\sigma(\Vec{S}_T)-\d\sigma(-\Vec{S}_T)$. Therefore, since we are neglecting
the intrinsic $\Vec{k}_T$ motion inside $A$, the spin density matrix elements
of our concern are $\rho^a_{+-}$ and $\rho^a_{-+}$, and the contributing
elementary cross-sections are $\d\hat\sigma_{+-+-}=\d\hat\sigma_{-+-+}$ and
$\d\hat\sigma_{+--+}=\d\hat\sigma_{-++-}$.

Using eqs.~(\ref{good17}) and (\ref{noncoll3b}--\ref{subnoncoll3e}) we find,
with our choices of axes
\begin{eqnarray}
  E_h \, \frac{\d\sigma( \Vec{S}_T)}{\d^3\Vec{P}_h} &-&
  E_h \, \frac{\d\sigma(-\Vec{S}_T)}{\d^3\Vec{P}_h} =
  \nonumber
\\
  && \null -
  2 | \Vec{S}_T | \, \sum_{abc} \int \! \d{x}_a \int \! \d{x}_b
  \int \! \d^2\Vec\kappa_T \, \frac1{\pi z} \, \DT{f}_a(x_a) \, f_b(x_b)
  \nonumber
\\
  &&  \hspace{1em} \null \times
  \left[
    \left( \frac{\d\hat\sigma}{\d\hat{t}} \right)_{+-+-} \hspace{-1em}
    \sin(\phi_\kappa + \phi_S) +
    \left( \frac{\d\hat\sigma}{\d\hat{t}} \right)_{+--+} \hspace{-1em}
    \sin(\phi_\kappa - \phi_S)
  \right]
  \nonumber
\\
  &&  \hspace{1em} \null \times
  \DT^0 D_{h/c}(z, \Vec\kappa_T^2) \, ,
  \label{hh14}
\end{eqnarray}
where $\phi_\kappa$ and $\phi_S$ are the azimuthal angles of $\Vec\kappa_T$ and
$\Vec{S}_T$, respectively, and $\DT^0D_{h/c}$ is the $T$-odd fragmentation
function related to $H_{1\perp}$, see (\ref{sicross14}).

In particular, if the spin of hadron $A$ is directed along $y$,
eq.~(\ref{hh14}) takes the form
\begin{eqnarray}
  E_h \, \frac{\d\sigma( \Vec{S}_T)}{\d^3\Vec{P}_h} &-&
  E_h \, \frac{\d\sigma(-\Vec{S}_T)}{\d^3\Vec{P}_h} =
  \nonumber
\\
  && \null -
  2 | \Vec{S}_T | \, \sum_{abc}
  \int \! \d{x}_a \int \! \frac{\d{x}_b}{\pi z} \int \! \d^2\Vec\kappa_T
  \DT{f}_a(x_a) \, f_b(x_b)
  \nonumber
\\
  && \null \times
  \Delta_{TT} \hat\sigma(x_a, x_b, \Vec\kappa_T) \,
  \DT^0 D_{h/c}(z, \Vec\kappa_T^2) \, ,
  \label{hh15}
\end{eqnarray}
where the elementary double-spin asymmetry $\Delta_{TT}\hat\sigma$ is given by
\begin{eqnarray}
  \Delta_{TT} \hat\sigma &=&
  \left(\frac{\d\hat\sigma}{\d\hat{t}} \right)_{+-+-} -
  \left(\frac{\d\hat\sigma}{\d\hat{t}} \right)_{+--+}
  \nonumber
\\
   &=&
  \frac{\d\hat\sigma( a^\uparrow b \to c^\uparrow d)}{\d\hat{t}} -
  \frac{\d\hat\sigma( a^\uparrow b \to c^{\downarrow}d)}{\d\hat{t}} \, .
  \label{hh16}
\end{eqnarray}

Equation (\ref{hh15}) gives the single-spin asymmetry under the hypothesis that
only the Collins mechanism (based on the existence of the $T$-odd fragmentation
functions $\DT^0D_{h/c}$ or $H_{1\perp}$) is at work. Another source of
single-spin asymmetries in hadron--hadron scattering is the Sivers effect
\cite{Sivers:1989cc, Anselmino:1995tv, Anselmino:1998yz, Anselmino:1999pw},
which relies on $T$-odd distribution functions. This mechanism predicts a
single-spin asymmetry of the form
\begin{eqnarray}
  E_h \, \frac{\d\sigma( \Vec{S}_T)}{\d^3\Vec{P}_h} &-&
  E_h \, \frac{\d\sigma(-\Vec{S}_T)}{\d^3\Vec{P}_h}
  \nonumber
\\
  &=&
  | \Vec{S}_T | \, \sum_{abc} \int \! \d{x}_a \int \! \frac{\d{x}_b}{\pi z}
  \int \! \d^2\Vec{k}_T \Delta_0^T f_a(x_a, \Vec{k}_T^2) \, f_b(x_b)
  \nonumber
\\
  && \null \times
  \frac{\d\hat\sigma}{\d\hat{t}}(x_a, x_b, \Vec{k}_T) \,
  D_{h/c}(z) \, ,
  \label{hh17}
\end{eqnarray}
where $\Delta_0^T f$, (related to $f_{1T}^\perp$) is the $T$-odd distribution
defined in (\ref{todd6b}).

Finally, the effect studied by Boer in \cite{Boer:1999mm} gives rise to an
asymmetry involving the other $T$-odd distribution, $\DT^0f$ (or $h_1^\perp$),
defined in (\ref{todd6c}). This asymmetry reads
\begin{eqnarray}
  E_h \, \frac{\d\sigma( \Vec{S}_T)}{\d^3\Vec{P}_h} &-&
  E_h \, \frac{\d\sigma(-\Vec{S}_T)}{\d^3\Vec{P}_h} =
  \nonumber
\\
  && \null -
  2 | \Vec{S}_T | \, \sum_{abc}
  \int \! \d{x}_a \int \! \frac{\d{x}_b}{\pi z} \int \! \d^2\Vec{k}'_T
  \DT{f}_a(x_a) \, \DT^0 f_b(x_b, \Vec{k'}_T^2) \,
  \nonumber
\\
  && \null \times
  \Delta_{TT} \hat\sigma'(x_a, x_b, \Vec{k}'_T) \, D_{h/c}(z) \, ,
  \label{hh18}
\end{eqnarray}
where the elementary asymmetry is
\begin{equation}
  \Delta_{TT} \hat\sigma' =
  \frac{\d\hat\sigma(a^{\uparrow}b^{\uparrow}\to c d)}{\d\hat{t}} -
  \frac{\d\hat\sigma(a^{\uparrow}b^{\downarrow}\to c d)}{\d\hat{t}} \, .
  \label{hh19}
\end{equation}

The \emph{caveat} of Sec.~\ref{todd} with regard to initial-state interaction
effects, which are assumed to generate the $T$-odd distributions, clearly
applies here and renders both the Sivers and the Boer mechanisms highly
conjectural. In the next section we shall see how single-spin asymmetries
emerge at higher twist.

\subsubsection{Single-spin asymmetries at twist three}
\label{singlespintwist3}

In the eighties Efremov and Teryaev \cite{Efremov:1981sh, Efremov:1984ip}
pointed out that non-vanishing single-spin asymmetries can be obtained in
perturbative \QCD if one resorts to higher twist. These asymmetries were later
evaluated in the context of \QCD factorisation by Qiu and Sterman, who studied
direct photon production \cite{Qiu:1991pp, Qiu:1992wg} and, more recently,
hadron production \cite{Qiu:1998ia}. This program has been extended to cover
the chirally-odd contributions by Kanazawa and Koike \cite{Kanazawa:2000hz,
Kanazawa:2000kp}. Here we limit ourselves to quoting the main general results
of these works (for a phenomenological analysis see Sec.~\ref{transvhh}).

At twist 3 the cross-section for the reaction (\ref{hh1}) can be formally
written as
\begin{eqnarray}
  \d\sigma &=&
  \sum_{abc}
  \left\{
    G_F^a(x_a, y_a) \otimes f_b(x_b) \otimes \d\hat\sigma \otimes D_{h/c}(z)
  \right.
  \nonumber
\\
  && \hspace{2.5em} \null +
  \DT{f}_a(x_a) \otimes E_F^b(x_b, y_b) \otimes \d\hat\sigma' \otimes D_{h/c}(z)
  \nonumber
\\[1ex]
  && \hspace{2.5em} \null +
  \left.
    \DT{f}_a(x_a) \otimes f_b(x_b) \otimes \d\hat\sigma'' \otimes D_{h/c}^{(3)}(z)
  \right\} ,
  \label{singtwist31}
\end{eqnarray}
where $G_F(x_a,x_b)$ and $E_F(x_a,x_b)$ are the quark--gluon correlation
functions introduced in Sec.~\ref{twist3drellyan}, $D_{h/c}^{(3)}$ is a
twist-three fragmentation function (that we do not specify further), and
$\d\hat\sigma$, $\d\hat\sigma'$ and $\d\hat\sigma''$ are cross-sections of hard
partonic subprocesses.

The first line in (\ref{singtwist31}), which does not contain the transversity
distributions, corresponds to the chirally-even mechanism studied in
\cite{Qiu:1998ia}. The second term in (\ref{singtwist31}) is the chirally-odd
contribution analysed in \cite{Kanazawa:2000hz}. The elementary cross-sections
can be found in the original papers. In Sec.~\ref{transvhh} we shall see how
the predictions based on eq.~\ref{singtwist31} compare with the available data
on single-spin asymmetries in hadron production. In practice, it turns out that
the transversity-dependent term is negligible \cite{Kanazawa:2000kp}.

\section{Model calculations of transverse polarisation distributions}
\label{models}

As we have no experimental information on the transversity distributions yet,
model calculations are presently the only way to acquire knowledge of these
quantities. This section is devoted to such calculations. We shall see how the
transverse polarisation distributions have been computed using various models
of the nucleon and other non-perturbative tools. In particular, three classes
of models will be discussed in detail:
\begin{enumerate}
\item
relativistic \emph{bag-like models}, such as the MIT bag model and the colour
dielectric model, which are dominated by the valence component;
\item
\emph{chiral soliton models}, in which the sea plays a more important r\^{o}le and
contributes significantly to various observables;
\item
\emph{light-cone models}, based on the Melosh rotation.
\end{enumerate}

Results obtained in other models, not included in the above list (for instance,
diquark spectator models), via QCD sum rules and from lattice calculations will
also be reported. Quite obviously, the presentation of all models will be
rather sketchy, our interest being essentially in their predictions for the
quark and antiquark transversity distributions.\footnote{Throughout this
section the transversity distributions will be denoted by $\DT{q}$ and the
tensor charges will be called $\delta{q}$.}

What models provide is the nucleon state (\ie, the wave functions and energy
spectrum), which appears in the field-theoretical expressions
(\ref{df16}--\ref{subdf18}) of quark distributions. In general, however, it is
impossible to solve the equations of motion \emph{exactly} for any realistic
model. Hence, one must resort to various approximations, which clearly affect
the results of the calculation. In order to test the validity of the
approximations (and of the models) one may check that the computed
distributions fulfill the valence-number sum rules
\begin{equation}
  \int_0^1 \d{x} \left[ \strut u(x) - \anti{u}(x) \right] = 2 \, , \qquad
  \int_0^1 \d{x} \left[        d(x) - \anti{d}(x) \right] = 1 \, ,
  \label{number}
\end{equation}
and that they satisfy other theoretical constraints, such as the Soffer
inequality (see Sec.~\ref{soffer})
\begin{equation}
  q(x) + \DL{q}(x) \le 2 | \DT{q}(x) |.
  \label{mod:soffer}
\end{equation}

As seen in Sec.~\ref{transvqcd}, the renormalisation of the operators in the
matrix elements of eqs.~(\ref{df16}--\ref{subdf18}) introduces a scale
dependence into the parton distributions. In contrast, when computed in any
model, the matrix elements of (\ref{df16}--\ref{subdf18}) are just numbers,
with no scale dependence. The problem thus arises as to how to reconcile QCD
perturbation theory with quark models. Since the early days of QCD various
authors \cite{Parisi:1976fz, Gluck:1977ah, Jaffe:1980ti} have proposed the
answer to this question: the twist-two matrix elements computed in quark models
should be interpreted as representing the nucleon at some \emph{fixed}, low
scale $Q_0^2$ (we shall call it the \emph{model scale}). In other terms, quark
models provide the \emph{initial condition} for QCD evolution. The experience
accumulated with the radiative generation models \cite{Gluck:1989xx,
Barone:1993ej, Barone:1993sb} has taught us that, in order to obtain a picture
of the nucleon at large $Q^2$ in agreement with experiment (at least in the
unpolarised case), the nucleon must contain a relatively large fraction of sea
and glue, even at low momentum scales. Purely valence models are usually unable
to fit the data at all well.

Although the model scale has the same order of magnitude in all models
($Q_0\sim0.3-0.8\GeV$), its precise value depends on the details of the model
and on the procedure adopted to determine it. The smallness of $Q_0^2$ clearly
raises another problem, namely to what extent one can apply perturbative
evolution to extrapolate the quark distributions from such low scales to large
$Q^2$. This problem is still unresolved (for an attempt to model a
non-perturbative evolution mechanism in an effective theory see
\cite{Ball:1994ke, Ball:1994ks}) although the success of fits based on
radiatively generated parton distributions \cite{Gluck:1989xx, Gluck:1990bm,
Gluck:1992ng, Gluck:1998xa} inspires some confidence that the realm of
perturbative QCD may extend to fairly small scales.

\subsection{Bag-like models}
\label{sec:bag}

In bag-like models (the MIT bag model \cite{Chodos:1974je, Chodos:1974pn,
DeGrand:1975cf} and the \CDM) confinement is implemented by situating the
quarks in a region characterised by a value of the colour dielectric constant
of order unity. The value of the dielectric is zero outside the nucleon, that
is, in the \emph{true vacuum}, from which the coloured degrees of freedom are
expelled. A certain amount of energy is associated with the excitation of
non-perturbative gluonic degrees of freedom. This energy is described by the
so-called \emph{vacuum pressure} in the MIT bag model and by excitations of a
phenomenological scalar field in the \CDM.

The two models (MIT bag and \CDM) differ in the following points. In the MIT
bag model the interior of the bag is supposedly described by perturbative QCD
and quarks have current masses; confinement is imposed by special boundary
conditions at the surface of the bag and the bag itself has no associated
dynamics. In contrast, in the \CDM chiral symmetry is broken both inside and
outside the nucleon, in a manner somewhat similar to the $\sigma$-model
\cite{Kahana:1984dx, Kahana:1984be}. Quark confinement is due to interaction
with the phenomenological scalar field which describes the non-perturbative
gluonic degrees of freedom. There is no rigid separation between ``inside'' and
``outside'', and confinement is implemented dynamically.

\subsubsection{Centre-of-mass motion}

A problem arising in many model calculations is the removal of spurious
contributions to physical observables due to the centre-of-mass motion (for a
comprehensive discussion of this matter, see the book by Ring and Schuck
\cite{Ring:1980b1}). The origin of the problem lies in the fact that the
solution of classical equations of motion (\ie, the mean-field approximation)
breaks translational invariance.

A way to restore this invariance is to first define the quantum state of the
nucleon at rest, and then minimise the normally-ordered Hamiltonian in this
state. In the evaluation of specific operators, boosted states of the nucleon
are also required. The difficulty of boosting the states renders the procedure
hard to fully implement.

\sloppypar A (non-relativistic) method frequently used to construct momentum
eigenstates is the so-called Peierls--Yoccoz projection \cite{Ring:1980b1}.
Writing $|\Vec{R},3q\rangle$ to denote a three-quark state centred at
$\Vec{R}$:
\begin{equation}
  | \Vec{R}, 3q \rangle
  =
  b_1^\dagger (\Vec{R}) \,
  b_2^\dagger (\Vec{R}) \,
  b_3^\dagger (\Vec{R}) \,
  | \Vec{R} , 0q \rangle \, ,
  \label{bag1}
\end{equation}
where $|\Vec{R},0q\rangle$ is the quantum state of the empty bag, a generic
nucleon eigenstate of momentum $\Vec{P}$ may be written as
\begin{equation}
  | \Vec{P} \rangle
  =
  \frac1{N_3 (P^0,\Vec{P})} \int \! \d^3\Vec{R} \,
  \e^{\I \Vec{P}{\cdot}\Vec{R}} | \Vec{R}, 3q \rangle \, ,
  \label{p-y}
\end{equation}
with normalisation
\begin{equation}
  {| N_3 (P^0,\Vec{P})|^2}
  =
  \frac1{2P^0} \int \! \d^3\Vec{R} \, \e^{\I \Vec{P}{\cdot}\Vec{R}} \,
  \langle \Vec{R}, 3q | \Vec0, 3q \rangle \, .
  \label{bag2}
\end{equation}

The expectation value of the normally-ordered Hamiltonian in the projected
zero-momentum eigenstate is
\begin{equation}
  E =
  \frac{\langle \Vec{P}=\Vec0 | \, {:}\hat{H}{:} \, | \Vec{P}=\Vec0 \rangle}
       {\langle \Vec{P}=\Vec0 | \Vec{P}=\Vec0 \rangle} \, .
  \label{bag3}
\end{equation}
Minimisation of this quantity amounts to solving a set of integro-differential
equations, for which a variational approach is generally adopted. In the
literature this procedure is known as \VAP, to be distinguished from the
simpler \VBP, which consists of minimising the unprojected Hamiltonian first
and then using the solutions in the Peierls--Yoccoz projection (\ref{p-y}). For
a detailed discussion of these techniques see \cite{Lubeck:1986if,
Lubeck:1987pj}.

\subsubsection{The quark distributions in bag models}
\label{sec:bagform}

\begin{figure}[htbp]
  \centering
  \begin{picture}(360,85)(0,20)
    \SetOffset(0,0)
    \SetWidth{2}
    \ArrowLine(25,60)(45,60)
    \ArrowLine(95,60)(115,60)
    \SetWidth{0.5}
    \ArrowLine(50,66)(90,66)
    \ArrowLine(50,54)(90,54)
    \ArrowLine(50,75)(50,95)
    \ArrowLine(90,95)(90,75)
    \GOval(50,60)(15,5)(0){0.5}
    \GOval(90,60)(15,5)(0){0.5}
    \Text(70,25)[]{(a)}
    \Text(45,95)[r]{$q$}
    \Text(95,95)[l]{$q$}
    \Text(70,80)[]{$2 q$}
    \SetOffset(110,0)
    \SetWidth{2}
    \ArrowLine(25,60)(45,60)
    \ArrowLine(95,60)(115,60)
    \SetWidth{0.5}
    \ArrowLine(50,51)(90,51)
    \ArrowLine(50,57)(90,57)
    \ArrowLine(50,63)(90,63)
    \ArrowLine(90,69)(50,69)
    \ArrowLine(50,75)(50,95)
    \ArrowLine(90,95)(90,75)
    \GOval(50,60)(15,5)(0){0.5}
    \GOval(90,60)(15,5)(0){0.5}
    \Text(70,25)[]{(b)}
    \Text(45,95)[r]{$q$}
    \Text(95,95)[l]{$q$}
    \Text(70,80)[]{$3 q \, 1 \anti{q}$}
    \SetOffset(220,0)
    \SetWidth{2}
    \ArrowLine(25,60)(45,60)
    \ArrowLine(95,60)(115,60)
    \SetWidth{0.5}
    \ArrowLine(50,51)(90,51)
    \ArrowLine(50,57)(90,57)
    \ArrowLine(50,63)(90,63)
    \ArrowLine(50,69)(90,69)
    \ArrowLine(50,95)(50,75)
    \ArrowLine(90,75)(90,95)
    \GOval(50,60)(15,5)(0){0.5}
    \GOval(90,60)(15,5)(0){0.5}
    \Text(70,25)[]{(c)}
    \Text(45,95)[r]{$\anti{q}$}
    \Text(95,95)[l]{$\anti{q}$}
    \Text(70,80)[]{$4 q$}
  \end{picture}
  \caption{Intermediate states in the parton model: (a) $2q$, (b) $3q1\bar{q}$
           and (c) $4q$.}
  \label{interm2}
\end{figure}
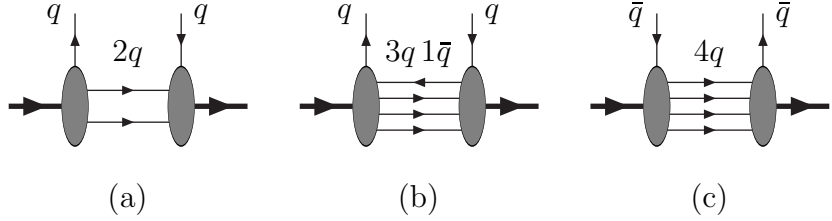

In the (projected) mean-field approximation, the matrix elements defining the
distribution functions can be rewritten in terms of single-particle (quark or
antiquark) wave functions, after inserting a complete set of states between the
two fermionic fields $\psi$ and $\anti\psi$ \cite{Signal:1988vf,
Schreiber:1988uw}. The intermediate states that contribute are $2q$ and
$3q1\anti{q}$ states for the quark distributions, and $4q$ states for the
antiquark distributions (see Fig.~\ref{interm2}).

The leading-twist quark distribution functions read ($f$ is the flavour)
\begin{subequations}
\begin{eqnarray}
  q_f(x) &=&
  \sum_\alpha \sum_m P(f,\alpha,m) \,
  F_\alpha(x) \, ,
  \label{eq5}
\\
  \DL{q}_f(x) &=&
  \sum_\alpha \sum_m P(f,\alpha,m) \, (-1)^{(m+\frac32+i_\alpha)} \,
  G_\alpha(x) \, ,
  \label{eq6}
\\
  \DT{q}_f(x) &=&
  \sum_\alpha \sum_m P(f,\alpha,m) \, (-1)^{(m+\frac32+i_\alpha)} \,
  H_\alpha(x) \, ,
  \label{eq7}
  \sublabel{subeq7}
\end{eqnarray}
\end{subequations}
where
\begin{eqnarray}
  \left.
    \begin{array}{r}
      F_\alpha(x) \\ G_\alpha(x) \\ H_\alpha(x)
    \end{array}
  \right\}
  &=&
  \int \! \frac{\d^3\Vec{p}_\alpha}{(2\pi)^3(2p_\alpha^0)} \,
  A_\alpha(p_\alpha) \, \delta \! \left( (1-x)P^+ - p^+_\alpha \right)
  \nonumber
\\
  &\times&
  \frac12
  \left\{
    \begin{array}{l}
      u^2(p_\alpha) +
      2u(p_\alpha) \, v(p_\alpha) \, \frac{p_\alpha^z}{| \Vec{p}_\alpha |} +
      v^2(p_\alpha) \, ,
    \\
      u^2(p_\alpha) +
      2u(p_\alpha) \, v(p_\alpha) \, \frac{p_\alpha^z}{| \Vec{p}_\alpha |} +
      v^2(p_\alpha)
      \left[
        2 \left( \frac{p_\alpha^z}{| \Vec{p}_\alpha |} \right)^2 - 1
      \right] , \qquad \STRUT
    \\
      u^2(p_\alpha) +
      2u(p_\alpha) \, v(p_\alpha) \, \frac{p_\alpha^z}{| \Vec{p}_\alpha |} +
      v^2(p_\alpha)
      \left[
        1- \left( \frac{p_{\alpha \perp}}{| \Vec{p}_\alpha |} \right)^2
      \right] .
    \end{array}
  \right.
  \label{correnti}
\end{eqnarray}

In (\ref{correnti}) $u$ and $v$ are respectively the upper and lower components
of the single-quark wave functions, $m$ is the projection of the quark spin
along the direction of the nucleon spin, and $P(f,\alpha,m)$ is the probability
of extracting a quark (or inserting an antiquark) of flavour $f$ and spin $m$,
leaving a state generically labelled by the quantum number $\alpha$. The index
$i_\alpha$ takes the value 0 when a quark is extracted and 1 when an antiquark
is inserted. The overlap function $A_\alpha(p_\alpha)$ contains the details of
the intermediate states. The antiquark distributions are obtained in a similar
manner (the index $i_\alpha$ is 1 for the $4q$ states).

When a quark (or an antiquark) is inserted, it can give rise to an infinite
number of states. Among all four-particle intermediate states, only that
corresponding to a quark or an antiquark inserted into the ground state is
usually considered because excited intermediate states have larger masses and,
as will be clear in the following, give a negligible contribution to the
distribution functions.

Concerning the antiquark distributions, we recall from Sec.~\ref{leading} that
the following formal expressions hold
\begin{subequations}
\begin{eqnarray}
  \anti{q}(x)    &=& -      q(-x) \, ,
  \label{neg1}
\\
  \DL\anti{q}(x) &=&   \DL{q}(-x) \, ,
  \label{neg2}
\\
  \DT\anti{q}(x) &=& - \DT{q}(-x) \, .
  \label{neg3}
  \sublabel{subneg3}
\end{eqnarray}
\end{subequations}
Although some authors use these relations to calculate the antiquark
distributions by extending the quark distributions to negative $x$, it should
be recalled that this is an incorrect procedure. The reason, explained in
Sec.~\ref{probab}, is that for $x<0$ there are \emph{semi-connected} diagrams
that contribute to the distributions whereas in computing the quark
distribution functions in the physical region one considers only
\emph{connected} diagrams (as stressed by Jaffe \cite{Jaffe:1983hp}, this
indeed \emph{defines} the parton model).

It is important to note that, in the non-relativistic limit, where the lower
components of the quark wave functions are neglected, the three currents in
eq.~(\ref{correnti}) coincide. This implies, in the light of
(\ref{eq6},~\ref{subeq7}), that ignoring relativistic effects the helicity
distributions are equal to the transversity distributions. This is obviously
only valid at the model scale (\ie, at very low $Q^2$) since, as shown in
Sec.~\ref{comparison}, QCD evolution discriminates between the two
distributions, in particular at small $x$.

In eqs.~(\ref{correnti}) energy--momentum conservation is enforced by the delta
function. This is also responsible for the correct support of the
distributions, which vanish for $x\ge1$. In fact, rewriting the integral in
(\ref{correnti}) as
\begin{equation}
  \int \d^3\Vec{p}_\alpha \, \delta \! \left( (1 - x)P^+ - p_\alpha^+ \right)
  =
  2\pi \int_{p_{\alpha_<}}^\infty \d{p}_\alpha \, p_\alpha \, ,
  \label{bag10}
\end{equation}
where ($m_\alpha$ is the mass of the intermediate state)
\begin{equation}
  p_{\alpha_<} =
  \left| \frac{M^2(1 - x)^2 - m_\alpha^2}{2M (1 - x)} \right| ,
  \label{bag11}
\end{equation}
one sees that for $x\to1$ the lower limit of integration $p_{\alpha<}$ tends to
infinity and (\ref{bag10}) gives zero.

The distributions are centred at the point $\anti{x}=1-(m_\alpha/M)$, which is
positive for the $2q$ term and negative for the $4q$ and $3q1\anti{q}$ terms.
In the latter case only the tails of the distributions (which are centred in
the non-physical region $x<0$) contribute to the physical region $0\le{x}\le1$.
More massive intermediate states would lead to distributions shifted towards
more negative $x$ values, and hence are negligible.

Let us now address the problem of the saturation of the Soffer inequality.
First of all, note that the three quantities $F_\alpha$, $G_\alpha$ and
$H_\alpha$ defined in (\ref{correnti}) satisfy
\begin{equation}
  F_\alpha(x) + G_\alpha(x) = 2 H_\alpha(x) \, .
  \label{bag12}
\end{equation}
This has led to the erroneous conclusion \cite{Soffer:1995ww} that the
inequality is saturated for a relativistic quark model, such as the MIT bag
model. It is clear from eqs.~(\ref{eq5}--\ref{subeq7}) that the spin--flavour
structure of the proton, which results in the appearance of the probabilities
$P(f,\alpha,m)$, spoils this argument and prevents in general the saturation of
the inequality.

Soffer's inequality is only saturated in very specific (and quite unrealistic)
cases. For instance, it is saturated when $P(f,\alpha,-\half)=0$, which happens
if the proton is modelled as a bound state of a quark and scalar diquark. It is
interesting to note that in SU(6) the $\Lambda^0$ hyperon is indeed a bound
state of a scalar--isoscalar $ud$ diquark and an $s$ quark: thus the
transversity distribution of the latter attains the maximal value compatible
with the inequality. Another instance of saturation occurs when
$F_\alpha=G_\alpha=H_\alpha$ and $P(f,\alpha,-\half)=2\,P(f,\alpha,\half)$. It
is easy to verify that this happens for the $d$ quark distribution in a
non-relativistic model of the proton with an SU(6) wavefunction. Apart from
these two special cases, Soffer's inequality should not generally be expected
to be saturated.

\subsubsection{Transversity distributions in the MIT bag model}

Calculations of structure functions within the MIT bag model have been
performed by various authors, using different versions of the model
\cite{Jaffe:1975nj, Benesh:1987ie, Signal:1988vf, Schreiber:1988uw,
Schreiber:1991tc}. The transversity distributions, however, have been evaluated
only in the simplest non-chiral version of the MIT bag.

In the first calculation of $\DT{q}$ \cite{Jaffe:1992ra}, the distributions
were estimated using the formalism developed in \cite{Jaffe:1975nj}, with no
attempt to restore translational invariance. The single-quark wave functions of
the non-translationally invariant bag are used directly in the evaluation of
the matrix elements. The single-quark contribution to the transversity
distributions, corresponding to $H_\alpha(x)$ in (\ref{correnti}), is then
\begin{equation}
  H(x) =
  \frac{\omega_n MR}{2\pi(\omega_n-1)j_0^2(\omega_n)}
  \int_{|y_\text{min}|}^\infty \d{y} \, y
  \left[
    t_0^2 + 2 t_0 t_1 \frac{y_\text{min}}{y} +
    t_1^2 \left( \frac{y_\text{min}}{y} \right)^2
  \right] ,
  \label{h1mit}
\end{equation}
where $\omega_n$ is the $n$-th root of the equation
$\tan\omega_n=-\omega_n/(\omega_n-1)$ and $y_\text{min}=xRM-\omega_n$. $R$ and
$M$ are the bag radius and nucleon mass, respectively, and $RM=4\,\omega_n$.
The functions $t_0$ and $t_1$ are defined as
\begin{equation}
  t_l(\omega_n,y) = \int_0^1 \d{u} \, u^2 j_l(u \,\omega_n) j_l(u y) \, ,
\end{equation}
where $j_l$ is the $l$-th order spherical Bessel function. For completeness, we
also give the unpolarised and helicity contributions
\begin{subequations}
\begin{eqnarray}
  F(x) &=&
  \frac{\omega_n MR}
       {2\pi(\omega_n-1)j_0^2(\omega_n)}
  \int_{|y_\text{min}|}^\infty \d{y} \, y
  \left[ t_0^2 + 2 t_0 t_1 \frac{y_\text{min}}{y} + t_1^2 \right] ,
  \label{f1mit}
\\
  G(x) &=&
  \frac{\omega_n MR}{2\pi(\omega_n-1)j_0^2(\omega_n)}
  \int_{|y_\text{min}|}^\infty \d{y} \, y
  \nonumber
\\
  &&
  \hspace{2.5em} \null \times
  \left[
    t_0^2 +
    2 t_0 t_1 \frac{y_\text{min}}{y} +
    t_1^2 \left( 2 \left( \frac{y_\text{min}}{y} \right)^2 - 1 \right)
  \right] .
  \label{g1mit}
\end{eqnarray}
\end{subequations}
To obtain the quark distributions one must insert $F(x)$, $G(x)$ and $H(x)$
into eqs.~(\ref{eq5}--\ref{subeq7}), along with the probabilities
$P(f,\alpha,n)$. In \cite{Jaffe:1975nj, Jaffe:1992ra} only valence quarks were
assumed to contribute to the distributions, hence the intermediate states
$|\alpha\rangle$ reduce to the diquark state alone. Therefore, the quark
distributions are just proportional to $F(x)$, $G(x)$ and $H(x)$. In
particular, with an SU(6) spin--flavour wave function one simply has
\begin{equation}
  \DL{u}(x) =  \frac43 \, G(x) \, , \qquad
  \DL{d}(x) = -\frac13 \, G(x) \, ,
  \label{eq20}
\end{equation}
and analogous relations for $\DT{u}$ and $\DT{d}$ with $G(x)$ replaced by
$H(x)$.

\begin{figure}[htbp]
  \centering
  \includegraphics*[width=12cm]{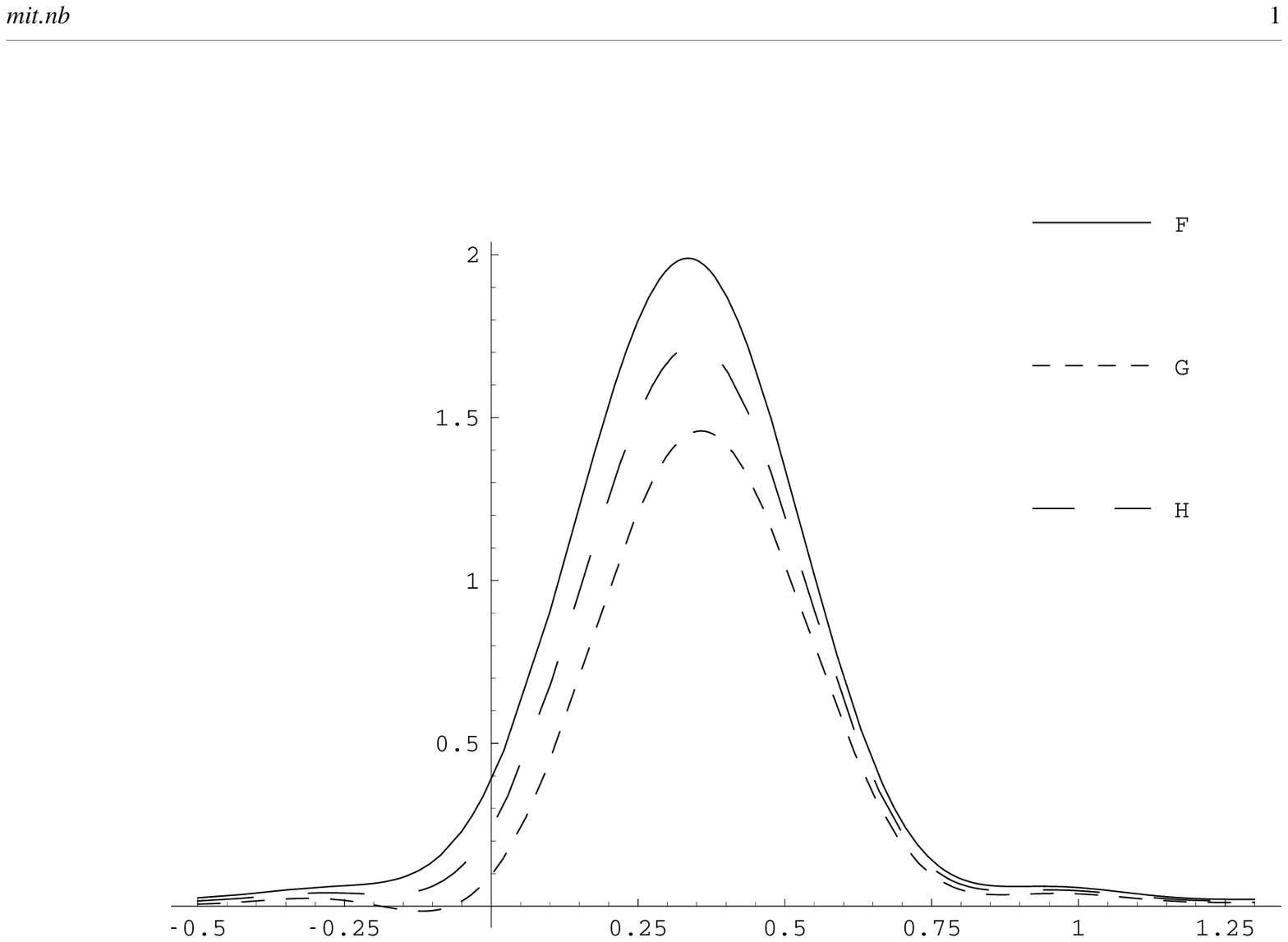}
  \caption{Single-quark contributions to the distribution functions in the
           MIT bag model of \cite{Jaffe:1992ra}.}
  \label{fig_h1mit}
\end{figure}

The quantities $F(x)$, $G(x)$ and $H(x)$ are plotted in Fig.~\ref{fig_h1mit}.
As one can see, the transversity distributions are not so different from those
for helicity. Since translational invariance is lost, the distributions do not
have the correct support. Thus, the integral of $F(x)$ over $x$ between 0 and 1
is not unity, as it should be. In particular, one has
$\int_0^1F(x)\,\d{x}=0.90$. This normalisation problem can be overcome (although
in a very \emph{ad hoc} manner) by integrating between $-\infty$ and $\infty$,
since $\int_{-\infty}^{\infty}F(x)\,\d{x}=1$. Proceeding in this manner, for
the tensor charges one obtains
\begin{equation}
  \delta{u} \equiv  \frac43 \int_{-\infty}^\infty H(x) \, \d{x} =  1.09 \, ,
  \qquad
  \delta{d} \equiv -\frac13 \int_{-\infty}^\infty H(x) \, \d{x} = -0.27 \, ,
\end{equation}
to be compared with the axial charges obtained similarly: $\DL{u}=0.87$ and
$\DL{d}=-0.22$.

Stratmann \cite{Stratmann:1993aw} recomputed $q(x)$, $\DL{q}(x)$ and
$\DT{q}(x)$ in the MIT bag model, introducing a Peierls--Yoccoz projection to
partially restore the translational invariance. However, he did not perform a
\VAP calculation. Since the masses of the intermediate states were not computed
within the model, the number sum rules turned out to be violated. Another
problem of the approach of \cite{Stratmann:1993aw} is that the antiquark
distribution functions were evaluated using eqs.~(\ref{neg1}--\ref{subneg3})
(we have already commented on the inconsistency of such a procedure in
Sec.~\ref{sec:bagform}).

\begin{figure}[htbp]
  \vspace{6cm}
  \includegraphics*[height=0.45\textwidth,angle=270]{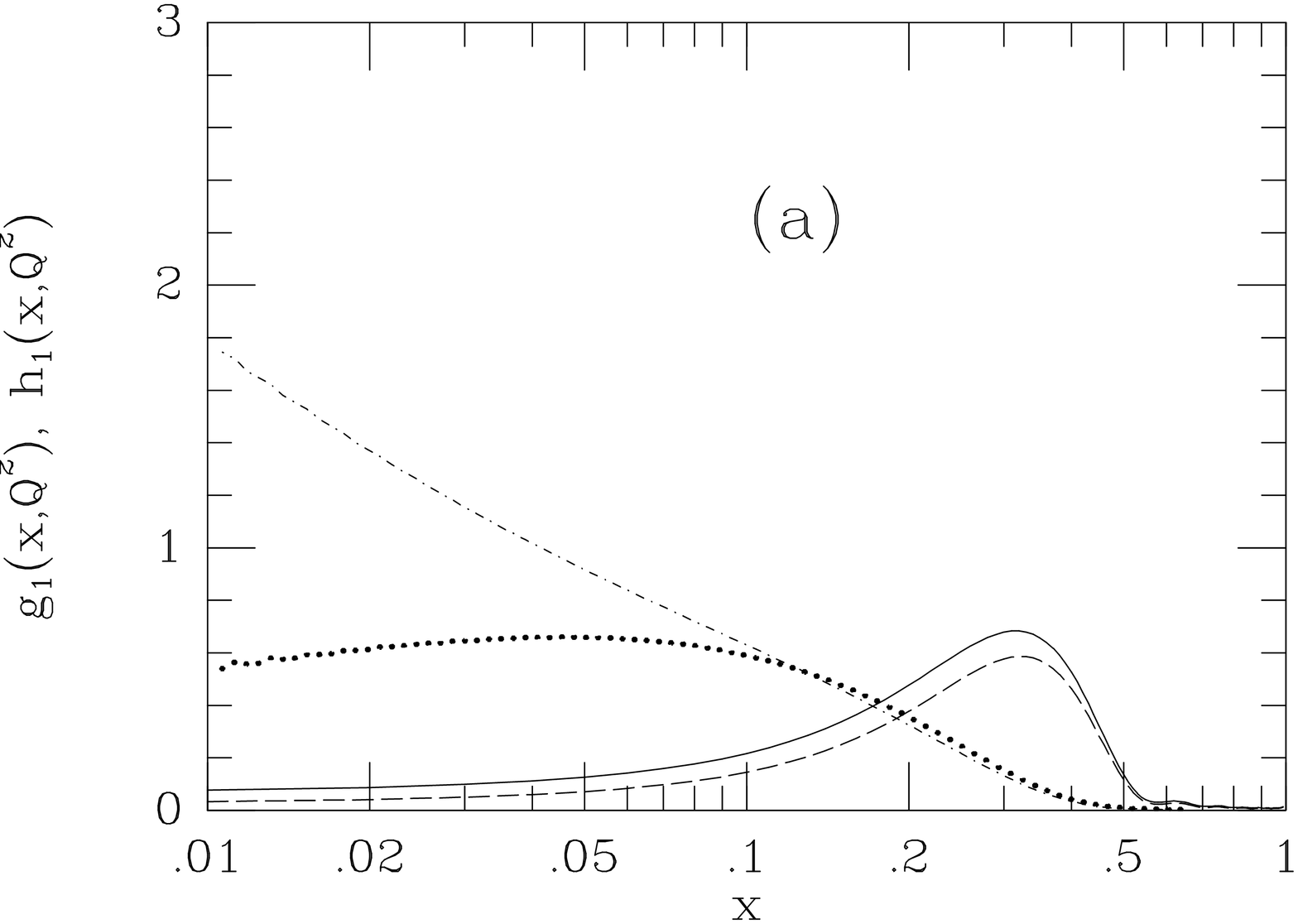}
  \hfill
  \includegraphics*[height=0.45\textwidth,angle=270]{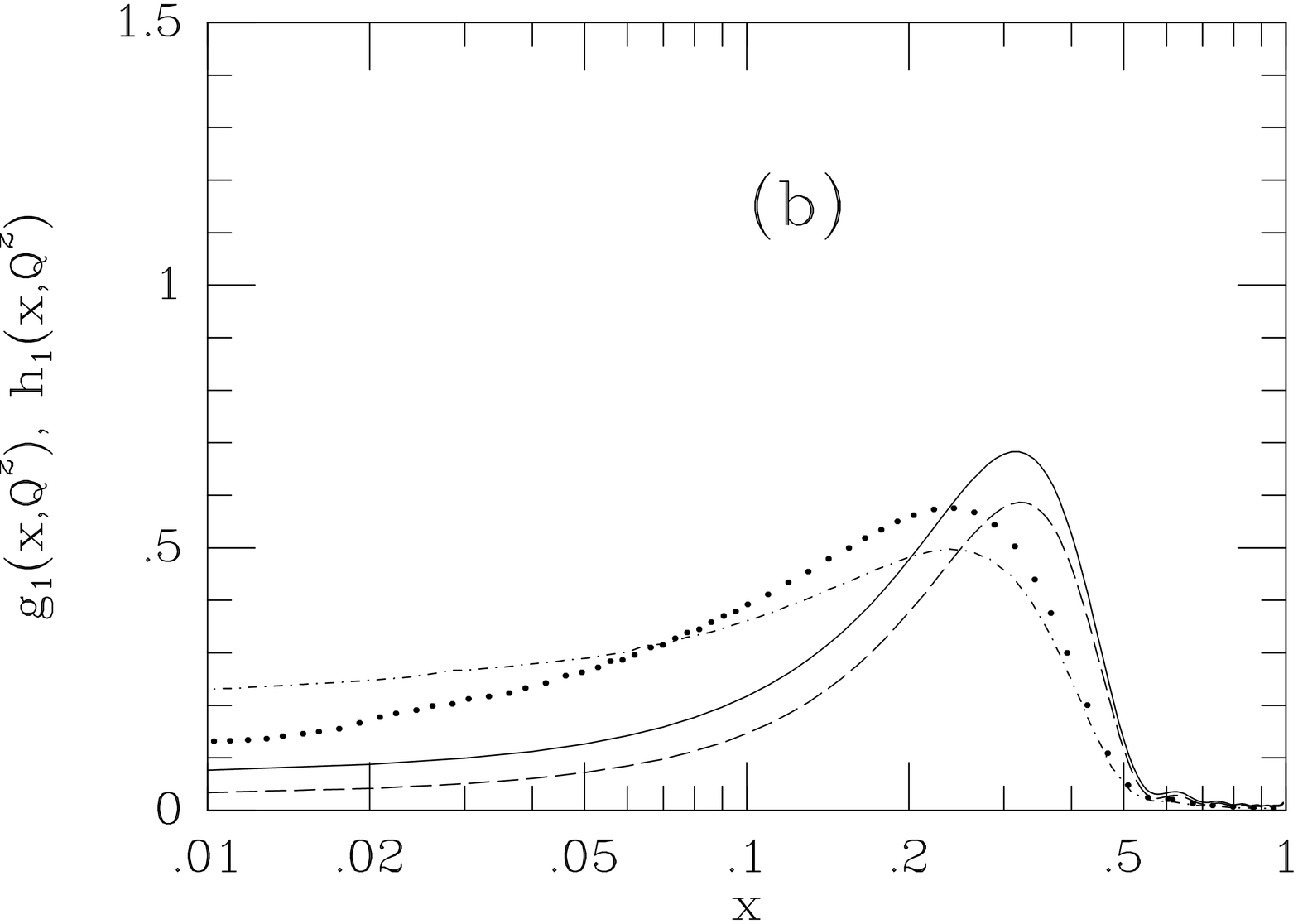}
  \caption{The transversity distribution $h_1(x,Q_0^2)$ (continuous line) and
           the spin distribution function $g_1(x,Q_0^2)$ (dashed line) for the
           MIT bag model at the initial scale $Q_0^2$. The corresponding evolved
           distributions $h_1(x,Q^2)$ (dotted line) and $g_1(x,Q^2)$ (dot--dashed
           line) are obtained starting from (a) $Q_0^2=0.079\GeV^2$ and (b)
           $Q_0^2=0.75\GeV^2$. From~\cite{Scopetta:1998qg}.}
  \label{fig:scopetta}
\end{figure}

The MIT bag model was also used to compute the transversity distributions in
\cite{Scopetta:1998qg}. The technique adopted in this work is rather different
from that discussed above and is based on a non-relativistic reduction of the
relativistic wave function (for a discussion of this non-covariant approach see
also \cite{Traini:1992ue}). The results of \cite{Scopetta:1998qg} for
$h_1=\half\sum_fe_f^2\DT{q}_f$ and $g_1=\half\sum_fe_f^2\DL{q}_f$ are shown in
Fig.~\ref{fig:scopetta}.

\subsubsection{Transversity distributions in the CDM}

The transversity distributions were calculated in the colour dielectric model
in \cite{Barone:1996un}. In particular, the chiral version of the \CDM was
used, in which the splitting between the masses of the nucleon and delta
resonance, or between the scalar and vector diquark, is due to the exchange of
pions, instead of perturbative gluons. Although this model suffers a number of
drawbacks, its main technical advantage with respect to the MIT bag model is
that it allows a full \VAP procedure to be performed, since confinement is
implemented by a dynamical field, not by a static bag surface. For the same
reason, the valence-number sum rules turn out to be fulfilled (to within a few
percent) if the masses of the intermediate states are consistently computed
within the model. As we shall see, the Soffer inequalities are also satisfied,
for both quarks and antiquarks.

The Lagrangian of the chiral \CDM is
\begin{eqnarray}
  \mathcal{L} &=&
  \I \anti\psi \gamma^\mu\partial_\mu \psi +
  \frac{g}{\chi} \, \anti\psi
  \left( \sigma + \I\gamma_5 \Vec\tau{\cdot}\Vec\pi \right) \psi
  \nonumber
\\
  &+&
  \half \left( \partial_\mu \chi \right)^2 -
  \half M^2\chi^2 +
  \half \left( \partial_\mu \sigma \right)^2 +
  \half \left( \partial_\mu \Vec\pi \right)^2 -
  U \left( \sigma, \Vec\pi \right) ,
  \label{eq:in1}
\end{eqnarray}
where $U(\sigma,\Vec\pi)$ is the usual Mexican-hat potential, see \eg,
\cite{Neuber:1993ah}. $\mathcal{L}$  describes a system of interacting quarks,
pions, sigma and a scalar--isoscalar chiral singlet field $\chi$. The
chromodielectric field $\chi$ incorporates non-perturbative gluonic degrees of
freedom. Through their interaction with the $\chi$ field, the quarks acquire a
mass that increases strongly at the boundary of the bag, hence leading to
absolute confinement.

The parameters of the model are: the chiral meson masses $m_\pi=0.14\GeV$,
$m_\sigma=1.2\GeV$ (the precise value of this parameter is actually
irrelevant), the pion decay constant $f_\pi=93\MeV$, the quark--meson coupling
constant $g$, and the mass $M$ of the $\chi$ field. The parameters $g$ and $M$,
which are the only free parameters of the model, can be uniquely fixed by
reproducing the average nucleon--delta mass and the isoscalar radius of the
proton.

The chiral \CDM Lagrangian (\ref{eq:in1}) contains a single-minimum potential
for the chromodielectric field $\chi$: $V(\chi)={\half}M^2\chi^2$. A
double-minimum version of the \CDM is also widely studied and used (see for
instance \cite{Barone:1994uc}). The structure functions computed in the two
versions of the chiral \CDM do not differ sensibly.\footnote{The single-minimum
\CDM seems to be preferable in the light of quark matter calculations
\cite{Drago:1995pr}.}

The solution of the field equations for the chiral \CDM proceeds through the
introduction of the so-called hedgehog ansatz, which corresponds to a
mean-field approximation \cite{Fiolhais:1985bt}. The technique used to compute
the physical nucleon state is based on a double projection of the mean-field
solution onto linear- and angular-momentum eigenstates. This technique has also
been used to compute the static properties of the nucleon \cite{Neuber:1993ah},
the unpolarised and the longitudinally polarised distribution functions
\cite{Barone:1994uc}, and the nucleon electromagnetic form factors
\cite{Drago:1997uj}. We refer the reader to these papers for more detail. A
different technique to obtain states of definite angular momentum and isospin,
based on the quantisation of the collective degrees of freedom associated with
the rotation of the hedgehog state, will be mentioned in Sec.~\ref{sec:CQSM}.
Let us simply remark that in the chiral \CDM the chiral field cannot develop a
non-zero winding number and its value is always very small. Thus, the choice of
a specific technique to obtain physical states from the hedgehog is less
critical in the chiral \CDM than in other models.

\begin{figure}[htbp]
  \centering
  \includegraphics*[width=0.37\textwidth,angle=90]{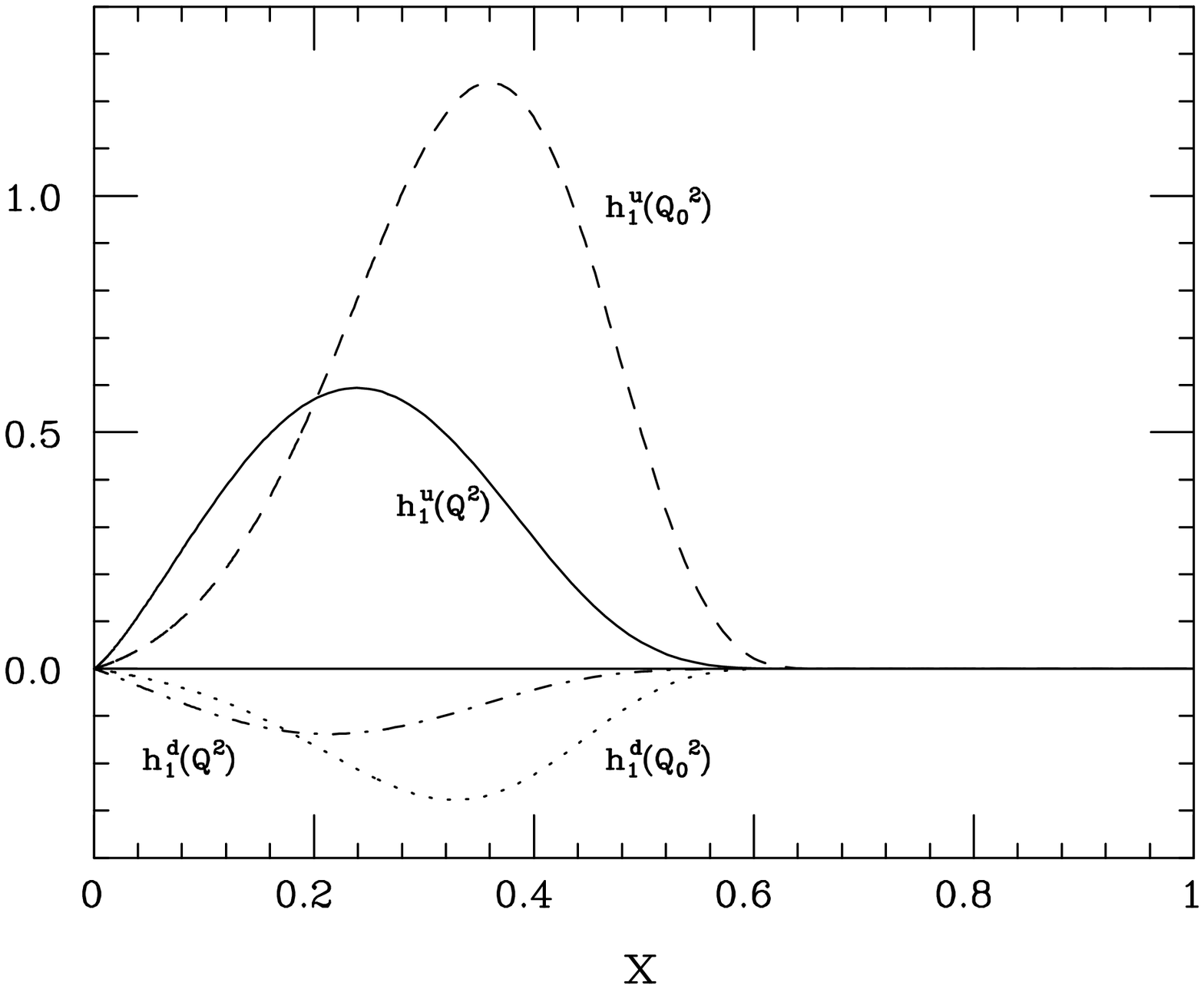}
  \quad
  \includegraphics*[width=0.37\textwidth,angle=90]{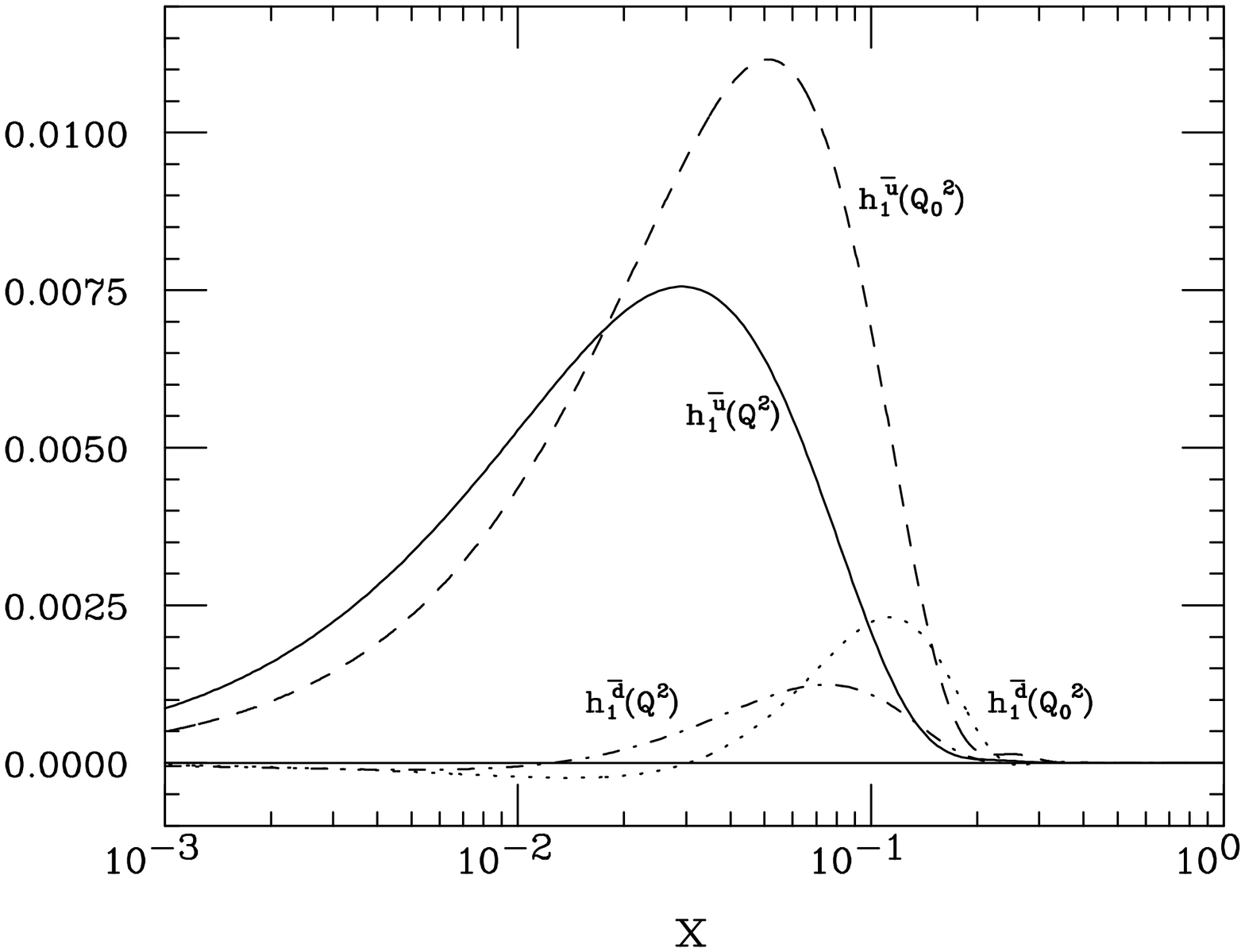}
  \caption{The transversity distributions $x\DT{q}(x)$ in the colour dielectric
           model (from~\cite{Barone:1996un}). Left: the quark distributions
           $x\DT{u}$ and $x\DT{d}$. Right: the antiquark distributions
           $x\DT\bar{u}$ and $x\DT\bar{d}$. The distributions are shown at the
           model scale $Q_0^2=0.16\GeV^2$ and at $Q^2=25\GeV^2$.}
  \label{figcdm}
\end{figure}

In Fig.~\ref{figcdm} we show the results of the calculation in
\cite{Barone:1996un}. One of the features of the distributions computed in the
\CDM is their rapid falloff and vanishing for $x>0.6$. This is due to the soft
confinement of quarks, which do not carry large momenta. It should also be
stressed that the Peierls--Yoccoz procedure, which is a non-relativistic
approximation, becomes unreliable at large $x$. Note also that the sea
contribution is rather small.

As for the model scale $Q_0^2$, in \cite{Barone:1993iq, Barone:1994uc,
Barone:1996un} it was determined by matching the value of the momentum fraction
carried by the valence, as computed in the model, with that obtained by
evolving backward the value experimentally determined at large $Q^2$. The result
is $Q_0^2=0.16\GeV^2$. Proceeding in a similar manner, Stratmann found
$Q_0^2=0.08\GeV^2$ \cite{Stratmann:1993aw} in the MIT bag. The \CDM distributions
evolved from $Q_0^2=0.16\GeV^2$ to $Q^2=25\GeV^2$ are also shown in
Fig.~\ref{figcdm}. Needless to say, perturbative evolution from such low $Q_0^2$
values should be taken with some caution. The tensor charges computed in the
\CDM are (at $Q_0^2=0.16\GeV^2$)
\begin{equation}
  \delta{u} = 1.22 \, , \qquad \delta{d} = -0.31 \, ,
\end{equation}
whereas the axial charges are $\DL{u}=1.08$ and $\DL{d}=-0.29$ (see
Table~\ref{tensc}).

\subsection{Chiral models}

In chiral models the $q\anti{q}$ excitations are described in terms of
effective degrees of freedom represented by chiral fields. There is now a huge
variety of models of this type and, as already seen in Sec.~\ref{sec:bag},
bag-like models also admit chiral versions. In this section we shall focus on
two models: the \CQSM \cite{Diakonov:1986eg, Diakonov:1988ty}, which can be
also derived from the Nambu--Jona-Lasinio model by imposing non-linear
constraints on the chiral fields\cite{Nambu:1961er, Nambu:1961fr, Vogl:1991qt,
Reinhardt:1988fz, Reinhardt:1989fv, Meissner:1988bg, Meissner:1989kq}, and the
\CQM \cite{Manohar:1984md}.

The main difference between these two models is that in the \CQSM chiral
symmetry is dynamically broken within the model itself, while in the \CQM
quarks have large dynamical masses arising from a process of spontaneous
chiral-symmetry breaking which is not described by the model. Another important
difference, reflected in the naming of the two models, is that in the \CQSM a
non-trivial topology is introduced, which is crucial for stabilising the
soliton, whereas in the \CQM chiral fields are treated as a perturbation.
Finally, while the \CQSM is a non-confining model, confinement may be
introduced into the \CQM, starting from a non-chiral confining model and
dressing the quarks with chiral fields.

As for the nucleon spin structure, chiral models are characterised by a
depolarisation of the valence quarks, due to a transfer of total angular
momentum of the nucleon into the orbital angular momentum of the sea, described
by the chiral fields. This feature has made the chiral models quite popular for
the study of the nucleon spin structure.

\subsubsection{Chiral quark--soliton model}
\label{sec:CQSM}

The basic idea of the chiral quark--soliton model is to describe the low-energy
behaviour of QCD by two effective degrees of freedom, Nambu--Goldstone pions
and quarks with a dynamical mass. The model is described by the following
vacuum functional \cite{Diakonov:1986eg, Diakonov:1988ty}
\begin{equation}
  \exp \left\{ \strut \I S_\text{eff}[ \pi(x) ] \right\} =
  \int D \psi \, D \anti\psi
  \exp
  \left\{
    \I\int \! \d^4 x \,
    \anti\psi \, (\I\gamma^\mu\partial_\mu - m U^{\gamma_5}) \, \psi
  \right\} ,
  \label{FI}
\end{equation}
with
\begin{subequations}
\begin{eqnarray}
  U &=& \exp\left[ \I \pi^a(x) \tau^a \right],
\\
  U^{\gamma_5} &=&
  \exp \left[ \I \pi^a(x) \tau^a\gamma_5 \right] =
  \half(1+\gamma_5) \, U + \half(1-\gamma_5) \, U^\dagger \, .
\end{eqnarray}
\end{subequations}

Here $\psi$ is the quark field, $m$ is the effective quark mass arising from
the spontaneous breakdown of chiral symmetry, and $U$ is the SU(2) chiral pion
field. A possible derivation of the effective action (\ref{FI}) is based on the
instanton model of QCD vacuum \cite{Diakonov:1986eg}.

The \CQSM describes the nucleon as a state of $\Nc$ valence quarks bound by a
self-consistent hedgehog-like pion field whose energy, in fact, coincides with
the aggregate energy of quarks from the negative-energy Dirac continuum. This
model differs from the $\sigma$-model \cite{Kahana:1984dx, Kahana:1984be} in
that no kinetic energy is associated to the chiral fields, which are effective
degrees of freedom, totally equivalent to the $q\anti{q}$ excitations of the
Dirac sea (the problem of double counting does not arise).

The \CQSM field equations are solved as follows. For a given time-inde\-pendent
pion field $U=\exp(\I\pi^a(\Vec{x})\tau^a)$, one finds the spectrum of the
Dirac Hamiltonian:
\begin{equation}
  H \Phi_n = E_n \Phi_n \, ,
  \label{Dirac-equation}
\end{equation}
which contains the upper and lower Dirac continua (distorted by the presence of
the external pion field), and may also contain discrete bound-state levels, if
the pion field is strong enough. If the pion field has winding number 1, there
is exactly one bound-state level which travels all the way from the upper to
the lower Dirac continuum as one increases the spatial size of the pion field
from zero to infinity. This level must be filled to obtain a non-zero
baryon-number state. Since the pion field is colour blind, in the discrete
level one may place $\Nc$ quarks in a state that is antisymmetric in colour.

Calling $E_\text{lev}$ (with $-M\leq{E}_\text{lev}\leq{M}$) the energy of the
discrete level, the nucleon mass is obtained by adding ${\Nc}E_\text{lev}$ to
the energy of the pion field (which coincides exactly with the overall energy
of the lower Dirac continuum) and subtracting the free continuum. The
self-consistent pion field is thus found by minimising the functional
\begin{equation}
  M =
  \min_U \Nc
  \left\{
    E_\text{lev}[U] +
    \sum_{E_n<0} \left( E_n[U] - E_n^{(0)} \right)
  \right\} .
  \label{nm}
\end{equation}
From symmetry considerations one looks for the minimum in a hedgehog
configuration
\begin{equation}
  U_c({\bf x}) =
  \exp \left[ \I \pi^a(\Vec{x}) \tau^a \right] =
  \exp\left[ \I (\Vec{n}{\cdot} \Vec\tau) P(r) \right], \quad
  r = |{\Vec{x}}| \, , \quad
  \Vec{n} = \frac{\Vec{x}}{r} \, ,
  \label{hedge}
\end{equation}
where $P(r)$ is the profile of the soliton. The latter is then obtained using a
variational procedure.

At lowest order in $1/\Nc$, the \CQSM essentially corresponds to a mean-field
picture. Some observables, however, vanish at zeroth order in $1/\Nc$ (this is
the case, as we shall see, of unpolarised isovector and polarised isoscalar
distribution functions) and for these quantities a calculation at first order
in $1/\Nc$ is clearly needed.


Within the \CQSM several calculations of distribution functions have been
performed \cite{Weigel:1996ef, Weigel:1996kw, Tanikawa:1996hc, Diakonov:1996sr,
Diakonov:1998ze}. In particular, the transversity distributions were computed
in \cite{Kim:1996bq, Pobylitsa:1996rs, Gamberg:1998vg, Wakamatsu:1998rx}. The
calculations mainly differ in the order of $1/\Nc$ expansion considered. We
shall see that this expansion is related to the expansion in the collective
angular velocity $\Omega$ of the hedgehog solution, and hence to the collective
quantisation of the hedgehog solitons.

\begin{figure}[htbp]
  \centering
  \includegraphics*[width=8cm,angle=-90]{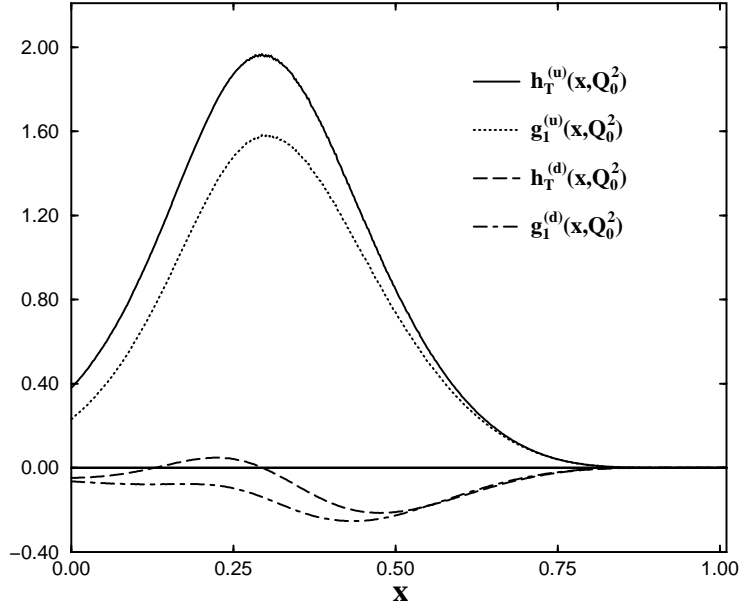}
  \caption{Longitudinal and transverse polarisation distributions in the
           valence-quark approximation of \cite{Gamberg:1998vg}.}
  \label{fig_h1grw}
\end{figure}

The first application of the \CQSM to transversity is contained in
\cite{Kim:1996bq}, where only the tensor charges were computed. In that paper
the \CQSM was extended from two to three flavours with a chiral
$\SU(3)_R\bigotimes\SU(3)_L$ symmetry (for a review see
\cite{Christov:1995vm}). Corrections of order $1/\Nc$ were taken into account
in building the quantised soliton. The procedure adopted in \cite{Kim:1996bq}
for constructing states with definite spin and flavour out of the hedgehog is
the so-called cranking procedure \cite{Ring:1980b1, Blaizot:1986b1}. The
parameters of the model are the constituent mass of the $u$ and $d$ quarks, the
explicit SU(3) symmetry breaking term for the mass of the $s$ quark, and a
cutoff needed to render the theory finite. These parameters are fixed from
hadronic spectroscopy. In particular the constituent mass of the quarks is
$m=420\MeV$ and the symmetry breaking term corresponds to an extra mass of
$180\MeV$ for the $s$ quark. The values of the tensor charges obtained are (the
model scale is taken to be $Q_0^2=0.36\GeV^2$, see Table~\ref{tensc})
\begin{equation}
  \delta{u} =  1.12 \, , \qquad
  \delta{d} = -0.42 \, , \qquad
  \delta{s} = -0.01 \, .
\end{equation}
These quantities are not much affected by the value of the constituent mass and
of the SU(3) symmetry breaking term.

We stress the importance of the $1/\Nc$ corrections. Without these corrections
the tensor (and also the axial) charge of the $u$ quark would be equal and
opposite to that of the $d$ quark. It is also important to notice that, while
the axial singlet charge is substantially reduced owing to the presence of the
chiral fields, the tensor charges are close to those obtained in other models
where the chiral fields are absent, or play a minor r\^{o}le.

A first attempt to compute the $x$ dependence of the transversity distributions
in the \CQSM was made in \cite{Pobylitsa:1996rs}. In this calculation, however,
no $1/\Nc$ corrections were taken into account, and hence they obtained
$\DT{u}+\DT{d}=0$, which is a spurious -- and unrealistic -- consequence of the
zeroth-order approximation adopted. The two most sophisticated calculations in
the \CQSM are those of \cite{Gamberg:1998vg} and \cite{Wakamatsu:1998rx}. In
\cite{Gamberg:1998vg}, both centre-of-mass motion corrections and $1/\Nc$
contributions were taken into account. The correct support for the
distributions is obtained by using a procedure that transforms the
distributions computed in the rest frame into the distributions in the infinite
momentum frame. This procedure essentially amounts to using the relation
\begin{equation}
  f_\text{IMF}(x) = \frac{\Theta (1-x)}{1-x} f_\text{RF}(-\ln(1-x)) \, .
\end{equation}

An important limitation of this work is that only the valence contribution to
the distribution functions is considered. The transversity distributions at the
scale of the model as computed in \cite{Gamberg:1998vg} are shown in
Fig.~\ref{fig_h1grw}.

Wakamatsu and Kubota \cite{Wakamatsu:1998rx} went beyond the valence-quark
approximation of \cite{Gamberg:1998vg} and included vacuum-polarisation
effects. Thus they were also able to compute the antiquark distributions, which
had only been previously evaluated in \cite{Barone:1996un}. A further
improvement is the treatment of the temporal non-locality of the bilinear
operators which appear in the distribution functions. The generic operator
$A^\dagger(0)O_aA(\xi^-)$ is expanded as (see \cite{Pobylitsa:1998tk})
\begin{equation}
  A^\dagger(0) O_a A(\xi^-)
  \simeq
  \widetilde{O}_a + \I \, \xi^- \,\half \left\{ \Omega , \widetilde{O}_a \right\} ,
  \label{tnonlocal}
\end{equation}
where $\widetilde{O}_a\equiv{A}^\dagger(0)O_a A(0)$ and
$\Omega=-\I\,A^\dagger(t)\,\dot{A}(t)$. Equation~(\ref{tnonlocal}) implies that
the non-locality of the operator $A^\dagger(0)O_a A(\xi^-)$ causes a rotational
correction proportional to the collective angular velocity $\Omega$.

\begin{figure}[htbp]
  \centering
  \includegraphics*[width=9cm,angle=-90]{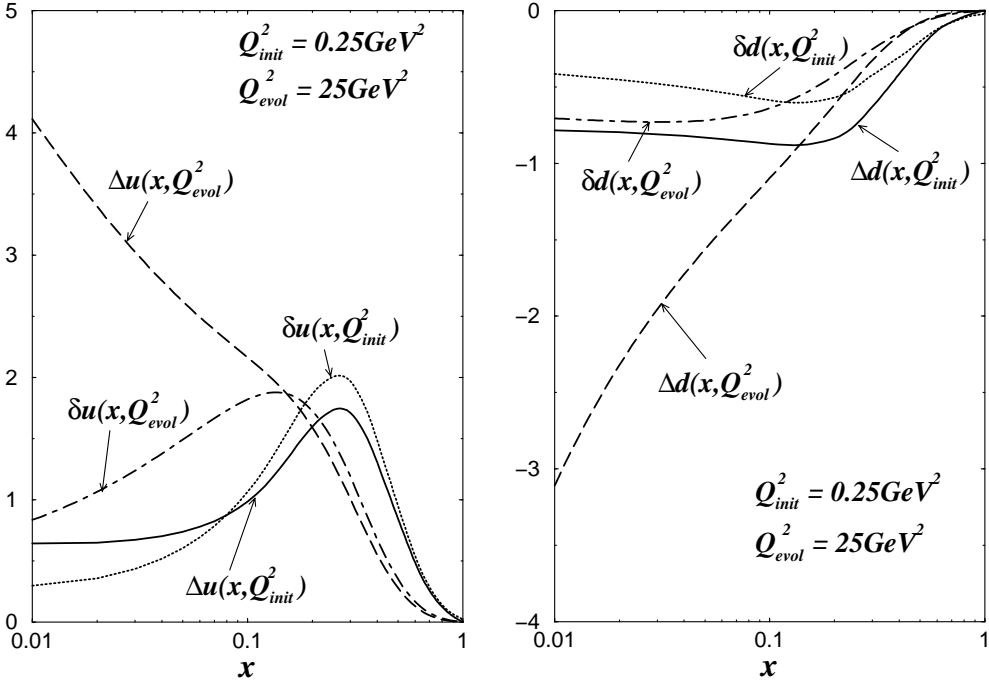}
  \caption{Longitudinal and transverse polarisation distributions of quarks, at
           the scale of the model and after the perturbative evolution.
           From~\cite{Wakamatsu:1998rx}.}
  \label{fig_h1wakamatsu1}
\end{figure}

\begin{figure}[htbp]
  \centering
  \includegraphics*[width=9cm,angle=-90]{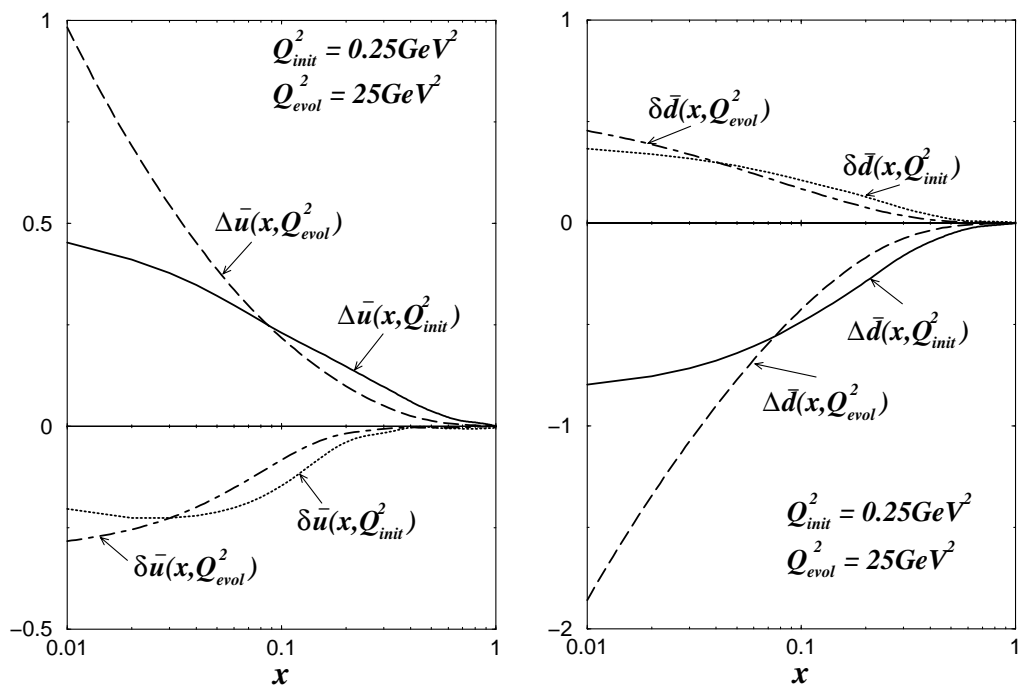}
  \caption{Longitudinal and transverse polarisation distributions of
           antiquarks, at the scale of the model and after the perturbative
           evolution. From~\cite{Wakamatsu:1998rx}.}
  \label{fig_h1wakamatsu2}
\end{figure}

In contrast to the calculation of \cite{Gamberg:1998vg}, in
\cite{Wakamatsu:1998rx} no centre-of-mass motion corrections are performed and
therefore the distributions do not have the correct support. The antiquark
distributions are obtained using the relations (\ref{neg1}--\ref{subneg3}), and
the \emph{caveat} concerning such a procedure thus applies. In
Figs.~(\ref{fig_h1wakamatsu1},\ref{fig_h1wakamatsu2}) the quark and antiquark
helicity and transversity distributions computed in \cite{Wakamatsu:1998rx} are
shown. While the quark distributions are not too different from those computed
in the \CDM (see Fig.~\ref{figcdm}), the $\delta_T\anti{u}$ distribution has a
different sign. This is a consequence of the different technique used in the
calculation of antiquark distributions in \cite{Barone:1996un} and
\cite{Wakamatsu:1998rx} (explicit evaluation of $\DT\anti{q}$ in
\cite{Barone:1996un} \vs use of (\ref{neg1}--\ref{subneg3}) in
\cite{Wakamatsu:1998rx}). The tensor charges obtained in
\cite{Wakamatsu:1998rx} are (again at $Q_0^2=0.36\GeV^2$, see Table~\ref{tensc})
\begin{equation}
  \delta{u} =  0.89 \, , \qquad
  \delta{d} = -0.33 \, .
\end{equation}

Very recently the technique adopted in \cite{Wakamatsu:1998rx} was criticised
in \cite{Schweitzer:2001sr}. In this work, however, the isoscalar and isovector
distributions are computed at a different order in $1/\Nc$ and cannot therefore
be combined to give the single-flavour distributions.

\subsubsection{Chiral quark model}
\label{cloud}

In the chiral quark model of Manohar and Georgi \cite{Manohar:1984md} the
relevant degrees of freedom at a scale below $1\GeV$ are constituent quarks and
Goldstone bosons. This model was used in \cite{Suzuki:1998wv} to compute the
quark and antiquark distribution functions. The \CQM is particularly
interesting for the study of the nucleon spin structure, as it predicts a
depolarisation of constituent quarks due to the emission of Goldstone bosons
into $P$-wave states.

In the \CQM model, the $u$, $d$ and $s$ quarks are assumed to develop large
dynamical masses as a consequence of a mechanism of spontaneous chiral-symmetry
breaking, which lies outside the model itself. We denote these ``bare'' massive
states by $\left|u_0\right\rangle$, $\left|d_0\right\rangle$ \etc. Once they
are dressed by Goldstone bosons, the constituent $u$ and $d$ quark states are
\begin{subequations}
\begin{eqnarray}
  | u \rangle  &=&
  \sqrt{Z} \, | u_0 \rangle + a_\pi \, | d \, \pi^+ \rangle +
  \frac{a_\pi}{2} \, | u \, \pi^0 \rangle + a_K \, | s \, K^+ \rangle +
  \frac{a_\eta}{6} \, | u \, \eta \rangle \, , \qquad
  \label{u-fock}
\\
  | d \rangle &=&
  \sqrt{Z} \, | d_0 \rangle + a_\pi \, | u \, \pi^- \rangle +
  \frac{a_\pi}{2} \, | d \, \pi^0 \rangle + a_K \, | s \, K^0 \rangle +
  \frac{a_\eta}{6} \, | d \, \eta \rangle \, , \qquad
  \label{d-fock}
\end{eqnarray}
\end{subequations}
where $Z$ is the renormalisation constant for a ``bare'' constituent quark (it
turns out to be about $0.7$) and $|a_i|^2$ are the probabilities of finding the
Goldstone bosons in the dressed constituent quark states. These probabilities
are related to each other by the underlying $\SU(3)_R\otimes\SU(3)_L$ symmetry
of the model. There is a single free parameter which may be fixed by computing
the axial coupling $g_A$. Thus the \CQM is a perturbative effective theory in
the Goldstone boson expansion.

\begin{figure}[htbp]
  \centering
  \includegraphics*[height=4.0cm]{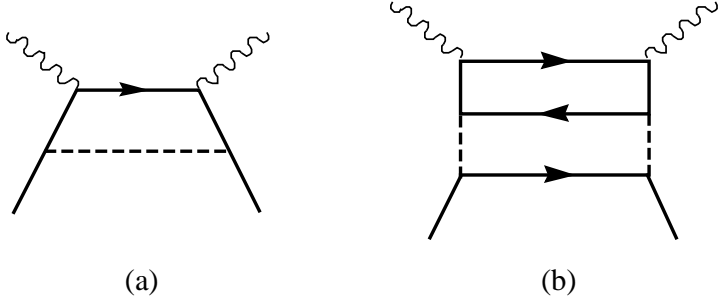}
  \caption{Diagrams contributing to the constituent quark structure: (a) the
           Goldstone boson spectator process, (b) the process probing the
           structure of the Goldstone boson. The solid lines represent quarks
           and dashed lines represent Goldstone bosons.
           From~\cite{Suzuki:1998wv}.
  \label{fig_CQM1}}
\end{figure}

There are three types of contributions to the quark distribution functions. The
first corresponds to the probability of finding a bare quark $f_0$ inside a
dressed quark $f$, and is the bare quark distribution renormalised by the $Z$
factor. The other two contributions correspond to diagrams (a) and (b) of
Fig.~\ref{fig_CQM1}. The spin-independent term corresponding to diagram (a) is
given by
\begin{equation}
  q_j (x) =
  \int_x^1 \! \frac{\d{y}}{y} \, P_{j \, \alpha/i}(y) \, q_i
  \left( \frac{x}{y} \right) .
  \label{suzuki1}
\end{equation}
Here the splitting function $P(y)_{j\,\alpha/i}$ is the probability of finding
a constituent quark $j$ carrying a momentum fraction $y$ and a (spectator)
Goldstone boson $\alpha$ ($\alpha=\pi,K,\eta$) inside a constituent quark $i$.
Diagram (b) of Fig.~\ref{fig_CQM1} corresponds to probing the internal
structure of the Goldstone bosons. This process gives the following
contribution
\begin{equation}
  q_k (x)  =
  \int \! \frac{\d{y}_1}{y_1}  \, \frac{\d{y}_2}{y_2} \, V_{k/\alpha}
  \left( \frac{x}{y_1} \right) P_{\alpha \, j/i}
  \left( \frac{y_1}{y_2} \right) q_i (y_2) \, ,
  \label{suzuki2}
\end{equation}
where $V_{k/\alpha}(x)$ is the distribution function of quarks of flavour $k$
inside the Goldstone boson $\alpha$. In analogy with (\ref{suzuki1},
\ref{suzuki2}), the longitudinal and transverse polarisation distributions
contain the splitting functions $\DL{P}(y)$ and $\DT{P}(y)$, respectively.

In \cite{Suzuki:1998wv} the splitting functions are computed within the \CQM,
the bare quark distributions are obtained from a covariant quark--diquark model
\cite{Meyer:1991fr, Artru:1990zv, Suzuki:1997vu}, and the quark distribution
functions in the Goldstone bosons are taken from phenomenological
parametrisations \cite{Gluck:1992ng} for pions and from models
\cite{Shigetani:1993dx, Shigetani:1994rp, Suzuki:1996vr} for kaons. The
following relation is found
\begin{equation}
  P(y) + \DL{P}(y) = 2 \DT{P}(y) \, ,
  \label{cloud3}
\end{equation}
which implies saturation of the Soffer inequality.

\begin{figure}[htbp]
  \centering
  \includegraphics*[width=\textwidth]{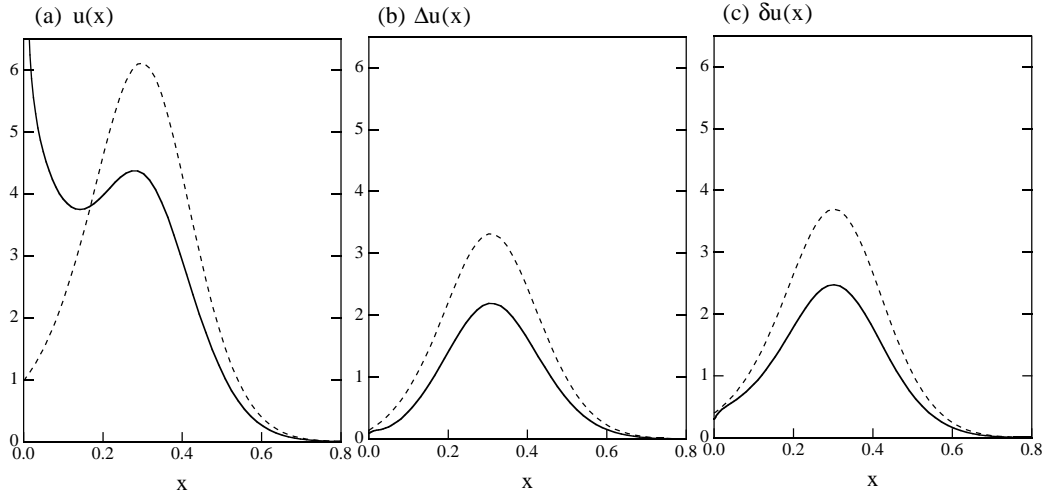}
  \caption{The $u$-quark distribution functions: (a) $u(x)$, (b) $\DL{u}(x)$
           and (c) $\DT{u}(x)$, respectively. In each figure, the result with
           dressed constituent quarks is shown by the solid curve and that
           without dressing by the dashed curve. The latter corresponds to the
           spectator model calculation of \cite{Suzuki:1997vu} (see
           Sec.~\ref{sec:spectator}). From~\cite{Suzuki:1998wv}.
  \label{figCQM4}}
\end{figure}

\begin{figure}[htbp]
  \centering
  \includegraphics*[width=0.7\textwidth]{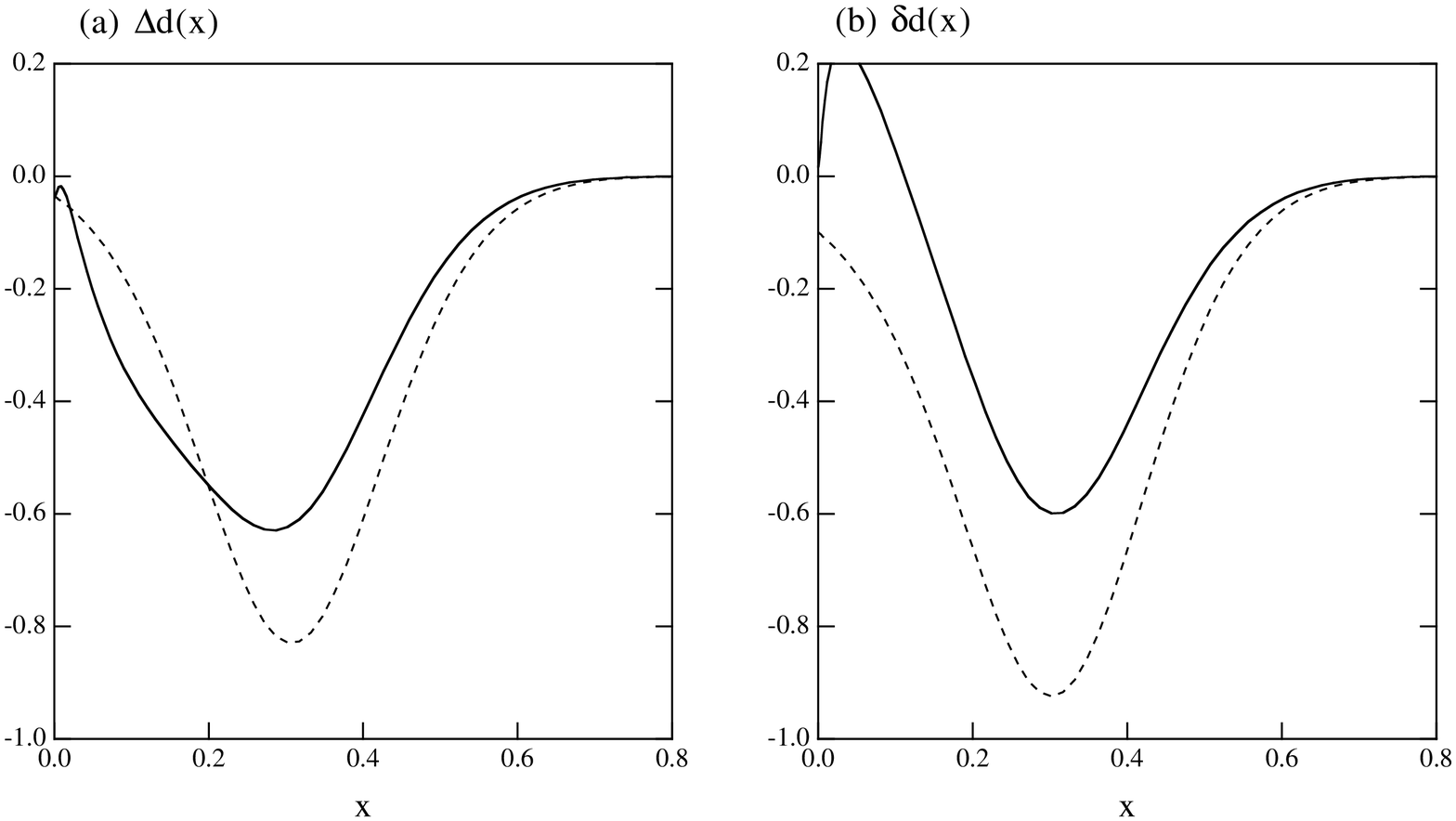}
  \caption{The $d$-quark distribution functions: (a) $\DL{d}(x)$ and (b)
           $\DT{d}(x)$, respectively. The notation is the same as in
           Fig.~\ref{figCQM4}. From~\cite{Suzuki:1998wv}.}
  \label{figCQM5}
\end{figure}

The dominant contribution to the dressing of the constituent quarks is due to
pion emission. The pion cloud affects the $u$ sector reducing both the helicity
and transversity distributions in a similar manner, as can be seen in
Fig.~\ref{figCQM4}. The situation is quite different in the $d$ sector. In
fact, while the renormalisation and meson cloud corrections approximately
cancel each other in the helicity distribution $\DL{d}$, for the transversity
distribution $\DT{d}$ these corrections are both positive. Thus, with respect
to the bare distributions, $\DL{d}$ is almost unmodified whereas $\DT{d}$ is
drastically reduced. The difference between $\DL{d}(x)$ and $\DT{d}(x)$ is an
important and peculiar feature of the model of \cite{Suzuki:1998wv}; the
results for the $d$ distributions are shown in Fig.~\ref{figCQM5}. One can see
that the meson cloud suppresses $\DT{d}$ much more than its helicity
counterpart. In terms of tensor charges, $\delta{d}$ is reduced by about 40\%
by the pion depolarisation effect while the corresponding axial charge is
almost unchanged. The tensor and axial charges are collected in
Table~\ref{tensc}. Notice that depolarisation due to Goldstone boson emission
is a significant effect, although not sufficient to reproduce the small value
of $\DL\Sigma$ observed experimentally.

\subsection{Light-cone models}
\label{sec:melosh}

This section is devoted to models using the so-called front-form dynamics to
describe the nucleon in the infinite momentum frame, and the Melosh rotation to
transform rest-frame quark states into infinite momentum states.

We start by recalling the general idea of front-form dynamics, originally due
to Dirac \cite{Dirac:1949cp}, and then present the calculations of the
transversity distributions based on this approach.

\subsubsection{Forms of dynamics and Melosh rotation}
\label{sec:FDMR}


As shown by Dirac \cite{Dirac:1949cp} (see also \cite{Leutwyler:1978vy}), we
have in general a certain freedom in describing the dynamics of a system.
Various choices of variables defining the initial conditions and of operators
generating the evolution of the system are possible. We shall refer to each of
these choices as a ``form'' of dynamics.

The state of a system is defined on a hypersurface $\Sigma$ in Minkowski space
that does not contain time-like directions. To characterise the state
unambiguously, $\Sigma$ must intersect every world-line once and only once. The
most familiar example of such a surface is, of course, the time instant
$x^0=0$.

Among the ten generators of the Poincar\'{e} algebra, there are some that map
$\Sigma$ into itself, not affecting the time evolution, and others that drive
the evolution of the system and contain the entire dynamics. The latter
generators are called Dirac ``Hamiltonians''.

Hereafter we shall only be interested in two forms of dynamics: the
\emph{instant-form} and the \emph{front-form} (for a more general discussion we
refer the reader to \cite{Leutwyler:1978vy}).

In the usual form of dynamics, the \emph{instant-form}, the initial conditions
are set at some instant of time and the hypersurfaces $\Sigma$ are flat
three-dimensional surfaces only containing directions that lie outside the
light-cone. The generators of rotations and space translations leave the
instant invariant and do not affect the dynamics. The remaining four generators
(boosts and time translations) are the ``Hamiltonians''.

In the \emph{front-form} dynamics one considers instead three-dimensional
surfaces in space--time formed by a plane-wave front advancing at the velocity
of light, \eg, the surface $x^+=0$. The quantities $P^1$, $P^2$, $P^+$, $M^{12}$,
$M^{-+}$, $M^{1+}$ and $M^{2+}$ are associated with transformations that leave
this front invariant. The remaining Poincar\'{e} generators, namely $P^-$, $M^{1-}$
and $M^{2-}$ are the ``Hamiltonians''. The advantage of using front-form dynamics
is that the number of Poincar\'{e} generators affecting the dynamics of the system
is reduced and there is one more Poincar\'{e} generator that transforms the states
without evolving them.

Working within front-form dynamics, there is an important transformation,
namely the \MW rotation \cite{Wigner:1939cj, Melosh:1974cu}, which relates the
spin wave functions $q_\text{RF}^{\uparrow\downarrow}$ in the \RF to the spin
wave functions $q_\text{IMF}^{\uparrow\downarrow}$ in the \IMF.\footnote{Note
that in literature the rest-frame wave functions are also called ``instant-form''
wave functions, and the infinite-momentum-frame wave functions are also called
``light-cone'' wave functions.} The Melosh--Wigner rotation is
\begin{subequations}
\begin{eqnarray}
  q_\text{IMF}^\uparrow &=&
  w
  \left[
    (k^++m) \, q_\text{RF}^\uparrow +
    (k^1 + \I k^2) \, q_\text{RF}^\downarrow
  \right]
  ,
  \label{melosh1}
\\
  q_\text{IMF}^{\downarrow} &=&
  w
  \left[
    -(k^1 - \I k^2) \, q_\text{RF}^\uparrow +
    (k^++m) \, q_\text{RF}^\downarrow
  \right] ,
  \label{melosh2}
  \sublabel{submelosh2}
\end{eqnarray}
\end{subequations}
where $w=[(k^++m)^2+\Vec{k}_\perp^2]^{-1/2}$ and $k^+=k^0+k^3$.

The reason that the \MW rotation is relevant in \DIS is that this process
probes quark dynamics on the light-cone rather than the constituent quarks in
the rest frame \cite{Ma:1990eq, Ma:1991ns}. As for the spin structure, in
front-form dynamics the spin of the proton is not simply the sum of the spins
of the individual quarks, but is the sum of the \MW-rotated spins of the
light-cone quarks \cite{Chung:1988my}.

Calculations of distribution functions using \MW have recently appeared in the
literature, see \eg, \cite{Faccioli:1998aq, Pace:1998di}.

\subsubsection{Transversity distributions in light-cone models}

B.-Q.~Ma has reconsidered the problem of the spin of the nucleon in the light
of the effects of the \MW rotation \cite{Ma:1991xq}. Applying the \MW rotation,
the quark contribution to the spin of the nucleon is reduced. In particular,
from eqs.(\ref{melosh1},~\ref{submelosh2}) one can show that the observed (\ie,
\IMF) axial charge $\DL{q}_\text{IMF}$ is related to the constituent quark
axial charge $\DL{q}_\text{RF}$ as follows
\begin{equation}
  \DL{q}_\text{IMF} = \langle M_q \rangle \, \DL{q}_\text{RF} \, ,
  \label{melosh3}
\end{equation}
where
\begin{equation}
  M_q = \frac{(k^+ + m)^2 - \Vec{k}_\perp^2}{(k^+ + m)^2 +
\Vec{k}_\perp^2} \, ,
  \label{melosh4}
\end{equation}
and $\langle{M_q}\rangle$ is its expectation value in the three-quark state
\begin{equation}
  \langle M_q \rangle = \int \! \d^3\Vec{k} \, M_q\ | \Psi (\Vec{k}) |^2 \, ,
  \label{melosh5}
\end{equation}
where $\Psi(\Vec{k})$ is the (normalised) momentum wavefunction of the
three-quark state. By choosing two different reasonable wave functions
(harmonic oscillator and power-law fall off) the calculation in
\cite{Brodsky:1994fz} gives $\langle{M_q}\rangle=0.75$ (both for $u$ and $d$ if
we assume $m_u=m_d$), which leads to a reduction of $\DL\Sigma$ from $1$ (the
constituent quark model value) down to $0.75$.

The effect of the \MW rotation on the tensor charges was discussed in
\cite{Schmidt:1997vm}. One finds that the \IMF tensor charge is related to the
constituent quark tensor charge by
\begin{equation}
  \delta{q}_\text{IMF} =
  \langle \widetilde{M}_q \rangle \, \delta{q}_\text{RF} \, ,
  \label{melosh6}
\end{equation}
where
\begin{equation}
  \widetilde{M}_q = \frac{(k^+ + m)^2}{(k^+ + m)^2 + \Vec{k}_\perp^2} \, ,
  \label{melosh7}
\end{equation}
and, again, $\langle\widetilde{M}_q\rangle$ is the expectation value of
$\widetilde{M}_q$. From (\ref{melosh4}) and (\ref{melosh7}) one finds an
important connection between the \MW rotation for the longitudinal and the
transverse polarisation, namely
\begin{equation}
  1 + M_q = 2 \widetilde{M}_q.
  \label{Melosh-long-transv}
\end{equation}
With the value $\langle{M_q}\rangle=0.75$, this implies
$\langle\widetilde{M}_q\rangle=0.88$ and, using the SU(6) values
$\delta_\text{RF}=\frac43$ and $\delta{d}_\text{RF}=-\frac13$, one obtains
\cite{Schmidt:1997vm} (we omit the ``IMF'' subscript)
\begin{equation}
  \delta{u} =  1.17 \, , \qquad
  \delta{d} = -0.29 \, .
\end{equation}

\begin{figure}[htbp]
  \centering
  \includegraphics[width=7cm]{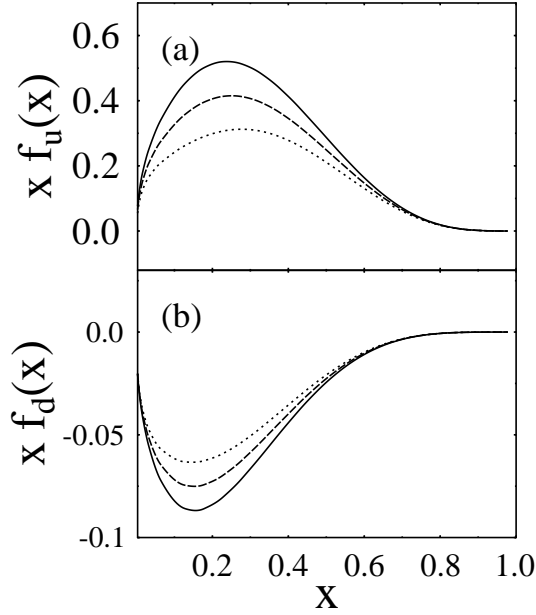}
  \caption{The quark spin distributions $x\DL{q}_\text{RF}(x)$ (solid
           curves), $x\DT{q}(x)$ (dashed curves), and $x\DL{q}(x)$ (dotted
           curves) in the light-cone SU(6) quark-spectator model, for (a) $u$
           quarks and (b) $d$ quarks. From~\cite{Ma:1998gy}.}
  \label{figma}
\end{figure}

The transverse polarisation distributions were computed using the \MW rotation
in \cite{Ma:1998gy}. In this paper a simple relation connecting the spin
distributions of quarks in the rest frame $\DL{q}_\text{RF}(x)$, the quark
helicity distributions $\DL{q}(x)$ and the quark transversity distributions
$\DT{q}(x)$ was derived. It reads
\begin{equation}
  \DL{q}_\text{RF}(x) + \DL{q}(x) = 2 \DT{q}(x) \, .
  \label{melosh-spin}
\end{equation}
Adopting a diquark spectator model \cite{Ma:1996sr} to compute the rest-frame
distributions, the authors of \cite{Ma:1998gy} obtain the curves shown in
Fig.~\ref{figma}.

From (\ref{melosh-spin}) and the measured values of the quantities
$\Gamma^{p,n}\equiv\int\d{x}\,g_1^{p,n}$, $g_A/g_V$ and $\DL{s}$, it is
possible to obtain predictions for the tensor charges, as shown by Ma and
Schmidt \cite{Ma:1998pm}. Taking $g_A/g_V=6\,(\Gamma^p-\Gamma^n)$ these authors
find (the ranges are determined by the experimental and theoretical
uncertainties)
\begin{equation}
  \delta{u} =   0.84 - 1.09  \, , \qquad
  \delta{d} = -(0.23 - 0.51) \, .
  \label{eq:Ma-results}
\end{equation}
Using the value $g_A/g_V=1.2573$ from neutron $\beta$ decay (denoted case 2),
they find instead
\begin{equation}
  \delta{u} =   0.89 - 1.11  \, , \qquad
  \delta{d} = -(0.29 - 0.53) \, .
  \tag{\ref{eq:Ma-results}$'$}
\end{equation}

Another calculation of the transversity distributions based on the \MW rotation
is presented in \cite{Cano:1999kx}. These authors use a three-quark wave
function obtained by solving the Schr\"{o}dinger equation with a hypercentral
phenomenological potential (for details see \cite{Faccioli:1998aq}). The effect
of the \MW rotation is to introduce a significant difference between
longitudinal and transverse polarisation already at the model scale.

\subsection{Spectator models}
\label{sec:spectator}

As we have seen, the main ingredients in model calculations of quark
distributions are the nucleon--quark vertices and the masses of the
intermediate states. In the spectator model the set of intermediate states is
reduced to only the diquark states and the vertex is parametrised in some
manner, for instance assuming an SU(6) spin--flavour structure. This model was
used in \cite{Meyer:1991fr} to estimate unpolarised and longitudinally
polarised distributions and, in \cite{Jakob:1997wg, Suzuki:1997vu} to compute
the transversity distributions. In \cite{Suzuki:1997vu, Suzuki:1998wv} it was
used as the starting point for the perturbative dressing of quarks by chiral
fields.

As already mentioned, the model contains the diquark masses as free parameters.
Typical values of these masses are in the range 600--800$\MeV$, with a
splitting of the scalar and vector diquark masses of the order of
100--200$\MeV$. The parameters entering the vertices are their Dirac structure
and form factors. The calculations in \cite{Jakob:1997wg} and in
\cite{Suzuki:1997vu} differ mainly in the choice of the parameters and in a
more-or-less simple form for the vertex.

The results of \cite{Suzuki:1997vu} for the distribution functions are those
already presented in Sec.~\ref{cloud} in Figs.~\ref{figCQM4} and \ref{figCQM5}
(they correspond to the undressed contributions, \ie, to the dashed curves).
Similar results were obtained in \cite{Jakob:1997wg}, where, fixing the
parameters so as to obtain the experimental value of the axial coupling, the
tensor charges were found to be
\begin{equation}
  \delta{u} =  1.22 \, , \qquad
  \delta{d} = -0.25 \, .
\end{equation}
In \cite{Meyer:1991fr} the scale of this model was estimated to be
$Q_0^2=0.063\GeV^2$.

\subsection{Non-perturbative QCD calculations}

\subsubsection{QCD sum rules}

In the QCD sum-rule approach (see for instance \cite{Reinders:1985sr}) one
considers correlation functions of the form
\begin{equation}
  \Pi (q^2) =
  \I \int \! \d^4\xi \, \e^{\I q{\cdot}\xi} \,
  \langle 0 | T \left( \strut j(\xi) j(0) \right) | 0 \rangle \, ,
  \label{sr1}
\end{equation}
where $j(x)=\anti{q}(x)\Gamma{q}(x)$ is a quark current (all indices are
omitted for simplicity). The vacuum polarisation (\ref{sr1}) is computed in two
different ways. On one hand, it is modelled by a dispersion relation,
expressing its imaginary part in terms of resonances exchanged in the
$s$-channel. On the other hand, in the limit of small light-cone separations,
\ie, large $Q^2\equiv-q^2$, one can make an operator-product expansion (OPE) of
$T\,(j(\xi)j(0))$, thus relating $\Pi(q^2)$ to quark condensates
$\langle\anti{q}q\rangle$. The two theoretical expressions of $\Pi(q^2)$ are
then equated, after performing a Borel transformation, which allows one to pick
up only the lowest-lying resonances in a particular channel. The result is an
expression for the matrix elements of certain quark currents in a hadron state,
in terms of quark and resonance parameters, condensates and the Borel mass
$M_B$. The generalisation of this procedure to three-point correlators is
straight-forward.

A QCD sum rule calculation of the tensor charges was reported by He and Ji in
\cite{He:1995gz}. Following the method presented in \cite{Balitsky:1983xk},
they consider the three-point correlation function
\begin{equation}
  \Pi^{\mu\nu} (q^2) =
  \I^2 \int \! \d^4\xi \, \d^4\zeta \, \e^{\I q{\cdot}\xi} \,
  \langle 0 |
    T \left( j^{\mu\nu}(\zeta) \eta(\xi) \anti\eta(0) \right)
  | 0 \rangle \, ,
  \label{sr2}
\end{equation}
where $j^{\mu\nu}$ is the quark tensor current
\begin{equation}
  j^{\mu\nu}(\zeta) = \anti{q}(\zeta) \, \sigma^{\mu\nu} \, q(\zeta) \, ,
  \label{sr3}
\end{equation}
and $\eta(\xi)$ is the nucleon interpolating field, \ie,
\begin{equation}
  \eta(\xi) =
  \varepsilon_{abc} \, [ u_a^T(\xi) C \gamma_\mu u_b(\xi) ]
  \gamma_5 \gamma^\mu d_c(\xi) \, .
  \label{sr4}
\end{equation}
(Here $a$, $b$ and $c$ are colour indices, the superscript $T$ indicates
transpose and $C$ is the charge conjugation matrix.) The interpolating current
$\eta$ is related to the nucleon spinor $U(P)$ by
\begin{equation}
  \langle 0 | \eta(0) | P \rangle = \lambda \, U(P) \, ,
  \label{sr5}
\end{equation}
where $\lambda$ is a coupling strength.

Computing (\ref{sr2}), using OPE on one hand and resonance saturation on the
other, for the tensor charges He and Ji obtained (at a scale $\mu^2=M^2$)
\begin{equation}
  \delta{u} = 1.0 \pm 0.5 \, , \qquad
  \delta{d} = 0.0 \pm 0.5 \, .
  \label{sr8}
\end{equation}
The uncertainty corresponds to a variation of the Borel mass $M_B^2$ from $M^2$
to $2M^2$.

In a subsequent paper \cite{He:1996wy} He and Ji presented a more refined QCD
sum rule calculation of $\delta{q}$ taking into account operators of higher
orders. Instead of the three-point function approach adopted in
\cite{He:1995gz}, they used the external-field approach. Their starting point
is the two-point correlation function in the presence of an external constant
tensor field $Z_{\mu\nu}$
\begin{equation}
  \Pi^{\mu\nu} (Z_{\mu\nu}, q^2) =
  \I \int \! \d^4 \xi \, \e^{\I q{\cdot}\xi} \,
  \langle 0 | T \, ( \eta(\xi) \anti\eta(0) ) | 0 \rangle_Z \, .
  \label{sr9}
\end{equation}
The coupling between quarks and $Z_{\mu\nu}$ is described by the additional
term
\begin{equation}
  \Delta\mathcal{L} = g_q \anti{q} \sigma^{\mu\nu} q Z_{\mu\nu} \, ,
\end{equation}
in the QCD Lagrangian.

Referring to the original paper for the details of the calculation, we give
here the results for $\delta{u}$ and $\delta{d}$ at the scale $\mu^2=M^2$
\begin{equation}
  \delta{u} = 1.33 \pm 0.53 \, , \qquad
  \delta{d} = 0.04 \pm 0.02 \, ,
  \label{sr10}
\end{equation}
where the error is an estimated theoretical one.

A similar study of tensor charges in the QCD sum rule framework was carried out
by Jin and Tang \cite{Jin:1997pe}, who discussed in detail various sources of
uncertainty, and in particular the dependence of the results on the vacuum
tensor susceptibility induced by the external field.

Finally, we recall that QCD sum rules have been also used to compute the
transversity distributions $\DT{q}(x)$. This was done in \cite{Ioffe:1995aa} by
considering a four-point correlator. The ranges of validity of the various
approximations adopted and the sensitivity of the results on the Borel mass
considerably restrict the interval of $x$ over which a reliable calculation can
be performed. The result of \cite{Ioffe:1995aa} is $\DT{u}\simeq0.5$ for
$0.3\lsim{x}\lsim0.5$, with no apparent variation in this range (the $Q^2$ scale
is estimated to be $Q^2\approx5-10\GeV^2$).

\subsubsection{Lattice}

Lattice evaluations of the tensor charges have  been presented by various
groups \cite{Aoki:1997pi, Brower:1997pe, Gockeler:1997es, Capitani:1999zd,
Dolgov:2000ca}. The lattice approach is based on a hypercubic discretisation of
the Euclidean path integral for QCD and a Monte Carlo computation of the
resulting partition function. The lattice size must be large enough  so that
finite size effects are small. An important parameter is the lattice spacing
$a$. The continuum limit corresponds to $a \to 0$. Usually, different values of
quark masses are used and a linear extrapolation is made to the chiral limit of
massless quarks.

Aoki \etal \cite{Aoki:1997pi} performed a simulation on a $16^3\times20$ lattice
with spacing $a\simeq0.14$\,fm, in the quenched approximation (which
corresponds to setting the fermion determinant in the partition function equal
to one). They obtained
\begin{equation}
  \delta{u} =  0.839(60) \, , \qquad
  \delta{d} = -0.231(55) \, ,
  \label{lattice1}
\end{equation}
at a scale $\mu=a^{-1}\simeq1.4\GeV$. For comparison, the axial charges
computed with the same lattice configuration are \cite{Fukugita:1995fh}
\begin{equation}
  \DL{u} =  0.638(54) \, , \qquad
  \DL{d} = -0.347(46) \, .
\end{equation}
Similar results were reported in \cite{Gockeler:1997es}. The continuum limit
was investigated by Capitani \etal \cite{Capitani:1999zd}, who found for the
difference $\delta{u}-\delta{d}$ the value
\begin{equation}
  \delta{u} - \delta{d} = 1.21(4) \, .
\end{equation}
Finally, a lattice calculation of the tensor charges in full QCD (that is, with
no quenching assumption) has been recently carried out by Dolgov \etal
\cite{Dolgov:2000ca}. Using $16^3 \times 32$ lattices, these authors obtain
\begin{equation}
  \delta{u} =  0.963(59) \, , \qquad
  \delta{d} = -0.202(36) \, .
\end{equation}

\subsection{Tensor charges: summary of results}

In Table~\ref{tensc} we compare the results for tensor charges computed in the
models discussed above. We also show the value of the axial
charges.\footnote{The only model where the polarised strange quark distribution
has been computed is the CQSM1 of \cite{Kim:1996bq}. They find $\DL{s}=-0.05$
and $\delta{s}=-0.01$.}

\begin{table}
  \centering
  \caption{Axial and tensor charges in various models. Tensor charges evolved
  in LO QCD from the intrinsic scale of the model ($Q_0^2$) to $Q^2=10\GeV^2$
  are also shown. See the text for details.}
  \label{tensc}
  \vspace{1ex}
  \tabcolsep=0.56\tabcolsep
  \begin{tabular}{lccccccccc}
  \hline
  Model [Ref.] &
  $\DL{u}$ & $\DL{d}$ & $\DL\Sigma$ & $\delta{u}$ & $\delta{d}$ &
  $|\delta{u}/\delta{d}|$ & $Q_0$[GeV] & $\delta{u}(Q^2)$ & $\delta{d}(Q^2)$ \\
  \hline\hline
  NRQM $\star$
  & 1.33 & -0.33 &   1  & 1.33 & -0.33 & 4.03 & 0.28 & 0.97 & -0.24 \\
  MIT \cite{Jaffe:1992ra} $\diamond$
  & 0.87 & -0.22 & 0.65 & 1.09 & -0.27 &  4.04 &0.87 & 0.99 & -0.25 \\
  CDM \cite{Barone:1996un} {\scriptsize $\oplus$}
  & 1.08 & -0.29 & 0.79 & 1.22 & -0.31 & 3.94 & 0.40 & 0.99 & -0.25 \\
  CQSM1 \cite{Kim:1996bq} {\scriptsize $\times$}
  & 0.90 & -0.48 & 0.37 & 1.12 & -0.42 & 2.67 & 0.60 & 0.97 & -0.37 \\
  CQSM2  \cite{Wakamatsu:1998rx} {\scriptsize $+$}
  & 0.88 & -0.53 & 0.35 & 0.89 & -0.33 & 2.70 & 0.60 & 0.77 & -0.29 \\
  CQM \cite{Suzuki:1998wv} {\scriptsize $\otimes$}
  & 0.65 & -0.22 & 0.43 & 0.80 & -0.15 & 5.33 & 0.80 & 0.72 & -0.13 \\
  LC \cite{Schmidt:1997vm} $\circ$
  & 1.00 & -0.25 & 0.75 & 1.17 & -0.29 & 4.03 & 0.28 & 0.85 & -0.21 \\
  Spect. \cite{Jakob:1997wg} {$\ast$}
  & 1.10 & -0.18 & 0.92 & 1.22 & -0.25 & 4.88 & 0.25 & 0.83 & -0.17 \\
  Latt. \cite{Fukugita:1995fh, Aoki:1997pi} $\triangleright$
  & 0.64 & -0.35 & 0.29 & 0.84 & -0.23 & 3.65 & 1.40 & 0.80 & -0.22 \\
  \hline
  \end{tabular}
  \vspace{1cm}
\end{table}

To allow a homogeneous comparison, we evolved the tensor charges from the model
scales $Q_0^2$ to $Q^2=10\GeV^2$ in \LO QCD. Given the very low input scales, the
result of this evolution should be taken with caution (it serves to give a
qualitative idea of the trend).

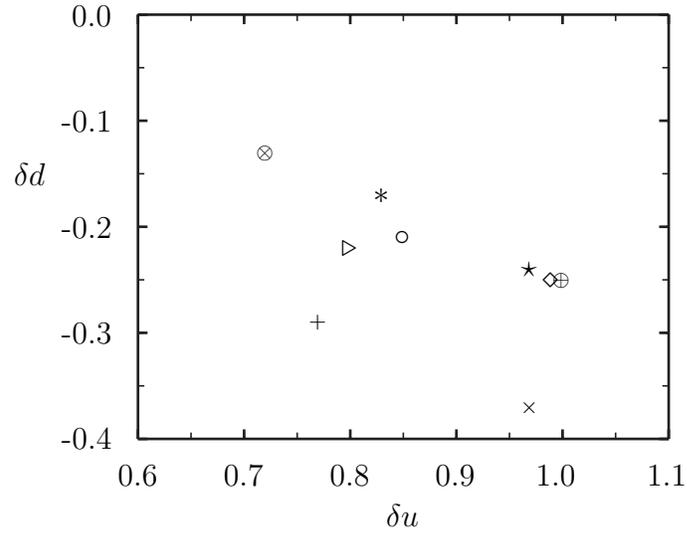
\begin{figure}[htbp]
  \centering
  \begin{picture}(280,210)(-60,-35)
    \LinAxis(0,0)(200,0)(5,2,3,0,1)
    \LinAxis(0,0)(0,160)(4,2,-3,0,1)
    \LinAxis(0,160)(200,160)(5,2,-3,0,1)
    \LinAxis(200,0)(200,160)(4,2,3,0,1)
    \Text(0,-10)[t]{0.6}
    \Text(40,-10)[t]{0.7}
    \Text(80,-10)[t]{0.8}
    \Text(120,-10)[t]{0.9}
    \Text(160,-10)[t]{1.0 }
    \Text(200,-10)[t]{1.1}
    \Text(100,-25)[t]{$\delta{u}$}
    \Text(-10,0)[r]{-0.4}
    \Text(-10,40)[r]{-0.3}
    \Text(-10,80)[r]{-0.2}
    \Text(-10,120)[r]{-0.1}
    \Text(-10,160)[r]{0.0}
    \Text(-35,100)[r]{$\delta{d}$}
    \Text(148,64)[]{$\star$}              
    \Text(156,60)[]{$\diamond$}           
    \Text(160,60)[]{\scriptsize$\oplus$}  
    \Text(148,12)[]{\scriptsize$\times$}  
    \Text(68,44)[]{\scriptsize$+$}        
    \Text(48,108)[]{\scriptsize$\otimes$} 
    \Text(100,76)[]{$\circ$}              
    \Text(80,72)[]{$\triangleright$}      
    \Text(92,92)[]{$\ast$}                
  \end{picture}
  \caption{The tensor charges as computed in various models and evolved to
           $Q^2=10\GeV^2$. For the symbols see Table~\ref{tensc}. The MIT and
           CDM points are slightly displaced for clarity.}
  \label{fig_tensor}
\end{figure}

Unfortunately, $Q_0$ has not been evaluated in the same manner in all models.
As discussed earlier, a possible way to estimate the model scale is to fix it
in such a manner that, starting from the computed value of the momentum
fraction carried by the valence, and evolving it to larger $Q^2$, one fits the
experimentally observed value. This procedure, with slight differences, has
been adopted to find the intrinsic scale in the \NRQM \cite{Scopetta:1998qg},
in the MIT bag model \cite{Jaffe:1980ti, Stratmann:1993aw}, in the \CDM
\cite{Barone:1993iq, Barone:1994uc, Barone:1996un} and in the spectator model
\cite{Meyer:1991fr}. For the light-cone (LC) model of \cite{Schmidt:1997vm} we
have taken the same scale as in the \NRQM, since the starting point of that
calculation is the rest-frame spin distributions of quarks. In other
calculations, in particular in chiral models, the authors have chosen $Q_0$ as
the scale up to which the model is expected to incorporate the relevant degrees
of freedom. The variety of procedures adopted to determine $Q_0$ adds a further
element of uncertainty in the results for $\delta{q}(Q^2)$ presented in
Table~\ref{tensc}. The evolved tensor charges are collected in
Fig.~\ref{fig_tensor}. As one can see, they span the ranges
\begin{equation}
  \delta{u} = 0.7 - 1 \, , \qquad
  \delta{d} = -(0.1 - 0.4) \qquad \text{at} \; Q^2 = 10\GeV^2 \,.
\end{equation}

It is important to notice that, since the evolution of the tensor charges is
multiplicative, the ratio $\delta{u}/\delta{d}$ does not depend on $Q^2$. As one
can see from Table~\ref{tensc}, most of the calculations give for
$|\delta{u}/\delta{d}|$ a value of the order of 4, or larger. Chiral soliton
models CQSM1 and CQSM2, in contrast, point to a considerably smaller value, of
the order of 2.7. An experimental measurement of the tensor charges may then
represent an important test of these models. Note also that the introduction of
chiral fields in a perturbative manner, as in the \CQM, actually has the effect
of increasing $|\delta{u}/\delta{d}|$ owing to a strong reduction in
$\delta{d}$. A possible way to extract the $u/d$ transversity ratio is to make
a precision measurement of the ratio of azimuthal asymmetries in $\pi^+/\pi^-$
leptoproduction. This could be done in the not so distant future (see
Sec.~\ref{experiments}).

\section{Phenomenology of transversity}
\label{phenomenology}

We now review some calculations of physical observables (typically, double-spin
and single-spin asymmetries) related to transversity.\footnote{In this section,
the transversity distributions will be denoted by $\DT{q}$.} Due to the
current lack of knowledge on $\DT{q}$ and the related fragmentation functions,
the available predictions are quite model-dependent and must be taken with a
grain of salt. They essentially provide an indication of the order of magnitude
of some phenomenological quantities.

We also discuss two recent results on azimuthal asymmetries in pion
leptoproduction that may find an explanation in the coupling of transversity to
a $T$-odd fragmentation function arising from final-state interactions.

\subsection{Transverse polarisation in hadron--hadron collisions}
\label{transvhh}

\subsubsection{Transverse double-spin asymmetries in Drell--Yan processes}
\label{transvdouble}

The Drell--Yan transverse double-spin asymmetries were calculated at \LO in
\cite{Barone:1996un, Barone:1997mj} and at \NLO in \cite{Martin:1996sn,
Martin:1998rz} (see also \cite{Hino:1998uq}). Earlier estimates
\cite{Ji:1992ev, Bourrely:1994sc} suffered a serious problem (they assumed the
same \QCD evolution for $\DL{q}$ and $\DT{q}$), which led to over-optimistic
values for $A_{TT}^\text{DY}$.

\begin{figure}[htbp]
  \centering
  \includegraphics[width=5cm,angle=90]{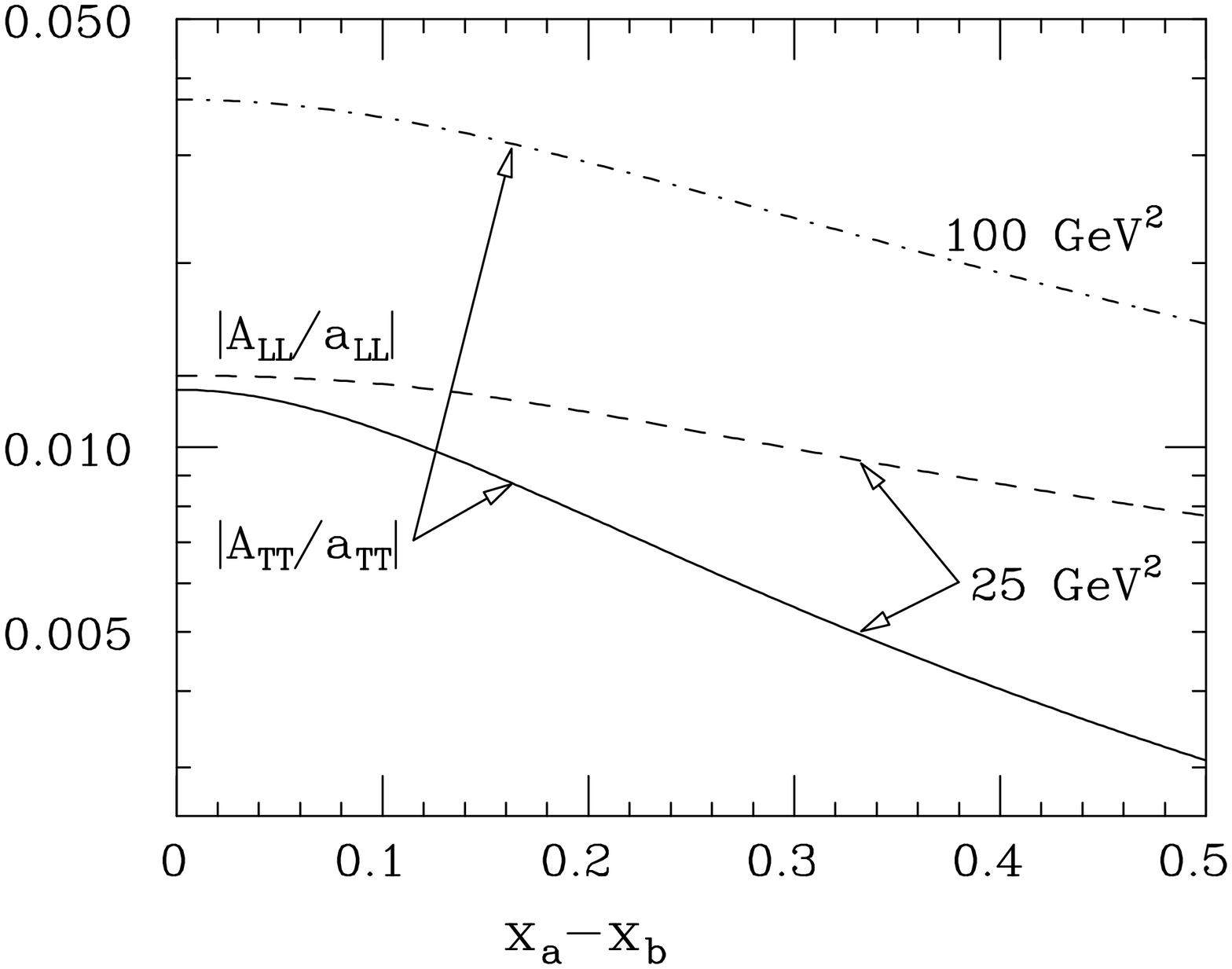}
  \quad
  \includegraphics[width=5cm,angle=90]{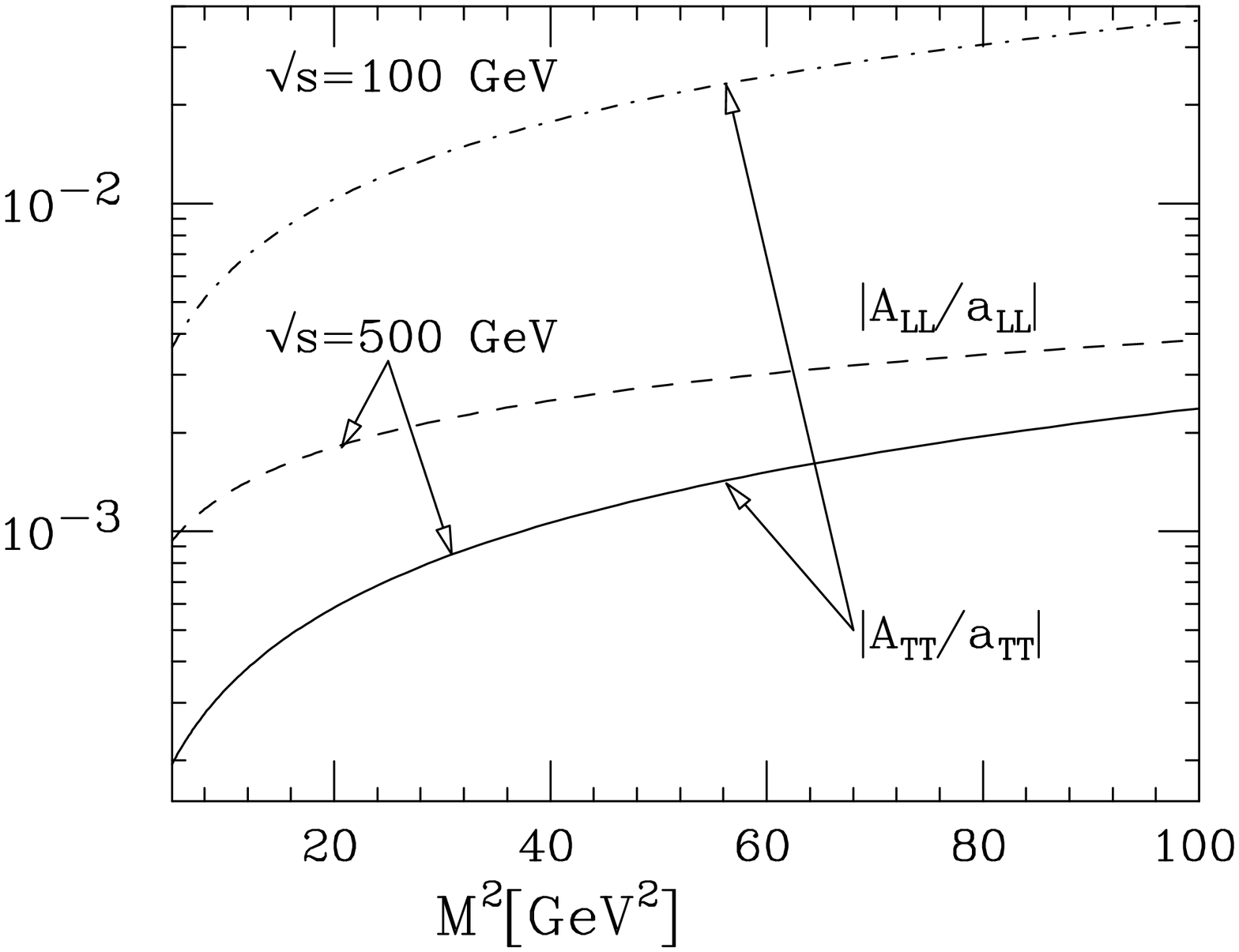}
  \caption{Drell--Yan longitudinal and transverse double-spin asymmetries
           normalised to the partonic asymmetry, as a function of $x_a-x_b$
           (\ie, $x_1-x_2$) for two values of the dilepton invariant mass
           (left), and as a function of the invariant mass of the dilepton pair
           $M^2$ for two values of the c.m.\ energy (right).
           From~\cite{Barone:1997mj}.}
  \label{fig2_att}
\end{figure}

\begin{figure}[htbp]
  \centering
  \includegraphics[width=6cm,angle=90]{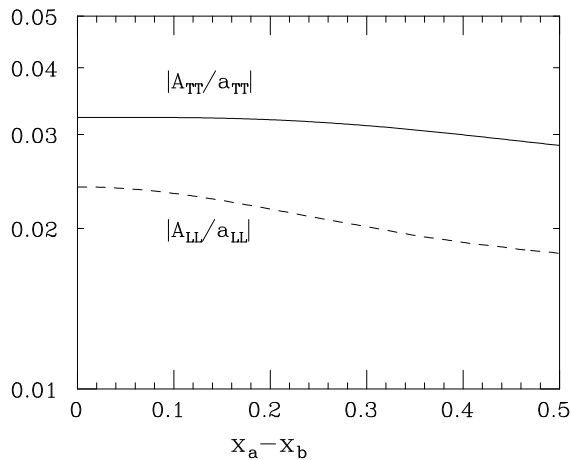}
  \caption{Longitudinal and transverse double-spin asymmetries (normalised to
           the partonic asymmetry) for the $Z^0$-mediated Drell--Yan process.
           From~\cite{Barone:1997mj}.}
  \label{fig6_att}
\end{figure}

In \cite{Barone:1997mj} the equality $\DT{q}(x,Q_0^2)=\DL{q}(x,Q_0^2)$ was
assumed to hold at a very low scale (the input $Q_0^2=0.23\GeV^2$ of the \GRV
distributions \cite{Gluck:1995yq}), as suggested by various non-perturbative
and confinement model calculations (see Sec.~\ref{models}). The transversity
distributions were then evolved according to their own Altarelli--Parisi
equation at \LO. The resulting asymmetry (divided by the partonic asymmetry) is
shown in Fig.~\ref{fig2_att}. Its value is just a few percent, which renders
the planned \RHIC measurement of $A_{TT}^\text{DY}$ rather difficult. The
asymmetry for the $Z^0$-mediated Drell--Yan process is plotted in
Fig.~\ref{fig6_att} and has the same order of magnitude as the electromagnetic
one.

\begin{figure}[htbp]
  \centering
  \includegraphics[width=6.5cm]{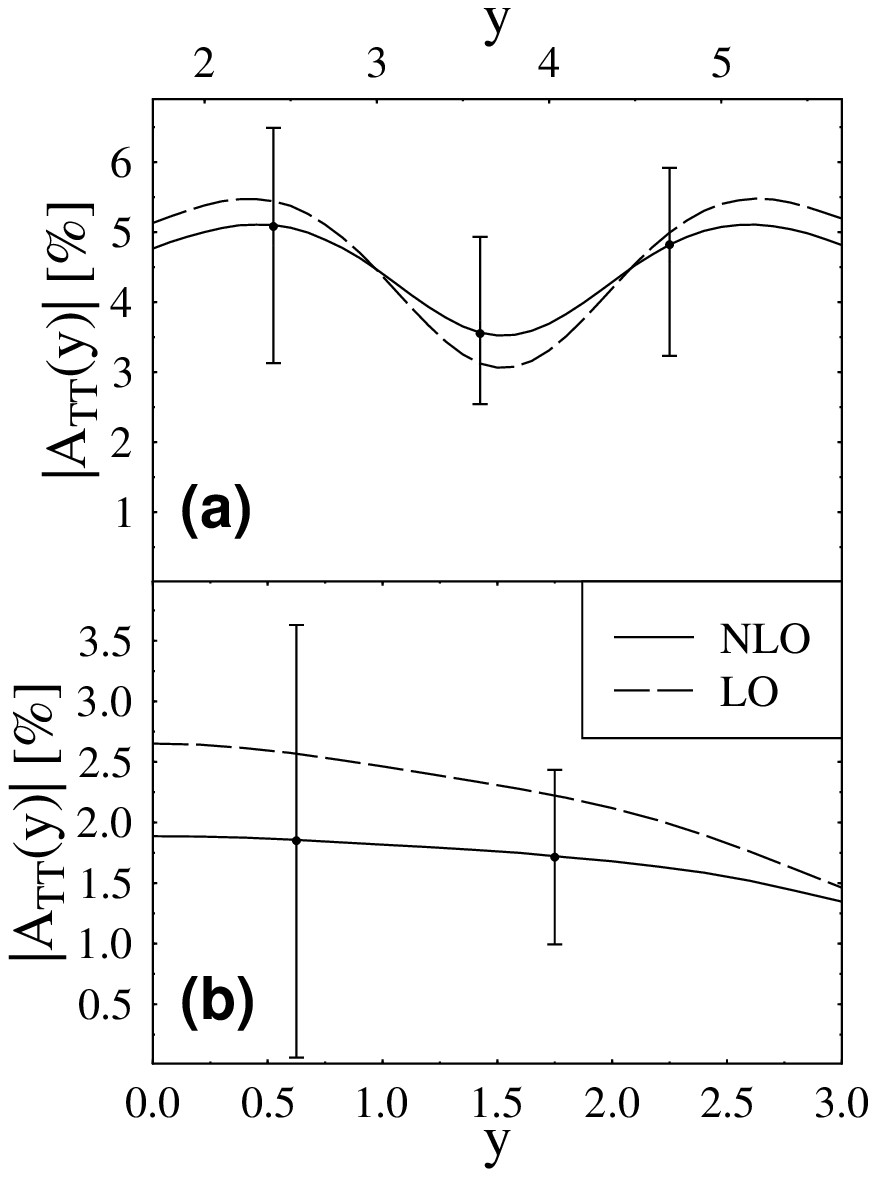}
  \includegraphics[width=6.5cm]{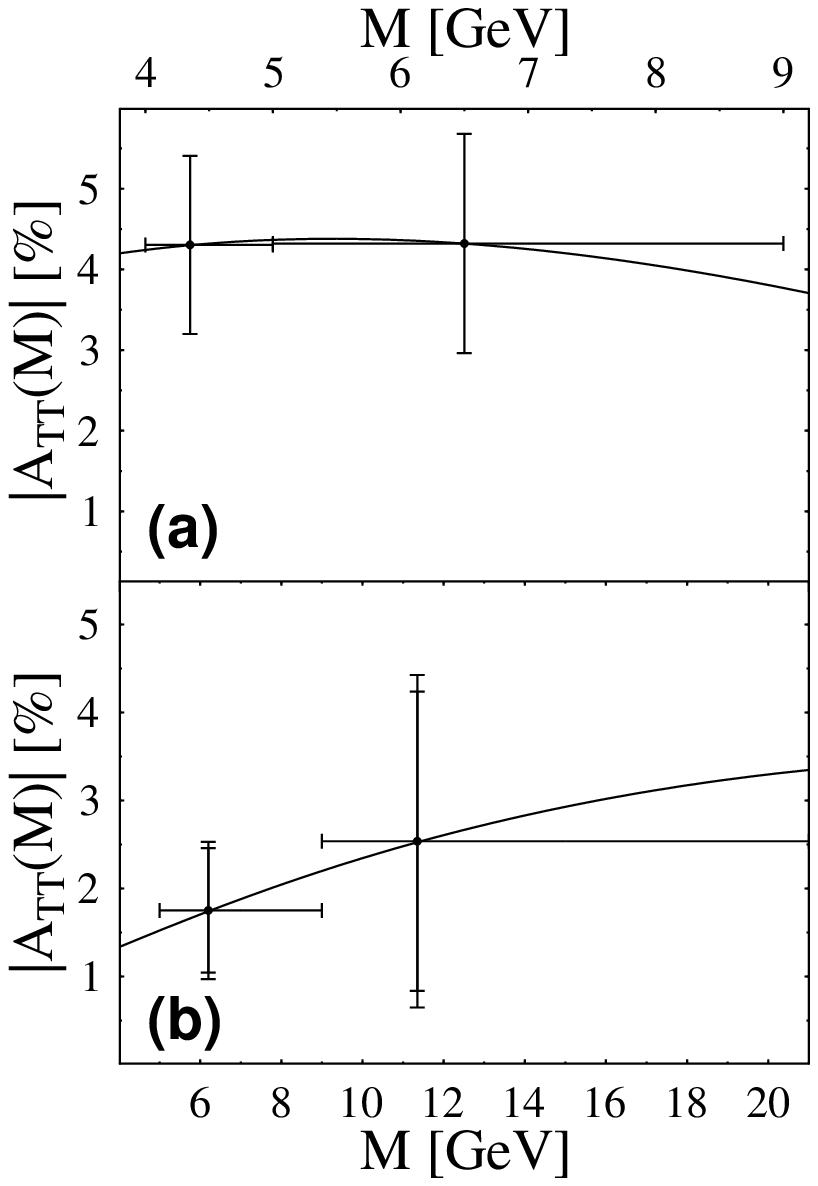}
  \caption{The Drell--Yan transverse double-spin asymmetry as a function of the
           virtual photon rapidity $y$ and of the dilepton invariant mass $M$,
           for two values of the c.m.\ energy: (a) $\sqrt{s}=40\GeV$
           (corresponding to HERA-$\vec{N}$) and (b) $\sqrt{s}=200\GeV$
           (corresponding to RHIC). The error bars represent the estimated
           statistical uncertainties of the two experiments.
           From~\cite{Martin:1998rz}.}
  \label{martin}
\end{figure}

The authors of \cite{Martin:1998rz} use a different procedure to estimate the
transversity distributions. They set $|\DT{q}|=2\,(q+\DL{q})$ at the \GRV
scale, thus imposing the saturation of Soffer's inequality. This yields the
maximal value for $A_{TT}^\text{DY}$. The transversity distributions are
evolved at \NLO. The \NLO corrections are found to be relatively small,
although non-negligible. The predicted curves for $A_{TT}^\text{DY}$ are shown
in Fig.~\ref{martin}.

Summarising the results of the calculations of $A_{TT}^\text{DY}$, we can say
that at the typical energies of the \RHIC experiments \cite{Enyo:2000nt,
Bland:1999gb} ($\sqrtno{s}>100\GeV$) one expects for the double-spin asymmetry,
integrated over the invariant mass $Q^2$ of the dileptons, a value
\begin{equation}
  A_{TT}^\text{DY} \sim (1-2)\%, \quad \text{at most}.
  \label{att2}
\end{equation}

It is interesting to note that as $\sqrtno{s}$ falls the asymmetry tends to
increase, as was first pointed out in \cite{Barone:1997mj}. Thus, at
$\sqrtno{s}=40\GeV$, which would correspond to the c.m.\ energy of the proposed
(but later cancelled) HERA-$\vec{N}$ experiment \cite{Anselmino:1996kg},
$A_{TT}^\text{DY}$ could reach a value of $\sim(3-4)\%$.

Model calculations of $A_{TT}^\text{DY}$ are reported in \cite{Barone:1996un,
Cano:1999kx}. The longitudinal--transverse Drell--Yan asymmetry
$A_{LT}^\text{DY}$ (see Sec.~\ref{twist3drellyan}) was estimated in
\cite{Kanazawa:1998rw, Kanazawa:1998ys} and found to be five to ten times
smaller than the double-transverse asymmetry. Polarised proton--deuteron
Drell--Yan processes were investigated in \cite{Hino:1998ww, Hino:1999qi,
Hino:1999fh, Kumano:1999bt}.

\subsubsection{Transverse single-spin asymmetries}
\label{transvsing}

In the early seventies data on single-spin asymmetries in inclusive pion
hadroproduction \cite{Dick:1975ty, Crabb:1977mg, Dragoset:1978gg} provoked a
theoretical certain interest, as it was widely held that large effects could
not be reproduced within the framework of perturbative \QCD \cite{Kane:1978nd}.
In 1991 the E704 experiment at Fermilab extended the results on large
single-spin asymmetries in inclusive pion hadroproduction with a transversely
polarised proton \cite{Adams:1991cs, Bravar:1996ki} to higher $p_T$. These
surprising results have prompted intensive theoretical work on the subject.

As a matter of fact, one year before the recent Fermilab measurements, Sivers
had suggested that single-spin asymmetries could originate, at leading twist,
from the intrinsic motion of quarks in the colliding protons
\cite{Sivers:1989cc, Sivers:1991fh}. This idea was pursued by the authors of
\cite{Anselmino:1995tv, Anselmino:1998yz}, who pointed out that the Sivers
effect is not forbidden by time-reversal invariance \cite{Collins:1993kk}
provided one takes into account soft interactions in the \emph{initial state}.
In so doing, $T$-odd distribution functions are introduced (see
Secs.~\ref{todd} and \ref{transvmotsinglespin}).

A different mechanism was proposed by Collins \cite{Collins:1993kk}. It relies
on the hypothesis of \emph{final-state} interactions, which would allow
polarised quarks with non-zero transverse momentum to fragment into an
unpolarised hadron (the Collins effect already discussed in
Secs.~\ref{sidisasymm} and \ref{transvmotsinglespin}).

Finally, as seen in Sec.~\ref{transvmotsinglespin}, another way to produce
single-spin asymmetries is to assume the existence of a $T$-odd transverse
polarisation distribution of quarks in the unpolarised initial-state hadron.

All the above effects manifest themselves at leading twist. We shall
concentrate on the Collins mechanism, which appears, among the three
hypothetical sources of single-spin asymmetries just mentioned, the likeliest
one (as repeatedly stressed, \emph{initial-state} interactions are definitely
harder to unravel than \emph{final-state} interactions).

\begin{figure}[htbp]
  \centering
  \includegraphics[width=8cm,angle=-90]{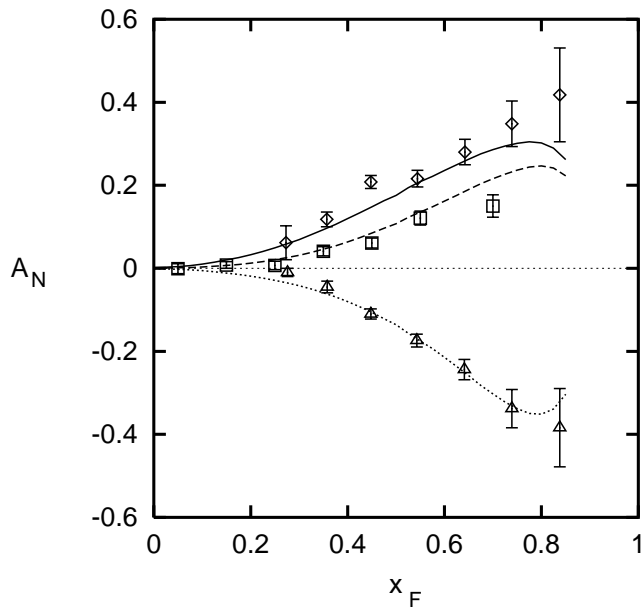}
  \caption{Fit to the data on $A_T^\pi$ for the process
           $p^{\uparrow}p\to\pi{X}$ \cite{Adams:1991cs, Bravar:1996ki} assuming
           that only the Collins effect is active; the upper, middle, and lower
           sets of data and curves refer to $\pi^+$, $\pi^0$ and $\pi^-$,
           respectively. From~\cite{Anselmino:1999pw}.}
  \label{pion}
\end{figure}

\begin{figure}[htbp]
  \centering
  \includegraphics[width=8.5cm,angle=-90]{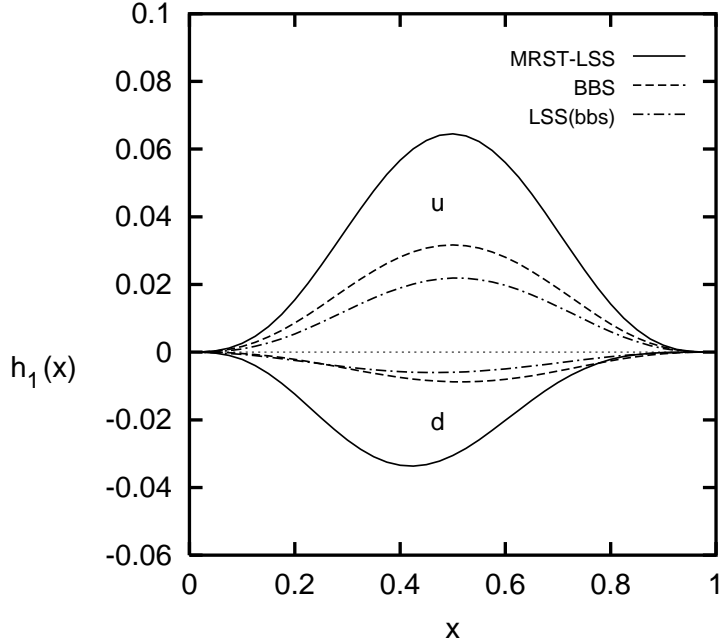}
  \caption{The transversity distributions obtained in \cite{Boglione:1999dq}
           from a fit to the E704 data. The curves correspond to different
           parametrisations of the helicity distributions (see
           \cite{Boglione:1999dq} for details). The figure is taken
           from~\cite{Boglione:2000jk}.}
  \label{boglione}
\end{figure}

The Collins effect was investigated phenomenologically in
\cite{Anselmino:1999pw}, under the hypothesis that it is the only mechanism
contributing to single-spin asymmetries. The authors of \cite{Anselmino:1999pw}
propose a simple parametrisation for the Collins fragmentation function
$\DT^0D_{\pi/q}(z,\langle\kappa_\perp\rangle)$ (see Sec.~\ref{sidisasymm} and
note that in \cite{Anselmino:1999pw} a function $\Delta^ND_{\pi/q^\uparrow}$ is
defined, which is related to our $\DT^0D_{\pi/q}$ by
$\Delta^ND_{\pi/q^\uparrow}=2\,\DT^0D_{\pi/q}$):
\begin{equation}
  \DT^0 D_{\pi/q}(z, \langle \kappa_\perp\rangle) =
  N \, \frac{\langle \kappa_\perp(z) \rangle}{M} \, z^\alpha (1-z)^\beta \, ,
  \label{single2}
\end{equation}
where $M=1\GeV$ and it is assumed that $\DT^0D_{\pi/q}$ is peaked around the
average value
$\langle\kappa_\perp\rangle\equiv\langle\Vec\kappa_\perp^2\rangle^{1/2}$. The
$z$ dependence of $\langle\kappa_\perp(z)\rangle$ is obtained from a fit to LEP
measurements of the transverse momentum of charged pions inside jets
\cite{Abreu:1996na} (remember that $\Vec\kappa_\perp\simeq-\Vec{P}_{h\perp}/z$
neglecting the intrinsic motion of quarks inside the target). Isospin and
charge-conjugation invariance allows one to reconstruct the entire flavour
structure of quark fragmentation into pions, giving the relations
\begin{multline}
  \DT^0 D_{\pi^+/u} =
  \DT^0 D_{\pi^-/d} =
  \DT^0 D_{\pi^+/\anti{d}} =
  \DT^0 D_{\pi^-/\anti{u}} =
\\
  2 \DT^0 D_{\pi^0/u} =
  2 \DT^0 D_{\pi^0/d} =
  2 \DT^0 D_{\pi^0/\anti{d}} =
  2 \DT^0 D_{\pi^0/\anti{u}} =
  \DT^0 D_{\pi/q} \, .
  \label{single4b}
\end{multline}
Only valence quarks in the incoming protons are considered in
\cite{Anselmino:1999pw}. Their transverse polarisation distributions are taken
to be proportional to the unpolarised distributions, according to
\begin{equation}
  \DT{u}(x) = P_T^{u/p} \, u(x) \, , \qquad
  \DT{d}(x) = P_T^{d/p} \, d(x) \, ,
  \label{single4}
\end{equation}
where the transverse polarisation $P_T^{u/p}$ of the $u$ quark is set equal to
$2/3$, as in the $\SU(6)$ model, whereas the transverse polarisation of the $d$
quark is left as a free parameter. The result of the fit to the single-spin
asymmetry data is shown in Fig.~\ref{pion}. Good agreement is obtained if
$\DT^0D_{\pi/q}$ the positivity constraint $|\DT^0D_{\pi/q}|\le{D}_{\pi/q}$
saturates at large $z$, otherwise the value of the single-spin asymmetry
$A_T^\pi$ is too small at large $x_F$. It also turns out that the resulting
transversity distribution of the $d$ quark violates the Soffer bound
$2\,|\DT{d}|\le{d}+\DL{d}$. Boglione and Leader pointed out
\cite{Boglione:1999dq} that, since $\DL{d}$ is negative in most
parametrisations, the Soffer constraint for the $d$ distributions is a rather
strict one. A fit to the $A_T^\pi$ data that satisfies the Soffer inequality
was performed in \cite{Boglione:1999dq}, with good results provided one allows
$\DL{d}$ to become positive at large $x$. In this case too, the positivity
constraint on $\DT^0D_{\pi/q}$ has to be saturated at large $z$. The inferred
transversity distributions are shown in Fig.~\ref{boglione}.

Another calculation of the single-spin asymmetry in pion hadroproduction, based
on the Collins effect, is presented in \cite{Artru:1995bh}. These authors
generate the $T$-odd fragmentation function by the Lund string mechanism and
obtain fair agreement with the E704 data by assuming the following behaviour
for the transversity distributions:
\begin{equation}
  \frac{\DT{u}(x)}{u(x)} \simeq - \frac{\DT{d}(x)}{d(x)} \to 1 \, ,
  \quad \text{as} \quad
  x \to 1 \, .
\end{equation}

A comment on the applicability of perturbative \QCD to the analysis of the E704
measurements is in order. First of all, we have already pointed out that
factorisation with intrinsic transverse momenta of quarks is not a proven
property but only a (plausible) hypothesis. Second, and more important, the
E704 data span a range of $|\Vec{P}_{\pi\perp}|$ that reaches $4\GeV$ for
$\pi^0$ in the central region, where the asymmetry is small, and up to only
$1.5\GeV$ for $\pi^{\pm},\pi^0$ in the forward region, where the asymmetry is
large. At such low values of transverse momenta perturbative \QCD is not
expected to be completely reliable, since cross-sections tend to rise very
steeply as $|\Vec{P}_{\pi\perp}|\to0$. What allows some confidence that a
perturbative \QCD treatment is nevertheless meaningful is the fact that both
intrinsic $\Vec\kappa_\perp$ effects and higher twists (see below) regularise
the cross-sections at $\Vec{P}_{\pi\perp}=0$.

A phenomenological analysis of the E704 results, based on the Sivers effect as
the only source of single-spin asymmetries, was carried out in
\cite{Anselmino:1995tv, Anselmino:1998yz}. For other (model) calculations of
$A_T^\pi$ see \cite{Suzuki:1999wc, Nakajima:2000th}.

\begin{figure}[htbp]
  \centering
  \includegraphics[width=6.5cm]{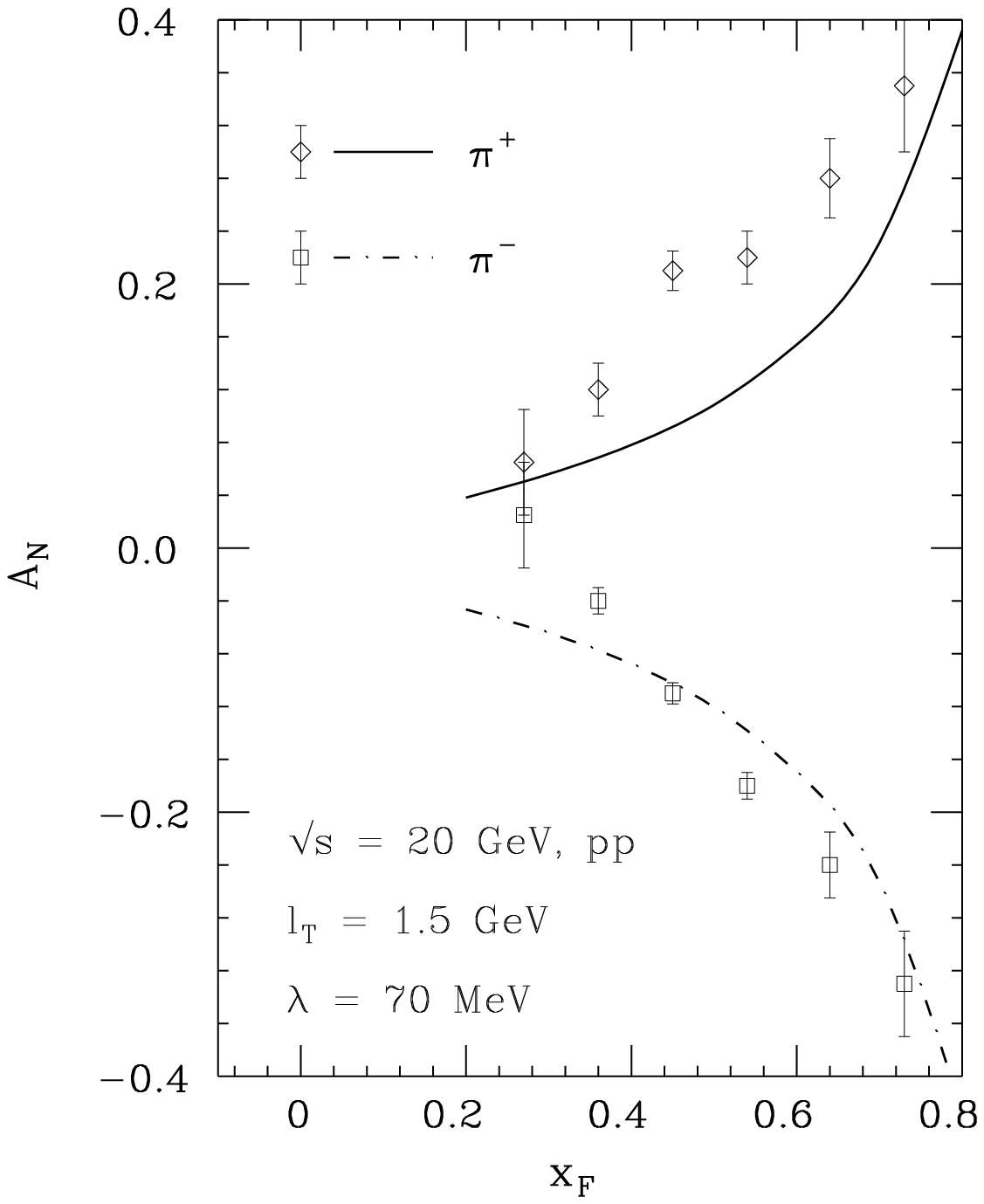}
  \hfil
  \includegraphics[width=6.5cm]{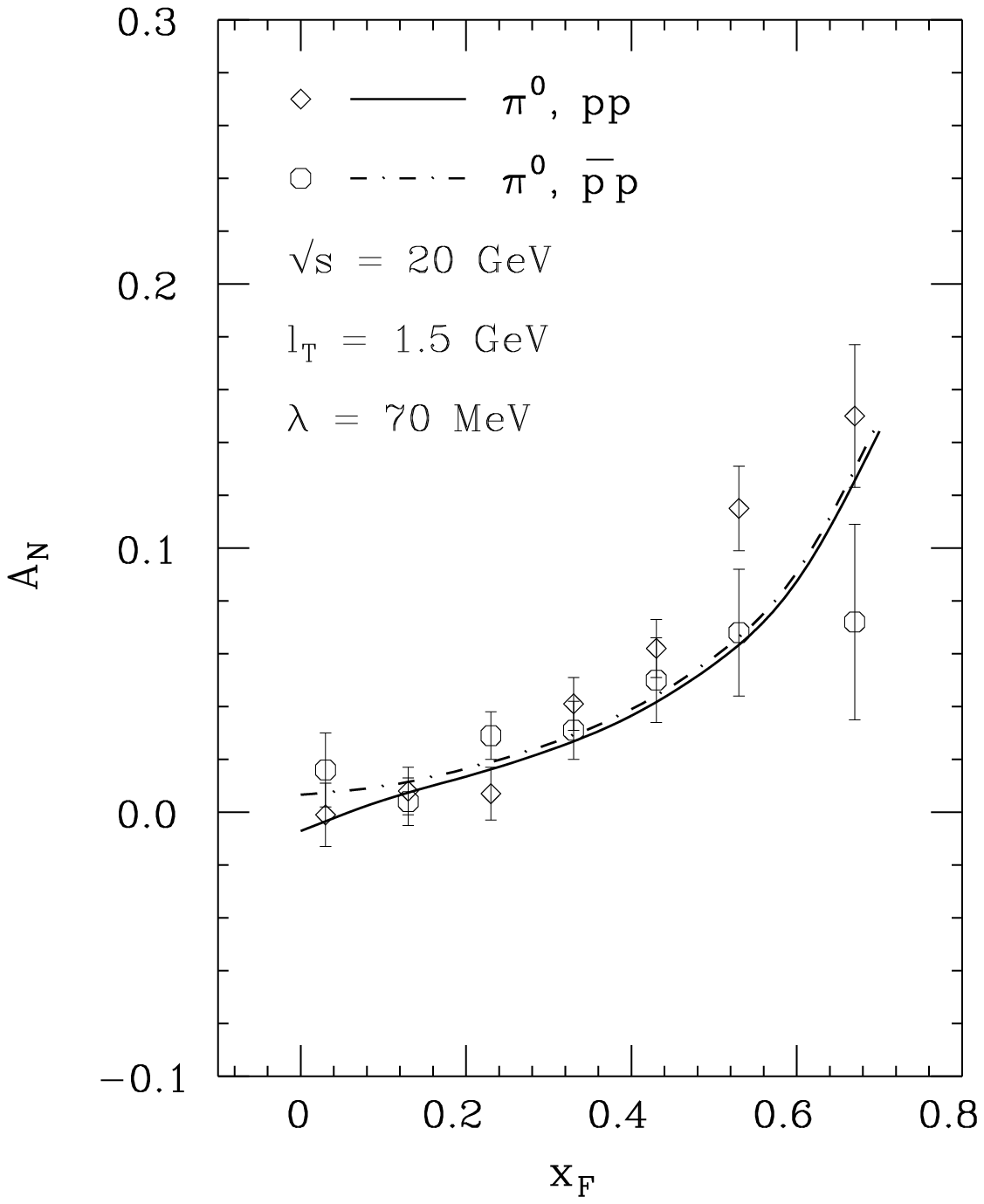}
  \caption{The fit to the single-spin asymmetry data (here $A_T^\pi$ is called
           $A_N$) performed in \cite{Qiu:1998ia}.}
  \label{qiu}
\end{figure}

As shown in Sec.~\ref{singlespintwist3}, single-spin asymmetries may also arise
as a result of twist-three effects \cite{Qiu:1991pp, Qiu:1998ia,
Kanazawa:2000hz, Kanazawa:2000kp}. Qiu and Sterman have used the first,
chirally even, term of eq.~(\ref{singtwist31}) to fit the E704 data on
$A_T^\pi$, setting
\begin{equation}
  G_F(x, x) = K \, q(x) \, ,
  \label{single5}
\end{equation}
where $K$ is a constant parameter. Their fit is shown in Fig.~\ref{qiu}.

Another twist-three contribution, the second term in eq.~(\ref{singtwist31}),
involves the transversity distributions. This term has been evaluated by
Kanazawa and Koike \cite{Kanazawa:2000hz, Kanazawa:2000kp} with an assumption
similar to (\ref{single5}) for the multiparton distribution $E_F$, \ie,
\begin{equation}
  E_F(x, x) = K'\, \DT{q}(x) \, .
  \label{single6}
\end{equation}
They found that, owing to the smallness of the hard partonic cross-sections,
this chirally-odd contribution to single-spin asymmetries turns out to be
negligible.

Clearly, in order to discriminate between leading-twist intrinsic
$\Vec\kappa_\perp$ effects and higher-twist mechanisms a precise measurement of
the $\Vec{P}_{\pi\perp}$ dependence of the asymmetry is needed, in particular
at large $\Vec{P}_{\pi\perp}$. Given the current experimental information on
$A_T^\pi$ it is just impossible to draw definite conclusions as to the
dynamical source of single-spin transverse asymmetries.

\subsection{Transverse polarisation in lepton--nucleon collisions}
\label{transvlepnuc}

Let us turn now to semi-inclusive \DIS on a transversely polarised proton. As
discussed at length in Sec.~\ref{semiinclusive}, there are three candidate
reactions for determining $\DT{q}$ at leading twist:
\begin{enumerate}
\item
semi-inclusive leptoproduction of a transversely polarised hadron with a
transversely polarised target;
\item
semi-inclusive leptoproduction of an unpolarised hadron with a transversely
polarised target;
\item
semi-inclusive leptoproduction of two hadrons with a transversely polarised
target.
\end{enumerate}

We shall review some calculations concerning the first two reactions.
Two-hadron production is more difficult to predict, as it involves interference
fragmentation functions for which we have at present no independent information
from other processes (a model calculation is presented in
\cite{Bianconi:1999uc}).

\subsubsection{$\Lambda^0$ hyperon polarimetry}
\label{polarimetry}

We have seen in Sec.~\ref{sidisasymm} that detecting a transversely polarised
hadron $h^\uparrow$ in the final state of a semi-inclusive \DIS process with a
transversely polarised target, $l\,p^\uparrow\to l'\,h^\uparrow\,X$, probes the
product $\DT{q}(x)\,\DT{D}_q(z)$ at leading twist. The relevant observable is
the polarisation of $h^\uparrow$, which at lowest order reads (we take the $y$
axis as the polarisation axis)
\begin{equation}
  \mathcal{P}_{y}^{h^\uparrow} =
  \hat{a}_T(t) \,
  \frac{\sum_{q, \anti{q}} e_q^2 \, \DT{q}(x,Q^2) \, \DT{D}_{h/q}(z,Q^2)}
       {\sum_{q, \anti{q}} e_q^2 \,      q(x,Q^2) \,      D_{h/q}(z,Q^2)} \, ,
  \label{polar1}
\end{equation}
where $\hat{a}_T(y)=2(1-y)/[1+(1-y)^2]$ is the elementary transverse asymmetry
(the QED depolarisation factor). In this class of reactions, the most promising
is $\Lambda^0$ production. The $\Lambda^0$ polarisation is, in fact, easily
measured by studying the angular distribution of the $\Lambda^0\to{p}\pi$
decay. The transverse polarisation of $\Lambda$'s produced in hard processes
was studied a long time ago in \cite{Craigie:1980tc, Baldracchini:1981uq} and
more recently in \cite{Jaffe:1996wp}. From the phenomenological viewpoint, the
main problem is that, in order to compute the quantity (\ref{polar1}), one
needs to know the fragmentation functions $\DT{D}_{h/q}(z,Q^2)$ besides the
transversity distributions.

\begin{figure}[htbp]
  \centering
  \includegraphics[width=7.5cm,angle=-90]{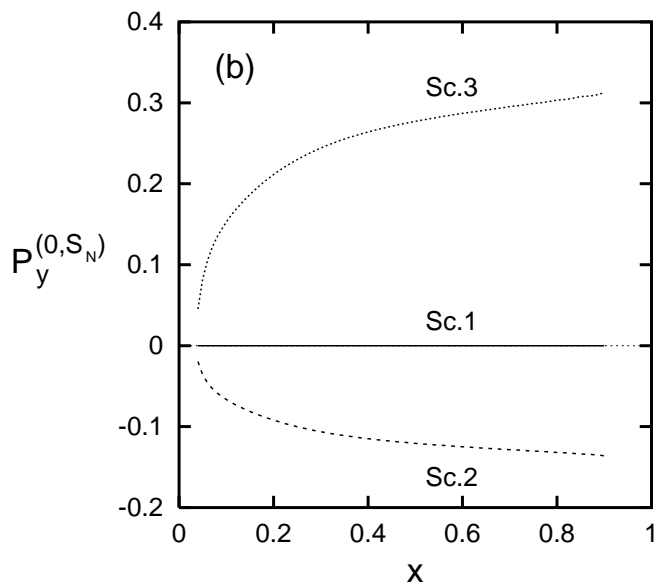}
  \caption{The polarisation of $\Lambda^0$ hyperons produced in semi-inclusive
           DIS, as predicted in~\cite{Anselmino:2000ga}.}
  \label{lambdapol}
\end{figure}

A prediction for $\mathcal{P}_y^{\Lambda^\uparrow}$ has been recently presented
by Anselmino, Boglione and Murgia \cite{Anselmino:2000ga}. These authors
assume, at some starting scale $Q_0^2$, the relations
\begin{subequations}
\begin{equation}
  D_{\Lambda/u} =
  D_{\Lambda/d} =
  D_{\Lambda/s} =
  D_{\Lambda/{\anti{u}}} =
  D_{\Lambda/{\anti{d}}} =
  D_{\Lambda/{\anti{s}}} \equiv
  D_{\Lambda/q} \, ,
  \label{polar2}
\end{equation}
\begin{equation}
  \DT{D}_{\Lambda/u} =
  \DT{D}_{\Lambda/d} =
  \DT{D}_{\Lambda/{\anti{u}}} =
  \DT{D}_{\Lambda/{\anti{d}}} =
  N \, \DT{D}_{\Lambda/s} =
  N \, \DL{D}_{\Lambda/s} \, ,
  \label{polar3}
\end{equation}
\end{subequations}
where $N$ is a free parameter. For $D_{\Lambda/q}$ and $\DL{D}_{\Lambda/s}$
they use the parametrisation of \cite{deFlorian:1998zj} at $Q_0^2=0.23\GeV^2$. As
for the transversity distributions, saturation of the Soffer bound is assumed
and sea densities are neglected. Leading-order \QCD evolution is applied. In
Fig.~\ref{lambdapol} we show the results of \cite{Anselmino:2000ga} for
$\mathcal{P}_y^{\Lambda^\uparrow}$, with three different choices of $N$ and
$\alpha$: the first scenario corresponds to the $\SU(6)$ non-relativistic quark
model (the entire spin of the $\Lambda^0$ is carried by the strange quark, \ie,
$N=0$); the second scenario corresponds to a negative $N$; and the third
scenario corresponds to all light quarks contributing equally to the
$\Lambda^0$ spin (\ie, $N=1$). Other predictions for the $\Lambda^0$
polarisation are offered by Ma \emph{et al}.~\cite{Ma:2001rm}.

\subsubsection{Azimuthal asymmetries in pion leptoproduction}
\label{pionlepto}

A potentially relevant reaction for the study of transversity is
leptoproduction of unpolarised hadrons (typically pions) with a transversely
polarised target, $l\,p^\uparrow\to{l'}\,h\,X$. In this case, as seen in
Sec.~\ref{sidisasymm}, $\DT{q}$ may be probed as a consequence of the Collins
effect (a $T$-odd contribution to quark fragmentation arising from final-state
interactions). In this case one essentially measures
$\DT{q}(x)\,H_1^{{\perp}q}(z,\Vec{P}_{h\perp}^2)$.

Preliminary results on single-spin transverse asymmetries in pion
leptoproduction have been recently reported by the SMC \cite{Bravar:1999rq} and
the HERMES collaboration \cite{Airapetian:1999tv}. Before presenting them, we
return to a kinematical problem already addressed in Sec.~\ref{deepinelastic}:
the definition of the target polarisation.

\begin{figure}[htbp]
  \centering
  \begin{picture}(240,215)(0,0)
    \LongArrow(120,90)(220,90)
    \Line(120,180)(120,210)
    \LongArrow(120,90)(20,40)
    \LongArrow(120,90)(120,5)
    \DashLine(160,195)(160,90){8}
    \SetWidth{1.0}
    \Photon(120,90)(120,176){-2}{4.5}
    \LongArrow(120,176)(120,182)
    \LongArrow(120,90)(160,105)
    \LongArrow(120,90)(160,195)
    \SetWidth{0.5}
    \DashLine(81,170)(80,50){8}
    \SetWidth{2}
    \LongArrow(120,90)(80,170)
    \DashLine(120,90)(84,54){5}
    \LongArrow(84,54)(80,50)
    \SetWidth{0.5}
    \DashCArc(120,90)(30,22,69){3}
    \DashCArc(120,90)(45,70,88){3}
    \DashCArc(140,50)(95,104,116){3}
    \Text(115,182)[r]{$\Vec{q}$}
    \Text(165,105)[l]{$\Vec\ell'$}
    \Text(165,195)[l]{$\Vec\ell$}
    \Text(75,170)[r]{$\Vec{S}$}
    \Text(77,45)[r]{$\Vec{S}_\perp$}
    \Text(220,85)[t]{$x$}
    \Text(125,5)[l]{$z$}
    \Text(20,35)[t]{$y$}
    \Text(147,123)[]{$\vartheta$}
    \Text(134,150)[]{$\vartheta_\gamma$}
    \Text(105,155)[]{$\vartheta_S$}
  \end{picture}
  \caption{The target spin and the lepton and photon momenta. Note that
           $\Vec{q}$ is directed along the negative $z$ axis.}
  \label{spinlep}
\end{figure}
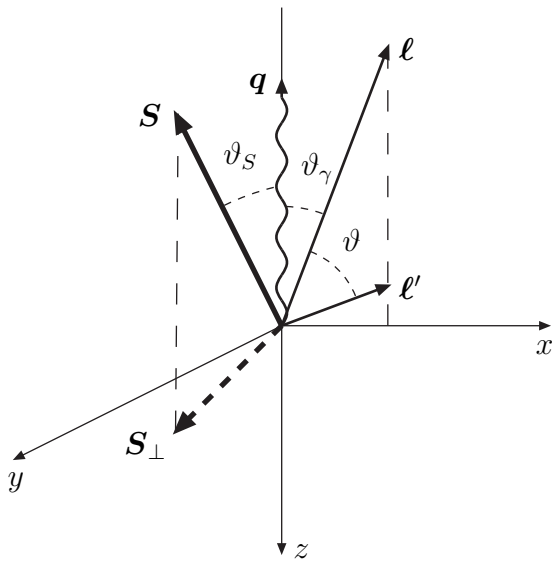

From the experimental point of view, the \DIS target is ``longitudinally''
(``transversely'') polarised when its spin $\Vec{S}$ is parallel (perpendicular)
to the initial lepton momentum $\Vec\ell$. If we parametrise $\Vec{S}$ as (see
Fig.~\ref{spinlep})
\begin{equation}
  \Vec{S} =
  | \Vec{S} | \,
  (
    \sin\vartheta_S \, \cos\phi_S, \,
   -\sin\vartheta_S \, \sin\phi_S, \,
   -\cos\vartheta_S
  ) \, ,
  \label{az1}
\end{equation}
and $\Vec\ell$ as
\begin{equation}
  \Vec\ell =
  E \,
  ( \sin\vartheta_\gamma, \, 0, \, -\cos\vartheta_\gamma ) \, ,
  \label{az2}
\end{equation}
the angle $\alpha$ between $\Vec{S}$ and $\Vec\ell$ is given by
\begin{equation}
  \cos\alpha =
  \sin\vartheta_\gamma \, \sin\vartheta_S \, \cos\phi_S +
  \cos\vartheta_\gamma \, \cos\vartheta_S \, .
  \label{az3}
\end{equation}
Thus, we have
\begin{displaymath}
  \begin{array}{l@{\quad}c@{\quad}l}
    \alpha = 0             &\Rightarrow& \text{``longitudinal'' polarisation} \, ,
  \\
    \alpha = \frac{\pi}{2} &\Rightarrow& \text{``transverse'' polarisation} \, .
  \end{array}
\end{displaymath}
We use quotation marks when adopting the \emph{experimental} terminology.

From the theoretical point of view, it is more convenient to focus on the
target and ignore the leptons. Thus the target is said to be longitudinal
(transverse) polarised when its spin is parallel (perpendicular) to the photon
momentum, \ie,
\begin{displaymath}
  \begin{array}{l@{\quad}c@{\quad}l}
  \vartheta_S = 0           &\Rightarrow& \text{longitudinal polarisation} \, ,
  \\
  \vartheta_S = \frac{\pi}{2} &\Rightarrow& \text{transverse polarisation} \, .
  \end{array}
\end{displaymath}
The absence of quotation marks signals the \emph{theoretical} terminology.

DIS kinematics gives
\begin{subequations}
\begin{eqnarray}
  \sin\vartheta_\gamma &=&
  \frac{2Mx}{Q} \, \sqrt{1-y} + \Ord(M^3/Q^3) \, ,
  \label{az4}
\\
  \cos\vartheta_\gamma &=&
  1 - \frac{2M^2x^2}{Q^2} \, (1-y) + \Ord(M^4/Q^4) \, ,
  \label{az5}
\end{eqnarray}
\end{subequations}
and inverting (\ref{az3}) we obtain
\begin{subequations}
\begin{eqnarray}
  \cos\vartheta_S &\simeq&
  \cos\alpha - \frac{2Mx}{Q} \, \sqrt{1-y} \, \sin\alpha \, \cos\phi_S \, ,
  \label{az6}
\\
  \sin\vartheta_S &\simeq&
  \sin\alpha + \frac{2Mx}{Q} \, \sqrt{1-y} \, \cos\alpha \, \cos\phi_S \, ,
  \label{az7}
\end{eqnarray}
\end{subequations}
where we have neglected $\Ord(M^2/Q^2)$ terms.

If the target is ``longitudinally'' polarised (\ie, $\alpha=0$), one has
\begin{subequations}
\begin{eqnarray}
  \cos\vartheta_S &\simeq& 1 \, ,
  \label{az8}
\\
  \sin\vartheta_S &\simeq& \frac{2Mx}{Q} \, \sqrt{1-y} \, \cos\phi_S \, ,
  \label{az9}
\end{eqnarray}
\end{subequations}
so that (setting $\phi_S=0$ since the lepton momenta lie in the $xz$ plane)
\begin{equation}
  | \Vec{S}_\perp | \simeq
  \frac{2Mx}{Q} \, \sqrt{1-y} \, | \Vec{S} | \, .
  \label{az10}
\end{equation}
Therefore, when the target is ``longitudinally'' polarised, its spin has a
non-zero transverse component, suppressed by a factor $1/Q$. This means that
there is a transverse single-spin asymmetry given by (see (\ref{sicross13}))
\begin{eqnarray}
  A_T^\pi &\simeq&
  \hat{a}_T(y) \,
  \frac{\sum_a e_a^2 \, \DT{f}_a(x) \, \DT^0 D_a(z,\Vec{P}_{h\perp}^2)}
       {\sum_a e_a^2 \,      f_a(x) \,       D_a(z,\Vec{P}_{h\perp}^2)}
  \nonumber
\\
  && \null
  \times | \Vec{S} | \, \frac{2Mx}{Q} \, \sqrt{1-y} \, \sin\phi_\pi \, .
  \label{az11}
\end{eqnarray}
We stress that the $1/Q$ factor in (\ref{az11}), which mimics a twist-three
contribution, has a purely kinematical origin. This is the situation explored
by HERMES \cite{Airapetian:1999tv}.

If the target is ``transversely'' polarised (\ie, $\alpha=\pi/2$), one has
\begin{subequations}
\begin{eqnarray}
  \cos\vartheta_S &\simeq& - \frac{2Mx}{Q} \, \sqrt{1-y} \, \cos\phi_S \, ,
  \label{az12}
\\
  \sin\vartheta_S &\simeq& 1 \, ,
  \label{az13}
\end{eqnarray}
\end{subequations}
and
\begin{equation}
  | \Vec{S}_\perp | \simeq | \Vec{S} | \, .
  \label{az14}
\end{equation}
In this case, neglecting $1/Q^2$ kinematical effects, the target is also
transversely polarised; the measured transverse single-spin asymmetry is
unsuppressed and is given by eq.~(\ref{sicross13}). This is the situation of
the SMC experiment \cite{Bravar:1999rq}.

\begin{figure}[htbp]
  \centering
  \includegraphics[width=10cm]{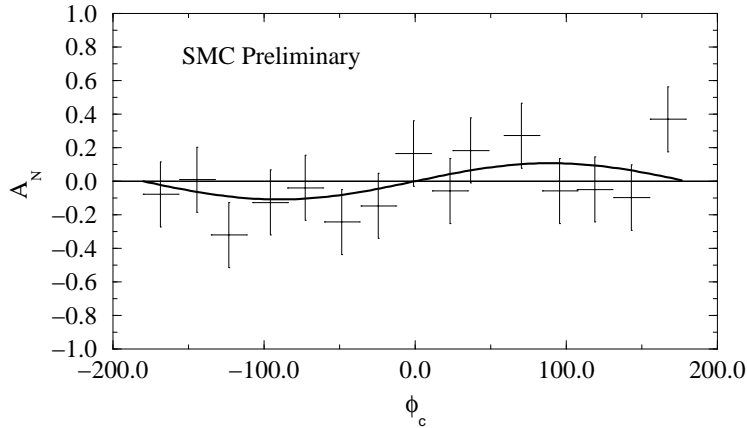}
  \caption{The SMC data \cite{Bravar:1999rq} on the transverse single-spin
           asymmetry in pion leptoproduction, as a function of the Collins
           angle.}
  \label{SMC}
\end{figure}

Let us now come to the data. The SMC \cite{Bravar:1999rq} presented a
preliminary measurement of $A_T^\pi$ for pion production in \DIS of unpolarised
muons off a transversely polarised proton target at $s=188.5\GeV^2$ and
\begin{equation}
  \langle x \rangle \simeq 0.08 \, ,
  \quad
  \langle y \rangle \simeq 0.33 \, ,
  \quad
  \langle z \rangle \simeq 0.45 \, ,
  \quad
  \langle Q^2 \rangle \simeq 5 \GeV^2 \, .
  \label{pion10}
\end{equation}
Two data sets, with $\langle\Vec{P}_{\pi\perp}\rangle=0.5\GeV$ and
$\langle\Vec{P}_{\pi\perp}\rangle=0.8\GeV$, are selected. The result for the
total amount of events is (note that SMC use a different choice of axes and,
moreover, their Collins angle has the opposite sign with respect to ours)
\begin{subequations}
\begin{eqnarray}
  A_T^{\pi^+}/\hat{a}_T &=& \hphantom{-}
  (0.11 \pm 0.06) \, \sin(\phi_\pi + \phi_S) \, ,
  \label{pion11}
\\
  A_T^{\pi^-}/\hat{a}_T &=& -
  (0.02 \pm 0.06) \, \sin(\phi_\pi + \phi_S) \, .
  \label{pion12}
  \sublabel{subpion12}
\end{eqnarray}
\end{subequations}
The SMC data are shown in Fig.~\ref{SMC}.

Evaluating the depolarisation factor $\hat{a}_T$ at
$\langle{y}\rangle\simeq0.33$, eqs.~(\ref{pion11},~\ref{subpion12}) imply
\begin{subequations}
\begin{eqnarray}
  A_T^{\pi^+} &=& \hphantom{-}
  (0.10 \pm 0.06) \, \sin(\phi_\pi + \phi_S),
  \label{pion13}
\\
  A_T^{\pi^-} &=& -
  (0.02 \pm 0.06) \, \sin(\phi_\pi + \phi_S) \, .
  \label{pion14}
\end{eqnarray}
\end{subequations}

\begin{figure}[htbp]
  \centering
  \includegraphics[width=6cm]{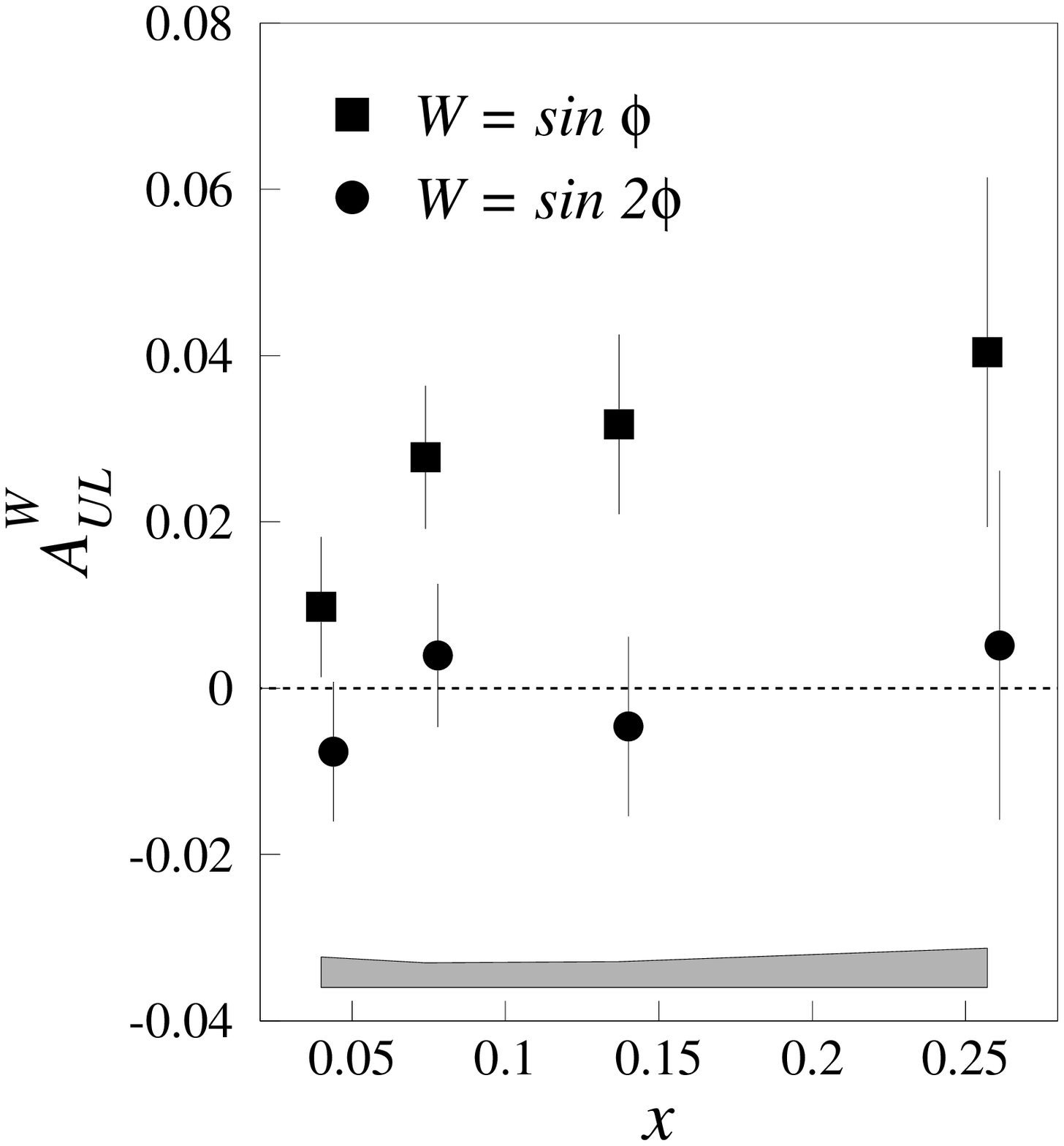}
  \includegraphics[width=6cm]{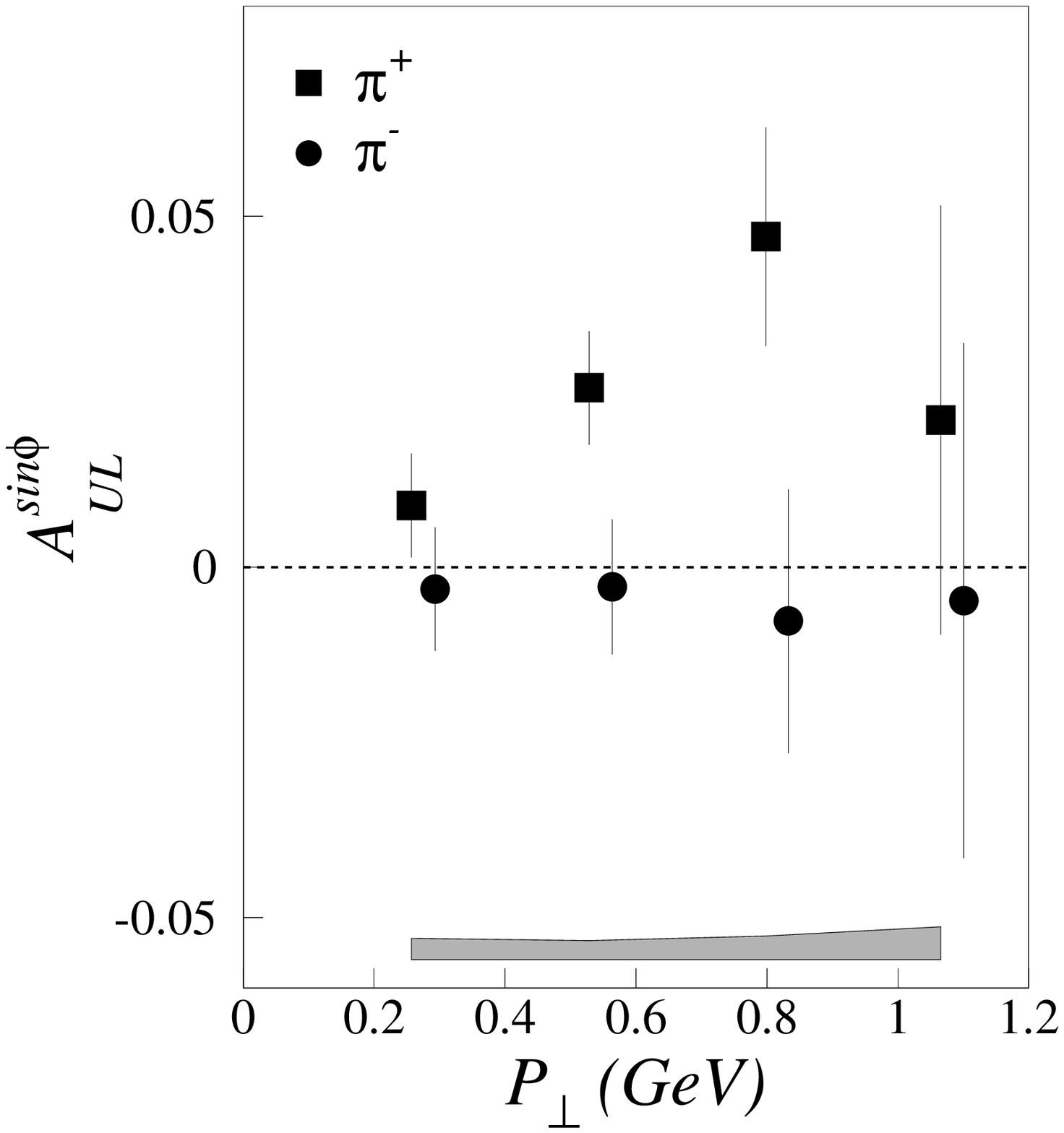}
  \caption{The HERMES data \cite{Airapetian:1999tv} on the transverse
           single-spin asymmetry in pion leptoproduction, as a function of $x$
           (left) and $|\Vec{P}_{\pi\perp}|$ (right).}
  \label{hermes}
\end{figure}

The HERMES experiment at HERA \cite{Airapetian:1999tv} has reported results on
$A_T^\pi$ for positive and negative pions produced in \DIS of unpolarised
positrons off a ``longitudinally'' polarised proton target at $s=52.6\GeV^2$ and
in the kinematical ranges:
\begin{equation}
  0.023 \le x \le 0.4  \, , \quad
  0.1   \le y \le 0.85 \, , \quad
  0.2   \le z \le 0.7  \, , \quad
  Q^2 \ge 1 \GeV^2 \, .
  \label{pion15}
\end{equation}
The transverse momentum of the produced pions is
$|\Vec{P}_{\pi\perp}|\lsim1\GeV$. The HERMES result is (see Fig.~\ref{hermes})
\begin{subequations}
\begin{eqnarray}
  A_T^{\pi^+} &=& +
  \left[ \strut
    0.022
    \pm 0.004 \, (\text{stat.})
    \pm 0.004 \, (\text{syst.})
  \right] \sin\phi_\pi \, ,
  \label{pion16}
\\
  A_T^{\pi^-} &=& -
  \left[ \strut
    0.001
    \pm 0.005 \, (\text{stat.})
    \pm 0.004 \, (\text{syst.})
  \right] \sin\phi_\pi \, .
  \label{pion17}
\end{eqnarray}
\end{subequations}
The conventions for the axes and the Collins angle used by HERMES are the same
as ours. There appears to be a sign difference between the SMC and HERMES
results. Unfortunately, the proliferation of conventions does not help to
settle sign problems. According to the discussion above, the HERMES data, which
are obtained with a ``longitudinally'' polarised target, gives a transverse
single-spin asymmetry suppressed by $1/Q$. Thus, higher-twist longitudinal
effects might be as relevant as the leading-twist Collins effect. The result
(\ref{pion16}), (\ref{pion17}) should be taken with this \emph{caveat} in mind.
Another, even deeper, reason to be very cautious in interpreting the SMC and
HERMES results is the low values of $\langle|\Vec{P}_{\pi\perp}|\rangle$
covered by the two experiments. This renders any perturbative \QCD analysis
rather problematic.

Anselmino and Murgia \cite{Anselmino:2000mb} have recently analysed the SMC and
HERMES data and extracted bounds on the Collins fragmentation function
$\DT^0D_{\pi/q}$ (for another analysis see \cite{Ma:2000ip}). They simplify the
expression of the single-spin transverse asymmetry (\ref{sicross13}) by
assuming that the transversity of the sea is negligible, \ie,
$\DT\anti{q}\simeq0$, using eqs.~(\ref{single4b}) for the fragmentation
functions into pions and similar relations for $D_{\pi/q}$, and ignoring the
non-valence quark contributions in pions. Thus, the single-spin transverse
asymmetries become
\begin{subequations}
\begin{eqnarray}
  A_T^{\pi^+} &\sim&
  \frac{4 \DT{u}(x)}
       {4 u(x) + \anti{d}(x)} \,
  \frac{\DT^0 D_{\pi/q}(z, P_{\pi\perp})}
       {D_{\pi/q}(z, P_{\pi\perp})} \, ,
  \label{pion20}
\\
  A_T^{\pi^-} &\sim&
  \frac{\DT{d}(x)}
       {d(x) + 4 \anti{u}(x)} \,
  \frac{\DT^0 D_{\pi/q}(z, P_{\pi\perp})}
       {D_{\pi/q}(z, P_{\pi\perp})} \, ,
  \label{pion21}
\\
  A_T^{\pi^0} &\sim&
  \frac{4 \DT{u}(x) + \DT{d}(x)}
       {4 u(x) +  d(x) + 4 \anti{u}(x) + \anti{d}(x)} \,
  \frac{\DT^0 D_{\pi/q}(z, P_{\pi\perp})}
       {D_{\pi/q}(z, P_{\pi\perp})} \, .
  \label{pion22}
\end{eqnarray}
\end{subequations}
Saturating the Soffer inequality, the authors of \cite{Anselmino:2000mb} derive
a lower bound for the quark analysing power $\DT^0D_{\pi/q}/D_{\pi/q}$ from the
data on $A_T^\pi$. From the SMC result they find
\begin{subequations}
\begin{equation}
  \frac{| \DT^0 D_{\pi/q}|}{D_{\pi/q}} \gsim 0.24 \pm 0.15 \, , \quad
  \langle z \rangle \simeq 0.45 \, , \quad
  \langle P_{\pi\perp} \rangle \simeq 0.65\GeV \, ,
  \label{pion23}
\end{equation}
and from the HERMES data
\begin{equation}
  \frac{| \DT^0 D_{\pi/q}|}{D_{\pi/q}} \gsim
  0.20 \pm 0.04 \, (\text{stat.}) \pm 0.04 \, (\text{syst.}) \, , \quad
   z \ge 0.2 \, .
  \label{pion24}
\end{equation}
\end{subequations}
These results, if confirmed, would indicate a large value of the Collins
fragmentation function and would therefore also point to a relevant
contribution of the Collins effect in other processes. More data at higher
$\Vec{P}_{\pi\perp}$ would clearly make a perturbative \QCD study definitely
safer. For another determination of the Collins analysing power see below,
Sec.~\ref{transvother}.

\begin{figure}[htbp]
  \centering
  \includegraphics[width=8.5cm]{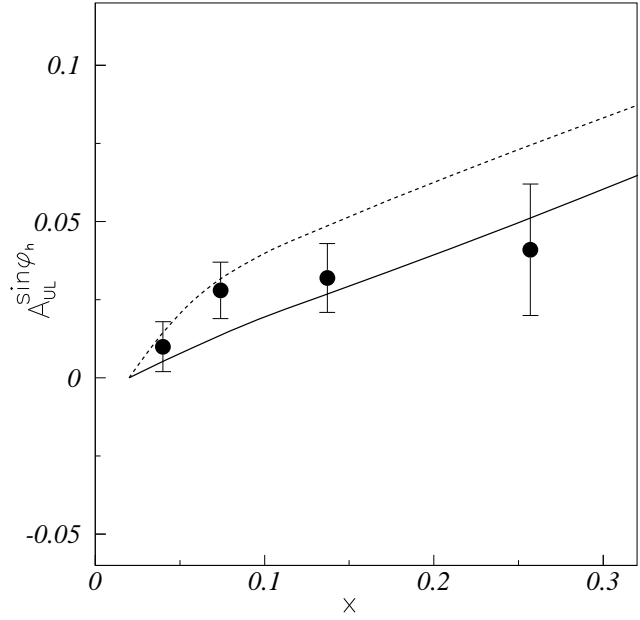}
  \caption{The single-spin azimuthal asymmetry in pion leptoproduction as
           computed in \cite{DeSanctis:2000fh}, compared to the HERMES data
           \cite{Airapetian:1999tv}. The solid line corresponds to
           $\DT{q}=\DL{q}$. The dashed line corresponds to saturation of the
           Soffer inequality.}
  \label{desanctis}
\end{figure}

As already recalled, the interpretation of the HERMES result is made difficult
by the fact that the target is ``longitudinally'' polarised (that is
$|\Vec{S}_\perp|\sim\frac{M}{Q}\,|\Vec{S}|$ and
$|\Vec{S}_\parallel|\sim|\Vec{S}|$). Thus, focusing on the dominant $1/Q$
effects, there are in principle two types of contributions to the
cross-section: 1) leading-twist contributions for a transversely polarised
target; 2) twist-three contributions for a longitudinally polarised target.
Type 1 is $\Ord(1/Q)$ owing to the kinematical relation (\ref{az10}); type 2 is
$\Ord(1/Q)$ owing to dynamical twist-three effects. The $\sin\phi_\pi$
asymmetry measured by HERMES receives the following contributions
\cite{Tangerman:1995hw, Mulders:1996dh, Kotzinian:1997wt, Kotzinian:2000td,
DeSanctis:2000fh, Oganessyan:2000um, Oganessyan:2000kr, Boglione:2000jk} (we
omit the factors in front of each term):
\begin{eqnarray}
  A_{\sin\phi_h}^\pi &\sim&
  | \Vec{S}_\perp | \, \DT{q} \otimes H_1^{\perp(1)} +
  \frac{M}{Q} \, | \Vec{S}_\parallel | \, h_L \otimes H_1^{\perp(1)} +
  \frac{M_h}{Q} \,
    | \Vec{S}_\parallel | \, h_{1L}^{\perp(1)} \otimes \widetilde{H}
  \nonumber
\\
  &\sim&
  \frac{M}{Q} \,   | \Vec{S} | \, \DT{q} \otimes H_1^{\perp(1)} +
  \frac{M}{Q} \,   | \Vec{S} | \, h_L \otimes H_1^{\perp(1)} +
  \frac{M_h}{Q} \, | \Vec{S} | \, h_{1L}^{\perp(1)} \otimes \widetilde{H} \, ,
  \nonumber
\\
  && \null
  \label{pion25}
\end{eqnarray}
where $h_{1L}^{\perp(1)}$ and $H_1^{\perp(1)}$ are defined in (\ref{higher6db})
and (\ref{pion26}) respectively. The first term in (\ref{pion25}) is the type 1
term described above and corresponds to the Collins effect studied in
\cite{Anselmino:2000mb}. The other two terms (type 2) were phenomenologically
investigated in \cite{Kotzinian:2000td, DeSanctis:2000fh, Oganessyan:2000um,
Oganessyan:2000kr, Boglione:2000jk}. In order to analyse the data by means of
(\ref{pion25}), extra input is needed, given the number of unknown quantities
involved. As we have seen in Sec.~\ref{sidisasymm}, when the target is
longitudinally polarised there is also a $\sin2\phi_h$ asymmetry, which appears
at leading twist and has the form
\begin{equation}
  A_{\sin2\phi_\pi}^\pi \sim
  | \Vec{S}_\parallel | \, h_{1L}^{\perp(1)} \otimes H_1^{\perp(1)} \, .
  \label{pion27}
\end{equation}
The smallness of $A_{\sin2\phi_\pi}^\pi$, as measured by HERMES (see
Fig.~\ref{hermes}), seems to be an indication in favour of
$h_{1L}^{\perp(1)}\simeq0$. If we make this assumption \cite{DeSanctis:2000fh},
recalling (\ref{sumrules1}) and (\ref{sumrules3}), we obtain
$\widetilde{h}_L(x)=h_L(x)=\DT{q}(x)$ and (\ref{pion25}) reduces to a single
term of the type $\DT{q}\otimes H_1^{\perp(1)}$. Using a simple parametrisation
\cite{Collins:1993kk} for the Collins fragmentation function, namely
\begin{equation}
  \frac{H_1^\perp(x, \Vec\kappa_\perp^2)}{D(x, \Vec\kappa_\perp^2)} =
  \frac{M_h \, M_C}{M_C^2 + \Vec\kappa_\perp^2} \, ,
  \label{pion28}
\end{equation}
with $M_C$ as a free parameter, the authors of \cite{DeSanctis:2000fh} fit the
HERMES data fairly well (see Fig.~\ref{desanctis}).

In \cite{Boglione:2000jk} it was pointed out that the HERMES data on the
$\sin2\phi_\pi$ asymmetry do not necessarily imply $h_{1L}^{\perp(1)}=0$. If
one assumes the interaction-dependent distribution $\widetilde{h}_L(x)$ to be
vanishing, so that from (\ref{higher6db}) and (\ref{sumrules3}) one has
(neglecting quark mass terms)
\begin{equation}
  h_{1L}^{\perp(1)}(x) =
  -\frac12 \, x h_L(x) =
  -x^2 \int_x^1 \! \frac{\d{y}}{y^2} \, \DT{q}(y) \, ,
  \label{pion29}
\end{equation}
then it is still possible to obtain a $\sin\phi_\pi$ asymmetry of the order of
few percent (as found by HERMES), with the $\sin2\phi_\pi$ asymmetry suppressed
by a factor~2.

The approximation (\ref{pion29}) was also adopted in \cite{Efremov:2000za},
where an analysis of the HERMES and SMC data in the framework of the chiral
quark--soliton model is presented.

In conclusion, we can say that the interpretation of the HERMES and SMC
measurements is far from clear. The experimental results seem to indicate that
transversity plays some r\^{o}le but the present scarcity of data, their errors,
our ignorance of most of the quantities involved in the process, and, last but
not least, uncertainty in the theoretical procedures make the entire matter
still rather vague. More, and more precise, data will be of great help in
settling the question.

\subsection{Transverse polarisation in $e^+e^-$ collisions.}
\label{transvother}

\begin{figure}[htbp]
  \centering
  \begin{picture}(340,140)(80,45)
    \Line(100,50)(300,50)
    \Line(250,100)(350,100) 
    \Line(100,50)(200,150)
    \Line(300,50)(400,150)
    \SetWidth{1}
    \LongArrow(210,130)(248,101.5)
    \LongArrow(290,70)(252,98.5)
    \SetWidth{0.5}
    \Text(210,135)[b]{$\Vec\ell'$}
    \Text(290,65)[t]{$\Vec\ell$}
    \Text(180,95)[t]{$\Vec{P}_2$}
    \SetWidth{0.5}
    \Line(250,100)(270,180)
    \Line(350,100)(370,180)
    \Line(270,180)(370,180)
    \SetWidth{0.5}
    \Text(315,145)[l]{$\Vec{P}_1$}
    \Text(265,145)[l]{$\Vec{P}_{1\perp}$}
    \SetWidth{2}
    \LongArrow(250,100)(310,140)
    \LongArrow(250,100)(260,140)
    \LongArrow(248,100)(180,100)
    \SetWidth{0.5}
    \Line(200,150)(260,150)
    \Line(400,150)(365,150)
    \CArc(350,100)(20,45,75)
    \CArc(250,100)(25,145,178)
    \Text(368,130)[]{$\phi_1$}
    \Text(222,112)[r]{$\theta_2$}
    \LongArrow(380,100)(365,100)
    \LongArrow(380,100)(380,85)
    \LongArrow(380,100)(390,110)
    \Text(365,97)[t]{$z$}
    \Text(377,83)[r]{$y$}
    \Text(395,110)[l]{$x$}
  \end{picture}
  \caption{Kinematics of two-hadron production in $e^+e^-$ annihilation.}
  \label{eeplane}
\end{figure}
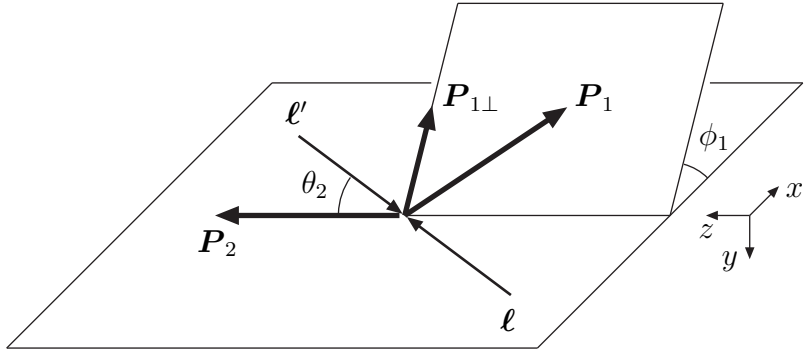

An independent source of information on the Collins fragmentation function
$H_1^\perp$ is inclusive two-hadron production in electron--positron collisions
(see Fig.~\ref{eeplane}):
\begin{equation}
  e^+ \, e^- \to h_1 \, h_2 \, X \, .
  \label{ee1}
\end{equation}
This process was studied in \cite{Chen:1995ar, Boer:1997mf, Boer:1998qn}. It
turns out that the cross-section has the following angular dependence (we
assume that two alike hadrons are produced, omit the flavour indices and refer
to Fig.~\ref{eeplane} for the kinematical variables)
\begin{eqnarray}
  \frac{\d\sigma}{\d\cos\theta_2 \d\phi_1} &\propto&
  (1 + \cos^2\theta_2) \, D(z_1) \anti{D}(z_2)
  \nonumber
\\
  && \null +
  C \, \sin^2\theta_2 \, \cos(2 \phi_1) \,
  H_1^{\perp(1)}(z_1) \anti{H}_1^{\perp(1)}(z_2) \, ,
  \label{ee2}
\end{eqnarray}
where $C$ is a constant containing the electroweak couplings.

Thus, the analysis of $\cos(2\phi_1)$ asymmetries in the process (\ref{ee1})
can shed light on the ratio between unpolarised and Collins fragmentation
functions. Efremov and collaborators \cite{Efremov:1998vd, Efremov:1998fu,
Efremov:1999nx} have carried out such a study using the DELPHI data on $Z^0$
hadronic decays. Under the assumption that all produced particles are pions and
that fragmentation functions have a Gaussian $\Vec\kappa_T$ dependence, they
find
\begin{equation}
  \left\langle \frac{\anti{H}_1^\perp}{D} \right\rangle =
  (6.3 \pm 1.7) \, \% \, ,
  \label{ee3}
\end{equation}
where the average is over flavours and the kinematical range covered by data.
The result (\ref{ee3}) is an indication of a non-zero fragmentation function of
transversely polarised quarks into unpolarised hadrons. The authors of
\cite{Efremov:1998vd, Efremov:1998fu, Efremov:1999nx} argue that a more careful
study of the $\theta_2$ dependence of the experimentally measured cross-section
could increase the value (\ref{ee3}) up to  $\sim10\%$. An analysing power of
this order of magnitude would make the possibility of observing the Collins
effect in the future experiments rather tangible.

\section{Experimental perspectives}
\label{experiments}

In this section, which completes the bulk of our report, we outline the present
experimental situation and the future prospects. The study of transversity
distributions is a more-or-less important fraction of the physics program of
many ongoing and forthcoming experiments in various laboratories (DESY, CERN
and Brookhaven). An overview of the experimental state of the field can be
found in the Proceedings of the RIKEN--BNL Workshop on ``Future Transversity
Measurements'' \cite{Boer:2000zy}.

\subsection{${\ell}N$ experiments}
\label{lnexp}

\subsubsection{HERMES}

The HERMES experiment uses the HERA $27.5\GeV$ positron (or electron) beam
incident on a longitudinally polarised $H$ or $D$ gas-jet target. The hydrogen
polarisation is approximately 85\%. Running since 1995, HERMES has already
provided a large amount of data on polarised inclusive and semi-inclusive DIS.
We have discussed (see Sec.~\ref{pionlepto}) their (preliminary) result of main
concern in this report, that is the observation of a relatively large azimuthal
spin asymmetry in semi-inclusive \DIS on a longitudinally polarised proton
target, which may involve the transversity distributions via the Collins effect
(as we have noted, similar findings have been reported by the SMC at CERN).

HERMES plans to continue data taking in the period 2001--2006, after the HERA
luminosity upgrade (which should increase the average luminosity by a factor
3). Two of the foreseen five years of running should be dedicated to a
transversely polarised target with an expected statistics of $7{\cdot}10^6$
reconstructed \DIS events. The foreseen target polarisation is $\sim75\%$. The
transverse polarisation program at HERMES includes \cite{Korotkov:2000ba}
(besides the extraction of the spin structure function $g_2$): \emph{i}) a
measurement of the twist-three azimuthal asymmetry in semi-inclusive pion
production with a longitudinally polarised lepton beam; \emph{ii}) the study of
the Collins effect in the scattering of an unpolarised lepton beam off a
transversely polarised target; and \emph{iii}) a measurement of the transverse
asymmetry in leptoproduction of two correlated mesons.

According to the estimates presented in \cite{Korotkov:1999jx,
Korotkov:2000ba}, HERMES should be able to determine both the transversity
distributions and the Collins fragmentation function (at least for the dominant
$u$ flavour) with good statistical accuracy.

\subsubsection{COMPASS}

COMPASS is a new fixed target experiment at CERN \cite{Baum:1996yv}, with two
main programs: the ``muon program'' and the ``hadron program''. The former (which
upgrades SMC) aims to study the nucleon spin structure with a high-energy muon
beam. COMPASS uses polarised muons of 100--200$\GeV$, scattering off polarised
proton and deuteron targets. Expected polarisations are 90\% and 50\% for the
proton and the deuteron, respectively. The transverse polarisation physics
program is similar to that of HERMES, but covers different kinematical regions.
In particular, single-spin asymmetries in hadron leptoproduction will be
measured. These will provide the transversity distributions via the Collins
effect. According to an estimate presented in \cite{Baum:1996yv} $\DT{q}$
should be determined with a $\sim10\%$ accuracy in the intermediate-$x$ region.
Data taking by COMPASS started in 2001.

\subsubsection{ELFE}

ELFE (Electron Laboratory for Europe) is a continuous electron-beam facility,
which has been discussed since the early nineties. The latest proposal is for
construction at CERN by exploiting the cavities and other components of LEP not
required for LHC \cite{Aulenbacher:1999hu}. The maximum energy of the electron
beam would be $25\GeV$. The very high luminosity (about three orders of
magnitude higher than HERMES and COMPASS) would allow accurate measurements of
semi-inclusive asymmetries with transversely polarised targets. In particular,
polarimetry in the final state should reach a good degree of precision (a month
of running time, with a luminosity of $10^{34}\cm^{-2}\secs^{-1}$, allows the
accumulation of about $10^6$ $\Lambda$'s with transverse momentum greater than
$1\GeV/c$.

\subsubsection{TESLA-N}

The TESLA-N project \cite{Anselmino:2000sm} is based on the idea of using one
of the arms of the $e^+e^-$ collider TESLA to produce collisions of
longitudinally polarised electrons on a fixed proton or deuteron target, which
may be either longitudinally or transversely polarised. The basic parameters
are: electron beam energy $250\GeV$, an integrated luminosity of 100\,fb$^{-1}$
per year, a target polarisation $\sim80$\% for protons, $\sim30$\% for
deuterium. The transversity program includes the measurement of single-spin
azimuthal asymmetries and two-pion correlations \cite{Korotkov:2000ba}. The
proposers of the project have estimated the statistical accuracy in the
extraction of the transversity distributions via the Collins effect and found
values comparable to the existing determinations of the helicity distributions.
They have also shown that the expected statistical accuracy in the measurement
of two-meson correlations is encouraging if the interference fragmentation
function is not much smaller that its upper bound.

\subsection{$pp$ experiments}
\label{ppexp}

\subsubsection{RHIC}

\begin{figure}[htbp]
  \centering
  \includegraphics[width=\textwidth]{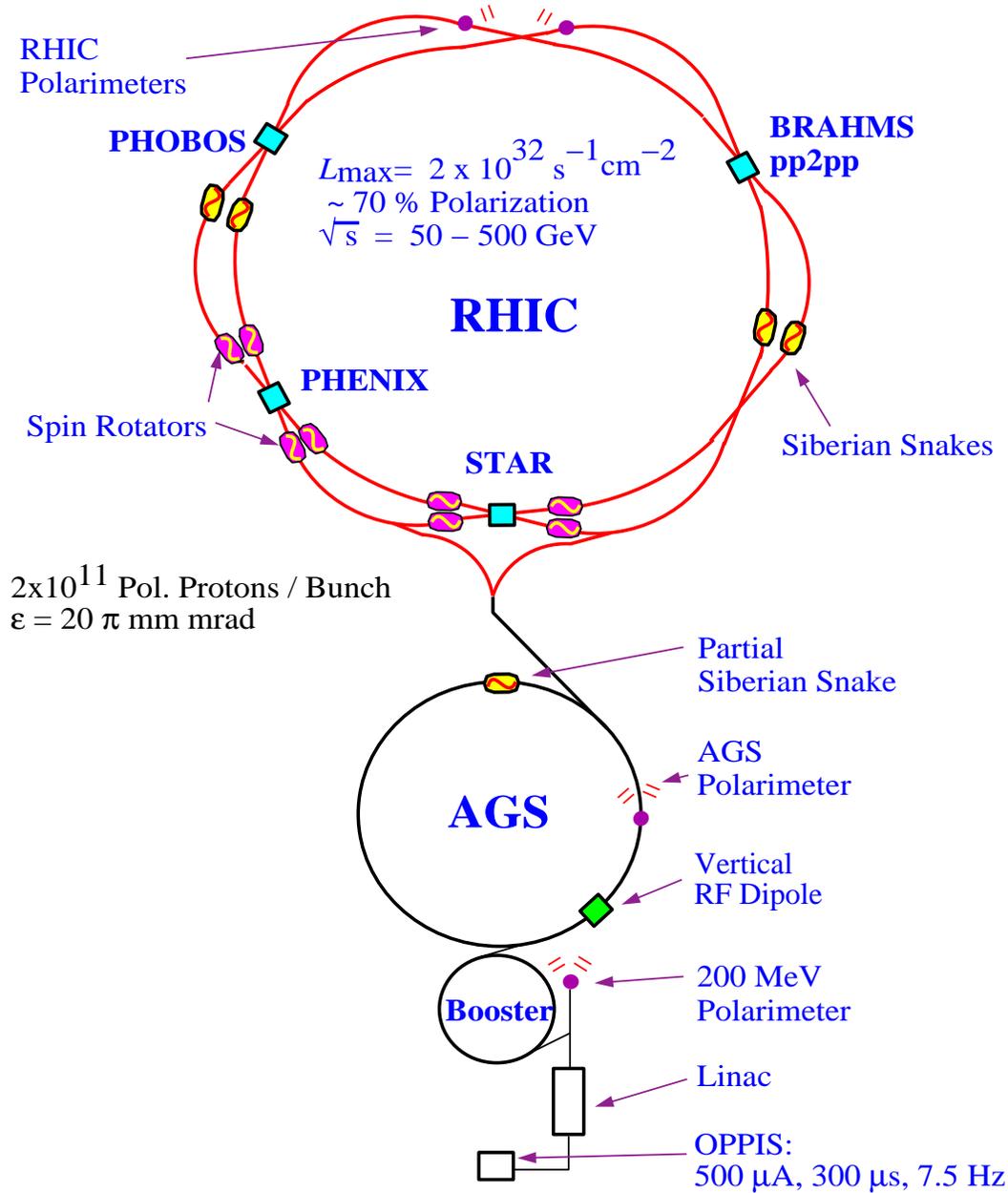}
  \caption{An overview of RHIC.}
  \label{rhicspin}
\end{figure}

The Relativistic Heavy Ion Collider (RHIC, Fig.~\ref{rhicspin}) at the
Brookhaven National Laboratory operates with gold ions and protons. With the
addition of Siberian snakes and spin rotators, there will be the possibility of
accelerating intense polarised proton beams up to energies of $250\GeV$ per
beam. The spin-physics program at \theRHIC will study reactions involving two
polarised proton beams with both longitudinal and transverse spin orientations,
at an average centre-of-mass energy of $500\GeV$ (for an overview of spin
physics at \theRHIC see \cite{Bunce:2000uv}). The expected luminosity is up to
$\sim2{\cdot}10^{32}\cm^{-2}\secs^{-2}$, with $70$\% beam polarisation. Two detectors
will be in operation: STAR (see, \eg, \cite{Bland:1999gb}) and PHENIX (see,
\eg, \cite{Enyo:2000nt}). The former is a general purpose detector with a large
solid angle; the latter is a dedicated detector mainly for leptons and photons.
Data taking with polarised protons will start in 2001. The most interesting
process involving transversity distributions to be studied at \theRHIC is
Drell--Yan lepton pair production mediated by $\gamma^*$ or $Z^0$. As seen in
Sec.~\ref{transvhh}, the expected double-spin asymmetry $A_{TT}^\text{DY}$ is
just few percent but may be visible experimentally, provided the transversity
distributions are not too small. Single-spin Drell--Yan measurements could be a
good testing ground for the existence of transversity in unpolarised hadrons
arising from $T$-odd initial-state interaction effects \cite{Boer:1999mm}.

\section{Conclusions}

The transverse polarisation of quarks represents an important piece of
information on the internal structure and dynamics of hadrons. In the previous
sections we have tried to substantiate this statement reviewing the current
state of knowledge. In conclusion, let us try to summarise what we have learned
so far about transversity.
\begin{itemize}
\item
The transverse polarisation (or transversity) distributions $\DT{q}$ are
chirally-odd leading-twist quantities that do not appear in fully inclusive
DIS, but do appear in semi-inclusive \DIS processes and in various
hadron-initiated reactions.
\item
The \QCD evolution of $\DT(x,Q^2)$ is known up to \NLO, and turns out to be
different from the evolution of the helicity counterpart.
\item
Many models (and other non-perturbative tools) have been used to calculate the
transversity distributions in the nucleon. These computations show that, at
least for the dominant $u$ sector, at low momentum scales $\DT{q}$ is not so
different from $\DL{q}$.
\item
The phenomenology of transversity is very rich. It includes transversely
polarised Drell--Yan processes, leptoproduction of polarised baryons and
mesons, correlated meson production and, via the Collins effect, lepto- and
hadro-production of pions.
\item
Only two preliminary results, that may have something to do with transversity,
are currently available. The HERMES and SMC collaborations have found
non-vanishing azimuthal asymmetries in pion leptoproduction, which may be
explained in terms of transverse polarisation distributions coupling to a
$T$-odd fragmentation function (Collins effect). However, no definite
conclusion on the physical explanation of these findings is possible as yet.
\end{itemize}

The intense theoretical effort developed over the last decade must now be put
to fruition by a vigorous experimental study of transversity. Many
collaborations around the world (at Brookhaven, DESY and CERN) aim at measuring
quark transverse polarisation in the nucleon. This is certainly a complex task
since the foreseen values of some of the relevant observables are close to the
sensitivity limits of the experiments. Nevertheless, the variety and accuracy
of the measurements planned for the coming years permit a certain confidence
that the veil of ignorance surrounding quark transversity will at last begin to
dissolve.

\ack We are grateful to Mauro Anselmino, Alessandro Bacchetta, Elena Boglione,
Umberto D'Alesio, Bo-Qiang Ma, Francesco Murgia, Sergio Scopetta, Oleg Teryaev
and Fabian Zomer for various discussions on the subject of this report.

\appendix
\nosubseceqnumbering                          
\section{Sudakov decomposition of vectors}
\label{sudakov}

We introduce two light-like vectors (the Sudakov vectors)
\begin{subequations}
\begin{eqnarray}
  p^\mu &=& \frac1{\sqrtno2} \, (\Lambda, 0, 0, \Lambda) \, ,
  \label{sud1}
\\
  n^\mu &=& \frac1{\sqrtno2} \, (\Lambda^{-1}, 0, 0, -\Lambda^{-1}) \, ,
  \label{sud2}
  \sublabel{subsud2}
\end{eqnarray}
\end{subequations}
where $\Lambda$ is arbitrary. These vectors satisfy
\begin{equation}
  p^2 = 0 = n^2   \, , \qquad
  p{\cdot}n = 1       \, , \qquad
  n^+ = 0 = p^- \, .
  \label{sud3}
\end{equation}
In light-cone components they read
\begin{subequations}
\begin{eqnarray}
  p^\mu &=& ( \Lambda, 0, \Vec{0}_\perp ) \, ,
  \label{sud4}
\\
  n^\mu &=& ( 0, \Lambda^{-1}, \Vec{0}_\perp ) \, .
  \label{sud5}
\end{eqnarray}
\end{subequations}
A generic vector $A^\mu$ can be parametrised as (a Sudakov decomposition)
\begin{eqnarray}
  A^\mu &=& \alpha \, p^\mu + \beta \, n^\mu + A^\mu_\perp
  \nonumber
\\
  &=& A{\cdot}n \, p^\mu + A{\cdot}p \, n^\mu + A^\mu_\perp \, ,
  \label{sud7}
\end{eqnarray}
with $A^\mu_\perp=(0,\Vec{A}_\perp,0)$. The modulus squared of $A^\mu$ is
\begin{equation}
  A^2 = 2\alpha\beta - \Vec{A}_\perp^2 \, .
\end{equation}

\section{Reference frames}
\label{frames}

\subsection{The $\gamma^*N$ collinear frames}
\label{gammaN}

In \DIS processes, we call the frames where the virtual photon and the target
nucleon move collinearly ``$\gamma^*N$ collinear frames''. If the motion takes
place along the $z$-axis, we can represent the nucleon momentum $P$ and the
photon momentum $q$ in terms of the Sudakov vectors $p$ and $n$ as
\begin{eqnarray}
  P^\mu &=& p^\mu + \half M^2 n^\mu \simeq p^\mu \, ,
  \label{sud9}
\\
  q^\mu &\simeq& P{\cdot}q \, n^\mu - x p^\mu = M\nu \, n^\mu - x p^\mu \, ,
  \label{sud11}
\end{eqnarray}
where the approximate equality sign indicates that we are neglecting $M^2$ with
respect to large scales such as $Q^2$, or $(P^+)^2$ in the infinite momentum
frame. Conventionally we always take the nucleon to be directed in the positive
$z$ direction.

With the identification (\ref{sud9}) the parameter $\Lambda$ appearing in the
definition of the Sudakov vectors (\ref{sud1}, \ref{subsud2}) coincides with
$P^+$ and fixes the specific frame. In particular:
\begin{itemize}
\item
in the \emph{target rest frame} (TRF) one has
\begin{eqnarray}
  P^\mu &=& (M, 0, 0, 0),
\\
  q^\mu &=&  \left (\nu, 0, 0, -\sqrt{\nu^2 + Q^2} \right ),
\end{eqnarray}
and $\Lambda\equiv{P}^+=M/\sqrt2$. The Bjorken limit in this frame corresponds
to $q^-=\sqrt2\,\nu\to\infty$ with $q^+=-Mx/\sqrt2$ fixed.
\item
in the \emph{infinite momentum frame} (IMF) the momenta are
\begin{eqnarray}
  P^\mu &\simeq&
  \frac1{\sqrtno2} \, (P^+, 0, 0, P^+) \, ,
  \label{imf1}
\\
  q^\mu &\simeq&
  \frac1{\sqrtno2}
  \left( \frac{M\nu}{P^+}-xP^+, 0, 0, -\frac{M\nu}{P^+}-xP^+ \right) ,
  \label{imf2}
\end{eqnarray}
Here we have $P^-\to0$ and $\Lambda\equiv{P}^+\to\infty$. In this frame the
vector $n^\mu$ is suppressed by a factor of $(1/P^+)^2$ with respect to $p^\mu$.
\end{itemize}

By means of the Sudakov vectors we can construct the \emph{perpendicular}
metric tensor $g_\perp^{\mu\nu}$ which projects onto the plane perpendicular to
$p$ and $n$, and to $P$ and $q$ (modulo $M^2/Q^2$ terms)
\begin{equation}
  g_\perp^{\mu\nu} =  g^{\mu\nu} - (p^\mu n^\nu + p^\nu n^\mu) \, .
  \label{gammaN4}
\end{equation}
Transverse vectors in the $\gamma^*N$ frame (or ``perpendicular'' vectors) will
be denoted by a $\perp$ subscript. Another projector onto the transverse plane
is
\begin{equation}
  \varepsilon_\perp^{\mu\nu}
  =
  \varepsilon^{\mu\nu\rho\sigma} p_\rho n_\sigma \, .
  \label{gammaN5}
\end{equation}
Consider now the spin vector of the nucleon. It may be written as
\begin{equation}
  S^\mu =
  \frac{\lambda_N}{M} \,
\left (p^\mu - \frac{M^2}{2} \, n^\mu \right ) + S_\perp^\mu
  \simeq
  \frac{\lambda_N}{M} \,p^\mu + S_\perp^\mu \, ,
  \label{gammaN5b}
\end{equation}
where $\lambda_N^2+\Vec{S}_\perp^2\le1$ (the equality sign applies to pure
states). The transverse spin vector $S_\perp^\mu$ is of order $\Ord(1)$, thus
it is suppressed by one power of $P^+$ with respect to longitudinal spin
$S_\parallel^\mu=\lambda_Np^\mu/M$.

Finally, in semi-inclusive \DIS the momentum $P_h$ of the produced hadron $h$
may be parametrised in the $\gamma^*N$ collinear frame as
\begin{equation}
  P_h^\mu \simeq z q^\mu + x z P^\mu + P_{h\perp}^\mu \, ,
  \label{gammaN6}
\end{equation}
where
\begin{equation}
  z = \frac{P{\cdot}P_h}{P{\cdot}q} = \frac{2x}{Q^2} \, P{\cdot}P_h \, .
  \label{gammaN7}
\end{equation}

\subsection{The $hN$ collinear frames}
\label{hN}

In polarised semi-inclusive \DIS it is often convenient to work in a frame
where the target nucleon $N$ and the produced hadron $h$ are collinear (a ``$hN$
collinear frame''). In the family of such frames the momenta of $N$ and $h$ are
parametrised, in terms of two Sudakov vectors $p'$ and $n'$, as
\begin{eqnarray}
  P^\mu &=& {p'}^\mu + \frac{M^2}{2} \, {n'}^\mu
  \simeq {p'}^\mu \, ,
  \label{hN1}
\\
  P_h^\mu &\simeq&
  \frac{M_h^2 x}{Q^2 z} \, {p'}^\mu +
  \frac{Q^2z}{2x} \, {n'}^\mu
  \simeq
  \frac{Q^2z}{2x} \, {n'}^\mu \, .
  \label{hN2}
\end{eqnarray}
The projectors onto the transverse plane (vectors lying in this plane will be
denoted by the subscript $T$) are
\begin{eqnarray}
  g_T^{\mu\nu} &=&
  g^{\mu\nu} - ({p'}^\mu {n'}^\nu + {p'}^\nu {n'}^\mu) \, ,
  \label{hN2b}
\\
  \varepsilon_T^{\mu\nu} &=&
  \varepsilon^{\mu\nu\rho\sigma} {p'}_\rho {n'}_\sigma \, .
  \label{hN2c}
\end{eqnarray}

In $hN$ collinear frames the photon acquires a transverse momentum
\begin{equation}
  q^\mu = - x {p'}^\mu + \frac{Q^2}{2x} \, {n'}^\mu + q_T^\mu \, .
  \label{hN3}
\end{equation}
Comparing this to the Sudakov decomposition (\ref{gammaN6}) of the momentum of
the produced hadron, we obtain
\begin{equation}
  P_{h\perp}^\mu \simeq - z q_T^\mu \, .
  \label{hN4}
\end{equation}
The spin vector of $h$ is
\begin{equation}
  S_h^\mu = \frac{\lambda_h}{M_h} \, P^\mu_h + S_{hT}^\mu \, .
  \label{hN5}
\end{equation}

The relation between transverse vectors in the $\gamma^*N$ frame
($\perp$-vectors) and transverse vectors in the $hN$ frame ($T$-vectors) is
\begin{equation}
  a_\perp^\mu =
  a_T^\mu - \frac{2 x}{Q^2} \,
  \left[ \strut a{\cdot}q_T \, P^\mu + a{\cdot}P \, q_T^\mu \right] \, .
  \label{hN6}
\end{equation}
Therefore, if we neglect order $1/Q$ corrections, that is if we ignore
higher-twist effects, we can identify transverse vectors in $\gamma^*N$
collinear frames with transverse vectors in $hN$ collinear frames (in other
terms, we have $g_\perp^{\mu\nu}\simeq{g}_T^{\mu\nu}$ and
$\varepsilon_\perp^{\mu\nu}\simeq\varepsilon_T^{\mu\nu}$).

\include{frames}
\section{Mellin moment identities}
\label{sec:mellin}

We first recall here the definition of the so-called plus regularisation,
necessary for the \IR singularities present in the \AP splitting kernels:
\begin{equation}
  \int_0^1 \d{x} \frac{f(x)}{(1-x)_+} =
  \int_0^1 \d{x} \frac{[f(x)-f(1)]}{(1-x)} \, .
  \label{eq:plus-def}
\end{equation}

A convenient identity regarding the above plus symbol is:
\begin{equation}
  \int_0^1 \d{x} \frac{x^{n-1}}{(1-x)_+} f(x) =
  \int_0^1 \d{x} \left\{\left[ \frac{x^{n-1}}{(1-x)} \right]_+
               - \delta(1-x) \sum_{j=1}^{n-1}\frac1j \right\} f(x) \, ,
  \label{eq:plus-id}
\end{equation}
This allows, for example, the following particularly compact expression for the
usual $qq$ \AP splitting kernels:
\begin{equation}
  \frac{(1+x^2)}{(1-x)_+} + \frac32 \delta(1-x) =
  \left[ \frac{1+x^2}{1-x} \right]_+ \, .
  \label{eq:plus-eq}
\end{equation}

\newpage
\bibliographystyle{pigroels} 
\bibliography{pigrostr,pigrotmp,pigrodbf,pigropgr,pigroxrf}
\end{document}